\documentclass[preprint,authoryear,12pt]{elsarticle}
\usepackage{amssymb}
\usepackage{float}

\newcommand{\mnras}{MNRAS }
\newcommand{\apjl}{ApJ }
\newcommand{\apj}{ApJ }

\newcommand{\nat}{Nature }
\newcommand{\aap}{A\&A }

\journal{New Astronomy Reviews}

\begin{document}

\begin{frontmatter}

\title{A Complete Reference of the Analytical Synchrotron
External Shock Models of Gamma-Ray Bursts}


\author{He Gao$^{1}$, Wei-Hua Lei$^{2,1}$, Yuan-Chuan Zou$^{2}$, Xue-Feng Wu$^3$ and Bing Zhang$^{1,4,5,*}$\\
1. Department of Physics \& Astronomy, University of Nevada, Las Vegas, NV 89154-4002, USA.\\
2. School of Physics, Huazhong University of Science and Technology,
Wuhan, 430074, China.\\
3. Purple Mountain Observatory, Chinese Academy of Sciences,
Nanjing, 210008, China.\\
4. Department of Astronomy, Peking University, Beijing 100871,
China\\
5. Kavli Institute of Astronomy and Astrophysics, Peking University,
Beijing 100871, China \\
$*$ Corresponding author: zhang@physics.unlv.edu}

\address{}

\begin{abstract}
Gamma-ray bursts are most luminous explosions in the universe. Their
ejecta are believed to move towards Earth with a relativistic speed.
The interaction between this ``relativistic jet'' and a circumburst
medium drives a pair of (forward and reverse) shocks. The electrons
accelerated in these shocks radiate synchrotron emission to power
the broad-band afterglow of GRBs. The external shock theory is an
elegant theory, since it invokes a limit number of model parameters,
and has well predicted spectral and temporal properties. On the
other hand, depending on many factors (e.g. the energy content,
ambient density profile, collimation of the ejecta, forward vs.
reverse shock dynamics, and synchrotron spectral regimes), there is
a wide variety of the models. These models have distinct predictions
on the afterglow decaying indices, the spectral indices, and the
relations between them (the so-called ``closure relations''), which
have been widely used to interpret the rich multi-wavelength
afterglow observations. This review article provides a complete
reference of all the analytical synchrotron external shock afterglow
models by deriving the temporal and spectral indices of all the models
in all spectral regimes, including some regimes that have not been
published before. The review article is designated to serve as a
useful tool for afterglow observers to quickly identify relevant
models to interpret their data. The limitations of the analytical
models are reviewed, with a list of situations summarized when
numerical treatments are needed.
\end{abstract}


\end{frontmatter}

\section{Introduction}

Gamma-ray bursts (GRBs) are the most luminous explosions in the
universe, which signify catastrophic events involving core
collapses of some rapidly spinning massive stars and mergers
of two compact objects (two neutron stars or one neutron star
and one stellar-mass black hole).
These events power an energetic, relativistic jet, which
beams towards Earth and gives rise to Doppler-boosted powerful
emission in $\gamma$-rays.

Although the nature of the progenitor and central engine as well as
the detailed physics of $\gamma$-ray emission are still rather
uncertain \citep[for reviews, see
e.g.][]{zhangmeszaros04,piran05,meszaros06,zhangcjaa07,kumarzhang13},
a generic synchrotron external shock model has been well
established to interpret the prompt emission and the
broad-band afterglow data
\citep{rees92,rees94,meszarosrees93,meszarosrees97,sari98,chevalier00}.
This model delineates the interaction between the relativistic GRB
jet and a circumburst medium. During the initial interaction, a pair
of shocks (forward and reverse) propagate into the ambient medium
and the ejecta, respectively. After the reverse shock crosses the
shell, the blastwave enters a self-similar phase described by the
Blandford-McKee self-similar solution \citep{blandford76}. In this
phase, the dynamics of the blastwave is solely determined by a few
parameters (e.g. the total energy of the system, the ambient
density and its profile).

Electrons are accelerated in both forward and reverse
shocks, which radiate synchrotron emission in the
magnetic fields behind the shocks that are believed to be generated
in situ due to plasma instabilities \citep[e.g.][]{medvedev99}.
Introducing several notations to parameterize micro-scopic processes,
i.e. the fractions of shock energy that go to electrons and magnetic
fields ($\epsilon_e$ and $\epsilon_B$), and the electron spectral
index $p$, one can then calculate the instantaneous synchrotron
spectrum at a given epoch, as well as the flux evolution with time (the
lightcurve) for a given observed frequency.

Since the simplest external shock theory does not invoke details of
a burst, and invokes only a limit number of model parameters, it is
an elegant theory with falsifiable predictions. It turned out that
the predicted power-law decay of lightcurves and broken power law
instantaneous spectra are well consistent with many late time
afterglow data in the pre-Swift era
\citep[e.g.][]{wijers97,waxman97a,wijers99,huang99,huang00,panaitescu01,panaitescu02,yost03},
suggesting that most of the observed multi-wavelength afterglows
indeed originate from jet-medium interaction, and that synchrotron
radiation is indeed the right radiation mechanism to power the
observed emission. Later observations showed more complicated
afterglow behaviors
\citep[e.g.][]{akerlof99,harrison99,berger03b,fox03,li03}, which
demand more complicated models \citep{meszaros98} that invoke joint
forward shock and reverse shock emission
\citep{meszarosrees97,meszarosrees99,saripiran99,saripiran99b,kobayashizhang03a,zhang03},
non-uniform density medium \citep{dailu98c,chevalier99,chevalier00},
continuous energy injection into the blastwave
\citep{dailu98b,rees98,zhangmeszaros01a}, collimation of the jet
\citep{rhoads99,sari99,zhangmeszaros02b,rossi02}, hard electron
injection spectrum \citep{daicheng01}, etc. Nonetheless, these more
complicated models, by introducing one or more additional
assumptions/parameters, still have clear testable predictions
regarding the afterglow decaying index $\alpha$, the spectral index
$\beta$, and the relation between them (the so-called ``closure
relations'') \citep[e.g.][for a collection of these
models]{zhangmeszaros04,zhang06}. The Swift mission
\citep{gehrels04} made it possible to systematically detect the
early phase of the GRB X-ray afterglow, which shows some
un-predicted features
\citep{tagliaferri05,burrows05,nousek06,obrien06,evans09} that
demand multiple physical processes that shape the observed
lightcurves \citep{zhang06}. Systematic data analyses
\citep{zhangbb07,liang07b,liang08,liang09,butlerkocevski07,kocevski07,chincarini07,chincarini10,margutti10}
suggest that the X-ray afterglow is a superposition of the
conventional external shock component and a radiation component that
is related to the late central engine activity
\citep[e.g.][]{zhangcjaa07,zhang11cr,zou13}. Nonetheless, the data
indicate that the low-energy (optical and radio) afterglows
\citep{kann10,kann11,chandra12} and the late-time X-ray afterglow is
more likely of the external shock origin. Recent Fermi observations
suggest that the GeV afterglow after the prompt emission phase is
also dominated by the emission from the external shock
\citep{kumar09,kumar10,ghisellini10,he11,liu11,maxham11}.
Observations with EVLA and ALMA start to reveal the early phase of
GRB afterglow in the radio and sub-mm regime, during which reverse
shock and self-absorption effects are important. These are the
regimes not fully covered by the already published materials. With
new data flooding in, it is essential to systematically survey a
complete list of external shock models in all possible temporal and
spectral regimes.

This review aims at providing a complete reference to the analytical
synchrotron external shock afterglow models. It includes all the
published models and spectral regimes, but also includes new
derivations in the previously not well-studied models or spectral
regimes. All the models are surveyed systematically, with typical
model parameters calculated, temporal and spectral indices and their
closure relations summarized in tables. It is designated as a
complete reference tool for GRB afterglow observers to quickly
identify the relevant models to interpret their broad-band data. In
Section 2, we provide a general description of the synchrotron
external shock models, which lay the foundation to derive any model
discussed later. Section 3 summarizes all the models in four
different phases: the reverse and forward shock models during the
reverse shock crossing phase (\S3.1), the forward shock models
during the isotropic self-similar deceleration phase (\S3.2), the
forward shock models in the post-jet break phase (\S3.3), and the
forward shock models in the non-relativistic (\S3.4) phase. For each
model, the expressions of key parameters, including the three
characteristic frequencies $\nu_a$ (self-absorption frequency),
$\nu_m$ (the characteristic synchrotron frequency of the electrons
at the minimum injection energy), and $\nu_c$ (the cooling
frequency), and the peak synchrotron flux density
$F_{\rm{\nu,max}}$, are presented. The spectral index $\beta$ and
the temporal index $\alpha$ (with the convention $F_\nu \propto
\nu^{-\beta} t^{-\alpha}$, as well as their closure relations are
presented in Tables 1-20. In Section 4, we describe how to make use
of the models to calculate lightcurves, and derive all possible
lightcurves (Fig.1-44) by allowing a wide range of parameters. We
also draw typical lightcurves in the radio, optical and X-ray bands
by adopting typical values of model parameters. Finally, we discuss
the limitations of these simple analytical models in Section 5.

\section{General description of the synchrotron external shock models of GRBs}

The synchrotron external shock models
\citep{rees92,meszarosrees93,meszarosrees97,sari98} describe the
interaction between the GRB outflow and the circum-burst hydrogen
medium (CBM). The physical parameters that enter the problem to
determine the dynamics of blastwave deceleration include the
``isotropic'' energy $E$ (the total energy assuming that the outflow
is isotropic), the initial Lorentz factor $\Gamma_0$, and the CBM
density and its profile $n(R)\propto R^{-k}, 0 \leq k<4$
\citep{blandford76} (and see \cite{sari06} for a discussion for the
cases with $k\geq 4$), where $R$ is the radius from the central
engine. As a result, these models are very generic, not depending on
the details of the central engine activity and prompt $\gamma$-ray
emission. There is another parameter, i.e. the magnetization of the
outflow $\sigma$, that would slightly affect the dynamics of the
system during the early phase of evolution
\citep[e.g.][]{zhangkobayashi05,mimica09}. In this review, we limit
ourselves to the regime of zero (or very low) magnetization. These
matter-dominated ejecta are also called ``fireballs''.

Assuming that a jet with opening angle $\theta_j$ is launched
from the central engine, which lasts a duration $T$ with constant
Lorentz factor $\Gamma_0$, the evolution of the a fireball jet
includes four phases\footnote{These simplified assumptions are
certainly not the case in reality, but may be a good approximation
after the prompt emission phase when the ejecta irregularities
are smoothed out after energy dissipation through internal shocks
\citep{rees94,kobayashi97,kumarpiran00a,maxham09}.}.
The first phase is when a pair of shocks (forward and reverse)
propagating into the CBM and the shell (with initial width
$\Delta_0 = c T$), respectively \citep{sari95}. After the
reverse shock crosses the shell, the blastwave quickly enters
a self-similar deceleration phase described by the Blandford-McKee
solution \citep{blandford76}. This is the second phase. Later,
as the blastwave is decelerated enough, the $1/\Gamma$ cone
becomes larger than the geometric cone defined by $\theta_j$,
the afterglow enters the post-jet-break phase. Finally, the
blastwave enters the Newtonian phase when the velocity is
much smaller than speed of light. The dynamics is then described
by the well-known Sedov solution widely used to study supernova
remnants.

During all the phases, particles are believed to be accelerated from
the forward shock front via the 1st-order Fermi acceleration
mechanism. For the reverse shock, particle acceleration occurs only
during the shock crossing phase. No new particles are accelerated in
the reverse-shocked region after the reverse shock crosses the
ejecta shell. Assume a power-law distribution of the electrons
$N(\gamma_e)d\gamma_e\propto \gamma_e^{-p}d\gamma_e$ (for $\gamma_m
\leq \gamma_e \leq \gamma_M$) and consider radiative cooling of
electrons and continuous injection of new electrons from the shock
front, one can obtain a broken power-law electron spectrum, which
leads to a multi-segment broken power-law photon spectrum at any
epoch \citep{sari98,meszaros98}.

Assuming that a constant
fraction $\epsilon_e$ of the shock energy is distributed to electrons,
one can derive the minimum injected electron Lorentz factor
\begin{eqnarray}
\gamma_m=g(p)\epsilon_e(\Gamma -1) \frac {m_p} {m_e},
\end{eqnarray}
where $\Gamma$ is the relative Lorentz factor between the unshocked
region and the shocked region, which is the Lorentz factor of
the blastwave for the forward shock, $m_p$ is proton mass, $m_e$
is electron mass, and the function $g(p)$ takes the form
\begin{eqnarray}
\label{gp} g(p) \simeq \left\{ \begin{array}{ll} \frac{p-2}{p-1}, & p>2;\\
\rm{ln}^{-1}(\gamma_{M}/\gamma_{m}), &
p=2; \\
\end{array} \right.
\end{eqnarray}
Here $\gamma_{M}$ is the maximum electron Lorentz factor, which
may be estimated by balancing the acceleration time scale and the
dynamical time scale, i.e.
\begin{eqnarray}
\gamma_{M}\sim \frac{\Gamma t q_e B}{\zeta m_p c},
\end{eqnarray}
where $\zeta$ is a parameter of order unity that describes the
details of acceleration, $t$ is the observational time, $q_e$ is the
electron charge, and $B$ is the comoving magnetic field strength. We
also assume that the magnetic energy density behind the shock is a
constant fraction $\epsilon_B$ of the shock energy density. This
gives
\begin{eqnarray}
B=(8 \pi e\epsilon_B)^{1/2},
\end{eqnarray}
where $e$ is the energy density in the shocked region. If the
electron energy has a harder spectral index
$1 < p < 2$, the minimum electron Lorentz factor would be derived as
\citep{daicheng01,bhattacharya01}
\begin{eqnarray}
\gamma_{m}=\left(\frac{2-p}{p-1}\frac{m_p}{m_e}\epsilon_e(\Gamma
-1)\gamma_{M}^{p-2}\right)^{1/(p-1)}
\end{eqnarray}

For synchrotron radiation, the observed radiation power and the
characteristic frequency of an electron with Lorentz factor $\gamma_e$
are given by \citep{rybicki79}
\begin{eqnarray}
\label{power} P(\gamma_e) \simeq \frac 4 3 \sigma_T c \Gamma^2
\gamma_e^2 \frac {B^2} {8\pi},
\end{eqnarray}
\begin{eqnarray}
\label{freq} \nu(\gamma_e) \simeq \Gamma \gamma_e^2 \frac {q_e B} {2
\pi m_e c},
\end{eqnarray}
where the factors of $\Gamma^2$ and $\Gamma$ are introduced to
transform the values from the rest frame of the shocked fluid to the
frame of the observer.

The spectral power of individual electron, $P_\nu$ (power per unit
frequency, in unit of ${\rm
erg\,Hz^{-1}\,s^{-1}}$), varies as $\nu^{1/3}$ for
$\nu<\nu(\gamma_e)$, and cuts off exponentially for
$\nu>\nu(\gamma_e)$ \citep{rybicki79}. The peak power occurs at
$\nu(\gamma_e)$, where it has the approximate value
\begin{eqnarray}
\label{flux} P_{\nu,max}\approx\frac{P(\gamma_e)}{\nu(\gamma_e)}=
\frac {m_e c^2 \sigma_T} {3 q_e} \Gamma B~.
\end{eqnarray}
Note that $P_{\nu,max}$ does not depend on $\gamma_e$.

The life time of a relativistic electron with Lorentz factor $\gamma_e$
in the observer frame can be estimated as
\begin{eqnarray}
\tau(\gamma_e)=\frac{\Gamma\gamma_e m_e c^2}{\frac 4 3 \sigma_T c
\Gamma^2 \gamma_e^2 \frac {B^2} {8\pi}}=\frac{6\pi m_e
c}{\Gamma\gamma_e \sigma_T B^2 }~.
\end{eqnarray}
One can define a critical electron Lorentz factor $\gamma_c$ by
setting $\tau(\gamma_e)=t$, i.e.,
\begin{eqnarray}
\label{cool} \gamma_c= \frac{6\pi m_e c}{\Gamma \sigma_T B^2t},
\end{eqnarray}
where $t$ refers to the time in the observer frame. Above $\gamma_c$,
cooling by synchrotron radiation becomes significant, so that the
electron distribution shape is modified in the $\gamma_e>\gamma_c$
regime.

The electron Lorentz factors $\gamma_m$ and $\gamma_c$ defines two
characteristic emission frequencies $\nu_m$ and $\nu_c$ in the
synchrotron spectrum. A third characteristic frequency $\nu_a$, is
defined by synchrotron self-absorption, below which the synchrotron
photons are self-absorbed. There are two methods to calculate this
frequency. The first one is to define $\nu_a$ by the condition that
the photon optical depth for self-absorption is unity \citep{rybicki79}.
A more convenient method \citep[e.g.][]{saripiran99b,kobayashizhang03a}
is to define
$\nu_a$ by equating the synchrotron flux and the flux of a blackbody, i.e.
\begin{eqnarray}
I_{\nu}^{syn}(\nu_a)=I_{\nu}^{bb}(\nu_a)\simeq2kT\cdot\frac{\nu_a^2}{c^2},
\end{eqnarray}
where the blackbody temperature is
\begin{eqnarray}
kT=\rm{max}[\gamma_{a},\rm{min}(\gamma_c,\gamma_m)]m_ec^2,
\end{eqnarray}
and $\gamma_{a}$ is the corresponding electron Lorentz factor of
$\nu_a$ for synchrotron radiation, i.e. $\gamma_a=(2\pi m_e
c\nu_a/\Gamma q_eB)^{1/2}$ (derived from Eq.[7]). One can prove
\citep{shen09} that the two methods are equivalent to each other,
even though the coefficient may slightly differ within a factor of
two.

In the afterglow phase, $\nu_a$ is usually the smallest among the
three frequencies. The broad-band synchrotron spectrum therefore
falls into two broad categories depending on the order of $\gamma_m$
and $\gamma_c$, namely the fast cooling regime ($\gamma_m>\gamma_c$)
or the slow cooling regime $\gamma_m<\gamma_c$ \citep{sari98}.

In the slow cooling regime, the electron energy distribution is
\begin{eqnarray}
\label{Ngam1} N(\gamma_e) = \left\{ \begin{array}{ll}
C_1(p-1)\gamma_{\rm m}^{p-1}\gamma_e^{-p}, &  \gamma_{\rm m} \leq \gamma_e \leq \gamma_{\rm c}, \\
C_1(p-1)\gamma_{\rm m}^{p-1}\gamma_{\rm c}\gamma_e^{-p-1}, &
\gamma_e
> \gamma_{\rm c}.
\end{array} \right.
\end{eqnarray}

In the fast cooling regime, usually one has the approximation
\begin{eqnarray}
\label{Ngam2} N(\gamma) = \left\{ \begin{array}{ll}
C_2 \gamma_{\rm c} \gamma_e^{-2}, &  \gamma_{\rm c} \leq \gamma_e \leq \gamma_{\rm m}, \\
C_2 \gamma_{\rm m}^{p-1}\gamma_{\rm c} \gamma_e^{-p-1}, &  \gamma_e
> \gamma_{\rm m}.
\end{array} \right.
\end{eqnarray}
where $C_1$ and $C_2$ are normalization factors\footnote{It is
realized that the fast-cooling spectrum below injection can be
harder than -2 in a decaying magnetic field, which is the case for
GRB afterglow emission \citep{uhm13a,uhm13b}. We will discuss this
more in Section 5.}.

For such an electron energy distribution, the observed synchrotron
radiation flux density $F_\nu$ can be expressed as

I. $\nu_a<\nu_m<\nu_c$:

\begin{eqnarray}
\label{gp} F_{\nu} = F_{\rm{\nu,max}}\left\{ \begin{array}{ll} \left(\frac{\nu_a}{\nu_m}\right)^{1/3}\left(\frac{\nu}{\nu_a}\right)^{2}, &\nu<\nu_a;\\
\left(\frac{\nu}{\nu_m}\right)^{1/3}, &\nu_a<\nu<\nu_m; \\
\left(\frac{\nu}{\nu_m}\right)^{-(p-1)/2}, &\nu_m<\nu<\nu_c; \\
\left(\frac{\nu_c}{\nu_m}\right)^{-(p-1)/2}\left(\frac{\nu}{\nu_c}\right)^{-p/2}, &\nu>\nu_c; \\
\end{array} \right.
\end{eqnarray}

II. $\nu_a<\nu_c<\nu_m$:

\begin{eqnarray}
\label{gp} F_{\nu} = F_{\rm{\nu,max}}\left\{ \begin{array}{ll} \left(\frac{\nu_a}{\nu_c}\right)^{1/3}\left(\frac{\nu}{\nu_a}\right)^{2}, &\nu<\nu_a;\\
\left(\frac{\nu}{\nu_c}\right)^{1/3}, &\nu_a<\nu<\nu_c; \\
\left(\frac{\nu}{\nu_c}\right)^{-1/2}, &\nu_c<\nu<\nu_m; \\
\left(\frac{\nu_m}{\nu_c}\right)^{-1/2}\left(\frac{\nu}{\nu_m}\right)^{-p/2}, &\nu>\nu_m. \\
\end{array} \right.
\end{eqnarray}

In general, there are six different orders among $\nu_a$, $\nu_m$
and $\nu_c$. Under extreme conditions they might be all possible.
When $\nu_a > \nu_c$, the electron energy distribution may be
significantly modified \citep{gao13}, so that analytical models are
no longer good approximations. Those cases are rare but not
impossible, and we will leave out from this review. A detailed
analysis can be found in \cite{kobayashi04} and \cite{gao13}.

For the $\nu_a < \nu_c$ regime, there is one more case, i.e.

III. $\nu_m<\nu_a<\nu_c$:

\begin{eqnarray}
\label{gp} F_{\nu} = F_{\rm{\nu,max}}\left\{ \begin{array}{ll} \left(\frac{\nu_m}{\nu_a}\right)^{(p+4)/2}\left(\frac{\nu}{\nu_m}\right)^{2}, &\nu<\nu_m;\\
\left(\frac{\nu_a}{\nu_m}\right)^{-(p-1)/2}\left(\frac{\nu}{\nu_a}\right)^{5/2}, &\nu_m<\nu<\nu_a; \\
\left(\frac{\nu}{\nu_m}\right)^{-(p-1)/2}, &\nu_a<\nu<\nu_c; \\
\left(\frac{\nu_c}{\nu_m}\right)^{-(p-1)/2}\left(\frac{\nu}{\nu_c}\right)^{-p/2}, &\nu>\nu_c. \\
\end{array} \right.
\end{eqnarray}

In all the above expressions,
$F_{\rm{\nu,max}}$ is the observed peak flux at luminosity
distance $D$ from the source, which can be estimated as
\citep{sari98}:
\begin{eqnarray}
F_{\rm{\nu,max}}\equiv N_eP_{\nu,max}/4\pi D^2,
\end{eqnarray}
where $N_e$ is the total number of electrons in the emission region.
For the forward shock emission, it is usually calculated as
$N_e\sim\int_{R_{\rm 0}}^{R}4\pi n(r)r^2dr$, where $R_{\rm 0}$ is
the central engine radius and $R$ is the radius from the center of
central engine.

The instantaneous spectra described above do not depend on the
hydrodynamical evolution of the shocks. However, in order to
calculate the light curve at a given frequency, we need to know the
temporal evolution of various quantities, such as the break
frequencies $\nu_a$, $\nu_m$ and $\nu_c$ and the peak flux density
$F_{\rm{\nu,max}}$, which depend on the dynamics of the system. For
the forward shock, the emission essentially depends on the temporal
evolution of three quantities $\Gamma$, $R$ and $B$ (or the energy
density $e$ if $\epsilon_B$ is assumed to be constant). In the next
section, we will derive how $\Gamma$, $R$ and $e$ evolve as a
function of $t$ for various systems and dynamical phases, and
quantify the evolutions of the break frequencies $\nu_m$, $\nu_c$,
$\nu_a$, as well as the peak flux density $F_{\rm{\nu,max}}$. We
will then present the spectral and temporal indices ($\beta$ and
$\alpha$) for all the spectral regimes of all the models, as well as
the closure relations between $\alpha$ and $\beta$.

\section{Analytical synchrotron external shock models}

There are many variations of the external shock synchrotron models.
First, during the reverse shock crossing phase, the dynamics of
the blastwave is complicated, and there are rich features in the
reverse shock and forward shock lightcurves. Second, even after
reverse shock crossing and when the blastwave is in the self-similar
deceleration phase, variations in the energy content of the blastwave
(e.g. radiative loss or energy injection) or in the profile of the
CBM (e.g. constant density ISM, a stratified wind, or a more general
profile) would give very different lightcurves. Next, the collimation
effect becomes important when the blastwave is decelerated enough
so that the relativistic beaming $1/\Gamma$ cone is large enough
to enclose a solid angle in which the anisotropic effect becomes
significant. Finally, the blastwave eventually enters the
Newtonian phase, when a different self-similar solution is reached.
For each dynamical model, there could be many possible lightcurves
in view of a range of initial spectral regime of the observing
frequency, and the complicated evolutions of three characteristic
frequencies and their relative orders.

In the following, we will discuss all these models based on the
four dynamical phases outlined above: Phase 1: reverse shock
crossing phase; Phase 2: relativistic, pre-jet-break self-similar
deceleration phase; Phase 3: post-jet-break phase; Phase 4:
Newtonian phase.

\subsection{Phase 1: Reverse shock crossing phase}

We consider a uniform and cold relativistic shell with isotropic
energy $E$, lab-frame width $\Delta_0 = cT$, coasting with an
initial Lorentz factor $\Gamma_0$. This shell sweeps into a
circumburst hydrogen medium (CBM) with a proton number density
profile $n=AR^{-k}$ ($0\leq k <4$).
A pair of shocks are developed: a forward shock
propagating into the CBM and a reverse shock propagating into the
shell. The two shocks and the contact discontinuity separate the
system into four regions: (1) the unshocked CBM (called Region 1
hereafter), (2) the shocked CBM (Region 2), (3) the shocked shell
(Region 3), and (4) the unshocked shell (Region 4). Using the
relativistic shock jump conditions \citep{blandford76} and assuming
equal pressure and velocity in the blastwave region (Regions 2 and
3)\footnote{Strictly speaking, such a situation cannot be achieved
since it violates energy conservation \citep{uhm11,uhm12}.
Nonetheless, for a short-lived reverse shock (finite width
$\Delta_0$ with constant $\Gamma_0$), such an approximation is good
enough to delineate the dynamical evolution of the system.}, i.e.,
$e_2=e_3$ and $\gamma_2=\gamma_3$, the values of the bulk Lorentz
factor $\Gamma$, the radius $R$, and the energy density $e$ in the
shocked regions can be estimated as functions of $n_1$, $n_4$, and
$\Gamma_0 = \gamma_4$, where $n_i$, $e_{i}$ and $\gamma_i$ are the
comoving number densities, energy density and Lorentz factors for
Region \emph{i}.

Analytical results can be obtained in both relativistic and
Newtonian reverse shock limits. These two cases are defined by
comparing a parameter $f\equiv n_4/n_1$
(ratio of the number densities between the unshocked shell and
the unshocked CBM) and $\gamma_4^2$ \citep{sari95}.
If $f\gg \gamma_4^2$, the reverse
shock is Newtonian (NRS, thin shell case), and if $f \ll
\gamma_4^2$, the reverse shock is relativistic (RRS, thick shell
case). The strength of the reverse shock depends on the
relative Lorentz factor between Region 3 and Region 4, i.e.
\begin{equation}
  \bar{\gamma}_{34}=\gamma_3\gamma_4(1-\sqrt{1-1/\gamma_3^2}\sqrt{1-1/\gamma_4^2}).
  \label{eq:gamma_34}
\end{equation}
For $\gamma_2, \gamma_4 \gg 1$ and assuming $\gamma_2 = \gamma_3$,
$\bar{\gamma}_{34}$ can be expressed as $\bar{\gamma}_{34} \simeq
\frac{1}{\sqrt{2}} \gamma_4^{1/2} f^{-1/4}$ for a RRS, while
$\bar{\gamma}_{34}-1 \simeq \frac{4}{7}\gamma_4^2 f^{-1}$ for a NRS.

The Phase 1 ends at the reverse shock crossing time
\begin{equation}
t_\times = {\rm max} (t_{\rm dec}, T),
\label{tcross}
\end{equation}
where $T$ is the duration of the burst, and
\begin{equation}
t_{\rm dec} = \left[\frac{(3-k)E}{2^{4-k}\pi
Am_p\Gamma_0^{8-2k}c^{5-k}}\right]^{\frac{1}{3-k}} \label{tdec}
\end{equation}
is the deceleration time of the ejecta for an impulsive injection
of fireball with energy $E$ and initial Lorentz factor $\Gamma_0$,
which corresponds to the time when the mass collected from the CBM
is about $1/\Gamma$ of the rest mass entrained in the ejecta.
For thin shells, one has $t_\times = t_{\rm dec}$, while for thick
shells, one has $t_\times = T$ \citep{kobayashi99}.

In the following, we discuss the synchrotron emission properties
for four models: thin shell forward shock model, thin
shell reverse shock model, thick shell forward shock model, and
thick shell reverse shock model.

\subsubsection{Thin Shell Forward Shock Model}

In the thin shell models, the reverse shock is Newtonian,
so that $\gamma_2 \simeq \gamma_4 = \Gamma_0$. We consider
the dynamics of Region 2, i.e.
\begin{eqnarray}
\gamma_2=\Gamma_0,~~~~~~~~~~~~~R_2=2c\Gamma_0^2t. \nonumber
\end{eqnarray}
In general, the expressions for an arbitrary density profile index
$k$ can be derived. The two most commonly used models are the
constant density interstellar medium (ISM) model ($k=0$) and the
free stratified wind model ($k=2$). Hereafter we will explicitly
derive the expressions for these two density profiles.

For the constant density case ($n_1=n_0$) with electron energy
spectral index $p > 2$, one has
\begin{eqnarray}
&&\nu_m=   3.1\times10^{16}~{\rm Hz}~\hat z^{-1}\frac{G(p)}{G(2.3)}\Gamma_{0,2}^{4}n_{0,0}^{1/2}\epsilon_{e,-1}^{2}\epsilon_{B,-2}^{1/2}           ,\nonumber\\
&&\nu_c=    4.1\times10^{16}~{\rm Hz}~\hat z\Gamma_{0,2}^{-4}n_{0,0}^{-3/2}\epsilon_{B,-2}^{-3/2}t_{2}^{-2},           \nonumber\\
&&F_{\rm{\nu,max}}=     1.1\times10^{4}~\mu {\rm
Jy}~\hat z^{-2}\Gamma_{0,2}^{8}n_{0,0}^{3/2}\epsilon_{B,-2}^{1/2}D_{28}^{-2}t_{2}^{3}          ,\nonumber\\
&&\nu_a =5.7\times10^{9}~{\rm Hz}~\hat z^{-8/5}\frac{g^{I}(p)}{g^{I}(2.3)}\Gamma_{0,2}^{8/5}n_{0,0}^{4/5}\epsilon_{e,-1}^{-1}\epsilon_{B,-2}^{1/5}t_{2}^{3/5}, ~~~~~~~~~~~~~~~~~~~~~~~~~ \nu_a < \nu_m < \nu_c\nonumber \\
&&\nu_a =8.3\times10^{12}~{\rm
Hz}~\hat z^{-\frac{p+6}{p+4}}\frac{g^{II}(p)}{g^{II}(2.3)}\Gamma_{0,2}^{\frac{4(p+2)}{p+4}}n_{0,0}^{\frac{p+6}{2(p+4)}}\epsilon_{e,-1}
^{\frac{2(p-1)}{p+4}}\epsilon_{B,-2}^{\frac{p+2}{2(p+4)}}t_{2}^{\frac{2}{p+4}},~~~~~~~~~~~~
\nu_m < \nu_a < \nu_c\nonumber\\
&&\nu_a =4.9\times10^{9}~{\rm Hz}~\hat z^{-13/5}\frac{g^{III}(p)}{g^{III}(2.3)}\Gamma_{0,2}^{28/5}n_{0,0}^{9/5}\epsilon_{B,-2}^{6/5}t_{2}^{8/5},               ~~~~~~~~~~~~~~~~~~~~~~~~~~~ \nu_a < \nu_c < \nu_m\nonumber \\
\end{eqnarray}
where $G(p)$ and $g^{i}(p)$ are numerical constants related to $p$,
and $\hat z=(1+z)/2$ is the redshift correction factor.
The explicit expressions of $G(p)$ and $g^{i}(p)$ are complicated,
and we present them (along with the $p$-dependent coefficients in
all other models) in Appendix A.

When $1<p<2$, expressions of $\nu_c$ and $F_{\rm{\nu,max}}$ remain
the same as the $p>2$ case (also apply to
other models discussed later). Other expressions are modified as follows
\begin{eqnarray}
&&\nu_m=3.2\times10^{14}~{\rm Hz}~\hat z^{-1}\frac{g^{IV}(p)}{g^{IV}(1.8)}\Gamma_{0,2}^{\frac{p+2}{p-1}}n_{0,0}^{\frac{1}{2(p-1)}}\zeta_{0}^{\frac{2-p}{p-1}}\epsilon_{e,-1}^{\frac{2}{p-1}}\epsilon_{B,-2}^{\frac{1}{2(p-1)}}           ,\nonumber\\
&&\nu_a=4.6\times10^{10}~{\rm Hz}~\hat
z^{-8/5}\frac{g^{V}(p)}{g^{V}(1.8)}\Gamma_{0,2}^{\frac{46-31p}{10(1-p)}}n_{0,0}^{\frac{26-21p}{20(1-p)}}\zeta_{0}^{\frac{p-2}{2(p-1)}}
\epsilon_{e,-1}^{\frac{1}{1-p}}\epsilon_{B,-2}^{\frac{14-9p}{20(1-p)}}t_{2}^{3/5},~~~~\nu_a < \nu_m < \nu_c\nonumber\\
&&\nu_a=2.0\times10^{10}~{\rm Hz}~\hat
z^{-\frac{p+6}{p+4}}\frac{g^{VI}(p)}{g^{VI}(1.8)}\Gamma_{0,2}^{\frac{p+14}{p+4}}n_{0,0}^{\frac{4}{p+4}}\zeta_{0}^{\frac{2-p}{p+4}}
\epsilon_{e,-1}^{\frac{2}{p+4}}\epsilon_{B,-2}^{\frac{2}{p+4}}t_{2}^{\frac{2}{p+4}}~~~~~~~~~~~~~~\nu_m < \nu_a < \nu_c\nonumber\\
&&\nu_a=4.0\times10^{9}~{\rm Hz}~\hat
z^{-13/5}\frac{g^{VII}(p)}{g^{VII}(1.8)}\Gamma_{0,2}^{28/5}n_{0,0}^{9/5}\epsilon_{B,-2}^{6/5}t_{2}^{8/5}~~~~~~~~~~~~~~~~~~~~~~~~~~~\nu_a<\nu_c<\nu_m\nonumber
\nonumber\\
\end{eqnarray}

For the wind model ($k=2$), one can express the density profile as
$n_1=AR^{-2}$, with $A=\dot{M}/4\pi
m_pv_w=3\times10^{35}A_*\rm{cm^{-1}}$,
$A_*=(\dot{M}/10^{-5}~\rm{M_{\odot}~yr^{-1}})(v_w/10^3\rm{~km~s^{-1}})^{-1}$
\citep{dailu98c,chevalier99,chevalier00}. For $p>2$, one has
\begin{eqnarray}
&&\nu_m=    8.7\times10^{16}~{\rm Hz}\frac{G(p)}{G(2.3)}A_{*,-1}^{1/2}\Gamma_{0,2}^{2}\epsilon_{e,-1}^{2}\epsilon_{B,-2}^{1/2}t_{2}^{-1}           ,\nonumber\\
&&\nu_c=    1.8\times10^{15}~{\rm Hz}~\hat z^{-2}\Gamma_{0,2}^{2}A_{*,-1}^{-3/2}\epsilon_{B,-2}^{-3/2}t_{2}^{}           \nonumber\\
&&F_{\rm{\nu,max}}=  7.5\times10^{5}~\mu {\rm
Jy}~\hat z A_{*,-1}^{3/2}\Gamma_{0,2}^{2}\epsilon_{B,-2}^{1/2}D_{28}^{-2}             ,\nonumber\\
&&\nu_a =5.9\times10^{10}~{\rm
Hz}\frac{g^{VIII}(p)}{g^{VIII}(2.3)}\Gamma_{0,2}^{-\frac{8}{5}}
A_{*,-1}^{\frac{4}{5}}\epsilon_{e,-1}^
{-1}\epsilon_{B,-2}^{\frac{1}{5}}t_{2}^{-1}, ~~~~~~~~~~~~~~~~~~~~~~~\nu_a < \nu_m < \nu_c\nonumber \\
&&\nu_a =4.7\times10^{13}~{\rm
Hz}\frac{g^{IX}(p)}{g^{IX}(2.3)}\Gamma_{0,2}^{\frac{2(p-2)}{p+4}}A_{*,-1}^{\frac{p+6}{2(p+4)}}
\epsilon_{e,-1}^{\frac{2(p-1)}{p+4}}\epsilon_{B,-2}^{\frac{p+2}{2(p+4)}}t_{2}^{-1}~~~~~~~~~~~~~~~~~~\nu_a < \nu_m < \nu_c\nonumber \\
&&\nu_a=4.1\times10^{11}~{\rm Hz}~\hat z\frac{g^{X}(p)}{g^{X}(2.3)}\Gamma_{0,2}^{-8/5}A_{*,-1}^{9/5}\epsilon_{B,-2}^{6/5}t_{2}^{-2},~~~~~~~~~~~~~~~~~~~~~~~~~~~~\nu_a < \nu_c < \nu_m\nonumber \\
\end{eqnarray}

For $1<p<2$, one has
\begin{eqnarray}
&&\nu_m=1.2\times10^{15}~{\rm Hz}~\hat z^{\frac{2-p}{p-1}}\frac{g^{XI}(p)}{g^{XI}(1.8)}A_{*,-1}^{\frac{1}{2(p-1)}}\Gamma_{0,2}^{\frac{p}{p-1}}\zeta_{0}^{\frac{2-p}{p-1}}\epsilon_{e,-1}^{\frac{2}{p-1}}\epsilon_{B,-2}^{\frac{1}{2(p-1)}}t_{2}^{\frac{1}{1-p}}           ,\nonumber\\
&&\nu_a=4.2\times10^{11}~{\rm Hz}~\hat z^{\frac{p-2}{2(p-1)}}\frac{g^{XII}(p)}{g^{XII}(1.8)}\Gamma_{0,2}^{\frac{11p-6}{10(1-p)}}A_{*,-1}^{\frac{26-21p}{20(1-p)}}\zeta_{0}^{\frac{p-2}{2(p-1)}}\epsilon_{e,-1}^{\frac{1}{1-p}}\epsilon_{B,-2}^{\frac{14-9p}{20(1-p)}}t_{2}^{\frac{4-3p}{2(p-1)}},\nu_a < \nu_m < \nu_c \nonumber\\
&&\nu_a=1.2\times10^{13}~{\rm Hz}~\hat
z^{\frac{2-p}{p+4}}\frac{g^{XIII}(p)}{g^{XIII}(1.8)}\Gamma_{0,2}^{\frac{p-2}{p+4}}A_{*,-1}^{\frac{4}{p+4}}\zeta_{0}^{\frac{2-p}{p+4}}
\epsilon_{e,-1}^{\frac{2}{p+4}}\epsilon_{B,-2}^{\frac{2}{p+4}}t_{2}^{-\frac{6}{p+4}},~~~~~~~~~~~~\nu_m
< \nu_a < \nu_m\nonumber\\
&&\nu_a =3.4\times10^{11}~{\rm Hz}~\hat z\frac{g^{XIV}(p)}{g^{XIV}(1.8)}\Gamma_{0,2}^{-8/5}A_{*,-1}^{9/5}\epsilon_{B,-2}^{6/5}t_{2}^{-2},~~~~~~~~~~~~~~~~~~~~~~~~~~~~~~\nu_a < \nu_c < \nu_m\nonumber \\
\end{eqnarray}

The $\alpha$ and $\beta$ values and their closure relations of the
models described in this section (with convention $F_{\nu} \propto
t^{-\alpha} \nu^{-\beta}$) are collected in Tables \ref{Thinfpre}
and \ref{Thinfpre1}.

We note that the temporal evolution of the characteristic
frequencies and the peak flux density are important to judge the
relevant models. Hereafter at the end of each subsection, we
summarize these dependences for easy identification.

For this regime (thin-shell forward shock model during shock
crossing) and for $p>2$, $\nu_m \propto t^0 ~(t^{-1})$, $\nu_c
\propto t^{-2}~ (t^{1})$, $F_{\rm \nu,max} \propto t^3~ (t^0)$ for
the ISM (wind) models, respectively. The temporal evolution of
$\nu_a$ depends on the relative orders between $\nu_a$, $\nu_m$ and
$\nu_c$. For $1<p<2$, $\nu_c$ and $F_{\rm{\nu,max}}$ evolutions are
the same as $p>2$ cases, while $\nu_m \propto
t^0~(t^{\frac{1}{1-p}})$ for the ISM (wind) models, respectively.

\begin{table}
\caption{The temporal decay index $\alpha$ and spectral index $\beta$ in thin
shell forward shock model with $\nu_a <
\rm{min}(\nu_m,\nu_c)$.\label{Thinfpre}}
\begin{tabular}{llllll}
\hline\hline
& & $p>2$ & &  $1<p<2$ & \\

& $\beta$ & $\alpha $  &  $\alpha (\beta)$ & $\alpha$  & $\alpha (\beta)$  \\
\hline
ISM & slow cooling  \\
\hline
$\nu<\nu_a$   &  $-2$  &   $-2$ & $\alpha={\beta}$  &$-2$ & $\alpha={\beta}$\\
$\nu_a<\nu<\nu_m$   &  $-{1 \over 3}$  &   $-3$ & $\alpha={3\beta}$ & $-3$  & $\alpha={3\beta}$\\
$\nu_m<\nu<\nu_c$   &  ${{p-1 \over 2}}$  &  $-3$
   &  $--$ & $-3$   & $--$\\
$\nu>\nu_c$   &  ${{p\over 2}}$   &   ${-2}$  & $--$ & $-2$   & $--$ \\

\hline
ISM & fast cooling \\
\hline
$\nu<\nu_a$   &  $-2$  &   $-1$ & $\alpha=\frac{\beta}{2}$ & $-1$  & $\alpha=\frac{\beta}{2}$\\
$\nu_a<\nu<\nu_c$   &  $-{1\over 3}$  &   $-{11\over 3}$ &  $\alpha={11\beta}$  &  $-{11 \over 3}$  & $\alpha={11\beta}$\\
$\nu_c<\nu<\nu_m$   &  ${1\over 2}$  &  $-2$  & $\alpha=-{4\beta}$ &  $-2$   &  $\alpha=-{4\beta}$ \\
$\nu>\nu_m$   &  ${p\over 2}$   &   $-2$      &  $--$  &  $-2$  & $--$ \\

\hline
Wind & slow cooling \\
\hline
$\nu<\nu_a$   &  $-2$   &   -2  &  $\alpha={\beta}$ & $\frac{5p-6}{2(1-p)}$  & $--$\\
$\nu_a<\nu<\nu_m$   &  $-{1\over 3}$   &   $-{1\over 3}$  &  $\alpha={\beta}$ & $-\frac{1}{3(p-1)}$  & $--$\\
$\nu_m<\nu<\nu_c$   &  ${p-1\over 2}$  &   ${p-1\over 2}$    &   $\alpha={\beta}$ & ${1\over 2}$ & $--$\\
$\nu>\nu_c$   &  ${p\over 2}$   &   ${p-2\over 2}$    &  $\alpha={\beta-1}$   &   $0$  &  $--$\\

\hline
Wind & fast cooling  \\
\hline
$\nu<\nu_a$   &  $-2$   &   $-3$ & $\alpha={3\beta \over 2}$ &  $-3$  & $\alpha={3\beta \over 2}$  \\
$\nu_a<\nu<\nu_c$   &  $-{1\over 3}$   &   ${1\over 3}$ & $\alpha=-\beta$ &  ${1 \over 3}$  & $\alpha=-\beta$  \\
$\nu_c<\nu<\nu_m$   &  ${1\over 2}$  &   $-{1\over 2}$  & $\alpha=-{\beta}$  & $-{1 \over 2}$ & $\alpha=-{\beta}$\\
$\nu>\nu_m$   &  ${p\over 2}$  &   ${p-2\over 2}$     & $\alpha={\beta-1}$  &  $0$  & $--$\\
\hline
\end{tabular}
\label{Tab:alpha-beta}
\end{table}

\begin{table}

\caption{The temporal decay index $\alpha$ and spectral index $\beta$ in thin
shell forward shock model in the $\nu_m<\nu_a <
\nu_c$ regime.\label{Thinfpre1}}
\begin{tabular}{llllll}
\hline\hline
& & $p>2$ & &  $1<p<2$ & \\

& $\beta$ & $\alpha $  &  $\alpha (\beta)$ & $\alpha$ & $\alpha (\beta)$  \\
\hline
ISM & slow cooling  \\
\hline
$\nu<\nu_m$   &  $-2$  &   $-2$ & $\alpha={\beta}$ & $-2$  & $\alpha={\beta}$\\
$\nu_m<\nu<\nu_a$   &  $-{5 \over 2}$  &   $-2$ & $\alpha={4\beta \over 5}$ & $-2$  & $\alpha={4\beta \over 5}$\\
$\nu_a<\nu<\nu_c$   &  ${{p-1 \over 2}}$   &  $-3$
  &  $--$ & $-3$  & $--$\\
$\nu>\nu_c$   &  ${{p\over 2}}$   &   $-2$  & $--$ & $-2$  & $--$ \\

\hline
Wind & slow cooling \\
\hline
$\nu<\nu_m$   &  $-2$   &   -2  &  $\alpha={\beta}$ & $\frac{6-5p}{2(p-1)}$  & $--$\\
$\nu_m<\nu<\nu_a$   &  $-{5\over 2}$   &   $-{5\over 2}$  &  $\alpha={\beta}$ & $-{5\over 2}$  & $\alpha={\beta}$\\
$\nu_a<\nu<\nu_c$   &  ${p-1\over 2}$   &   ${p-1\over 2}$    &   $\alpha={\beta}$ & ${1\over 2}$ & $--$\\
$\nu>\nu_c$   &  ${p\over 2}$   &   ${p-2\over 2}$    &  $\alpha={\beta-1}$   &   $0$  &  $--$\\

\hline
\end{tabular}
\label{Tab:alpha-beta}
\end{table}

\subsubsection{Thin Shell Reverse Shock Model}

The scalings of this regime have been derived by \cite{kobayashi00}.
During the reverse shock crossing phase, the blastwave dynamics is
same as the thin-shell forward shock case. However, the emission
properties of the reverse shock depend on $\bar{\gamma}_{34}$ and
$n_4$, while those of the forward shock depend on $\gamma_{2}$ and
$n_1$. Following the similar procedure described above, one can
derive the expressions of various parameters of this model. For the
ISM model ($k=0$) and $p>2$, one has
\begin{eqnarray}
&&\nu_m=1.9\times10^{12}~{\rm Hz}~\hat z^{-7}\frac{G(p)}{G(2.3)}E_{52}^{-2}\Gamma_{0,2}^{18}n_{0,0}^{5/2}\epsilon_{e,-1}^{2}\epsilon_{B,-2}^{1/2}t_{2}^{6}           ,\nonumber\\
&&\nu_c=4.1\times10^{16}~{\rm Hz}~\hat z\Gamma_{0,2}^{-4}n_{0,0}^{-3/2}\epsilon_{B,-2}^{-3/2}t_{2}^{-2}           \nonumber\\
&&F_{\rm{\nu,max}}=9.1\times10^{5}~\mu {\rm
Jy}~\hat z^{-1/2}E_{52}^{1/2} \Gamma_{0,2}^{5}n_{0,0}^{}\epsilon_{B,-2}^{1/2}D_{28}^{-2}t_{2}^{3/2}          ,\nonumber\\
&&\nu_a =1.0\times10^{13}~{\rm
Hz}~\hat z^{23/10}\frac{g^{I}(p)}{g^{I}(2.3)}E_{52}^{13/10}\Gamma_{0,2}^{-36/5}n_{0,0}^{-1/2}
\epsilon_{e,-1}^{-1}\epsilon_{B,-2}^{1/5}t_{2}^{-33/10},~~~~~~~~
\nu_a < \nu_m < \nu_c\nonumber \\
&&\nu_a =4.7\times10^{12}~{\rm
Hz}~\hat z^{\frac{3-7p}{p+4}}\frac{g^{II}(p)}{g^{II}(2.3)}E_{52}^{\frac{3-2p}{p+4}}\Gamma_{0}^{\frac{18p-12}{p+4}}n_{0,0}^{\frac{5p}{2(p+4)}}
\epsilon_{e,-1}^{\frac{2(p-1)}{p+4}}\epsilon_{B,-2}^{\frac{p+2}{2(p+4)}}t_{2}^{\frac{6p-7}{p+4}},~~~~~
\nu_m < \nu_a < \nu_c\nonumber \\
&&\nu_a =7.0\times10^{10}~{\rm
Hz}~\hat z^{-17/10}\frac{g^{III}(p)}{g^{III}(2.3)}E_{52}^{3/10}\Gamma_{0,2}^{19/5}n_{0,0}^{3/2}
\epsilon_{B,-2}^{6/5}t_{2}^{7/10}. ~~~~~~~~~~~~~~~~~~
\nu_a < \nu_c < \nu_m\nonumber \\
\end{eqnarray}

For $1<p<2$, one has
\begin{eqnarray}
&&\nu_m=1.8\times10^{9}~{\rm Hz}~\hat z^{\frac{p+5}{1-p}}\frac{g^{IV}(p)}{g^{IV}(1.8)}E_{52}^{-\frac{2}{p-1}}n_{0,0}^{\frac{5}{2(p-1)}}\Gamma_{0,2}^{\frac{p+16}{p-1}}\zeta_{0}^{\frac{2-p}{p-1}}\epsilon_{e,-1}^{\frac{2}{p-1}}\epsilon_{B,-2}^{\frac{1}{2(p-1)}}t_{2}^{\frac{6}{p-1}}            ,\nonumber\\
&&\nu_a =2.7\times10^{14}~{\rm Hz}~\hat
z^{\frac{37-7p}{10(p-1)}}\frac{g^{V}(p)}{g^{V}(1.8)}E_{52}^{\frac{3p+7}{10(p-1)}}\Gamma_{0,2}^{\frac{98-13p}{10(1-p)}}n_{0,0}^{\frac{8-3p}{4(1-p)}}\zeta_{0}^{\frac{p-2}{2(p-1)}}
\epsilon_{e,-1}^{\frac{1}{1-p}}\epsilon_{B,-2}^{\frac{14-9p}{20(1-p)}}t_{2}^{-\frac{3(p+9)}{10(p-1)}},\nonumber \\
&&~~~~~~~~~~~~~~~~~~~~~~~~~~~~~~~~~~~~~~~~~~~~~~~~~~~~~~~~~~~~~~~~~~~~~~~~~~~~~~~~~~~~~~~~~~~~~~~
\nu_a < \nu_m < \nu_c\nonumber \\
&&\nu_a =1.7\times10^{12}~{\rm Hz}~\hat
z^{-\frac{p+9}{p+4}}\frac{g^{VI}(p)}{g^{VI}(1.8)}E_{52}^{-\frac{1}{p+4}}\Gamma_{0,2}^{\frac{p+22}{p+4}}n_{0,0}^{\frac{5}{p+4}}\zeta_{0}^{\frac{2-p}{p+4}}
\epsilon_{e,-1}^{\frac{2}{p+4}}\epsilon_{B,-2}^{\frac{2}{p+4}}t_{2}^{\frac{5}{p+4}},
~~~~~~
\nu_m < \nu_a < \nu_c\nonumber \\
&&\nu_a =5.8\times10^{10}~{\rm Hz}~\hat
z^{-17/10}\frac{g^{VII}(p)}{g^{VII}(1.8)}E_{52}^{3/10}\Gamma_{0,2}^{19/5}n_{0,0}^{3/2}
\epsilon_{B,-2}^{6/5}t_{2}^{7/10}, ~~~~~~~~~~~~~~~~~
\nu_a < \nu_c < \nu_m\nonumber \\
\end{eqnarray}

For the wind model ($k=2$) and $p>2$, one has
\begin{eqnarray}
&&\nu_m=3.3\times10^{15}~{\rm Hz}~\hat z^{-2}\frac{G(p)}{G(2.3)}E_{52}^{-2}A_{*,-1}^{5/2}\Gamma_{0,2}^{8}\epsilon_{e,-1}^{2}\epsilon_{B,-2}^{1/2}t_{2}^{}           ,\nonumber\\
&&\nu_c=1.8\times10^{15}~{\rm Hz}~\hat z^{-2}\Gamma_{0,2}^{2}A_{*,-1}^{-3/2}\epsilon_{B,-2}^{-3/2}t_{2}^{}           \nonumber\\
&&F_{\rm{\nu,max}}=1.3\times10^{7}~\mu {\rm
Jy}~\hat z^{3/2}E_{52}^{1/2}A_{*,-1}\Gamma_{0,2}\epsilon_{B,-2}^{1/2}D_{28}^{-2}t_{2}^{-1/2}              ,\nonumber\\
&&\nu_a =1.7\times10^{12}~{\rm
Hz}~\hat z^{13/10}\frac{g^{VIII}(p)}{g^{VIII}(2.3)}E_{52}^{13/10}\Gamma_{0,2}^{-26/5}
A_{*,-1}^{-1/2}\epsilon_{e,-1}^{-1}\epsilon_{B,-2}^{1/5}t_{2}^{-23/10},
~~~~~
\nu_a < \nu_m < \nu_c\nonumber \\
&&\nu_a =5.9\times10^{13}~{\rm
Hz}~\hat z^{\frac{3-2p}{p+4}}\frac{g^{IX}(p)}{g^{IX}(2.3)}E_{52}^{\frac{3-2p}{p+4}}\Gamma_{0,2}^{\frac{8p-12}{p+4}}
A_{*,-1}^{\frac{5p}{2(p+4)}}\epsilon_{e,-1}^{\frac{2(p-1)}{p+4}}
\epsilon_{B,-2}^{\frac{p+2}{2(p+4)}}t_{2}^{\frac{p-7}{p+4}}, ~~~~~~
\nu_m < \nu_a < \nu_c\nonumber \\
&&\nu_a =2.3\times10^{12}~{\rm Hz}~\hat z^{13/10}
\frac{g^{X}(p)}{g^{X}(2.3)}E_{52}^{3/10}\Gamma_{0,2}^{-11/5}A_{*,-1}^{3/2}\epsilon_{B,-2}^{6/5}t_{2}^{-23/10},
~~~~~~~~~~~~~~~
\nu_a < \nu_c < \nu_m\nonumber \\
\end{eqnarray}

For $1<p<2$, one has
\begin{eqnarray}
&&\nu_m=2.0\times10^{13}~{\rm Hz}~\hat z^{\frac{p}{1-p}}\frac{g^{XI}(p)}{g^{XI}(1.8)}E_{52}^{-\frac{2}{p-1}}A_{*,-1}^{\frac{5}{2(p-1)}}\Gamma_{0,2}^{\frac{p+6}{p-1}}\zeta_{0}^{\frac{2-p}{p-1}}\epsilon_{e,-1}^{\frac{2}{p-1}}\epsilon_{B,-2}^{\frac{1}{2(p-1)}}t_{2}^{\frac{1}{p-1}}            ,\nonumber\\
&&\nu_a=1.8\times10^{13}~{\rm Hz}~\hat
z^{\frac{8p-3}{10(p-1)}}\frac{g^{XII}(p)}{g^{XII}(1.8)}E_{52}^{\frac{3p+7}{10(p-1)}}\Gamma_{0,2}^{\frac{17p+18}{10(1-p)}}A_{*,-1}^{\frac{8-3p}{4(1-p)}}\zeta_{0}^{\frac{p-2}{2(p-1)}}
\epsilon_{e,-1}^{\frac{1}{1-p}}\epsilon_{B,-2}^{\frac{14-9p}{20(1-p)}}t_{2}^{\frac{13-18p}{10(p-1)}},\nonumber \\
&&~~~~~~~~~~~~~~~~~~~~~~~~~~~~~~~~~~~~~~~~~~~~~~~~~~~~~~~~~~~~~~~~~~~~~~~~~~~~~~~~~~~~~~~~~~~~~~~
\nu_a < \nu_m < \nu_c\nonumber \\
&&\nu_a=1.9\times10^{13}~{\rm Hz}~\hat
z^{\frac{1-p}{p+4}}\frac{g^{XIII}(p)}{g^{XIII}(1.8)}E_{52}^{-\frac{1}{p+4}}\Gamma_{0,2}^{\frac{p+2}{p+4}}A_{*,-1}^{\frac{5}{p+4}}\zeta_{0}^{\frac{2-p}{p+4}}
\epsilon_{e,-1}^{\frac{2}{p+4}}\epsilon_{B,-2}^{\frac{2}{p+4}}t_{2}^{-\frac{5}{p+4}},~~~
\nu_m < \nu_a < \nu_c\nonumber \\
&&\nu_a=1.9\times10^{12}~{\rm Hz}~\hat
z^{13/10}\frac{g^{XIV}(p)}{g^{XIV}(1.8)}E_{52}^{3/10}\Gamma_{0,2}^{-11/5}A_{*,-1}^{3/2}
\epsilon_{B,-2}^{6/5}t_{2}^{-23/10}, ~~~~~~~~~~~~
\nu_a < \nu_c < \nu_m\nonumber \\
\end{eqnarray}

After the NRS crosses the shell, the Lorentz factor of the shocked
shell may be assumed to have a general power-law decay behavior $\gamma_3
\propto r^{-g}$ \citep{meszarosrees99,kobayashisari00}. The dynamical
behavior in Region 3 could be expressed with some scaling-laws:
\begin{eqnarray}
  \gamma_3 &\propto& t^{-g/(1+2g)}, n_3 \propto t^{-6(3+g)/7(1+2g)}, \nonumber \\
  e_3 &\propto& t^{-8(3+g)/7(1+2g)},
  r \propto t^{1/(1+2g)}, N_{e,3} \propto t^0,
\end{eqnarray}

For the ISM case ($k=0$), one may adopt $g\simeq2$
\citep{kobayashi00,zou05}. For $p>2$, one has
\begin{eqnarray}
&&\nu_m=8.5\times10^{11}~{\rm Hz}~\hat z^{19/35}\frac{G(p)}{G(2.3)}E_{52}^{18/35}\Gamma_{0,2}^{-74/35}n_{0,0}^{-1/70}\epsilon_{e,-1}^{2}\epsilon_{B,-2}^{1/2}t_{2}^{-54/35}           ,\nonumber\\
&&\nu_{\rm cut}=4.3\times10^{16}~{\rm Hz}~\hat z^{19/35}E_{52}^{-16/105}\Gamma_{0,2}^{-292/105}n_{0,0}^{-283/210}\epsilon_{B,-2}^{-3/2}t_{2}^{-54/35}           \nonumber\\
&&F_{\rm{\nu,max}}=7.0\times10^{5}~\mu {\rm
Jy}~\hat z^{69/35}E_{52}^{139/105} \Gamma_{0,2}^{-167/105}n_{0,0}^{37/210}\epsilon_{B,-2}^{1/2}D_{28}^{-2}t_{2}^{-34/35}          ,\nonumber\\
&&\nu_a=1.4\times10^{13}~{\rm
Hz}~\hat z^{-73/175}\frac{g^{XV}(p)}{g^{XV}(2.3)}E_{52}^{69/175}\Gamma_{0,2}^{8/175}n_{0,0}^{71/175}\epsilon_{e,-1}^{-1}\epsilon_{B,-2}^{1/5}t_{2}^{-102/175},
\nu_a < \nu_m < \nu_c\nonumber \\
&&\nu_a=3.7\times10^{12}~{\rm
Hz}~\hat z^{\frac{19p-36}{35(p+4)}}\frac{g^{XVI}(p)}{g^{XVI}(2.3)}E_{52}^{\frac{2(9p+29)}{35(p+4)}}\Gamma_{0,2}^{\frac{-74p-44}{35(p+4)}}n_{0,0}^{\frac{94-p}{70(p+4)}}
\epsilon_{e,-1}^{\frac{2(p-1)}{p+4}}\epsilon_{B,-2}^{\frac{p+2}{2(p+4)}}t_{2}^{-\frac{54p+104}{35(p+4)}},\nonumber\\
&&~~~~~~~~~~~~~~~~~~~~~~~~~~~~~~~~~~~~~~~~~~~~~~~~~~~~~~~~~~~~~~~~~~~~~~~~~~~~~~~~~~~~~~~~~~~~~~~
\nu_m < \nu_a < \nu_c\nonumber \\
\end{eqnarray}

Here $\nu_{\rm cut}$ is the cut-off frequency of the synchrotron
spectrum, which is different from the traditional $\nu_c$. After
reverse shock crossing, no new electrons are accelerated. The
maximum electron energy is defined by $\nu_{\rm cut}$, which is
calculated by $\nu_c$ at the shock crossing time with correction due
to adiabatic expansion \citep{kobayashi00}. In this case, fast
cooling is not relevant, so there are only two regimes, i.e.,
$\nu_a<\nu_m<\nu_{\rm cut}$ and $\nu_m<\nu_a<\nu_{\rm cut}$.

For $1<p<2$, again the expressions of $\nu_{\rm cut}$ and $F_{\rm \nu,max}$ remain the same, and other parameters are
\begin{eqnarray}
&&\nu_m=6.8\times10^{11}~{\rm Hz}~\hat z^{19/35}\frac{g^{XVII}(p)}{g^{XVII}(1.8)}E_{52}^{\frac{18}{35}}\Gamma_{0,2}^{\frac{109p-144}{35(1-p)}}n_{0,0}^{\frac{71-36p}{70(p-1)}}\zeta_{0}^{\frac{2-p}{p-1}}\epsilon_{e,-1}^{\frac{2}{p-1}}\epsilon_{B,-2}^{\frac{1}{2(p-1)}}t_{2}^{-\frac{54}{35}}          ,\nonumber\\
&&\nu_a=1.3\times10^{13}~{\rm Hz}~\hat
z^{-73/175}\frac{g^{XVIII}(p)}{g^{XVIII}(1.8)}E_{52}^{\frac{69}{175}}\Gamma_{0,2}^{\frac{191p-366}{350(p-1)}}n_{0,0}^{\frac{459p-634}{700(p-1)}}\zeta_{0}^{\frac{p-2}{2(p-1)}}
\epsilon_{e,-1}^{\frac{1}{1-p}}\nonumber\\
&&~~~~~~~~~~\epsilon_{B,-2}^{\frac{14-9p}{20(1-p)}}t_{2}^{-\frac{102}{175}},~~~~~~~~~~~~~~~~~~~~~~~~~~~~~~~~~~~~~~~~~~~~~~~~~~~~~~~~~~\nu_a < \nu_m < \nu_{\rm cut}\nonumber \\
&&\nu_a=3.7\times10^{12}~{\rm Hz}~\hat
z^{\frac{19p-36}{35(p+4)}}\frac{g^{XIX}(p)}{g^{XIX}(1.8)}E_{52}^{\frac{2(9p+29)}{35(p+4)}}\Gamma_{0,2}^{\frac{26-109p}{35(p+4)}}n_{0,0}^{-\frac{2(9p-41)}{35(p+4)}}\zeta_{0}^{\frac{2-p}{p+4}}
\epsilon_{e,-1}^{\frac{2}{p+4}}\epsilon_{B,-2}^{\frac{2}{p+4}}t_{2}^{-\frac{54p+104}{35(p+4)}},
\nonumber \\
&&~~~~~~~~~~~~~~~~~~~~~~~~~~~~~~~~~~~~~~~~~~~~~~~~~~~~~~~~~~~~~~~~~~~~~~~~~~~~~~~~~~~~~~~~\nu_m < \nu_a < \nu_{\rm cut}\nonumber \\
\end{eqnarray}

For the wind model ($k=2$), one could adopt $g\simeq1$ \citep{zou05}.
For $p>2$, one has
\begin{eqnarray}
&&\nu_m=1.4\times10^{11}~{\rm Hz}~\hat z^{6/7}\frac{G(p)}{G(2.3)}E_{52}^{6/7}A_{*,-1}^{-5/14}\Gamma_{0,2}^{-24/7}\epsilon_{e,-1}^{2}\epsilon_{B,-2}^{1/2}t_{2}^{-13/7}           ,\nonumber\\
&&\nu_{\rm cut}=7.4\times10^{10}~{\rm Hz}~\hat z^{6/7}E_{52}^{20/7}\Gamma_{0,2}^{-66/7}A_{*,-1}^{-61/14}\epsilon_{B,-2}^{-3/2}t_{2}^{-13/7}           \nonumber\\
&&F_{\rm{\nu,max}}=1.6\times10^{6}~\mu {\rm
Jy}~\hat z^{44/21}E_{52}^{23/21}A_{*,-1}^{17/42}\Gamma_{0,2}^{-29/21}\epsilon_{B,-2}^{1/2}D_{28}^{-2}t_{2}^{-23/21}              ,\nonumber\\
&&\nu_a=5.5\times10^{14}~{\rm
Hz}~\hat z^{-8/35}\frac{g^{XX}(p)}{g^{XX}(2.3)}E_{52}^{-12/35}\Gamma_{0,2}^{48/35}A_{*,-1}^{8/7}
\epsilon_{e,-1}^{-1}\epsilon_{B,-2}^{1/5}t_{2}^{-23/35}, ~~~
\nu_a < \nu_m < \nu_{\rm cut} \nonumber\\
&&\nu_a=5.5\times10^{14}~{\rm
Hz}~\hat z^{\frac{6p-4}{7(p+4)}}\frac{g^{XXI}(p)}{g^{XXI}(2.3)}E_{52}^{\frac{6p-4}{7(p+4)}}\Gamma_{0,2}^{\frac{16-24p}{7(p+4)}}
A_{*,-1}^{\frac{50-5p}{14(p+4)}}\epsilon_{e,-1}^{\frac{2(p-1)}{p+4}}
\epsilon_{B,-2}^{\frac{p+2}{2(p+4)}}t_{2}^{-\frac{13p+24}{7(p+4)}},\nonumber\\
&&~~~~~~~~~~~~~~~~~~~~~~~~~~~~~~~~~~~~~~~~~~~~~~~~~~~~~~~~~~~~~~~~~~~~~~~~~~~~~~~~~~~~~~~~~~~~~\nu_m < \nu_a < \nu_{\rm cut} \nonumber\\
\end{eqnarray}

For $1<p<2$, $\nu_{\rm cut}$ and $F_{\rm \nu,max}$ remain the same, and
\begin{eqnarray}
&&\nu_m=3.5\times10^{11}~{\rm Hz}~\hat z^{6/7}\frac{g^{XXII}(p)}{g^{XXII}(1.8)}E_{52}^{\frac{13p-20}{7(p-1)}} \Gamma_{0,2}^{\frac{45p-66}{7(1-p)}}A_{*,-1}^{\frac{47-26p}{14(p-1)}}\zeta_{0}^{\frac{2-p}{p-1}}\epsilon_{e,-1}^{\frac{2}{p-1}}\epsilon_{B,-2}^{\frac{1}{2(p-1)}} t_{2}^{-\frac{13}{7}}          ,\nonumber\\
&&\nu_a=2.8\times10^{14}~{\rm Hz}~\hat
z^{-8/35}\frac{g^{XXIII}(p)}{g^{XXIII}(1.8)}E_{52}^{\frac{94-59p}{70(p-1)}}\Gamma_{0,2}^{\frac{3(67p-102)}{70(p-1)}}A_{*,-1}^{\frac{74-53p}{28(1-p)}}\zeta_{0}^{\frac{p-2}{2(p-1)}}
\epsilon_{e,-1}^{\frac{1}{1-p}}\nonumber \\
&&~~~~~~~~~\epsilon_{B,-2}^{\frac{14-9p}{20(1-p)}}t_{2}^{-\frac{23}{35}},
~~~~~~~~~~~~~~~~~~~~~~~~~~~~~~~~~~~~~~~~~~~~~~~~~~~~~~~~~~~~~~~~~~\nu_a < \nu_m < \nu_{\rm cut}\nonumber \\
&&\nu_a=1.6\times10^{13}~{\rm Hz}~\hat
z^{\frac{6p-4}{7(p+4)}}\frac{g^{XXIV}(p)}{g^{XXIV}(1.8)}E_{52}^{\frac{13p-18}{7(p+4)}}\Gamma_{0,2}^{\frac{58-45p}{7p+28}}A_{*,-1}^{\frac{46-13p}{7p+28}}\zeta_{0}^{\frac{2-p}{p+4}}
\epsilon_{e,-1}^{\frac{2}{p+4}}\epsilon_{B,-2}^{\frac{2}{p+4}}t_{2}^{-\frac{13p+24}{7(p+4)}},
\nonumber \\
&&~~~~~~~~~~~~~~~~~~~~~~~~~~~~~~~~~~~~~~~~~~~~~~~~~~~~~~~~~~~~~~~~~~~~~~~~~~~~~~~~~~~~~~~~~~~~~\nu_m < \nu_a < \nu_{\rm cut}\nonumber \\
\end{eqnarray}

The $\alpha$ and $\beta$ values and their
closure relations for the thin shell reverse shock models are
presented in Tables \ref{ThinRPre} and \ref{ThinRPre1}
(for pre-shock-crossing), and Tables \ref{ThinRPost}
and \ref{ThinRPost2} (for post-shock-crossing).

For this regime (thin-shell reverse shock model during shock
crossing), for $p>2$, one has $\nu_m \propto t^6~(t^1)$, $\nu_c
\propto t^{-2}~(t^1)$, $F_{\rm \nu,max} \propto t^{3/2} ~(t^{-1/2})$
for the ISM (wind) models, respectively.
For $1<p<2$, $\nu_c$ and
$F_{\rm{\nu,max}}$ evolutions are the same as $p>2$ cases, while
$\nu_m \propto t^{\frac{6}{p-1}}~ (t^{\frac{1}{p-1}})$ for the ISM
(wind) models, respectively.

After shock crossing, $\nu_m \propto \nu_{\rm cut} \propto
t^{-54/35} ~(t^{-13/7})$, $F_{\rm \nu,max} \propto t^{-34/35}
~(t^{-23/21})$  for the ISM (wind) models, respectively.

\begin{table}
\caption{Temporal decay index $\alpha$ and spectral index $\beta$ in the thin
shell reverse shock model during the reverse shock crossing phase in the
$\nu_a < \rm{min}(\nu_m,\nu_c)$ spectral regime.\label{ThinRPre}}
\begin{tabular}{llllll}
\hline\hline
& & $p>2$ & &  $1<p<2$ & \\

& $\beta$ & $\alpha $  &  $\alpha (\beta)$ & $\alpha$  & $\alpha (\beta)$  \\
\hline
ISM & slow cooling  \\
\hline
$\nu<\nu_a$   &  $-2$  &   $-5$ & $\alpha={5\beta \over 2}$  &$-\frac{2p+1}{p-1}$ & $--$\\
$\nu_a<\nu<\nu_m$   &  $-{1 \over 3}$  &   ${1 \over 2}$ & $\alpha={3\beta \over 2}$ & $-\frac{3p-7}{2(p-1)}$  & $--$\\
$\nu_m<\nu<\nu_c$   &  ${{p-1 \over 2}}$  &  $-{6p-3 \over 2}$
   &  $\alpha=-{3(4\beta+1) \over 2}$ & $-{9 \over 2}$   & $--$\\
$\nu>\nu_c$   &  ${{p\over 2}}$   &   $-{6p-5 \over 2}$  & $-{11\beta+1 \over 2}$ & $-{7 \over 2}$   & $--$ \\

\hline
ISM & fast cooling \\
\hline
$\nu<\nu_a$   &  $-2$  &   $-1$ & $\alpha=\frac{\beta}{2}$ & $-1$  & $\alpha=\frac{\beta}{2}$\\
$\nu_a<\nu<\nu_c$   &  $-{1\over 3}$  &   $-{13\over 6}$ &  $\alpha={13\beta \over 2}$  &  $-{13 \over 6}$  & $\alpha={13\beta \over 2}$\\
$\nu_c<\nu<\nu_m$   &  ${1\over 2}$  &  $-{1\over 2}$  & $\alpha=-{\beta}$ &  $-{1 \over 2}$   &  $\alpha=-{\beta}$ \\
$\nu>\nu_m$   &  ${p\over 2}$   &   $-{6p-5\over 2}$      &  $-{12\beta-5 \over 2}$  &  $-{7 \over 2}$  & $--$ \\

\hline
Wind & slow cooling \\
\hline
$\nu<\nu_a$   &  $-2$   &   -3  &  $\alpha={3\beta \over 2}$ & $-\frac{5p-4}{2(p-1)}$  & $--$\\
$\nu_a<\nu<\nu_m$   &  $-{1\over 3}$   &   ${5\over 6}$  &  $\alpha={5\beta \over 2}$ & ${3p-1\over 6(p-1)}$  & $--$\\
$\nu_m<\nu<\nu_c$   &  ${p-1\over 2}$  &   $-{p-2\over 2}$    &   $\alpha={1-2\beta \over 2}$ & $0$ & $--$\\
$\nu>\nu_c$   &  ${p\over 2}$   &   $-{p-1\over 2}$    &  $\alpha={1-2\beta \over 2}$   &   $-{1\over 2}$  &  $--$\\

\hline
Wind & fast cooling  \\
\hline
$\nu<\nu_a$   &  $-2$   &   $-3$ & $\alpha={3\beta \over 2}$ &  $-3$  & $\alpha={3\beta \over 2}$  \\
$\nu_a<\nu<\nu_c$   &  $-{1\over 3}$   &   ${5\over 6}$ & $\alpha=-{5\beta \over 2}$ &  ${5 \over 6}$  & $\alpha=-{5\beta \over 2}$  \\
$\nu_c<\nu<\nu_m$   &  ${1\over 2}$  &   $0$  & $--$  & $0$ & $--$\\
$\nu>\nu_m$   &  ${p\over 2}$  &   $-{p-1\over 2}$     & $\alpha={1-2\beta \over 2}$  &  $-{1\over 2}$  & $--$\\
\hline
\end{tabular}
\label{Tab:alpha-beta}
\end{table}

\begin{table}

\caption{Temporal decay index $\alpha$ and
spectral index $\beta$ in the thin
shell reverse shock model during the reverse shock crossing phase in the
$\nu_m <\nu_a < \nu_c$ spectral regime.\label{ThinRPre1}}
\begin{tabular}{llllll}
\hline\hline
& & $p>2$ & &  $1<p<2$ & \\

& $\beta$ & $\alpha $  &  $\alpha (\beta)$ & $\alpha$ & $\alpha (\beta)$  \\
\hline
ISM & slow cooling  \\
\hline
$\nu<\nu_m$   &  $-2$  &   $-5$ & $\alpha={5\beta \over 2}$ & $-\frac{2p+1}{p-1}$  & $--$\\
$\nu_m<\nu<\nu_a$   &  $-{5 \over 2}$  &   $-2$ & $\alpha={4\beta \over 5}$ & $-2$  & $\alpha={4\beta \over 5}$\\
$\nu_a<\nu<\nu_c$   &  ${p-1\over 2}$   &  $-{{6p-3 \over 2}}$
  &  $\alpha=-{3(4\beta+1) \over 2}$ & $-{9 \over 2}$  & $--$\\
$\nu>\nu_c$   &  ${{p\over 2}}$   &   $-{{6p-5 \over 2}}$  & $-{12\beta-5 \over 2}$ & $-{7\over 2}$  & $--$ \\

\hline
Wind & slow cooling \\
\hline
$\nu<\nu_m$   &  $-2$   &   -3  &  $\alpha={3\beta \over 2}$ & $-\frac{5p-4}{2(p-1)}$  & $--$\\
$\nu_m<\nu<\nu_a$   &  $-{5\over 2}$   &   $-{5\over 2}$  &  $\alpha={\beta}$ & $-{5 \over 2}$  & $\alpha={\beta}$\\
$\nu_a<\nu<\nu_c$   &  ${p-1\over 2}$   &   $-{p-2\over 2}$    &   $\alpha={1-2\beta \over 2}$ & $0$ & $--$\\
$\nu>\nu_c$   &  ${p\over 2}$   &   $-{p-1\over 2}$    &  $\alpha={1-2\beta \over 2}$   &   $-{1\over 2}$  &  $--$\\

\hline
\end{tabular}
\label{Tab:alpha-beta}
\end{table}

\begin{table}
\caption{Temporal decay index $\alpha$ and spectral index $\beta$ in
thin shell reverse shock model after reverse shock crossing in the
$\nu_a < \rm{min}(\nu_m,\nu_{\rm cut})$ spectral
regime.\label{ThinRPost}}
\begin{tabular}{llllll}
\hline\hline
& & $p>2$ & &  $1<p<2$ & \\

& $\beta$ & $\alpha $  &  $\alpha (\beta)$ & $\alpha$  & $\alpha (\beta)$  \\
\hline
ISM & slow cooling  \\
\hline
$\nu<\nu_a$   &  $-2$  &   $-{18 \over 35}$ & $\alpha={9\beta \over 35}$  &$-{18 \over 35}$ & $\alpha={9\beta \over 35}$\\
$\nu_a<\nu<\nu_m$   &  $-{1 \over 3}$  &   ${16 \over 35}$ & $\alpha=-{16\beta \over 105}$ & ${16 \over 35}$ & $\alpha=-{16\beta \over 105}$\\
$\nu_m<\nu<\nu_{\rm cut}$   &  ${{p-1 \over 2}}$  &  ${27p+7 \over
35}$
   &  $\alpha={54\beta+34 \over 35}$ & ${27p+7 \over
35}$
   &  $\alpha={54\beta+34 \over 35}$\\
\hline
Wind & slow cooling \\
\hline
$\nu<\nu_a$   &  $-2$   &   $-{13 \over 21}$  &  $\alpha={13\beta \over 42}$ & $-{13 \over 21}$  &  $\alpha={13\beta \over 42}$\\
$\nu_a<\nu<\nu_m$   &  $-{1\over 3}$   &   ${10\over 21}$  &  $\alpha={10\beta \over 7}$ & ${10\over 21}$  &  $\alpha={10\beta \over 7}$\\
$\nu_m<\nu<\nu_{\rm cut}$   &  ${p-1\over 2}$  &   ${39p+7\over 42}$    &   $\alpha={78\beta+46 \over 2}$ & ${39p+7\over 42}$    &   $\alpha={78\beta+46 \over 2}$\\
\hline
\end{tabular}
\label{Tab:alpha-beta}
\end{table}

\begin{table}

\caption{Temporal decay index $\alpha$ and spectral index $\beta$ in
thin shell reverse shock model after reverse shock crossing in the $
\nu_m <\nu_a < \nu_{\rm cut}$ spectral regime.\label{ThinRPost2}}
\begin{tabular}{llllll}
\hline\hline
& & $p>2$ & &  $1<p<2$ & \\

& $\beta$ & $\alpha $  &  $\alpha (\beta)$ & $\alpha$ & $\alpha (\beta)$  \\
\hline
ISM & slow cooling  \\
\hline
$\nu<\nu_m$   &  $-2$  &   $-{18 \over 35}$ & $\alpha={9\beta \over 35}$ & $-{18 \over 35}$ & $\alpha={9\beta \over 35}$\\
$\nu_m<\nu<\nu_a$   &  $-{5 \over 2}$  &   $-{9 \over 7}$ & $\alpha={18\beta \over 35}$ & $-{9 \over 7}$ & $\alpha={18\beta \over 35}$\\
$\nu_a<\nu<\nu_{\rm cut}$   &  ${p-1\over 2}$   &  ${27p+7\over 35}$
  &  $\alpha={54\beta+34 \over 35}$ & ${27p+7\over 35}$
  &  $\alpha={54\beta+34 \over 35}$\\

\hline

Wind & slow cooling \\
\hline
$\nu<\nu_m$   &  $-2$   &   $-{13\over 21}$  &  $\alpha={13\beta \over 42}$ & $-{13\over 21}$  &  $\alpha={13\beta \over 42}$\\
$\nu_m<\nu<\nu_a$   &  $-{5\over 2}$   &   $-{65\over 42}$  &  $\alpha={13\beta \over 24}$ & $-{65\over 42}$  &  $\alpha={13\beta \over 24}$\\
$\nu_a<\nu<\nu_{\rm cut}$   &  ${p-1\over 2}$   &   ${39p+7\over 42}$    &   $\alpha={78\beta+46 \over 2}$ & ${39p+7\over 42}$    &   $\alpha={78\beta+46 \over 2}$\\

\hline
\end{tabular}
\label{Tab:alpha-beta}
\end{table}

\subsubsection{Thick Shell Forward Shock Model}

For the thick shell case, the reverse shock becomes relativistic early on during shock crossing.
In this relativistic shock crossing phase, the blastwave dynamics can be
characterized as
\begin{eqnarray}
\gamma_2 = \gamma_3=\frac{1}{\sqrt{2}}\left(\frac{l^{3-k}}{\Delta_0}\right)^{\frac{1}{2(4-k)}}
\left(\frac{t}{T}\right)^{\frac{k-2}{2(k-4)}}\Delta_0^{\frac{k-2}{2(k-4)}},~~~~~~~~~~R=2c\gamma_2^2t\nonumber\\
\label{thick-dynamics}
\end{eqnarray}
where $l=\left(\frac{(3-k)E}{4\pi Am_pc^2}\right)^{\frac{1}{3-k}}$
is the Sedov length, and $T=\frac{\Delta_0}{c}$ is the shock
crossing time \citep{yi13}.

For the ISM model and when $p>2$, the forward shock emission can be
characterized by
\begin{eqnarray}
&&\nu_m=   1.0\times10^{16}~{\rm Hz}\frac{G(p)}{G(2.3)}E_{52}^{1/2}\Delta_{0,13}^{-1/2}\epsilon_{e,-1}^{2}\epsilon_{B,-2}^{1/2}t_{2}^{-1}           ,\nonumber\\
&&\nu_c=    1.2\times10^{17}~{\rm Hz}E_{52}^{-1/2}\Delta_{0,13}^{1/2}n_{0,0}^{-1}\epsilon_{B,-2}^{-3/2}t_{2}^{-1}           \nonumber\\
&&F_{\rm{\nu,max}}=     1.2\times10^{3}~\mu {\rm
Jy}~\hat z E_{52}^{} \Delta_{0,13}^{-1}n_{0,0}^{1/2}\epsilon_{B,-2}^{1/2}D_{28}^{-2}          ,\nonumber\\
&&\nu_a=3.6\times10^{9}~{\rm
Hz}~\hat z^{-6/5}\frac{g^{I}(p)}{g^{I}(2.3)}E_{52}^{1/5}\Delta_{0,13}^{-1/5}n_{0,0}^{3/5}\epsilon_{e,-1}^{-1}
\epsilon_{B,-2}^{1/5}t_{2}^{1/5}, ~~~~~~~~~~~~~~~~~~
\nu_a < \nu_m < \nu_c\nonumber \\
&&\nu_a=3.9\times10^{12}~{\rm
Hz}~\hat z^{-\frac{4}{p+4}}\frac{g^{II}(p)}{g^{II}(2.3)}E_{52}^{\frac{p+2}{2(p+4)}}\Delta_{0,13}^{-\frac{p+2}{2(p+4)}}n_{0,0}^{\frac{2}{p+4}}\epsilon_{e,-1}^{\frac{2(p-1)}{p+4}}
\epsilon_{B,-2}^{\frac{p+2}{2(p+4)}}t_{2}^{-\frac{p}{p+4}}, ~~~
\nu_m < \nu_a < \nu_c\nonumber \\
&&\nu_a=1.0\times10^{9}~{\rm
Hz}~\hat z^{-6/5}\frac{g^{III}(p)}{g^{III}(2.3)}E_{52}^{7/10}\Delta_{0,13}^{-7/10}n_{0,0}^{11/10}
\epsilon_{B,-2}^{6/5}t_{2}^{1/5}, ~~~~~~~~~~~~~~~~~~
\nu_a < \nu_c < \nu_m\nonumber \\
\end{eqnarray}

For $1<p<2$, one has ($\nu_c$ and $F_{\rm \nu,max}$ remain the same)
\begin{eqnarray}
&&\nu_m=   8.6\times10^{13}~{\rm Hz}~\hat z^{\frac{6-3p}{4(p-1)}}\frac{g^{IV}(p)}{g^{IV}(1.8)}E_{52}^{\frac{p+2}{8(p-1)}}n_{0,0}^{\frac{2-p}{8(p-1)}}\Delta_{0,13}^{\frac{p+2}{8(1-p)}}\zeta_{0}^{\frac{2-p}{p-1}}\epsilon_{e,-1}^{\frac{2}{p-1}}\epsilon_{B,-2}^{\frac{1}{2(p-1)}} t_{2}^{\frac{p+2}{4(1-p)}}           ,\nonumber\\
&&\nu_a= 3.2\times10^{10}~{\rm Hz}\frac{g^{V}(p)}{g^{V}(1.8)}\hat
z^{\frac{18-33p}{40(p-1)}}E_{52}^{\frac{46-31p}{80(1-p)}}
\Delta_{0,13}^{\frac{46-31p}{80(p-1)}}
n_{0,0}^{\frac{58-53p}{80(1-p)}}\zeta_{0}^{\frac{p-2}{2(p-1)}}
\epsilon_{e,-1}^{\frac{1}{1-p}}\epsilon_{B,-2}^{\frac{14-9p}{20(1-p)}}t_{2}^{\frac{22-7p}{40(p-1)}},,\nonumber \\
&&~~~~~~~~~~~~~~~~~~~~~~~~~~~~~~~~~~~~~~~~~~~~~~~~~~~~~~~~~~~~~~~~~~~~~~~~~~~~~~~~~~~~~~~~~~~~~~~~~~~
\nu_a < \nu_m < \nu_c\nonumber \\
&&\nu_a= 9.3\times10^{11}~{\rm Hz}~\hat
z^{-\frac{3p+10}{4(p+4)}}\frac{g^{VI}(p)}{g^{VI}(1.8)}E_{52}^{\frac{p+14}{8(p+4)}}
\Delta_{0,13}^{-\frac{p+14}{8(p+4)}}
n_{0,0}^{\frac{18-p}{8(p+4)}}\zeta_{0}^{\frac{2-p}{p+4}}
\epsilon_{e,-1}^{\frac{2}{p+4}}\epsilon_{B,-2}^{\frac{2}{p+4}}t_{2}^{-\frac{p+6}{4(p+4)}},\nonumber \\
&&~~~~~~~~~~~~~~~~~~~~~~~~~~~~~~~~~~~~~~~~~~~~~~~~~~~~~~~~~~~~~~~~~~~~~~~~~~~~~~~~~~~~~~~~~~~~~~~~~~~
\nu_m < \nu_a < \nu_c\nonumber \\
&&\nu_a= 8.5\times10^{8}~{\rm Hz}~\hat
z^{-6/5}\frac{g^{VII}(p)}{g^{VII}(1.8)}E_{52}^{7/10}
\Delta_{0,13}^{-7/10}
n_{0,0}^{11/10}\epsilon_{B,-2}^{6/5}t_{2}^{1/5},
~~~~~~~~~~~~~~~~~~~~~
\nu_a < \nu_c < \nu_m\nonumber \\
\end{eqnarray}

For the wind model and $p>2$, one has
\begin{eqnarray}
&&\nu_m=   5.8\times10^{15}~{\rm Hz}\frac{G(p)}{G(2.3)}E_{52}^{1/2}\Delta_{0,13}^{-1/2}\epsilon_{e,-1}^{2}\epsilon_{B,-2}^{1/2}t_{2}^{-1}           ,\nonumber\\
&&\nu_c=    1.2\times10^{14}~{\rm Hz}~\hat z^{-2}E_{52}^{1/2}\Delta_{0,13}^{-1/2}A_{*,-1}^{-2}\epsilon_{B,-2}^{-3/2}t_{2}^{}           \nonumber\\
&&F_{\rm{\nu,max}}=     5.0\times10^{4}~\mu {\rm
Jy}~\hat z E_{52}^{1/2} \Delta_{0,13}^{-1/2}A_{*,-1}^{}\epsilon_{B,-2}^{1/2}D_{28}^{-2}          ,\nonumber\\
&&\nu_a=5.1\times10^{11}~{\rm
Hz}\frac{g^{VIII}(p)}{g^{VIII}(2.3)}E_{52}^{-2/5}\Delta_{0,13}^{2/5}A_{*,-1}^{6/5}
\epsilon_{e,-1}^{-1}\epsilon_{B,-2}^{1/5}t_{2}^{-1},
~~~~~~~~~~~~~~~~
\nu_a < \nu_m < \nu_c\nonumber \\
&&\nu_a=4.2\times10^{13}~{\rm
Hz}\frac{g^{IX}(p)}{g^{IX}(2.3)}E_{52}^{\frac{p-2}{2(p+4)}}\Delta_{0,13}^{\frac{2-p}{2(p+4)}}A_{*,-1}^{\frac{4}{p+4}}
\epsilon_{e,-1}^{\frac{2(p-1)}{p+4}}\epsilon_{B,-2}^{\frac{p+2}{2(p+4)}}t_{2}^{-1},
~~~~~~~~~~
\nu_m < \nu_a < \nu_c\nonumber \\
&&\nu_a=3.6\times10^{12}~{\rm Hz}~\hat
z\frac{g^{X}(p)}{g^{X}(2.3)}E_{52}^{-2/5}\Delta_{0,13}^{2/5}A_{*,-1}^{11/5}\epsilon_{B,-2}^{6/5}t_{2}^{-2},
~~~~~~~~~~~~~~~~~~~~~
\nu_a < \nu_c < \nu_m\nonumber \\
\end{eqnarray}

For $1<p<2$, one has ($\nu_c$ and $F_{\rm \nu,max}$ remain the same)
\begin{eqnarray}
&&\nu_m=   5.6\times10^{13}~{\rm Hz} \hat z^{\frac{2-p}{p-1}}\frac{g^{XI}(p)}{g^{XI}(1.8)}E_{52}^{\frac{p}{4(p-1)}}A_{*,-1}^{\frac{2-p}{4(p-1)}}\Delta_{0,13}^{\frac{p}{4(1-p)}}\zeta_{0}^{\frac{2-p}{p-1}}\epsilon_{e,-1}^{\frac{2}{p-1}}\epsilon_{B,-2}^{\frac{1}{2(p-1)}}t_{2}^{\frac{1}{1-p}}           ,\nonumber\\
&&\nu_a =  4.3\times10^{12}~{\rm Hz}~\hat
z^{\frac{p-2}{2(p-1)}}\frac{g^{XII}(p)}{g^{XII}(1.8)}
E_{52}^{\frac{6-11p}{40(p-1)}} \Delta_{0,13}^{\frac{11-6p}{40(p-1)}}
A_{*,-1}^{\frac{58-53p}{40(1-p)}}\zeta_{0}^{\frac{p-2}{2(p-1)}}
\epsilon_{e,-1}^{\frac{1}{1-p}}\epsilon_{B,-2}^{\frac{14-9p}{20(1-p)}}
t_{2}^{\frac{4-3p}{2(p-1)}},,\nonumber \\
&&~~~~~~~~~~~~~~~~~~~~~~~~~~~~~~~~~~~~~~~~~~~~~~~~~~~~~~~~~~~~~~~~~~~~~~~~~~~~~~~~~~~~~~~~~~~~~~~~~~~
\nu_a < \nu_m < \nu_c\nonumber \\
&&\nu_a =  1.3\times10^{13}~{\rm Hz}~\hat
z^{\frac{2-p}{p+4}}\frac{g^{XIII}(p)}{g^{XIII}(1.8)}
E_{52}^{\frac{p-2}{4(p+4)}} \Delta_{0,13}^{\frac{2-p}{4(p+4)}}
A_{*,-1}^{\frac{18-p}{4(p+4)}}\zeta_{0}^{\frac{2-p}{p+4}}
\epsilon_{e,-1}^{\frac{2}{p+4}}\epsilon_{B,-2}^{\frac{2}{p+4}}
t_{2}^{-\frac{6}{p+4}}, ~~
\nu_m < \nu_a < \nu_c\nonumber \\
&&\nu_a= 3.0\times10^{12}~{\rm Hz}~\hat
z\frac{g^{XIV}(p)}{g^{XIV}(1.8)} E_{52}^{-2/5} \Delta_{0,13}^{2/5}
A_{*,-1}^{11/5}\epsilon_{B,-2}^{6/5}t_{2}^{-2},
~~~~~~~~~~~~~~~~~~~~~~~~~~\nu_a < \nu_c < \nu_m\nonumber \\
\end{eqnarray}

The $\alpha$ and $\beta$ values and their
closure relations for the thick shell forward shock models
are presented in Tables \ref{ThickF} and \ref{ThickF1}.

For this regime (thick-shell forward shock model during shock
crossing), for $p>2$, one has $\nu_m \propto t^{-1}~(t^{-1})$,
$\nu_c \propto t^{-1}~(t^1)$, $F_{\rm \nu,max} \propto t^{0}
~(t^{0})$ for the ISM (wind) models, respectively. For $1<p<2$,
$\nu_c$ and $F_{\rm{\nu,max}}$ evolutions are the same as $p>2$
cases, while $\nu_m \propto t^{\frac{p+2}{4(1-p)}}~
(t^{\frac{1}{1-p}})$ for the ISM (wind) models, respectively.

\begin{table}
\caption{The temporal decay index $\alpha$ and spectral index $\beta$ of
the thick shell forward shock model in the $\nu_a <
\rm{min}(\nu_m,\nu_c)$ spectral regime.\label{ThickF}}
\begin{tabular}{llllll}
\hline\hline
& & $p>2$ & &  $1<p<2$ & \\

& $\beta$ & $\alpha $  &  $\alpha (\beta)$ & $\alpha$  & $\alpha (\beta)$  \\
\hline
ISM & slow cooling  \\
\hline
$\nu<\nu_a$   &  $-2$  &   $-1$ & $\alpha={\beta \over 2}$  &${11p-14 \over 8(1-p)}$ & $--$\\
$\nu_a<\nu<\nu_m$   &  $-{1 \over 3}$  &   $-{4 \over 3}$ & $\alpha={4\beta}$ & ${13p-10 \over 12(1-p)}$  & $--$\\
$\nu_m<\nu<\nu_c$   &  ${p-1 \over 2}$  &  ${p-3 \over 2}$
   &  $\alpha={\beta-1}$ & ${p-6 \over 8}$   & $\alpha={2\beta-5 \over 8}$\\
$\nu>\nu_c$   &  ${p\over 2}$   &   ${p-2 \over 2}$  & $\alpha={\beta-1}$ & ${p-2\over 8}$   & $\alpha={\beta-1 \over 4}$ \\

\hline
ISM & fast cooling \\
\hline
$\nu<\nu_a$   &  $-2$  &   $-1$ & $\alpha=\frac{\beta}{2}$ & $-1$  & $\alpha=\frac{\beta}{2}$\\
$\nu_a<\nu<\nu_c$   &  $-{1\over 3}$  &   $-{4\over 3}$ &  $\alpha={4\beta}$  &  $-{4 \over 3}$  & $\alpha={4\beta}$\\
$\nu_c<\nu<\nu_m$   &  ${1\over 2}$  &  $-{1 \over 2}$  & $\alpha=-{\beta}$ &  $-{1 \over 2}$   &  $\alpha=-{\beta}$ \\
$\nu>\nu_m$   &  ${p\over 2}$   &   ${p-2 \over 2}$      &  $\alpha={\beta-1}$  &  $0$  & $--$ \\

\hline
Wind & slow cooling \\
\hline
$\nu<\nu_a$   &  $-2$   &   $-2$  &  $\alpha={\beta}$ & $\frac{5p-6}{2(1-p)}$  & $--$\\
$\nu_a<\nu<\nu_m$   &  $-{1\over 3}$   &   $-{1\over 3}$  &  $\alpha={\beta}$ & $-{1\over 3(p-1)}$  & $--$\\
$\nu_m<\nu<\nu_c$   &  ${p-1\over 2}$  &   ${p-1\over 2}$    &   $\alpha={\beta}$ & ${1 \over 2}$ & $--$\\
$\nu>\nu_c$   &  ${p\over 2}$   &   ${p-2\over 2}$    &  $\alpha={\beta-1}$   &   $0$  &  $--$\\

\hline
Wind & fast cooling  \\
\hline
$\nu<\nu_a$   &  $-2$   &   $-3$ & $\alpha={3\beta \over 2}$ &  $-3$  & $\alpha={3\beta \over 2}$  \\
$\nu_a<\nu<\nu_c$   &  $-{1\over 3}$   &   ${1\over 3}$ & $\alpha=-\beta$ &  ${1 \over 3}$  & $\alpha=-{\beta}$  \\
$\nu_c<\nu<\nu_m$   &  ${1\over 2}$  &   $-{1\over 2}$  & $\alpha=-{\beta}$  & $-{1 \over 2}$ & $\alpha=-{\beta}$\\
$\nu>\nu_m$   &  ${p\over 2}$  &   ${p-2\over 2}$     & $\alpha={\beta-1}$  &  $0$  & $--$\\
\hline
\end{tabular}
\label{Tab:alpha-beta}
\end{table}

\begin{table}
\caption{The temporal decay index $\alpha$ and spectral index $\beta$
of the thick shell forward shock model in the $\nu_m < \nu_a
<\nu_c$ spectral regime.\label{ThickF1}}
\begin{tabular}{llllll}
\hline\hline
& & $p>2$ & &  $1<p<2$ & \\

& $\beta$ & $\alpha $  &  $\alpha (\beta)$ & $\alpha$  & $\alpha (\beta)$  \\
\hline
ISM & slow cooling  \\
\hline
$\nu<\nu_a$   &  $-2$  &   $-1$ & $\alpha={\beta \over 2}$  &${11p-14 \over 8(1-p)}$ & $--$\\
$\nu_a<\nu<\nu_m$   &  $-{1 \over 3}$  &   $-{3 \over 2}$ & $\alpha={9\beta \over 2}$ & $-{3 \over 2}$  & $\alpha={9\beta \over 2}$\\
$\nu_m<\nu<\nu_c$   &  ${p-1 \over 2}$  &  ${p-3 \over 2}$
   &  $\alpha={\beta-1}$ & ${p-6 \over 8}$   & $\alpha={2\beta-5 \over 8}$\\
$\nu>\nu_c$   &  ${p\over 2}$   &   ${p-2 \over 2}$  & $\alpha={\beta-1}$ & ${p-2\over 8}$   & $\alpha={\beta-1 \over 4}$ \\

\hline
Wind & slow cooling \\
\hline
$\nu<\nu_a$   &  $-2$   &   $-2$  &  $\alpha={\beta}$ & $\frac{5p-6}{2(1-p)}$  & $--$\\
$\nu_a<\nu<\nu_m$   &  $-{1\over 3}$   &   $-{5\over 2}$  &  $\alpha={15\beta \over 2}$ & $-{5\over 2}$  & $\alpha={15\beta \over 2}$\\
$\nu_m<\nu<\nu_c$   &  ${p-1\over 2}$  &   ${p-1\over 2}$    &   $\alpha={\beta}$ & ${1 \over 2}$ & $--$\\
$\nu>\nu_c$   &  ${p\over 2}$   &   ${p-2\over 2}$    &  $\alpha={\beta-1}$   &   $0$  &  $--$\\

\hline
\end{tabular}
\label{Tab:alpha-beta}
\end{table}

\subsubsection{Thick Shell Reverse Shock Model}

Using the same dynamics in Eq.(\ref{thick-dynamics}), one can characterize
the reverse shock emission during the shock crossing phase.

For the ISM model and $p>2$, the reverse shock emission can be characterized by
\begin{eqnarray}
&&\nu_m=   7.6\times10^{11}~{\rm Hz}~\hat z^{-1}\frac{G(p)}{G(2.3)}\Gamma_{0,2}^{2}n_{0,0}^{1/2}\epsilon_{e,-1}^{2}\epsilon_{B,-2}^{1/2}           ,\nonumber\\
&&\nu_c=    1.2\times10^{17}~{\rm Hz}E_{52}^{-1/2}\Delta_{0,13}^{1/2}n_{0,0}^{-1}\epsilon_{B,-2}^{-3/2}t_{2}^{-1}           \nonumber\\
&&F_{\rm{\nu,max}}=     1.3\times10^{5} ~\mu {\rm
Jy}~\hat z^{1/2}E_{52}^{5/4} \Delta_{0,13}^{-5/4}\Gamma_{0,2}^{-1}n_{0,0}^{1/4}\epsilon_{B,-2}^{1/2}D_{28}^{-2}t_{2}^{1/2}          ,\nonumber\\
&&\nu_a =7.2\times10^{12}~{\rm
Hz}~\hat z^{-2/5}\frac{g^{I}(p)}{g^{I}(2.3)}E_{52}^{3/5}\Gamma_{0,2}^{-8/5}\Delta_{0,13}^{-3/5}n_{0,0}^{1/5}
\epsilon_{e,-1}^{-1}\epsilon_{B,-2}^{1/5}t_{2}^{-3/5}, ~~~~~~
\nu_a < \nu_m < \nu_c\nonumber \\
&&\nu_a =2.5\times10^{12}~{\rm
Hz}~\hat z^{-\frac{p+2}{p+4}}\frac{g^{II}(p)}{g^{II}(2.3)}E_{52}^{\frac{2}{p+4}}\Gamma_{0,2}^{\frac{2(p-2)}{p+4}}
\Delta_{0,13}^{-\frac{2}{p+4}}n_{0,0}^{\frac{p+2}{2(p+4)}}
\epsilon_{e,-1}^{\frac{2(p-1)}{p+4}}\epsilon_{B,-2}^{\frac{p+2}{2(p+4)}}t_{2}^{-\frac{2}{p+4}},\nonumber \\
&&~~~~~~~~~~~~~~~~~~~~~~~~~~~~~~~~~~~~~~~~~~~~~~~~~~~~~~~~~~~~~~~~~~~~~~~~~~~~~~~~~~~~~~~~~~~~~~~~
\nu_m < \nu_a < \nu_c\nonumber \\
&&\nu_a =1.8\times10^{10}~{\rm
Hz}~\hat z^{-9/10}\frac{g^{III}(p)}{g^{III}(2.3)}E_{52}^{17/20}\Gamma_{0,2}^{-3/5}\Delta_{0,13}^{-17/20}n_{0,0}^{19/20}\epsilon_{B,-2}^{6/5}t_{2}^{-1/10},
~~~~
\nu_a < \nu_c < \nu_m\nonumber \\
\end{eqnarray}

For $1<p<2$, one has ($\nu_c$ and $F_{\rm \nu,max}$ remain the same)
\begin{eqnarray}
&&\nu_m=   6.1\times10^{8}~{\rm Hz}~\hat z^{\frac{2-3p}{4(p-1)}}\frac{g^{IV}(p)}{g^{IV}(1.8)}E_{52}^{\frac{p-2}{8(p-1)}}n_{0,0}^{\frac{6-p}{8(p-1)}}\Gamma_{0,2}^{\frac{2}{p-1}}\Delta_{0,13}^{\frac{p-2}{8(1-p)}}\zeta_{0}^{\frac{2-p}{p-1}}\epsilon_{e,-1}^{\frac{2}{p-1}}\epsilon_{B,-2}^{\frac{1}{2(p-1)}}t_{2}^{\frac{2-p}{4(p-1)}}            ,\nonumber\\
&&\nu_a=2.1\times10^{14}~{\rm Hz}~\hat
z^{\frac{26-21p}{40(p-1)}}\frac{g^{V}(p)}{g^{V}(1.8)}E_{52}^{\frac{38-43p}{80(1-p)}}\Gamma_{0,2}^{\frac{3p+2}{5(1-p)}}\Delta_{0,13}^{\frac{38-43p}{80(p-1)}}n_{0,0}^{\frac{66-41p}{80(1-p)}}\zeta_{0}^{\frac{p-2}{2(p-1)}}
\epsilon_{e,-1}^{\frac{1}{1-p}}\epsilon_{B,-2}^{\frac{14-9p}{20(1-p)}}t_{2}^{\frac{14-19p}{40(p-1)}},\nonumber\\
&&~~~~~~~~~~~~~~~~~~~~~~~~~~~~~~~~~~~~~~~~~~~~~~~~~~~~~~~~~~~~~~~~~~~~~~~~~~~~~~~~~~~~~~~~~~~~~~~~~~\nu_a < \nu_m < \nu_c\nonumber \\
&&\nu_a =9.3\times10^{11}~{\rm Hz}~\hat
z^{-\frac{3p+10}{4(p+4)}}\frac{g^{VI}(p)}{g^{VI}(1.8)}E_{52}^{\frac{p+14}{8(p+4)}}
\Delta_{0,13}^{-\frac{p+14}{8(p+4)}}n_{0,0}^{-\frac{p-18}{8(p+4)}}\zeta_{0}^{\frac{2-p}{p+4}}
\epsilon_{e,-1}^{\frac{2}{p+4}}\epsilon_{B,-2}^{\frac{2}{p+4}}t_{2}^{-\frac{p+6}{4(p+4)}},\nonumber \\
&&~~~~~~~~~~~~~~~~~~~~~~~~~~~~~~~~~~~~~~~~~~~~~~~~~~~~~~~~~~~~~~~~~~~~~~~~~~~~~~~~~~~~~~~~~~~~~~~~~~
\nu_m < \nu_a < \nu_c\nonumber \\
&&\nu_a=1.5\times10^{10}~{\rm Hz}~\hat
z^{-9/10}\frac{g^{VII}(p)}{g^{VII}(1.8)}E_{52}^{17/20}\Gamma_{0,2}^{-3/5}\Delta_{0,13}^{-17/20}n_{0,0}^{19/20}\epsilon_{B,-2}^{6/5}t_{2}^{-1/10},
~~~~~~
\nu_a < \nu_c < \nu_m\nonumber \\
\end{eqnarray}

For the wind model and $p>2$, one has
\begin{eqnarray}
&&\nu_m=    3.3\times10^{13}~{\rm Hz}\frac{G(p)}{G(2.3)}E_{52}^{-1/2}A_{*,-1}^{}\Gamma_{0,2}^{2}\Delta_{0,13}^{1/2}\epsilon_{e,-1}^{2}\epsilon_{B,-2}^{1/2}t_{2}^{-1}           ,\nonumber\\
&&\nu_c=    1.2\times10^{14}~{\rm Hz}~\hat z^{-2}E_{52}^{1/2}\Delta_{0,13}^{-1/2}A_{*,-1}^{-2}\epsilon_{B,-2}^{-3/2}t_{2}^{}           \nonumber\\
&&F_{\rm{\nu,max}}=  6.7\times10^{5}~\mu {\rm
Jy}~\hat zE_{52}^{}A_{*,-1}^{1/2}\Gamma_{0,2}^{-1}\Delta_{0,13}^{-1}\epsilon_{B,-2}^{1/2}D_{28}^{-2}              ,\nonumber\\
&&\nu_a=3.2\times10^{13}~{\rm
Hz}\frac{g^{VIII}(p)}{g^{VIII}(2.3)}E_{52}^{2/5}\Gamma_{0,2}^{-8/5}A_{*,-1}^{2/5}\Delta_{0,13}^{-2/5}\epsilon_{e,-1}^{-1}\epsilon_{B,-2}^{1/5}t_{2}^{-1},
~~~~~~~~~~~~~~
\nu_a < \nu_m < \nu_c\nonumber \\
&&\nu_a =3.3\times10^{13}~{\rm
Hz}\frac{g^{IX}(p)}{g^{IX}(2.3)}E_{52}^{\frac{2-p}{2(p+4)}}\Gamma_{0,2}^{\frac{2(p-2)}{p+4}}
\Delta_{0,13}^{\frac{p-2}{2(p+4)}}A_{*,-1}^{\frac{p+2}{p+4}}
\epsilon_{e,-1}^{\frac{2(p-1)}{p+4}}\epsilon_{B,-2}^{\frac{p+2}{2(p+4)}}t_{2}^{-1},
~~~~~~~
\nu_m < \nu_a < \nu_c\nonumber \\
&&\nu_a=1.7\times10^{13}~{\rm Hz}~\hat
z\frac{g^{X}(p)}{g^{X}(2.3)}E_{52}^{-1/10}\Gamma_{0,2}^{-3/5}\Delta_{0,13}^{1/10}A_{*,-1}^{19/10}\epsilon_{B,-2}^{6/5}t_{2}^{-2},
~~~~~~~~~~~~~~~~~~
\nu_a < \nu_c < \nu_m\nonumber \\
\end{eqnarray}

For $1<p<2$, one has ($\nu_c$ and $F_{\rm \nu,max}$ remain the same)
\begin{eqnarray}
&&\nu_m=  8.7\times10^{10}~{\rm Hz}~\hat z^{\frac{2-p}{p-1}}\frac{g^{XI}(p)}{g^{XI}(1.8)}E_{52}^{\frac{p-4}{4(p-1)}}A_{*,-1}^{\frac{6-p}{4(p-1)}}\Gamma_{0,2}^{\frac{2}{p-1}}\Delta_{0,13}^{\frac{p-4}{4(1-p)}}\zeta_{0}^{\frac{2-p}{p-1}}\epsilon_{e,-1}^{\frac{2}{p-1}}\epsilon_{B,-2}^{\frac{1}{2(p-1)}}t_{2}^{\frac{1}{1-p}}           ,\nonumber\\
&&\nu_a=5.2\times10^{14}~{\rm Hz}~\hat
z^{\frac{p-2}{2(p-1)}}\frac{g^{XII}(p)}{g^{XII}(1.8)}E_{52}^{\frac{p+14}{40(p-1)}}\Gamma_{0,2}^{\frac{3p+2}{5(1-p)}}\Delta_{0,13}^{\frac{p+14}{40(p-1)}}A_{*,-1}^{\frac{66-41p}{40(1-p)}}\zeta_{0}^{\frac{p-2}{2(p-1)}}
\epsilon_{e,-1}^{\frac{1}{1-p}}\nonumber\\
&&~~~~~~~~\epsilon_{B,-2}^{\frac{14-9p}{20(1-p)}}t_{2}^{\frac{4-3p}{2(p-1)}},~~~~~~~~~~~~~~~~~~~~~~~~~~~~~~~~~~~~~~~~~~~~~~~~~~~~~~~~~~~~~~~~~\nu_a < \nu_m < \nu_c\nonumber \\
&&\nu_a =1.3\times10^{13}~{\rm Hz}~\hat
z^{\frac{2-p}{p+4}}\frac{g^{XIII}(p)}{g^{XIII}(1.8)}E_{52}^{\frac{p-2}{4(p+4)}}\Delta_{0,13}^{\frac{2-p}{4(p+4)}}A_{*,-1}^{\frac{18-p}{4(p+4)}}\zeta_{1}^{\frac{2-p}{p+4}}
\epsilon_{e,-1}^{\frac{2}{p+4}}\epsilon_{B,-2}^{\frac{2}{p+4}}t_{2}^{-\frac{6}{p+4}},\nonumber \\
&&~~~~~~~~~~~~~~~~~~~~~~~~~~~~~~~~~~~~~~~~~~~~~~~~~~~~~~~~~~~~~~~~~~~~~~~~~~~~~~~~~~~~~~~~~~~\nu_m < \nu_a < \nu_c\nonumber \\
&&\nu_a=1.4\times10^{13}~{\rm Hz}~\hat
z\frac{g^{XIV}(p)}{g^{XIV}(1.8)}E_{52}^{-1/10}\Gamma_{0,2}^{-3/5}\Delta_{0,13}^{1/10}A_{*,-1}^{19/10}\epsilon_{B,-2}^{6/5}t_{2}^{-2},
~~~~~~~~~
\nu_a < \nu_c < \nu_m\nonumber \\
\end{eqnarray}

After the reverse shock crosses the shell, the shocked shell can be
roughly described by the BM solution
\citep{kobayashisari00,wu03,kobayashizhang03b,kobayashi04},
{\setlength\arraycolsep{2 pt}
\begin{eqnarray}
  \gamma_3 &\propto& t^{(2k-7)/4(4-k)}, e_3 \propto t^{(2k-13)/3(4-k)},
  r \propto t^{1/(8-2k)}, N_{e,3} \propto t^0,
\end{eqnarray}}

For the ISM case, one has
\begin{eqnarray}
\gamma=\gamma_{3,\times}\left(\frac{t}{T}\right)^{-\frac{7}{16}},~~~~~~~~~~~~~~~~~~~~~~~R=R_\times \left(\frac{t}{T}\right)^{\frac{1}{8}}\nonumber
\end{eqnarray}
where $\gamma_{3,\times}$ and $R_\times$ are the Lorentz factor and radius of
Region 3 at the shock crossing time.

For $p>2$, one has
\begin{eqnarray}
&&\nu_m=   4.8\times10^{12}~{\rm Hz}~\hat z^{25/48}\frac{G(p)}{G(2.3)}\Gamma_{0,2}^{2}\Delta_{0,13}^{73/48}n_{0,0}^{1/2}\epsilon_{e,-1}^{2}\epsilon_{B,-2}^{1/2}t_{2}^{-73/48}           ,\nonumber\\
&&\nu_{\rm cut}=    2.3\times10^{17}~{\rm Hz}~\hat z^{25/48}E_{52}^{-1/2}\Delta_{0,13}^{49/48}n_{0,0}^{-1}\epsilon_{B,-2}^{-3/2}t_{2}^{-73/48}           \nonumber\\
&&F_{\rm{\nu,max}}=     7.9\times10^{5} ~\mu {\rm
Jy}~\hat z^{95/48}E_{52}^{5/4} \Gamma_{0,2}^{-1}\Delta_{0,13}^{11/48}n_{0,0}^{1/4}\epsilon_{B,-2}^{1/2}D_{28}^{-2}t_{2}^{-47/48}          ,\nonumber\\
&&\nu_a=6.6\times10^{12}~{\rm
Hz}~\hat z^{-7/15}\frac{g^{XV}(p)}{g^{XV}(2.3)}E_{52}^{3/5}\Gamma_{0,2}^{-8/5}\Delta_{0,13}^{-2/3}n_{0,0}^{1/5}\epsilon_{e,-1}^{-1}\epsilon_{B,-2}^{1/5}t_{2}^{-8/15},
~~~~
\nu_a < \nu_m < \nu_{\rm cut} \nonumber \\
&&\nu_a =5.7\times10^{12}~{\rm
Hz}~\hat z^{\frac{25p-58}{48(p+4)}}\frac{g^{XVI}(p)}{g^{XVI}(2.3)}E_{52}^{\frac{2}{p+4}}\Gamma_{0,2}^{\frac{2(p-2)}{p+4}}\Delta_{0,13}^{\frac{73p-58}{48(p+4)}}n_{0,0}^{\frac{p+2}{2(p+4)}}
\epsilon_{e,-1}^{\frac{2(p-1)}{p+4}}\epsilon_{B,-2}^{\frac{p+2}{2(p+4)}}t_{2}^{-\frac{73p+134}{48(p+4)}},\nonumber \\
&&~~~~~~~~~~~~~~~~~~~~~~~~~~~~~~~~~~~~~~~~~~~~~~~~~~~~~~~~~~~~~~~~~~~~~~~~~~~~~~~~~~~~~~~~~~~~~~~
\nu_m < \nu_a < \nu_{\rm cut} \nonumber \\
\end{eqnarray}

For $1<p<2$, one has ($\nu_c$ and $F_{\rm \nu,max}$ remain the same)
\begin{eqnarray}
&&\nu_m=  4.2\times10^{12}~{\rm Hz} \hat z^{25/48}\frac{g^{XVII}(p)}{g^{XVII}(1.8)}E_{52}^{\frac{p-2}{8(p-1)}}\Gamma_{0,2}^{\frac{2}{p-1}}n_{0,0}^{\frac{6-p}{8(p-1)}}\Delta_{0,13}^{\frac{55p-37}{48(p-1)}}\zeta_{0}^{\frac{2-p}{p-1}}\epsilon_{e,-1}^{\frac{2}{p-1}}\epsilon_{B,-2}^{\frac{1}{2(p-1)}}t_{2}^{-73/48}         ,\nonumber\\
&&\nu_a=5.8\times10^{12}~{\rm Hz}~\hat
z^{-7/15}\frac{g^{XVIII}(p)}{g^{XVIII}(1.8)}E_{52}^{\frac{43p-38}{80(p-1)}}\Gamma_{0,2}^{\frac{3p+2}{5(1-p)}}\Delta_{0,13}^{\frac{23p-14}{48(1-p)}}n_{0,0}^{\frac{66-41p}{80(1-p)}}\zeta_{0}^{\frac{p-2}{2(p-1)}}
\epsilon_{e,-1}^{\frac{1}{1-p}}\nonumber\\
&&~~~~~~~~~~~~\epsilon_{B,-2}^{\frac{14-9p}{20(1-p)}}t_{2}^{-8/15},
~~~~~~~~~~~~~~~~~~~~~~~~~~~~~~~~~~~~~~~~~~~~~~~~~~~~~~~~~~~~\nu_a < \nu_m < \nu_{\rm cut} \nonumber \\
&&\nu_a =5.1\times10^{12}~{\rm Hz}~\hat
z^{\frac{25p-58}{48(p+4)}}\frac{g^{XIX}(p)}{g^{XIX}(1.8)}E_{52}^{\frac{p+14}{8(p+4)}}\Delta_{0,13}^{\frac{11(5p-2)}{48(p+4)}}n_{0,0}^{\frac{18-p}{8(p+4)}}\zeta_{0}^{\frac{2-p}{p+4}}
\epsilon_{e,-1}^{\frac{2}{p+4}}\epsilon_{B,-2}^{\frac{2}{p+4}}t_{2}^{-\frac{73p+134}{48(p+4)}},\nonumber \\
&&~~~~~~~~~~~~~~~~~~~~~~~~~~~~~~~~~~~~~~~~~~~~~~~~~~~~~~~~~~~~~~~~~~~~~~~~~~~~~~~~~~~~~~~~~~~~
\nu_m < \nu_a < \nu_{\rm cut} \nonumber \\
\end{eqnarray}

For the wind model and $p>2$, one has
\begin{eqnarray}
&&\nu_m=    9.4\times10^{13}~{\rm Hz}~\hat z^{7/8}\frac{G(p)}{G(2.3)}E_{52}^{-1/2}A_{*,-1}^{}\Gamma_{0,2}^{2}\Delta_{0,13}^{11/8}\epsilon_{e,-1}^{2}\epsilon_{B,-2}^{1/2}t_{2}^{-15/8}           ,\nonumber\\
&&\nu_{\rm cut}=    3.7\times10^{15}~{\rm Hz}~\hat z^{7/8}E_{52}^{1/2}\Delta_{0,13}^{19/8}A_{*,-1}^{-2}\epsilon_{B,-2}^{-3/2}t_{2}^{-15/8}           \nonumber\\
&&F_{\rm{\nu,max}}=  2.6\times10^{6}~\mu {\rm
Jy}~\hat z^{17/8}E_{52}^{}A_{*,-1}^{1/2}\Gamma_{0,2}^{-1}\Delta_{0,13}^{1/8}\epsilon_{B,-2}^{1/2}D_{28}^{-2} t_{2}^{-9/8}             ,\nonumber\\
&&\nu_a=1.9\times10^{13}~{\rm Hz}~\hat
z^{-2/5}\frac{g^{XX}(p)}{g^{XX}(2.3)}E_{52}^{2/5}\Gamma_{0,2}^{-8/5}A_{*,-1}^{2/5}\Delta_{0,13}^{-4/5}\epsilon_{e,-1}^{-1}\epsilon_{B,-2}^{1/5}t_{2}^{-3/5},
~~~~~~~~~~
\nu_a < \nu_m < \nu_{\rm cut} \nonumber \\
&&\nu_a =4.1\times10^{13}~{\rm
Hz}~\hat z^{\frac{7p-6}{8(p+4)}}\frac{g^{XXI}(p)}{g^{XXI}(2.3)}E_{52}^{\frac{2-p}{2(p+4)}}\Gamma_{0,2}^{\frac{2(p-2)}{p+4}}\Delta_{0,13}^{\frac{11p-14}{8(p+4)}}
A_{*,-1}^{\frac{p+2}{p+4}}\epsilon_{e,-1}^{\frac{2(p-1)}{p+4}}\epsilon_{B,-2}^{\frac{p+2}{2(p+4)}}t_{2}^{-\frac{15p+26}{8(p+4)}},\nonumber \\
&&~~~~~~~~~~~~~~~~~~~~~~~~~~~~~~~~~~~~~~~~~~~~~~~~~~~~~~~~~~~~~~~~~~~~~~~~~~~~~~~~~~~~~~~~~~~~~~~
\nu_m < \nu_a < \nu_{\rm cut} \nonumber \\
\end{eqnarray}

For $1<p<2$, one has ($\nu_c$ and $F_{\rm \nu,max}$ remain the same)
\begin{eqnarray}
&&\nu_m= 1.9\times10^{14}~{\rm Hz}~\hat z^{7/8}\frac{g^{XXII}(p)}{g^{XXII}(1.8)}E_{52}^{\frac{p-4}{4(p-1)}}\Gamma_{0,2}^{\frac{2}{p-1}}A_{*,-1}^{\frac{6-p}{4(p-1)}}\Delta_{0,13}^{\frac{13p-15}{8(p-1)}}\zeta_{0}^{\frac{2-p}{p-1}}\epsilon_{e,-1}^{\frac{2}{p-1}}\epsilon_{B,-2}^{\frac{1}{2(p-1)}}t_{2}^{-15/8}          ,\nonumber\\
&&\nu_a=1.2\times10^{13}~{\rm Hz}~\hat
z^{-2/5}\frac{g^{XXIII}(p)}{g^{XXIII}(1.8)}
E_{52}^{\frac{p+14}{40(p-1)}}\Gamma_{0,2}^{\frac{3p+2}{5(1-p)}}A_{*,-1}^{\frac{66-41p}{40(1-p)}}\Delta_{0,13}^{\frac{42-37p}{40(p-1)}}\zeta_{0}^{\frac{2-p}{p-1}}\epsilon_{e,-1}^{\frac{2}{p-1}}\nonumber \\
&&~~~~~~~~~~\epsilon_{B,-2}^{\frac{1}{2(p-1)}}t_{2}^{-3/5},~~~~~~~~~~~~~~~~~~~~~~~~~~~~~~~~~~~~~~~~~~~~~~~~~~~~~~~~~~~~~~~~~~~\nu_a < \nu_m < \nu_{\rm cut} \nonumber \\
&&\nu_a =3.8\times10^{13}~{\rm Hz}~\hat
z^{\frac{7p-6}{8(p+4)}}\frac{g^{XXIV}(p)}{g^{XXIV}(1.8)}E_{52}^{\frac{p-2}{4(p+4)}}\Delta_{0,13}^{\frac{13p-18}{8(p+4)}}A_{*,-1}^{\frac{18-p}{4(p+4)}}\zeta_{0}^{\frac{2-p}{p+4}}
\epsilon_{e,-1}^{\frac{2}{p+4}}\epsilon_{B,-2}^{\frac{2}{p+4}}t_{2}^{-\frac{15p+26}{8(p+4)}},\nonumber\\
&&~~~~~~~~~~~~~~~~~~~~~~~~~~~~~~~~~~~~~~~~~~~~~~~~~~~~~~~~~~~~~~~~~~~~~~~~~~~~~~~~~~~~~~~~~~~~~~~~
\nu_m < \nu_a < \nu_{\rm cut} \nonumber \\
\end{eqnarray}

The $\alpha$ and $\beta$ values and their closure relations
for the thick shell reverse shock models are
presented in Tables \ref{ThickRPre} and \ref{ThickRPre1} (for
pre-shock-crossing), and Tables \ref{ThickRPost} and \ref{ThickRPost2}
(for post-shock-crossing).

For this regime (thick-shell reverse shock model during shock
crossing), for $p>2$, one has $\nu_m \propto t^{0}~(t^{-1})$, $\nu_c
\propto t^{-1}~(t^1)$, $F_{\rm \nu,max} \propto t^{1/2} ~(t^{0})$
for the ISM (wind) models, respectively. For $1<p<2$, $\nu_c$ and
$F_{\rm{\nu,max}}$ evolutions are the same as $p>2$ cases, while
$\nu_m \propto t^{\frac{2-p}{4(p-1)}}~ (t^{\frac{1}{1-p}})$ for the
ISM (wind) models, respectively.

After shock crossing, $\nu_m \propto \nu_{\rm cut} \propto
t^{-73/48} ~(t^{-15/8})$, $F_{\rm \nu,max} \propto t^{-47/48}
~(t^{-9/8})$ for the ISM (wind) models, respectively.

Notice that in the above treatment, a relativistic reverse shock has
been assumed. In reality, there is a brief epoch before the reverse
shock becomes relativistic. There should be an additional dynamical
change at $R_{\rm N}$ (the transition radius from Newtonian to
relativistic reverse shock), which is much smaller than $R_{\times}$
\citep{sari95}. The light curves may show an additional break at
this epoch, before which the thin shell scalings discussed in
\S3.1.1 and \S3.1.2 apply.

\begin{table}
\caption{The temporal decay index $\alpha$ and spectral index $\beta$ of
the thick shell reverse shock model during the shock crossing phase
in the $\nu_a < \rm{min}(\nu_m,\nu_c)$ spectral regime.\label{ThickRPre}}
\begin{tabular}{llllll}
\hline\hline
& & $p>2$ & &  $1<p<2$ & \\

& $\beta$ & $\alpha $  &  $\alpha (\beta)$ & $\alpha$  & $\alpha (\beta)$  \\
\hline
ISM & slow cooling  \\
\hline
$\nu<\nu_a$   &  $-2$  &   $-{3 \over 2}$ & $\alpha={3\beta \over 4}$  &${11p-10 \over 8(1-p)}$ & $--$\\
$\nu_a<\nu<\nu_m$   &  $-{1 \over 3}$  &   $-{1 \over 2}$ & $\alpha={3\beta \over 2}$ & ${7p-8 \over 12(1-p)}$  & $--$\\
$\nu_m<\nu<\nu_c$   &  ${{p-1 \over 2}}$  &  $-{1 \over 2}$
   &  $--$ & ${p-6 \over 8}$   & $\alpha={2\beta-5 \over 8}$\\
$\nu>\nu_c$   &  ${{p\over 2}}$   &   $0$  & $--$ & ${p-2\over 8}$   & ${\beta-1 \over 4}$ \\

\hline
ISM & fast cooling \\
\hline
$\nu<\nu_a$   &  $-2$  &   $-1$ & $\alpha=\frac{\beta}{2}$ & $-1$  & $\alpha=\frac{\beta}{2}$\\
$\nu_a<\nu<\nu_c$   &  $-{1\over 3}$  &   $-{5\over 6}$ &  $\alpha={5\beta \over 2}$  &  $-{5 \over 6}$  & $\alpha={5\beta \over 2}$\\
$\nu_c<\nu<\nu_m$   &  ${1\over 2}$  &  $0$  & $--$ &  $0$   &  $--$ \\
$\nu>\nu_m$   &  ${p\over 2}$   &   $0$      &  $--$  &  ${p-2 \over 8}$  & $\alpha={\beta-1 \over 4}$ \\

\hline
Wind & slow cooling \\
\hline
$\nu<\nu_a$   &  $-2$   &   $-2$  &  $\alpha={\beta}$ & $\frac{5p-6}{2(1-p)}$  & $--$\\
$\nu_a<\nu<\nu_m$   &  $-{1\over 3}$   &   $-{1\over 3}$  &  $\alpha={\beta}$ & $-{1\over 3(p-1)}$  & $--$\\
$\nu_m<\nu<\nu_c$   &  ${p-1\over 2}$  &   ${p-1\over 2}$    &   $\alpha={\beta}$ & ${1 \over 2}$ & $--$\\
$\nu>\nu_c$   &  ${p\over 2}$   &   ${p-2\over 2}$    &  $\alpha={\beta-1}$   &   $0$  &  $--$\\

\hline
Wind & fast cooling  \\
\hline
$\nu<\nu_a$   &  $-2$   &   $-3$ & $\alpha={3\beta \over 2}$ &  $-3$  & $\alpha={3\beta \over 2}$  \\
$\nu_a<\nu<\nu_c$   &  $-{1\over 3}$   &   ${1\over 3}$ & $\alpha=-\beta$ &  ${1 \over 3}$  & $\alpha=-\beta$  \\
$\nu_c<\nu<\nu_m$   &  ${1\over 2}$  &   $-{1\over 2}$  & $\alpha=-\beta$  & $-{1 \over 2}$ & $\alpha=-\beta$\\
$\nu>\nu_m$   &  ${p\over 2}$  &   ${p-2\over 2}$     & $\alpha={\beta-1}$  &  $0$  & $--$\\
\hline
\end{tabular}
\label{Tab:alpha-beta}
\end{table}

\begin{table}

\caption{The temporal decay index $\alpha$ and spectral index $\beta$ for
the thick shell reverse shock model during the reverse shock crossing phase
in the $ \nu_m <\nu_a < \nu_c$ spectral regime.\label{ThickRPre1}}
\begin{tabular}{llllll}
\hline\hline
& & $p>2$ & &  $1<p<2$ & \\

& $\beta$ & $\alpha $  &  $\alpha (\beta)$ & $\alpha$ & $\alpha (\beta)$  \\
\hline
ISM & slow cooling  \\
\hline
$\nu<\nu_m$   &  $-2$  &   $-{3 \over 2}$ & $\alpha={3\beta \over 4}$ & ${11p-10 \over 8(1-p)}$  & $--$\\
$\nu_m<\nu<\nu_a$   &  $-{5 \over 2}$  &   $-{3 \over 2}$ & $\alpha={3\beta \over 5}$ & $-{3 \over 2}$  & $\alpha={3\beta \over 5}$\\
$\nu_a<\nu<\nu_c$   &  ${p-1\over 2}$   &  $-{1 \over 2}$
  &  $--$ & ${p-6 \over 8}$  & $\alpha={2\beta-5 \over 8}$\\
$\nu>\nu_c$   &  ${{p\over 2}}$   &   $0$  & $--$ & ${p-2 \over 8}$  & ${\beta-1 \over 4}$ \\

\hline
Wind & slow cooling \\
\hline
$\nu<\nu_m$   &  $-2$   &   $-2$  &  $\alpha={\beta}$ & $\frac{6-5p}{2(p-1)}$  & $--$\\
$\nu_m<\nu<\nu_a$   &  $-{5\over 2}$   &   $-{5\over 2}$  &  $\alpha={\beta}$ & $-{5 \over 2}$  & $\alpha={\beta}$\\
$\nu_a<\nu<\nu_c$   &  ${p-1\over 2}$   &   ${p-1\over 2}$    &   $\alpha={\beta}$ & ${1 \over 2}$ & $--$\\
$\nu>\nu_c$   &  ${p\over 2}$   &   ${p-2\over 2}$    &  $\alpha={\beta-1}$   &   $0$  &  $--$\\

\hline
\end{tabular}
\label{Tab:alpha-beta}
\end{table}

\begin{table}
\caption{The temporal decay index $\alpha$ and spectral index
$\beta$ of the thick shell reverse shock model in the post-shock
crossing phase in the $\nu_a < \rm{min}(\nu_m,\nu_{\rm cut})$
spectral regime.\label{ThickRPost}}
\begin{tabular}{llllll}
\hline\hline
& & $p>2$ & &  $1<p<2$ & \\

& $\beta$ & $\alpha $  &  $\alpha (\beta)$ & $\alpha$  & $\alpha (\beta)$  \\
\hline
ISM & slow cooling  \\
\hline
$\nu<\nu_a$   &  $-2$  &   $-{5 \over 12}$ & $\alpha={5\beta \over 24}$  &$-{5 \over 12}$ & $\alpha={5\beta \over 24}$\\
$\nu_a<\nu<\nu_m$   &  $-{1 \over 3}$  &   ${17 \over 36}$ & $-\alpha={17\beta \over 12}$ & ${17 \over 36}$ & $-\alpha={17\beta \over 12}$\\
$\nu_m<\nu<\nu_{\rm cut}$   &  ${{p-1 \over 2}}$  &  ${73p+21 \over
96}$
   &  $\alpha={73\beta+47 \over 48}$ & ${73p+21 \over
96}$
   &  $\alpha={73\beta+47 \over 48}$\\

\hline
Wind & slow cooling \\
\hline
$\nu<\nu_a$   &  $-2$   &   $-{1\over 2}$  &  $\alpha={\beta \over 4}$ & $-{1\over 2}$  &  $\alpha={\beta \over 4}$\\
$\nu_a<\nu<\nu_m$   &  $-{1\over 3}$   &   ${1\over 2}$  &  $\alpha=-{3\beta \over 2}$ & ${1\over 2}$  &  $\alpha=-{3\beta \over 2}$\\
$\nu_m<\nu<\nu_{\rm cut}$   &  ${p-1\over 2}$  &   ${3(5p+1) \over 16}$    &   $\alpha={3(5\beta+3) \over 8}$ & ${3(5p+1) \over 16}$    &   $\alpha={3(5\beta+3) \over 8}$\\

\hline
\end{tabular}
\label{Tab:alpha-beta}
\end{table}

\begin{table}

\caption{The temporal decay index $\alpha$ and spectral index
$\beta$ of the thick shell reverse shock model in the post-shock
crossing phase in the $\nu_m <\nu_a < \nu_{\rm cut}$ spectral
regime.\label{ThickRPost2}}
\begin{tabular}{llllll}
\hline\hline
& & $p>2$ & &  $1<p<2$ & \\

& $\beta$ & $\alpha $  &  $\alpha (\beta)$ & $\alpha$ & $\alpha (\beta)$  \\
\hline
ISM & slow cooling  \\
\hline
$\nu<\nu_m$   &  $-2$  &   $-{5 \over 12}$ & $\alpha={5\beta \over 24}$ & $-{5 \over 12}$ & $\alpha={5\beta \over 24}$\\
$\nu_m<\nu<\nu_a$   &  $-{5 \over 2}$  &   $-{113\over 96}$ & $\alpha={226\beta \over 480}$ & $-{113\over 96}$ & $\alpha={226\beta \over 480}$\\
$\nu_a<\nu<\nu_{\rm cut}$   &  ${p-1\over 2}$   &  ${73p+21 \over
96}$
  &  $\alpha={73\beta+47 \over 48}$ & ${73p+21 \over
96}$
  &  $\alpha={73\beta+47 \over 48}$\\

\hline
Wind & slow cooling \\
\hline
$\nu<\nu_m$   &  $-2$   &   $-{1\over 2}$  &  $\alpha={\beta \over 4}$ & $-{1\over 2}$  &  $\alpha={\beta \over 4}$\\
$\nu_m<\nu<\nu_a$   &  $-{5\over 2}$   &   $-{23\over 16}$  &  $\alpha={23\beta \over 40}$ & $-{23\over 16}$  &  $\alpha={23\beta \over 40}$\\
$\nu_a<\nu<\nu_{\rm cut}$   &  ${p-1\over 2}$   &   ${3(5p+1)\over 16}$    &   $\alpha={3(5\beta+3) \over 8}$ & ${3(5p+1)\over 16}$    &   $\alpha={3(5\beta+3) \over 8}$\\

\hline

\end{tabular}
\label{Tab:alpha-beta}
\end{table}

\subsection{Phase 2: Relativistic, pre-jet-break, self-similar deceleration phase}

After reverse shock crosses the shell, the blastwave would quickly
adjusts itself to a self-similar deceleration phase
\citep{blandford76}\footnote{This is the case for the idealized
situation. In reality, there might be irregulatities in the system
(e.g. ambient density fluctuations or non-power-law energy
injection). The blastwave is no longer self-similar. We limit
ourselves to the self-similar assumption and derive the scalings in
this subsection, and discuss more complicated simulations in \S5.}.
Early on, the blastwave is ultra-relativistic with $1/\Gamma \ll
\theta_j$. The closure relations in this phase have been reviewed
previously \citep[e.g.][]{zhangmeszaros04,zhang06}.
\subsubsection{Adiabatic deceleration without energy injection}

The simplest model invokes a constant energy in the blastwave. This
requires that the blastwave is adiabatic (no radiative loss), and
that there is no energy injection into the blastwave. The adiabatic
approximation usually gives a reasonable description to the
blastwave evolution. This is because the radiative loss fraction is
at most $\epsilon_e$ (for fast cooling), which is constrained to be
around 0.1 and lower
\citep{panaitescu01,panaitescu02,yost03}\footnote{Note that since
the blast-wave energy is given again and again to newly heated
material, the radiative energy loss may become important after
several orders of magnitude of deceleration time \citep{sari97b}.}.

For an arbitrary $k$ density profile, the dynamics of the blast wave
in the constant energy regime can be described as
\begin{eqnarray}
\gamma=\left(\frac{(17-4k)E}{4^{5-k}(4-k)^{3-k}\pi
Am_pc^{5-k}t^{3-k}}\right)^\frac{1}{2(4-k)},~~~~~~~~R=\left(\frac{(17-4k)(4-k)Et}{4\pi
Am_pc}\right)^\frac{1}{4-k},\nonumber
\end{eqnarray}

For the ISM model ($k=0$) and $p>2$, one has
\begin{eqnarray}
&&\nu_m=   4.3\times10^{10}~{\rm Hz}~\hat z^{1/2}\frac{G(p)}{G(2.3)}E_{52}^{1/2}\epsilon_{e,-1}^{2}\epsilon_{B,-2}^{1/2}t_{5}^{-3/2}           ,\nonumber\\
&&\nu_c=    2.9\times10^{16}~{\rm Hz}~\hat z^{-1/2}E_{52}^{-1/2}n_{0,0}^{-1}\epsilon_{B,-2}^{-3/2}t_{5}^{-1/2}           \nonumber\\
&&F_{\rm{\nu,max}}=     1.1\times10^{4}~\mu {\rm
Jy}~\hat z E_{52}^{}n_{0,0}^{1/2}\epsilon_{B,-2}^{1/2}D_{28}^{-2}          ,\nonumber\\
&&\nu_a=5.7\times10^{9}~{\rm Hz}~\hat
z^{-1}\frac{g^{I}(p)}{g^{I}(2.3)}E_{52}^{1/5}n_{0,0}^{3/5}\epsilon_{e,-1}^{-1}\epsilon_{B,-2}^{1/5},
~~~~~~~~~~~~~~~~~~~~~~~~~~~~~~~~~~~
\nu_a < \nu_m < \nu_c\nonumber \\
&&\nu_a=1.5\times10^{10}~{\rm Hz}~\hat
z^{\frac{p-6}{2(p+4)}}\frac{g^{II}(p)}{g^{II}(2.3)}E_{52}^{\frac{p+2}{2(p+4)}}n_{0,0}^{\frac{2}{p+4}}\epsilon_{e,-1}^
{\frac{2(p-1)}{p+4}}\epsilon_{B,-2}^{\frac{p+2}{2(p+4)}}t_{5}^{-\frac{3p+2}{2(p+4)}},
~~~~~~~~~~~~~
\nu_m < \nu_a < \nu_c\nonumber \\
&&\nu_a=6.9\times10^{6}~{\rm Hz}~\hat
z^{-1/2}\frac{g^{III}(p)}{g^{III}(2.3)}E_{52}^{7/10}n_{0,0}^{11/10}\epsilon_{B,-2}^{6/5}t_{5}^{-1/2},
~~~~~~~~~~~~~~~~~~~~~~~~~~~~
\nu_a < \nu_c < \nu_m\nonumber \\
\end{eqnarray}

For $1<p<2$, one has ($\nu_c$ and $F_{\rm \nu.max}$ remain the same)
\begin{eqnarray}
&&\nu_m=   3.6\times10^{7}~{\rm Hz}~\hat z^{\frac{14-5p}{8(p-1)}}\frac{g^{IV}(p)}{g^{IV}(1.8)}E_{52}^{\frac{p+2}{8(p-1)}}n_{0,0}^{\frac{2-p}{8(p-1)}}\zeta_{0}^{\frac{2-p}{p-1}}\epsilon_{e,-1}^{\frac{2}{p-1}}\epsilon_{B,-2}^{\frac{1}{2(p-1)}}t_{5}^{\frac{3p+6}{8(1-p)}}           ,\nonumber\\
&&\nu_a=1.6\times10^{11}~{\rm Hz}~\hat
z^{-\frac{7p+2}{16(p-1)}}\frac{g^{V}(p)}{g^{V}(1.8)}E_{52}^{\frac{46-31p}{80(1-p)}}n_{0,0}^{\frac{58-53p}{80(1-p)}}\zeta_{0}^{\frac{p-2}{2(p-1)}}
\epsilon_{e,-1}^{\frac{1}{1-p}}\epsilon_{B,-2}^{\frac{14-9p}{20(1-p)}}t_{5}^{-\frac{9(p-2)}{16(p-1)}},
~~
\nu_a < \nu_m < \nu_c\nonumber \\
&&\nu_a=4.5\times10^{9}~{\rm Hz}~\hat
z^{-\frac{5p+6}{8(p+4)}}\frac{g^{VI}(p)}{g^{VI}(1.8)}E_{52}^{\frac{p+14}{8(p+4)}}n^{\frac{18-p}{8(p+4)}}\zeta^{\frac{2-p}{p+4}}
\epsilon_{e}^{\frac{2}{p+4}}\epsilon_{B}^{\frac{2}{p+4}}t_{d}^{-\frac{3p+26}{8(p+4)}},
~~~~~~~~~~~
\nu_m < \nu_a < \nu_c\nonumber \\
&&\nu_a=5.7\times10^{6}~{\rm Hz}~\hat
z^{-1/2}\frac{g^{VII}(p)}{g^{VII}(1.8)}E_{52}^{7/10}n_{0,0}^{11/10}\epsilon_{B,-2}^{6/5}t_{5}^{-1/2},
~~~~~~~~~~~~~~~~~~~~~~~~~~~~~
\nu_a < \nu_c < \nu_m\nonumber \\
\end{eqnarray}

For the wind model ($k=2$) and $p>2$, one has
\begin{eqnarray}
&&\nu_m=    2.2\times10^{10}~{\rm Hz}~\hat z^{1/2}\frac{G(p)}{G(2.3)}E_{52}^{1/2}\epsilon_{e,-1}^{2}\epsilon_{B,-2}^{1/2}t_{5}^{-3/2}           ,\nonumber\\
&&\nu_c=    1.8\times10^{18}~{\rm Hz}~\hat z^{-3/2}E_{52}^{1/2}A_{*,-1}^{-2}\epsilon_{B,-2}^{-3/2}t_{5}^{1/2}           \nonumber\\
&&F_{\rm{\nu,max}}=  1.5\times10^{3}~\mu {\rm
Jy}~\hat z^{3/2} E_{52}^{1/2}A_{*,-1}^{}\epsilon_{B,-2}^{1/2}D_{28}^{-2}t_{5}^{-1/2}             ,\nonumber\\
&&\nu_a=1.0\times10^{9}~{\rm Hz}~\hat
z^{-2/5}\frac{g^{VIII}(p)}{g^{VIII}(2.3)}E_{52}^{-2/5}A_{*,-1}^{6/5}\epsilon_{e,-1}^{-1}\epsilon_{B,-2}^{1/5}t_{5}^{-3/5},
~~~~~~~~~~~~~~~~~~~~
\nu_a < \nu_m < \nu_c\nonumber \\
&&\nu_a=4.4\times10^{9}~{\rm Hz}~\hat
z^{\frac{p-2}{2(p+4)}}\frac{g^{IX}(p)}{g^{IX}(2.3)}E_{52}^{\frac{p-2}{2(p+4)}}A_{*,-1}^{\frac{4}{p+4}}\epsilon_{e,-1}^
{\frac{2(p-1)}{p+4}}\epsilon_{B,-2}^{\frac{p+2}{2(p+4)}}t_{5}^{-\frac{3(p+2)}{2(p+4)}},
~~~~~~~~~~~~~
\nu_m < \nu_a < \nu_c\nonumber \\
&&\nu_a=1.2\times10^{5}~{\rm Hz}~\hat
z^{3/5}\frac{g^{X}(p)}{g^{X}(2.3)}E_{52}^{-2/5}A_{*,-1}^{11/5}\epsilon_{B,-2}^{6/5}t_{5}^{-8/5},
~~~~~~~~~~~~~~~~~~~~~~~~~~~~~~
\nu_a < \nu_c < \nu_m\nonumber \\
\end{eqnarray}

For $1<p<2$, one has ($\nu_c$ and $F_{\rm \nu.max}$ remain the same)
\begin{eqnarray}
&&\nu_m=   1.5\times10^{7}~{\rm Hz} \hat z^{\frac{8-3p}{4(p-1)}}\frac{g^{XI}(p)}{g^{XI}(1.8)}E_{52}^{\frac{p}{4(p-1)}}A_{*,-1}^{\frac{2-p}{4(p-1)}}\zeta_{0}^{\frac{2-p}{p-1}}\epsilon_{e,-1}^{\frac{2}{p-1}}\epsilon_{B,-2}^{\frac{1}{2(p-1)}}t_{5}^{\frac{p+4}{4(1-p)}}           ,\nonumber\\
&&\nu_a=3.3\times10^{10}~{\rm Hz}~\hat
z^{\frac{9p-34}{40(p-1)}}\frac{g^{XII}(p)}{g^{XII}(1.8)}E_{52}^{\frac{6-11p}{40(p-1)}}A_{*,-1}^{\frac{58-53p}{40(1-p)}}\zeta_{0}^{\frac{p-2}{2(p-1)}}
\epsilon_{e,-1}^{\frac{1}{1-p}}\epsilon_{B,-2}^{\frac{14-9p}{20(1-p)}}t_{5}^{\frac{74-49p}{40(p-1)}},
~~
\nu_a < \nu_m < \nu_c\nonumber \\
&&\nu_a=1.3\times10^{9}~{\rm Hz}~\hat
z^{\frac{6-3p}{4(p+4)}}\frac{g^{XIII}(p)}{g^{XIII}(1.8)}E_{52}^{\frac{p-2}{4(p+4)}}A_{*,-1}^{\frac{18-p}{4(p+4)}}\zeta_{0}^{\frac{2-p}{p+4}}
\epsilon_{e,-1}^{\frac{2}{p+4}}\epsilon_{B,-2}^{\frac{2}{p+4}}t_{5}^{-\frac{22+p}{4(p+4)}},
~~~~~~~~
\nu_m < \nu_a < \nu_c\nonumber \\
&&\nu_a=9.5\times10^{4}~{\rm Hz}~\hat
z^{3/5}\frac{g^{XIV}(p)}{g^{XIV}(1.8)}E_{52}^{-2/5}A_{*,-1}^{11/5}\epsilon_{B,-2}^{6/5}t_{5}^{-8/5},
~~~~~~~~~~~~~~~~~~~~~~~~~~~~~~
\nu_a < \nu_c < \nu_m\nonumber \\
\end{eqnarray}

The $\alpha$ and $\beta$ values and their
closure relations of these models are presented in
Tables \ref{adia_table} to \ref{BMpl22}.

For this model (adiabatic deceleration without energy injection),
for $p>2$, one has $\nu_m \propto t^{-3/2}~(t^{-3/2})$, $\nu_c
\propto t^{-1/2}~(t^{1/2})$, $F_{\rm \nu,max} \propto t^{0}
~(t^{-1/2})$ for the ISM (wind) models, respectively. For $1<p<2$,
$\nu_c$ and $F_{\rm{\nu,max}}$ evolutions are the same as $p>2$
cases, while $\nu_m \propto t^{\frac{3p+6}{8(1-p)}}~
(t^{\frac{p+4}{4(1-p)}})$ for the ISM (wind) models, respectively.

\subsubsection{Adiabatic deceleration with energy injection}

In some central engines models, such as the millisecond magnetar
model \citep{dailu98a,zhangmeszaros01a}, significant energy
injection into the blastwave is possible.
Assume that the central engine has a power-law
luminosity history $L(t)=L_0\left(\frac{t}{t_0}\right)^{-q}$,
the injected energy is $E_{inj}=\frac{L_0t_0^q}{1-q}t^{1-q}$.
If the injected energy is in the form of a Poynting flux
so that a reverse shock does not exist or is weak, one can
approximately treat the blastwave as a system with continuous
energy increase. The energy injection effect becomes significant
when $E_{\rm inj} > E_{\rm imp}$, where $E_{\rm imp}$ is the impulsively
injected energy during the prompt emission phase
\citep{zhangmeszaros01a}. The dynamics of the system can
be described by
\begin{eqnarray}
\gamma=\left(\frac{(17-4k)E}{4^{5-k}(4-k)^{3-k}\pi
Am_pc^{5-k}t^{q+2-k}}\right)^\frac{1}{2(4-k)},~~~~~~~~R=\left(\frac{(17-4k)(4-k)Et^{2-q}}{4\pi
Am_pc}\right)^\frac{1}{4-k}.\nonumber
\end{eqnarray}

There is an alternative type of energy injection, which does
not invoke a long lasting central engine, but rather invokes a
Lorentz factor stratification of the ejecta \citep{rees98,sarimeszaros00}, e.g.
\begin{equation}
M(>\gamma) \propto \gamma^{-s}
\end{equation}
As the blastwave decelerates, ejecta with lower $\gamma$ gradually
piles up onto the blastwave so that the energy of the blastwave
is increased. Since energy is injected when
$\Gamma \sim \gamma$, the reverse shock is very weak, one can
treat the blastwave as a system with continuous energy injection.

The two energy injection mechanisms can be considered equivalent
when bridging the two injection parameter $s$ and $q$, i.e.,
\begin{equation}
 s = \frac{10-3k-7q+2kq}{2+q-k}, ~~~ q=\frac{10-2s-3k+ks}{7+s-2k}
\end{equation}
for general density profile $n_1=AR^{-k}$. For the ISM model and
wind model, it becomes $s = \frac{10-7q}{2+q}, ~~~
q=\frac{10-2s}{7+s}$ and $s = \frac{4-3q}{q}, ~~~ q=\frac{4}{3+s}$
respectively \citep{zhang06}.

In the following, we derive all the expressions using the parameter $q$.
For the ISM model ($k=0$) and $p>2$, one has
\begin{eqnarray}
&&\nu_m= 1.37\times10^{18}~{\rm Hz}~\hat z^{q/2}E_{52}^{1/2}\epsilon_{e,-1}^{2}\epsilon_{B,-2}^{1/2}t_{}^{-1-q/2}           ,\nonumber\\
&&\nu_c=  9.2\times10^{18}~{\rm Hz}~\hat z^{-q/2}E_{52}^{-1/2}n_{0,0}^{-1}\epsilon_{B,-2}^{-3/2}t_{}^{-1+q/2} ,          \nonumber\\
&&F_{\rm{\nu,max}}=   1.1\times10^{4}~\mu {\rm
Jy}~\hat z^{q}E_{52}^{}n_{0,0}^{1/2}\epsilon_{B,-2}^{1/2}D_{28}^{-2}t_{}^{1-q}          ,\nonumber\\
&&\nu_a=5.7\times10^{9}~{\rm Hz}~\hat
z^{\frac{q-6}{5}}\frac{g^{I}(p)}{g^{I}(2.3)}E_{52}^{1/5}n_{0,0}^{3/5}\epsilon_{e,-1}^{-1}\epsilon_{B,-2}^{1/5}t_{}^{\frac{1-q}{5}},
~~~~~~~~~~~~~~~~~~~~~~~~~~
\nu_a < \nu_m < \nu_c\nonumber  \\
&&\nu_a=5.0\times10^{13}~{\rm Hz}~\hat
z^{\frac{(p+2)q-8}{2(p+4)}}\frac{g^{II}(p)}{g^{II}(2.3)}E_{52}^{\frac{p+2}{2(p+4)}}n_{0,0}^{\frac{2}{p+4}}\epsilon_{e,-1}^
{\frac{2(p-1)}{p+4}}\epsilon_{B,-2}^{\frac{p+2}{2(p+4)}}t_{}^{-\frac{2p+(p+2)q}{2(p+4)}},
~~~~
\nu_m < \nu_a < \nu_c\nonumber  \\
&&\nu_a=2.2\times10^{9}~{\rm Hz}~\hat
z^{\frac{7q-12}{10}}\frac{g^{III}(p)}{g^{III}(2.3)}E_{52}^{7/10}n_{0,0}^{11/10}\epsilon_{B,-2}^{6/5}t_{}^{-\frac{7q-2}{10}},
~~~~~~~~~~~~~~~~~~~~~~~~
\nu_a < \nu_c < \nu_m\nonumber \\
\end{eqnarray}

For $1<p<2$, one has ($\nu_c$ and $F_{\rm \nu.max}$ remain the same)
\begin{eqnarray}
&&\nu_m=  2.9\times10^{16}~{\rm Hz}~\hat z^{\frac{pq-6p+2q+12}{8(p-1)}}\frac{g^{IV}(p)}{g^{IV}(1.8)}E_{52}^{\frac{p+2}{8(p-1)}}n_{0,0}^{\frac{2-p}{8(p-1)}}\zeta_{0}^{\frac{2-p}{p-1}}\epsilon_{e,-1}^{\frac{2}{p-1}}\epsilon_{B,-2}^{\frac{1}{2(p-1)}}t_{}^{-\frac{(q+2)(p+2)}{8(p-1)}}           ,\nonumber\\
&&\nu_a=3.2\times10^{10}~{\rm Hz}~\hat
z^{\frac{31pq-66p-46q+36}{8(p-1)}}\frac{g^{V}(p)}{g^{V}(1.8)}E_{52}^{\frac{46-31p}{80(1-p)}}n_{0,0}^{\frac{58-53p}{80(1-p)}}\zeta_{0}^{\frac{p-2}{2(p-1)}}
\epsilon_{e,-1}^{\frac{1}{1-p}}\epsilon_{B,-2}^{\frac{14-9p}{20(1-p)}}t_{}^{\frac{44-14p+46q-31pq}{80(p-1)}},\nonumber\\
&&~~~~~~~~~~~~~~~~~~~~~~~~~~~~~~~~~~~~~~~~~~~~~~~~~~~~~~~~~~~~~~~~~~~~~~~~~~~~~~~~~~~~~~~~~~~~~~~~
\nu_a < \nu_m < \nu_c\nonumber \\
&&\nu_a=1.1\times10^{13}~{\rm Hz}~\hat
z^{\frac{pq-6p+14q-20}{8(p+4)}}\frac{g^{VI}(p)}{g^{VI}(1.8)}E_{52}^{\frac{p+14}{8(p+4)}}n_{0,0}^{\frac{18-p}{8(p+4)}}\zeta_{0}^{\frac{2-p}{p+4}}
\epsilon_{e,-1}^{\frac{2}{p+4}}\epsilon_{B,-2}^{\frac{2}{p+4}}t_{}^{-\frac{p(q+2)+2(7q+6)}{8(p+4)}},\nonumber\\
&&~~~~~~~~~~~~~~~~~~~~~~~~~~~~~~~~~~~~~~~~~~~~~~~~~~~~~~~~~~~~~~~~~~~~~~~~~~~~~~~~~~~~~~~~~~~~~~~~\nu_m < \nu_a < \nu_c\nonumber \\
&&\nu_a=1.8\times10^{9}~{\rm Hz}~\hat
z^{\frac{7q-12}{10}}\frac{g^{VII}(p)}{g^{VII}(1.8)}E_{52}^{7/10}n_{0,0}^{11/10}\epsilon_{B,-2}^{6/5}t_{}^{-\frac{7q-2}{10}},
~~~~~~~~~~~~~~~~~~~~~~
\nu_a < \nu_c < \nu_m\nonumber \\
\end{eqnarray}

For the wind model ($k=2$) and $p>2$, one has
\begin{eqnarray}
&&\nu_m= 7.0\times10^{17}~{\rm Hz}~\hat z^{q/2}E_{52}^{1/2}\epsilon_{e,-1}^{2}\epsilon_{B,-2}^{1/2}t_{}^{-1-q/2}           ,\nonumber\\
&&\nu_c= 5.8\times10^{15}~{\rm Hz}~\hat z^{q/2-2}E_{52}^{1/2}A_{*,-1}^{-2}\epsilon_{B,-2}^{-3/2}t_{}^{1-q/2}           \nonumber\\
&&F_{\rm{\nu,max}}=  4.9\times10^{5}~\mu {\rm
Jy}~\hat z^{\frac{q+2}{2}}E_{52}^{1/2}A_{*,-1}^{}\epsilon_{B,-2}^{1/2}D_{28}^{-2}t_{}^{-q/2}             ,\nonumber\\
&&\nu_a=1.0\times10^{12}~{\rm Hz}~\hat
z^{-\frac{2q}{5}}\frac{g^{VIII}(p)}{g^{VIII}(2.3)}E_{52}^{-2/5}A_{*,-1}^{6/5}\epsilon_{e,-1}^{-1}\epsilon_{B,-2}^{1/5}t_{}^{-1+2q/5},
~~~~~~~~~~~~~~~~
\nu_a < \nu_m < \nu_c\nonumber \\
&&\nu_a=5.8\times10^{14}~{\rm Hz}~\hat
z^{\frac{(p-2)q}{2(p+4)}}\frac{g^{IX}(p)}{g^{IX}(2.3)}E_{52}^{\frac{p-2}{2(p+4)}}A_{*,-1}^{\frac{4}{p+4}}\epsilon_{e,-1}^{\frac{2(p-1)}{p+4}}
\epsilon_{B,-2}^{\frac{p+2}{2(p+4)}}t_{}^{-1-\frac{(p-2)q}{2(p+4)}},
~~~~~~~~~
\nu_m < \nu_a < \nu_c\nonumber \\
&&\nu_a=1.2\times10^{13}~{\rm Hz}~\hat
z^{\frac{5-2q}{5}}\frac{g^{X}(p)}{g^{X}(2.3)}E_{52}^{-2/5}A_{*,-1}^{11/5}\epsilon_{B,-2}^{6/5}t_{}^{2q/5-2},
~~~~~~~~~~~~~~~~~~~~~~~~~~
\nu_a < \nu_c < \nu_m\nonumber \\
\end{eqnarray}

For $1<p<2$, one has ($\nu_c$ and $F_{\rm \nu.max}$ remain the same)
\begin{eqnarray}
&&\nu_m= 1.7\times10^{16}~{\rm Hz}~\hat z^{\frac{pq-4p+8}{4(p-1)}}\frac{g^{XI}(p)}{g^{XI}(1.8)}E_{52}^{\frac{p}{4(p-1)}}A_{*,-1}^{\frac{2-p}{4(p-1)}}\zeta_{0}^{\frac{2-p}{p-1}}\epsilon_{e,-1}^{\frac{2}{p-1}}\epsilon_{B,-2}^{\frac{1}{2(p-1)}}t_{}^{\frac{4+pq}{4(1-p)}}           ,\nonumber\\
&&\nu_a= 5.5\times10^{12}~{\rm Hz}~\hat
z^{\frac{120-100p-6q+11pq}{40(p-1)}}\frac{g^{XII}(p)}{g^{XII}(1.8)}
E_{52}^{\frac{6-11p}{40(p-1)}}A_{*,-1}^{\frac{58-53p}{40(1-p)}}\zeta_{0}^{\frac{p-2}{2(p-1)}}
\epsilon_{e,-1}^{\frac{1}{1-p}}\epsilon_{B,-2}^{\frac{14-9p}{20(1-p)}}t_{}^{\frac{20p-40+6q-11pq}{40(p-1)}},\nonumber\\
&&~~~~~~~~~~~~~~~~~~~~~~~~~~~~~~~~~~~~~~~~~~~~~~~~~~~~~~~~~~~~~~~~~~~~~~~~~~~~~~~~~~~~~~~~~~~~~~~~~
\nu_a < \nu_m < \nu_c\nonumber \\
&&\nu_a= 1.7\times10^{14}~{\rm Hz}~\hat
z^{\frac{(p-2)q-4p+8}{4(p+4)}}\frac{g^{XIII}(p)}{g^{XIII}(1.8)}
E_{52}^{\frac{p-2}{4(p+4)}}A_{*,-1}^{\frac{18-p}{4(p+4)}}\zeta_{0}^{\frac{2-p}{p+4}}
\epsilon_{e,-1}^{\frac{2}{p+4}}\epsilon_{B,-2}^{\frac{2}{p+4}}t_{}^{-\frac{(p-2)q+24}{4(p+4)}},\nonumber\\
&&~~~~~~~~~~~~~~~~~~~~~~~~~~~~~~~~~~~~~~~~~~~~~~~~~~~~~~~~~~~~~~~~~~~~~~~~~~~~~~~~~~~~~~~~~~~~~~~~~
\nu_m < \nu_a < \nu_c\nonumber \\
&&\nu_a=9.5\times10^{12}~{\rm Hz}~\hat z^{\frac{5-2q}{5}}
\frac{g^{XIV}(p)}{g^{XIV}(1.8)}
E_{52}^{-2/5}A_{*,-1}^{11/5}\epsilon_{B,-2}^{6/5}t_{}^{2q/5-2},
~~~~~~~~~~~~~~~~~~~~~~
\nu_a < \nu_c < \nu_m\nonumber \\
\end{eqnarray}

The $\alpha$ and $\beta$ values and their
closure relations for these models are also presented in Tables
\ref{adia_table} to \ref{BMpl22}.

For this model (adiabatic deceleration without energy injection),
for $p>2$, one has $\nu_m \propto t^{-1-q/2}~(t^{-1-q/2})$, $\nu_c
\propto t^{-1+q/2}~(t^{1-q/2})$, $F_{\rm \nu,max} \propto t^{1-q}
~(t^{-q/2})$ for the ISM (wind) models, respectively. For $1<p<2$,
$\nu_c$ and $F_{\rm{\nu,max}}$ evolutions are the same as $p>2$
cases, while $\nu_m \propto t^{\frac{(q+2)(p+2)}{8(1-p)}}~
(t^{\frac{4+pq}{4(1-p)}})$ for the ISM (wind) models, respectively.

\begin{table}
\caption{The temporal decay index $\alpha$ and spectral index $\beta$ in
relativistic, isotropic, self-similar deceleration phase for
$\nu_a < \rm{min}(\nu_m,\nu_c)$ and $p>2$.\label{adia_table}}
\begin{tabular}{llllll}
\hline\hline
& & no injection & &  injection & \\

& $\beta$ & $\alpha $  &  $\alpha (\beta)$ & $\alpha$  & $\alpha (\beta)$  \\
\hline
ISM & slow cooling  \\
\hline
$\nu<\nu_a$   &  $-2$  &   $-{1\over 2}$ & $\alpha={\beta \over 4}$  &$\frac{q}{2}-1$ & $--$\\
$\nu_a<\nu<\nu_m$   &  $-{1 \over 3}$  &   $-{1\over 2}$ & $\alpha={3\beta \over 2}$ & ${5q-8 \over 6}$  & $--$\\
$\nu_m<\nu<\nu_c$   &  ${{p-1 \over 2}}$  &  ${3(p-1)\over 4}$
   &  $\alpha={3\beta \over 2}$ & ${(2p-6)+(p+3)q \over 4}$   & $\alpha=(q-1)+\frac{(2+q)\beta}{2}$\\
$\nu>\nu_c$   &  ${{p\over 2}}$   &   ${3p-2 \over 4}$  & $\alpha={3\beta-1 \over 2}$ & ${(2p-4)+(p+2)q\over 4}$   & $\alpha=\frac{q-2}{2}+\frac{(2+q)\beta}{2}$ \\

\hline
ISM & fast cooling \\
\hline
$\nu<\nu_a$   &  $-2$  &   $-1$ & $\alpha=\frac{\beta}{2}$ & $-1$  & $\alpha=\frac{\beta}{2}$\\
$\nu_a<\nu<\nu_c$   &  $-{1\over 3}$  &   $-{1\over 6}$ &  $\alpha={\beta \over 2}$  &  ${7q-8 \over 6}$  & $--$\\
$\nu_c<\nu<\nu_m$   &  ${1\over 2}$  &  ${1\over 4}$  & $\alpha={\beta \over 2}$ &  ${3q-2 \over 4}$   &  $--$ \\
$\nu>\nu_m$   &  ${p\over 2}$   &   ${3p-2\over 4}$      &  $\alpha={3\beta-1 \over 2}$  &  ${(2p-4)+(p+2)q\over 4}$  & $\alpha=\frac{q-2}{2}+\frac{(2+q)\beta}{2}$ \\

\hline
Wind & slow cooling \\
\hline
$\nu<\nu_a$   &  $-2$   &   -1  &  $\alpha={\beta \over 2}$ & $q-2$  & $--$\\
$\nu_a<\nu<\nu_m$   &  $-{1\over 3}$   &   0  &  0 & $--$  & $--$\\
$\nu_m<\nu<\nu_c$   &  ${p-1\over 2}$  &   ${3p-1\over 4}$    &   $\alpha={3\beta+1 \over 2}$ & ${(2p-2)+(p+1)q \over 4}$ & $\alpha=\frac{q}{2}+\frac{(2+q)\beta}{2}$\\
$\nu>\nu_c$   &  ${p\over 2}$   &   ${3p-2\over 4}$    &  $\alpha={3\beta-1 \over 2}$   &   ${(2p-4)+(p+2)q\over 4}$  &  $\alpha=\frac{q-2}{2}+\frac{(2+q)\beta}{2}$\\

\hline
Wind & fast cooling  \\
\hline
$\nu<\nu_a$   &  $-2$   &   $-2$ & $\alpha=\beta$ &  $q-3$  & $--$  \\
$\nu_a<\nu<\nu_c$   &  $-{1\over 3}$   &   ${2\over 3}$ & $\alpha=-2\beta$ &  ${(1+q) \over 3}$  & $--$  \\
$\nu_c<\nu<\nu_m$   &  ${1\over 2}$  &   ${1\over 4}$  & $\alpha={\beta \over 2}$  & ${3q-2 \over 4}$ & $--$\\
$\nu>\nu_m$   &  ${p\over 2}$  &   ${3p-2\over 4}$     & $\alpha={3\beta-1 \over 2}$  &  ${(2p-4)+(p+2)q\over 4}$  & $\alpha=\frac{q-2}{2}+\frac{(2+q)\beta}{2}$\\
\hline
\end{tabular}
\label{Tab:alpha-beta}
\end{table}

\begin{table}

\caption{The temporal decay index $\alpha$ and spectral index $\beta$ in
relativistic, isotropic, self-similar deceleration phase
for $\nu_m <\nu_a < \nu_c$ and $p>2$.\label{adia_table2}}
\begin{tabular}{llllll}
\hline\hline
& & no injection & &  injection & \\

& $\beta$ & $\alpha $  &  $\alpha (\beta)$ & $\alpha$ & $\alpha (\beta)$  \\
\hline
ISM & slow cooling  \\
\hline
$\nu<\nu_m$   &  $-2$  &   $-{1\over 2}$ & $\alpha={\beta \over 4}$ & $\frac{q}{2}-1$  & $--$\\
$\nu_m<\nu<\nu_a$   &  $-{5 \over 2}$  &   $-{5\over 4}$ & $\alpha={\beta \over 2}$ & ${q-6 \over 4}$  & $--$\\
$\nu_a<\nu<\nu_c$   &  ${{p-1 \over 2}}$   &  ${3(p-1)\over 4}$
  &  $\alpha={3\beta \over 2}$ & ${(2p-6)+(p+3)q \over 4}$  & $\alpha=(q-1)+\frac{(2+q)\beta}{2}$\\
$\nu>\nu_c$   &  ${{p\over 2}}$   &   ${3p-2 \over 4}$  & $\alpha={3\beta-1 \over 2}$ & ${(2p-4)+(p+2)q\over 4}$  & $\alpha=\frac{q-2}{2}+\frac{(2+q)\beta}{2}$ \\

\hline
Wind & slow cooling \\
\hline
$\nu<\nu_m$   &  $-2$   &   -1  &  $\alpha={\beta \over 2}$ & ${q-2}$  & $--$\\
$\nu_m<\nu<\nu_a$   &  $-{5\over 2}$   &   $-{7\over 4}$  &  $\alpha={7\beta \over 10}$ & ${3q-10 \over 4}$  & $--$\\
$\nu_a<\nu<\nu_c$   &  ${p-1\over 2}$   &   ${3p-1\over 4}$    &   $\alpha={3\beta+1 \over 2}$ & ${(2p-2)+(p+1)q \over 4}$ & $\alpha=\frac{q}{2}+\frac{(2+q)\beta}{2}$\\
$\nu>\nu_c$   &  ${p\over 2}$   &   ${3p-2\over 4}$    &  $\alpha={3\beta-1 \over 2}$   &   ${(2p-4)+(p+2)q\over 4}$  &  $\alpha=\frac{q-2}{2}+\frac{(2+q)\beta}{2}$\\

\hline
\end{tabular}
\label{Tab:alpha-beta}
\end{table}

\begin{table}
\caption{The temporal decay index $\alpha$ and spectral index $\beta$ in
relativistic, isotropic, self-similar deceleration phase
for $\nu_a < \rm{min}(\nu_m,\nu_c)$ and $1<p<2$.\label{BMpl2}}
\begin{tabular}{llllll}
\hline\hline
& & no injection & &  injection & \\

& $\beta$ & $\alpha $  &  $\alpha (\beta)$ & $\alpha$  & $\alpha (\beta)$  \\
\hline
ISM & slow cooling  \\
\hline
$\nu<\nu_a$   &  $-2$  &   ${26-17p\over 16(p-1)}$ & $--$  &$\frac{28-22p-2q+5pq}{16(p-1)}$ & $--$\\
$\nu_a<\nu<\nu_m$   &  $-{1 \over 3}$  &   $-{p+2\over 8(p-1)}$ & $--$ & $\frac{20-26p-26q+23pq}{24(p-1)}$  & $--$\\
$\nu_m<\nu<\nu_c$   &  ${{p-1 \over 2}}$  &  ${3(p+2)\over 16}$
   &  $\alpha={6\beta+9 \over 16}$ & $-{12-18q-p(q+2) \over 16}$   & $\alpha=\frac{19q-10}{16}+\frac{(2+q)\beta}{8}$\\
$\nu>\nu_c$   &  ${{p\over 2}}$   &   ${3p+10 \over 16}$  & $\alpha={3\beta+5 \over 8}$ & ${14q+p(q+2)-4\over 16}$   & $\alpha=\frac{7q-2}{8}+\frac{(2+q)\beta}{8}$ \\

\hline
ISM & fast cooling \\
\hline
$\nu<\nu_a$   &  $-2$  &   $-1$ & $\alpha=\frac{\beta}{2}$ & $-1$  & $\alpha=\frac{\beta}{2}$\\
$\nu_a<\nu<\nu_c$   &  $-{1\over 3}$  &   $-{1\over 6}$ &  $\alpha={\beta \over 2}$  &  ${7q-8 \over 6}$  & $--$\\
$\nu_c<\nu<\nu_m$   &  ${1\over 2}$  &  ${1\over 4}$  & $\alpha={\beta \over 2}$ &  ${3q-2 \over 4}$   &  $--$ \\
$\nu>\nu_m$   &  ${p\over 2}$   &   ${3p+10 \over 16}$      &  $\alpha={3\beta+5 \over 8}$  &  ${14q+p(q+2)-4\over 16}$  & $\alpha=\frac{7q-2}{8}+\frac{(2+q)\beta}{8}$ \\

\hline
Wind & slow cooling \\
\hline
$\nu<\nu_a$   &  $-2$   &   $\frac{13p-18}{8(1-p)}$  &  $--$ & $\frac{20p+6q-7pq-24}{8(1-p)}$  & $--$\\
$\nu_a<\nu<\nu_m$   &  $-{1\over 3}$   &   $\frac{5(p-2)}{12(p-1)}$  &  $--$ & $\frac{4+6q-5pq}{12(1-p)}$  & $--$\\
$\nu_m<\nu<\nu_c$   &  ${p-1\over 2}$  &   ${p+8\over 8}$    &   $\alpha={2\beta+9 \over 8}$ & ${4+(p+4)q \over 8}$ & $\alpha=\frac{5q+4}{8}+\frac{\beta q}{4}$\\
$\nu>\nu_c$   &  ${p\over 2}$   &   ${p+6\over 8}$    &  $\alpha={2\beta+7 \over 8}$   &   ${(6+p)q\over 8}$  &  $\alpha=\frac{(\beta+3)q}{4}$\\

\hline
Wind & fast cooling  \\
\hline
$\nu<\nu_a$   &  $-2$   &   $-2$ & $\alpha=\beta$ &  $q-3$  & $--$  \\
$\nu_a<\nu<\nu_c$   &  $-{1\over 3}$   &   ${2\over 3}$ & $\alpha=-2\beta$ &  ${1+q \over 3}$  & $--$  \\
$\nu_c<\nu<\nu_m$   &  ${1\over 2}$  &   ${1\over 4}$  & $\alpha={\beta \over 2}$  & ${3q-2 \over 4}$ & $--$\\
$\nu>\nu_m$   &  ${p\over 2}$  &   ${p+6\over 8}$     & $\alpha={2\beta+7 \over 8}$  &  ${(6+p)q\over 8}$  & $\alpha=\frac{(\beta+3)q}{4}$\\
\hline
\end{tabular}
\label{Tab:alpha-beta}
\end{table}

\begin{table}

\caption{The temporal decay index $\alpha$ and spectral index $\beta$ in
relativistic, isotropic, self-similar deceleration phase for
$\nu_m <\nu_a < \nu_c$ and $1<p<2$.
\label{BMpl22}}
\begin{tabular}{llllll}
\hline\hline
& & no injection & &  injection & \\

& $\beta$ & $\alpha $  &  $\alpha (\beta)$ & $\alpha$ & $\alpha (\beta)$  \\
\hline
ISM & slow cooling  \\
\hline
$\nu<\nu_m$   &  $-2$  &   ${26-17p \over 16(p-1)}$ & $--$ & $\frac{28-22p-2q+5pq}{16(p-1)}$  & $--$\\
$\nu_m<\nu<\nu_a$   &  $-{5 \over 2}$  &   $-{5\over 4}$ & $\alpha={\beta \over 2}$ & ${q-6 \over 4}$  & $--$\\
$\nu_a<\nu<\nu_c$   &  ${{p-1 \over 2}}$   &  ${3(p+2)\over 16}$
  &  $\alpha={6\beta+9 \over 16}$ & ${18q+p(q+2)-12 \over 16}$  & $\alpha=\frac{19q-10}{16}+\frac{(2+q)\beta}{8}$\\
$\nu>\nu_c$   &  ${{p\over 2}}$   &   ${3p+10 \over 16}$  & $\alpha={3\beta+5 \over 8}$ & ${14q+p(q+2)-4\over 16}$  & $\alpha=\frac{7q-2}{8}+\frac{(2+q)\beta}{8}$ \\

\hline
Wind & slow cooling \\
\hline
$\nu<\nu_m$   &  $-2$   &  $\frac{13p-18}{8(1-p)}$  &  $--$ & $\frac{20p+6q-7pq-24}{8(1-p)}$  & $-$\\
$\nu_m<\nu<\nu_a$   &  $-{5\over 2}$   &   $-{7\over 4}$  &  $\alpha={7\beta \over 10}$ & ${3q-10 \over 4}$  & $--$\\
$\nu_a<\nu<\nu_c$   &  ${p-1\over 2}$   &   ${p+8\over 8}$    &   $\alpha={2\beta+9 \over 8}$ & ${4+(p+4)q \over 8}$ & $\alpha=\frac{5q+4}{8}+\frac{\beta q}{4}$\\
$\nu>\nu_c$   &  ${p\over 2}$   &   ${6+p\over 8}$    &  $\alpha={2\beta+7 \over 8}$   &   ${(6+p)q\over 8}$  &  $\alpha=\frac{(\beta+3)q}{4}$\\

\hline
\end{tabular}
\label{Tab:alpha-beta}
\end{table}

\subsection{Phase 3: Post Jet Break Phase}

The above calculations are based on the assumption
of a spherical expansion. However, achromatic breaks seen in many
afterglow lightcurves suggest that GRB outflows are collimated.
For a simplified conical jet model with an opening angle $\theta_j$,
the jet effect becomes important when $1/\Gamma > \theta_j$.
The lightcurve shows a steepening break around this time.

In the literature, two effects have been discussed to steepen the
lightcurve. The first is the pure edge effect
\citep[e.g.][]{panaitescu98}. Since an observer sees emission within
the $1/\Gamma$ cone for a blastwave moving with bulk Lorentz factor
$\Gamma$, he/she would feel the deficit of flux outside the
$\theta_j$ cone when $1/\Gamma > \theta_j$ is satisfied. Assuming
that the dynamics does not change, the flux reduction factor would
be $\theta_j^{2}/(1/\Gamma)^2 = \Gamma^2 \theta_j^2$. This defines
the degree of steepening at the jet break.

The second effect discussed in the literature is the sideway
expansion effect. According to \citep{rhoads99,sari99},
when $\Gamma\sim\theta_j^{-1}$ is satisfied, sound
waves in the jet would cross the jet in the transverse direction and
lead to its sideways expansion. This leads to a exponentially
deceleration of the jet. However, later numerical simulations,
and more sophisticated analytical
treatments suggest that sideways expansion is not important
until $\Gamma$ drops below a few
\citep{kumar03,cannizzo04,zhangmacfadyen09,granotpiran12}.
We therefore do not discuss this effect.

For the edge effect only, in the post-jet-break phase the
expressions of the break frequencies $\nu_a$, $\nu_m$ and $\nu_c$
and the peak flux density $ F_{\rm{\nu,max}}$ all remain the same as
the isotropic phase. The temporal decay indices are changed with the
extra steepening correction factor. In rare cases, continuous energy
injection may extend to the post-jet-break phase. For completeness,
we also discuss these cases. After shock crossing, the reverse
shocked region decelerates with a different dynamics from the
forward shocked region. Given a same jet opening angle, it
corresponds to an earlier jet break time. In Table \ref{tjet}, we
present the expressions of jet break time and the temporal indices
changes ($\Delta \alpha$ defined as post-jet-break $\alpha_2$ minus
pre-jet-break $\alpha_1$) for all the models in different regimes.

In Tables \ref{Edgeeffect} and \ref{Edgeeffect2}, we present
$\alpha$ and $\beta$ values and their closure relations for the jet
model. Since the reverse shock jet break is usually undetectable,
only forward shock models are presented.

\begin{table}[tbhp]
 \caption{Collection of jet break time and temporal indices changes $\Delta \alpha= \alpha_2 - \alpha_1$ for different regimes.}
 \begin{center}{\scriptsize
 \begin{tabular}{c|c|c} \hline\hline
                                                &     $t_{jet}$                                                                                                       & $\Delta\alpha $     \\
 \hline
 $\rm Thin RS_{post}$ (ISM)                 \ \ & \ \ $2.8\times10^4~{\rm s}~\hat z E_{52}^{1/3}\theta_{j,-1}^{5/2}n_0^{-1/3}\Gamma_{0,2}^{-1/6}$\ \                                  & \ \ $4/5$\ \         \\
\hline
 $\rm Thin RS_{post}$ (wind)                \ \ & \ \ $2.9\times10^3~{\rm s}~\hat z E_{52}^{}\theta_{j,-1}^{3}A_{*,-1}^{-1}\Gamma_{0,2}^{-1}$\ \                                      & \ \ $2/3$\ \         \\
     \hline
 $\rm Thick RS_{post}$  (ISM)               \ \ & \ \ $1.2\times10^4~{\rm s}~\hat z E_{52}^{2/7}\theta_{j,-1}^{16/7}n_0^{-2/7}\Delta_{0,12}^{1/7}$\ \                                 & \ \ $7/8$\ \           \\
  \hline
 $\rm Thick RS_{post}$ (Wind)               \ \ & \ \ $1.9\times10^3~{\rm s}~\hat z E_{52}^{2/3}\theta_{j,-1}^{8/3}A_{*,-1}^{-2/3}\Delta_{0,12}^{1/3}$\ \                             & \ \ $3/4$\ \          \\
\hline
 $\rm FS$  (ISM,  no injection)             \ \ & \ \ $5.8\times10^3~{\rm s}~\hat z E_{52}^{1/3}\theta_{j,-1}^{8/3}n_0^{-1/3}$\ \                                                     & \ \ $3/4$\ \           \\
     \hline
 $\rm FS$  (wind, no injection)             \ \ & \ \ $1.7\times10^4~{\rm s}~\hat z E_{52}\theta_{j,-1}^{4}A_{*,-1}^{-1}$\ \                                                          & \ \ $1/2$\ \           \\
  \hline
 $\rm FS$  (ISM,  injection)                \ \ & \ \ $2.0\times10^{\frac{11}{2+q}}~{\rm s}~\hat z E_{52}^{\frac{1}{2+q}}\theta_{j,-1}^{\frac{8}{2+q}}n_0^{-\frac{1}{2+q}}$\ \        & \ \ $(2+q)/4$\ \           \\
     \hline
 $\rm FS$  (wind, injection)                \ \ & \ \ $1.7\times10^{\frac{4}{q}}~{\rm s}~\hat z E_{52}^{\frac{1}{q}}\theta_{j,-1}^{\frac{4}{q}}A_{*,-1}^{-\frac{1}{q}}$\ \                                            & \ \ $q/2$\ \           \\
   \hline\hline
 \end{tabular}
 }
 \end{center}
 \label{tjet}
 \end{table}

\begin{table}
\caption{The temporal decay index $\alpha$ and spectral index $\beta$ after
jet break for $\nu_a < \rm{min}(\nu_m,\nu_c)$, considering edge
effect only.\label{Edgeeffect}}
\begin{tabular}{llllll}
\hline\hline
& & $p>2$ & &  $1<p<2$ & \\

& $\beta$ & $\alpha $  &  $\alpha (\beta)$ & $\alpha$  & $\alpha (\beta)$  \\
\hline
ISM & no injection\\
\hline
$\nu<\nu_a$   &  $-2$  &   ${1 \over 4}$ & $\alpha={\beta \over 8}$  &${14-5p \over 16(p-1)}$ & $--$\\
$\nu_a<\nu<\nu_m$   &  $-{1 \over 3}$  &   ${1 \over 4}$ & $\alpha={3\beta \over 4}$ & ${5p-8 \over 8(p-1)}$  & $--$\\
$\nu_m<\nu<\nu_c$   &  ${{p-1 \over 2}}$  &  ${3p \over 4}$
   &  $\alpha={6\beta+3 \over 4}$ & ${3(p+6) \over 16}$   & $\alpha={3(2\beta+7) \over 16}$\\
$\nu>\nu_c$   &  ${{p\over 2}}$   &   ${3p+1 \over 4}$  & $\alpha={6\beta+1 \over 4}$ & ${3p+22 \over 16}$   & $\alpha={3\beta+11 \over 8}$ \\

\hline Wind & no injection\\
\hline
$\nu<\nu_a$   &  $-2$   &   $-{1 \over 2}$  &  $\alpha={\beta \over 4}$ & $\frac{14-9p}{8(p-1)}$  & $--$\\
$\nu_a<\nu<\nu_m$   &  $-{5\over 2}$   &   ${1\over 2}$  &  $\alpha={\beta \over 5}$ & ${11p-16 \over 12(p-1)}$  & $--$\\
$\nu_m<\nu<\nu_c$   &  ${p-1\over 2}$   &   ${3p+1\over 4}$    &   $\alpha={3\beta+2 \over 2}$ & ${p+12 \over 8}$ & $\alpha={2\beta+13 \over 8}$\\
$\nu>\nu_c$   &  ${p\over 2}$   &   ${3p\over 4}$    &  $\alpha={3\beta \over 2}$   &   ${p+10 \over 8}$  &  $\alpha={\beta+5 \over 4}$\\

\hline
ISM & injection\\
\hline
$\nu<\nu_a$   &  $-2$  &   ${3q-2 \over 4}$ & $--$  &${20-14p-6q+9pq \over 16(p-1)}$ & $--$\\
$\nu_a<\nu<\nu_m$   &  $-{1 \over 3}$  &   ${13q-10 \over 12}$ & $--$ & ${8-14p-32q+29pq \over 24(p-1)}$  & $--$\\
$\nu_m<\nu<\nu_c$   &  ${{p-1 \over 2}}$  &  ${p(q+2)-4(1-q) \over
4}$
   &  $\alpha=\frac{5q-2}{4}+\frac{(2+q)\beta}{2}$ & ${22q-4+p(q+2) \over 16}$   & $\alpha=\frac{11q-2}{8}+\frac{(2+q)\beta}{8}$\\
$\nu>\nu_c$   &  ${{p\over 2}}$   &   ${3q-2+p(q+2) \over 4}$  & $\alpha={3q-2+2\beta(q+2) \over 4}$ & ${18q+4+p(q+2) \over 16}$   & $\alpha={9q+2+\beta(q+2) \over 8}$ \\

\hline
Wind & injection\\
\hline
$\nu<\nu_a$   &  $-2$   &   ${3q-4 \over 2}$  &  $--$ & $\frac{24-20p-10q+11pq}{8(p-1)}$  & $--$\\
$\nu_a<\nu<\nu_m$   &  $-{5\over 2}$   &   ${5q-2\over 6}$  &  $--$ & ${11pq-12q-4 \over 12(p-1)}$  & $--$\\
$\nu_m<\nu<\nu_c$   &  ${p-1\over 2}$   &   ${3q-2+p(q+2)\over 4}$    &   $\alpha=q+\frac{(2+q)\beta}{2}$ & ${pq+8q+4 \over 8}$ & $\alpha=\frac{1}{2}+\frac{2\beta+9}{8}$\\
$\nu>\nu_c$   &  ${p\over 2}$   &   ${p(q+2)-4(1-q)\over 4}$    &  $\alpha={\beta(q+2)-2(1-q)\over 2}$   &   ${(p+10)q \over 8}$  &  $\alpha={(\beta+5)q \over 4}$\\

\hline
\end{tabular}
\label{Tab:alpha-beta}
\end{table}

\begin{table}
\caption{The temporal decay index $\alpha$ and spectral index $\beta$ after
jet break for $\nu_m < \nu_a <\nu_c$, considering edge effect
only.\label{Edgeeffect2}}
\begin{tabular}{llllll}
\hline\hline
& & $p>2$ & &  $1<p<2$ & \\

& $\beta$ & $\alpha $  &  $\alpha (\beta)$ & $\alpha$  & $\alpha (\beta)$  \\
\hline
ISM & no injection\\
\hline
$\nu<\nu_m$   &  $-2$  &   ${1 \over 4}$ & $\alpha={\beta \over 8}$  &${14-5p \over 16(p-1)}$ & $--$\\
$\nu_m<\nu<\nu_a$   &  $-{1 \over 3}$  &   $-{1 \over 2}$ & $\alpha={3\beta \over 2}$ & $-{1 \over 2}$  & $\alpha={3\beta \over 2}$\\
$\nu_a<\nu<\nu_c$   &  ${{p-1 \over 2}}$  &  ${3p \over 4}$
   &  $\alpha={6\beta+3 \over 4}$ & ${3(p+6) \over 16}$   & $\alpha={3(2\beta+7) \over 16}$\\
$\nu>\nu_c$   &  ${{p\over 2}}$   &   ${3p+1 \over 4}$  & $\alpha={6\beta+1 \over 4}$ & ${3p+22 \over 16}$   & $\alpha={3\beta+11 \over 8}$ \\

\hline Wind & no injection\\
\hline
$\nu<\nu_m$   &  $-2$   &   $-{1 \over 2}$  &  $\alpha={\beta \over 4}$ & $\frac{14-9p}{8(p-1)}$  & $--$\\
$\nu_m<\nu<\nu_a$   &  $-{5\over 2}$   &   $-{5\over 4}$  &  $\alpha={\beta \over 2}$ & $-{5\over 4}$  & $\alpha={\beta \over 2}$\\
$\nu_a<\nu<\nu_c$   &  ${p-1\over 2}$   &   ${3p+1\over 4}$    &   $\alpha={3\beta+2 \over 2}$ & ${p+12 \over 8}$ & $\alpha={2\beta+13 \over 8}$\\
$\nu>\nu_c$   &  ${p\over 2}$   &   ${3p\over 4}$    &  $\alpha={3\beta \over 2}$   &   ${p+10 \over 8}$  &  $\alpha={\beta+5 \over 4}$\\

\hline
ISM & injection\\
\hline
$\nu<\nu_m$   &  $-2$  &   ${3q-2 \over 4}$ & $--$  &${20-14p-6q+9pq \over 16(p-1)}$ & $--$\\
$\nu_m<\nu<\nu_a$   &  $-{1 \over 3}$  &   ${q-2 \over 2}$ & $--$ & ${q-2 \over 2}$  & $--$\\
$\nu_a<\nu<\nu_c$   &  ${{p-1 \over 2}}$  &  ${p(q+2)-4(1-q) \over
4}$
   &  $\alpha=\frac{5q-2}{4}+\frac{(2+q)\beta}{2}$ & ${22q-4+p(q+2) \over 16}$   & $\alpha=\frac{11q-2}{8}+\frac{(2+q)\beta}{8}$\\
$\nu>\nu_c$   &  ${{p\over 2}}$   &   ${3q-2+p(q+2) \over 4}$  & $\alpha={3q-2+2\beta(q+2) \over 4}$ & ${18q+4+p(q+2) \over 16}$   & $\alpha={9q+2+\beta(q+2) \over 8}$ \\

\hline
Wind & injection\\
\hline
$\nu<\nu_m$   &  $-2$   &   ${3q-4 \over 2}$  &  $--$ & $\frac{24-20p-10q+11pq}{8(p-1)}$  & $--$\\
$\nu_m<\nu<\nu_a$   &  $-{5\over 2}$   &   ${5(q-2)\over 4}$  &  $--$ & ${5(q-2)\over 4}$  & $--$\\
$\nu_a<\nu<\nu_c$   &  ${p-1\over 2}$   &   ${3q-2+p(q+2)\over 4}$    &   $\alpha=q+\frac{(2+q)\beta}{2}$ & ${pq+8q+4 \over 8}$ & $\alpha=\frac{1}{2}+\frac{2\beta+9}{8}$\\
$\nu>\nu_c$   &  ${p\over 2}$   &   ${p(q+2)-4(1-q)\over 4}$    &  $\alpha={\beta(q+2)-2(1-q)\over 2}$   &   ${(p+10)q \over 8}$  &  $\alpha={(\beta+5)q \over 4}$\\

\hline
\end{tabular}
\label{Tab:alpha-beta}
\end{table}

\subsection{Phase 4: Newtonian Phase}

The blastwave eventually enters the Newtonian phase when it has swept
up a CBM mass comparable to the initial mass entrained in the ejecta.
In the deep Newtonian phase, the dynamics is
described by the well known Sedov-Taylor solution:
\begin{eqnarray}
R=\left(\frac{5-k}{2}\right)^{\frac{2}{5-k}}\left[\frac{(3-k)E}{2\pi
Am_p}\right]^{\frac{1}{5-k}}t^{\frac{2}{5-k}},~~~~~~~v=\left(\frac{5-k}{2}\right)^{\frac{k-3}{5-k}}\left[\frac{(3-k)E}{2\pi
Am_p}\right]^{\frac{1}{5-k}}t^{\frac{k-3}{5-k}}
\end{eqnarray}
This phase has been studied extensively in the literature
\citep{wijers97,dailu99,huang99,huang00,livio00,huang03}.

In this phase, for an ISM medium and $p>2$, one has
\begin{eqnarray}
&&\nu_m=   2.0\times10^{14}~{\rm Hz}~\hat z^{2}\frac{G(p)}{G(2.3)}E_{52}^{}n_{0,0}^{-1/2}\epsilon_{e,-1}^{2}\epsilon_{B,-2}^{1/2}t_{5}^{-3}           ,\nonumber\\
&&\nu_c=    7.0\times10^{15}~{\rm Hz}~\hat z^{-4/5}E_{52}^{-3/5}n_{0,0}^{-9/10}\epsilon_{B,-2}^{-3/2}t_{5}^{-1/5}           \nonumber\\
&&F_{\rm{\nu,max}}=     2.3\times10^{2}~\mu {\rm
Jy}~\hat z^{2/5} E_{52}^{4/5}n_{0,0}^{7/10}\epsilon_{B,-2}^{1/2}D_{28}^{-2}t_{5}^{3/5}          ,\nonumber\\
&&\nu_a=1.4\times10^{7}~{\rm Hz}~\hat
z^{-11/5}\frac{g^{I}(p)}{g^{I}(2.3)}E_{52}^{-1/5}n_{0,0}^{}\epsilon_{e,-1}^{-1}\epsilon_{B,-2}^{1/5}t_{5}^{6/5},
~~~~~~~~~~~~~~~~~~~~~~~
\nu_a < \nu_m < \nu_c\nonumber \\
&&\nu_a=3.3\times10^{10}~{\rm Hz}~\hat
z^{\frac{2p-6}{p+4}}\frac{g^{II}(p)}{g^{II}(2.3)}E_{52}^{\frac{p}{p+4}}n_{0,0}^{\frac{6-p}{2(p+4)}}
\epsilon_{e,-1}^{\frac{2(p-1)}{p+4}}\epsilon_{B,-2}^{\frac{p+2}{2(p+4)}}t_{5}^{-\frac{3p-2}{p+4}},
~~~~~~~~~~~~
\nu_m < \nu_a < \nu_c\nonumber \\
\end{eqnarray}

For $1<p<2$, one has ($\nu_c$ and $F_{\rm \nu.max}$ remain the same)
\begin{eqnarray}
&&\nu_m=   1.9\times10^{12}~{\rm Hz}~\hat z^{\frac{4-p}{p-1}}\frac{g^{III}(p)}{g^{III}(1.8)}E_{52}^{\frac{1}{p-1}}n_{0,0}^{\frac{1}{2(1-p)}}\zeta_{0}^{\frac{2-p}{p-1}}\epsilon_{e,-1}^{\frac{2}{p-1}}\epsilon_{B,-2}^{\frac{1}{2(p-1)}}t_{5}^{-\frac{3}{p-1}}           ,\nonumber\\
&&\nu_a=1.2\times10^{8}~{\rm Hz}\frac{g^{IV}(p)}{g^{IV}(1.8)}\hat
z^{\frac{7p+8}{10(p-1)}}E_{52}^{\frac{8-3p}{10(1-p)}}n_{0,0}^{\frac{2-3p}{4(1-p)}}\zeta_{0}^{\frac{p-2}{2(p-1)}}\epsilon_{e,-1}^{\frac{1}{1-p}}\epsilon_{B,-2}^{\frac{14-9p}{20(1-p)}}t_{5}^{-\frac{3(p-6)}{10(p-1)}},
~~~~
\nu_a < \nu_m < \nu_c\nonumber \\
&&\nu_a=7.4\times10^{9}~{\rm Hz}~\hat
z^{-\frac{p}{p+4}}\frac{g^{V}(p)}{g^{V}(1.8)}E_{52}^{\frac{2}{p+4}}n_{0,0}^{\frac{2}{p+4}}\zeta_{0}^{\frac{2-p}{p+4}}
\epsilon_{e,-1}^{\frac{2}{p+4}}\epsilon_{B,-2}^{\frac{2}{p+4}}t_{5}^{-\frac{4}{p+4}},
~~~~~~~~~~~~~~~~~~~~
\nu_m < \nu_a < \nu_c\nonumber \\
\end{eqnarray}

For the wind model and $p>2$, one has
\begin{eqnarray}
&&\nu_m=   1.6\times10^{14}~{\rm Hz}~\hat z^{4/3}\frac{G(p)}{G(2.3)}E_{52}^{4/3}A_{*,-1}^{-5/6}\epsilon_{e,-1}^{2}\epsilon_{B,-2}^{1/2}t_{5}^{-7/3}           ,\nonumber\\
&&\nu_c=    1.7\times10^{15}~{\rm Hz}~\hat z^{-2}A_{*,1}^{-3/2}\epsilon_{B,-2}^{-3/2}t_{5}^{}           \nonumber\\
&&F_{\rm{\nu,max}}=     5.3\times10^{2}~\mu {\rm
Jy}~\hat z^{4/3} E_{52}^{1/3}A_{*,-1}^{7/6}\epsilon_{B,-2}^{1/2}D_{28}^{-2}t_{5}^{-1/3}          ,\nonumber\\
&&\nu_a=6.9\times10^{7}~{\rm Hz}~\hat
z^{-13/15}\frac{g^{VI}(p)}{g^{VI}(2.3)}E_{52}^{-13/15}A_{*,-1}^{5/3}\epsilon_{e,-1}^{-1}\epsilon_{B,-2}^{1/5}t_{5}^{-2/15},
~~~~~~~~~~~~~~~~~~
\nu_a < \nu_m < \nu_c\nonumber \\
&&\nu_a=6.9\times10^{10}~{\rm Hz}~\hat
z^{\frac{4p-6}{3(p+4)}}\frac{g^{VII}(p)}{g^{VII}(2.3)}E_{52}^{\frac{2(2p-3)}{3(p+4)}}A_{*,-1}^{\frac{5(6-p)}{6(p+4)}}
\epsilon_{e,-1}^{\frac{2(p-1)}{p+4}}\epsilon_{B,-2}^{\frac{p+2}{2(p+4)}}t_{5}^{-\frac{7p+6}{3(p+4)}},
~~~~~~~~
\nu_m < \nu_a < \nu_c\nonumber \\
\end{eqnarray}

For $1<p<2$, one has ($\nu_c$ and $F_{\rm \nu.max}$ remain the same)
\begin{eqnarray}
&&\nu_m=  1.4\times10^{12}~{\rm Hz}~\hat z^{\frac{10-3p}{3(p-1)}}\frac{g^{VIII}(p)}{g^{VIII}(1.8)}E_{52}^{\frac{4}{3(p-1)}}A_{*,-1}^{\frac{5}{6(1-p)}}\zeta_{0}^{\frac{2-p}{p-1}}\epsilon_{e,-1}^{\frac{2}{p-1}}\epsilon_{B,-2}^{\frac{1}{2(p-1)}}t_{5}^{-\frac{7}{3(p-1)}}           ,\nonumber\\
&&\nu_a=6.0\times10^{8}~{\rm Hz}~\hat
z^{\frac{9p-44}{30(p-1)}}\frac{g^{IX}(p)}{g^{IX}(1.8)}E_{52}^{\frac{3p+7}{15(1-p)}}A_{*,-1}^{\frac{5(3p-2)}{12(p-1)}}\zeta_{0}^{\frac{p-2}{2(p-1)}}\epsilon_{e,-1}^{\frac{1}{1-p}}\epsilon_{B,-2}^{\frac{14-9p}{20(1-p)}}t_{5}^{\frac{74-39p}{30(p-1)}},
~~~
\nu_a < \nu_m < \nu_c\nonumber \\
&&\nu_a=1.6\times10^{10}~{\rm Hz}~\hat
z^{\frac{8-3p}{3p+4}}\frac{g^{X}(p)}{g^{X}(1.8)}E_{52}^{\frac{2}{3(p+4)}}A_{*,-1}^{\frac{10}{3(p+4)}}\zeta_{0}^{\frac{2-p}{p+4}}
\epsilon_{e,-1}^{\frac{2}{p+4}}\epsilon_{B,-2}^{\frac{2}{p+4}}t_{5}^{-\frac{20}{3(p+4)}},
~~~~~~~~~~~~~
\nu_m < \nu_a < \nu_c\nonumber \\
\end{eqnarray}

The $\alpha$ and $\beta$ values and their closure relations
in this phase are presented in Tables
\ref{Newtonian} and \ref{Newtonian1}.

For this model (newtonian Phase), for $p>2$, one has $\nu_m \propto
t^{-3}~(t^{-7/3})$, $\nu_c \propto t^{-1/5}~(t^{1})$, $F_{\rm
\nu,max} \propto t^{3/5} ~(t^{-1/3})$ for the ISM (wind) models,
respectively. For $1<p<2$, $\nu_c$ and $F_{\rm{\nu,max}}$ evolutions
are the same as $p>2$ cases, while $\nu_m \propto t^{\frac{3}{1-p}}~
(t^{\frac{7}{3(1-p)}})$ for the ISM (wind) models, respectively.

\begin{table}
\caption{The temporal decay index $\alpha$ and spectral index $\beta$ in
the Newtonian phase for $\nu_a <
\rm{min}(\nu_m,\nu_c)$.\label{Newtonian}}
\begin{tabular}{llllll}
\hline\hline
& & no injection & &  injection & \\

& $\beta$ & $\alpha $  &  $\alpha (\beta)$ & $\alpha$  & $\alpha (\beta)$  \\
\hline
ISM & slow cooling  \\
\hline
$\nu<\nu_a$   &  $-2$  &   ${2 \over 5}$ & $\alpha={\beta \over 5}$  &$\frac{26-11p}{10(p-1)}$ & $--$\\
$\nu_a<\nu<\nu_m$   &  $-{1 \over 3}$  &   $-{8 \over 5}$ & $\alpha={24\beta \over 5}$ & $-{3p+2 \over 5(p-1)}$  & $--$\\
$\nu_m<\nu<\nu_c$   &  ${{p-1 \over 2}}$  &  ${3(5p-7) \over 10}$
   &  $\alpha={3(5\beta-1) \over 5}$ & ${9 \over 10}$   & $--$\\
$\nu>\nu_c$   &  ${{p\over 2}}$   &   ${3p-4 \over 2}$  & $\alpha={3\beta-2}$ & $1$   & $--$ \\

\hline
Wind & slow cooling  \\
\hline
$\nu<\nu_a$   &  $-2$  &   $-{2 \over 3}$ & $\alpha={\beta \over 3}$  &$\frac{18-11p}{6(p-1)}$ & $--$\\
$\nu_a<\nu<\nu_m$   &  $-{1 \over 3}$  &   $-{4 \over 9}$ & $\alpha={4\beta \over 3}$ & ${3p-10 \over 9(p-1)}$  & $--$\\
$\nu_m<\nu<\nu_c$   &  ${{p-1 \over 2}}$  &  ${7p-5 \over 6}$
   &  $\alpha={7\beta+1 \over 3}$ & ${3 \over 2}$   & $--$\\
$\nu>\nu_c$   &  ${{p\over 2}}$   &   ${7p-8 \over 6}$  & $\alpha={7\beta-4 \over 3}$ & $1$   & $--$ \\

\hline
\end{tabular}
\label{Tab:alpha-beta}
\end{table}

\begin{table}
\caption{The temporal decay index $\alpha$ and spectral index $\beta$ in
the Newtonian phase for $\nu_m < \nu_a <\nu_c$.\label{Newtonian1}}
\begin{tabular}{llllll}
\hline\hline
& & no injection & &  injection & \\

& $\beta$ & $\alpha $  &  $\alpha (\beta)$ & $\alpha$  & $\alpha (\beta)$  \\
\hline
ISM & slow cooling  \\
\hline
$\nu<\nu_m$   &  $-2$  &   ${2 \over 5}$ & $\alpha={\beta \over 5}$  &$\frac{26-11p}{10(p-1)}$ & $--$\\
$\nu_m<\nu<\nu_a$   &  $-{1 \over 3}$  &   $-{11 \over 10}$ & $\alpha={33\beta \over 10}$ & $-{11 \over 10}$  & $\alpha={33\beta \over 10}$\\
$\nu_a<\nu<\nu_c$   &  ${{p-1 \over 2}}$  &  ${3(5p-7) \over 10}$
   &  $\alpha={3(5\beta-1) \over 5}$ & ${9 \over 10}$   & $--$\\
$\nu>\nu_c$   &  ${{p\over 2}}$   &   ${3p-4 \over 2}$  & $\alpha={3\beta-2}$ & $1$   & $--$ \\

\hline
Wind & slow cooling  \\
\hline
$\nu<\nu_m$   &  $-2$  &   $-{2 \over 3}$ & $\alpha={\beta \over 3}$  &$\frac{18-11p}{6(p-1)}$ & $--$\\
$\nu_m<\nu<\nu_a$   &  $-{1 \over 3}$  &   $-{11 \over 6}$ & $\alpha={11\beta \over 2}$ & $-{11 \over 6}$  & $\alpha={11\beta \over 2}$\\
$\nu_a<\nu<\nu_c$   &  ${{p-1 \over 2}}$  &  ${7p-5 \over 6}$
   &  $\alpha={7\beta+1 \over 3}$ & ${3 \over 2}$   & $--$\\
$\nu>\nu_c$   &  ${{p\over 2}}$   &   ${7p-8 \over 6}$  & $\alpha={7\beta-4 \over 3}$ & $1$   & $--$ \\

\hline
\end{tabular}
\label{Tab:alpha-beta}
\end{table}

\section{Applications of the models}

Section 3 gives a complete reference of all the possible analytical
synchrotron external shock models. There are two opposite ways of
applying this reference tool. First, one can fit the observational
data to get both temporal decay index $\alpha$ and spectral index
$\beta$, and then identify which spectral regime the observational
frequency lies in. One can then constrain related afterglow
parameters. To fully determine the parameters, one needs
multi-wavelength, multi-epoch observational data. In any case, for
the relativistic deceleration phase before the jet break, from which
most data are collected, usually a closure relation study could give
a quick judgement about the possible spectral regime and medium
type. Alternatively, one can start to assign reasonable ranges of a
set of model parameters, and apply the models to draw predicted
light curves. By varying parameters, one can use the model to fit
the observational data.

Since the three characteristic frequencies $\nu_a$, $\nu_m$, and $\nu_c$
all evolve with time, the order among them may change during the
evolution. The characteristic frequencies may also pass the observed
band, so that the observational spectral regime may also change.
These factors introduce complications in drawing theoretical lightcurves.
First, one needs to estimate how spectral regimes evolve with time,
using the related expressions of the characteristic frequencies;
Second, one needs to use the closure relation tables to find out
the temporal decay index for each segment of the light curve, and
then connect all the segments. Lightcurves can differ for different
dynamical models, different initial ordering of the characteristic
frequencies, and different spectral regimes.

\begin{figure}
  \includegraphics[angle=0 ,width=1.0\textwidth]{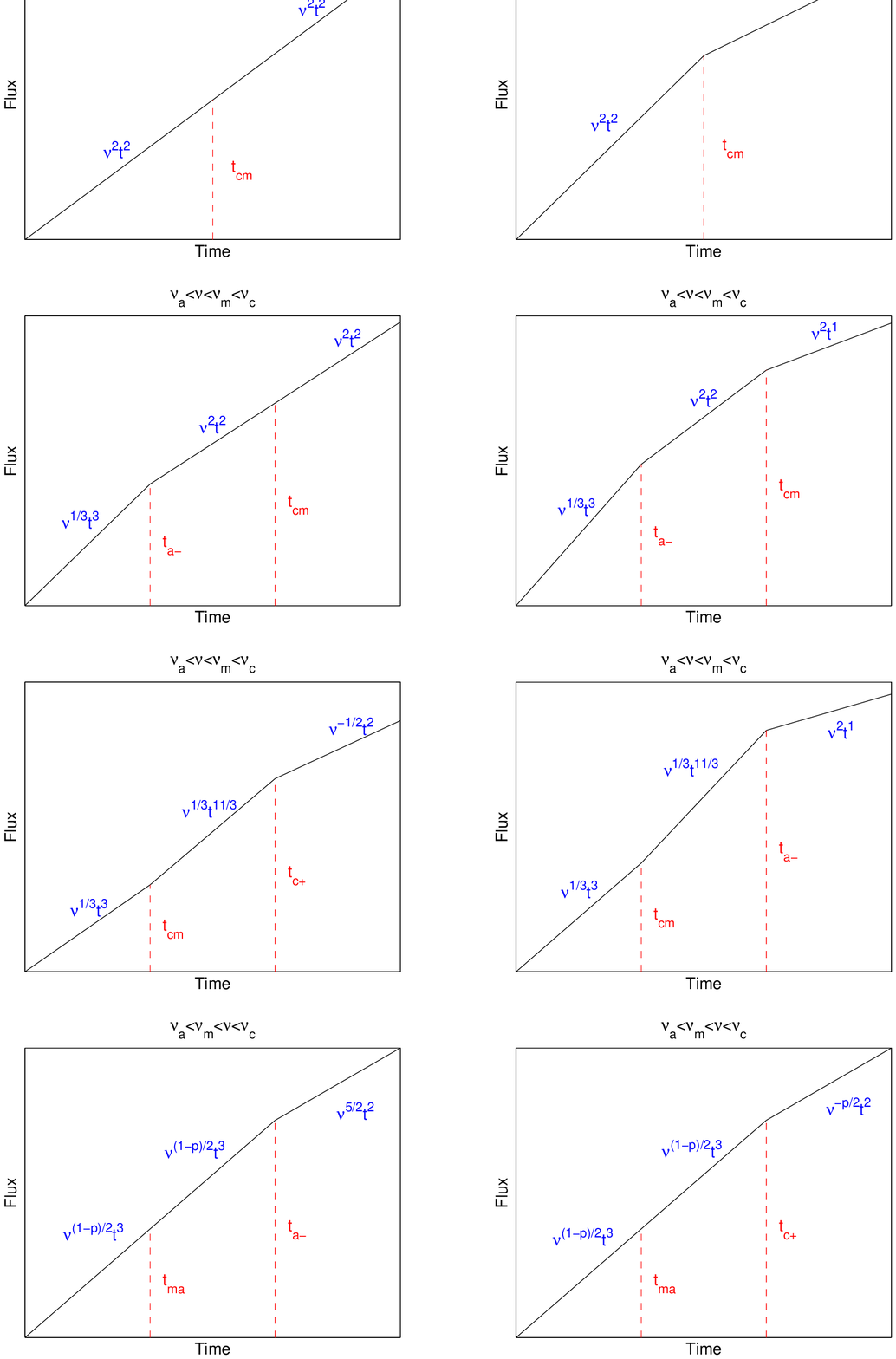}
    \caption{All possible forward shock lightcurves during Phase 1
(reverse shock crossing phase), for thin shell ISM model
and the initial characteristic frequency order $\nu_a < \nu_m < \nu_c$.
The notations $t_{i+}, i=a,m,c$ denote frequency regime change
from $\nu_i>\nu$ to $\nu_i<\nu$; $t_{i-}, i=a,m,c$ denote
frequency regime change from $\nu_i<\nu$ to $\nu_i>\nu$; $t_{ij},
\{i,j\}=a,m,c$ denote frequency regime change from $\nu_i>\nu_j$ to
$\nu_i<\nu_j$. The title for each sub-figure is the initial
spectral regime of the observed frequency $\nu$. }
  \label{fig11}
\end{figure}

\begin{figure}
  \includegraphics[angle=0,width=1.0\textwidth]{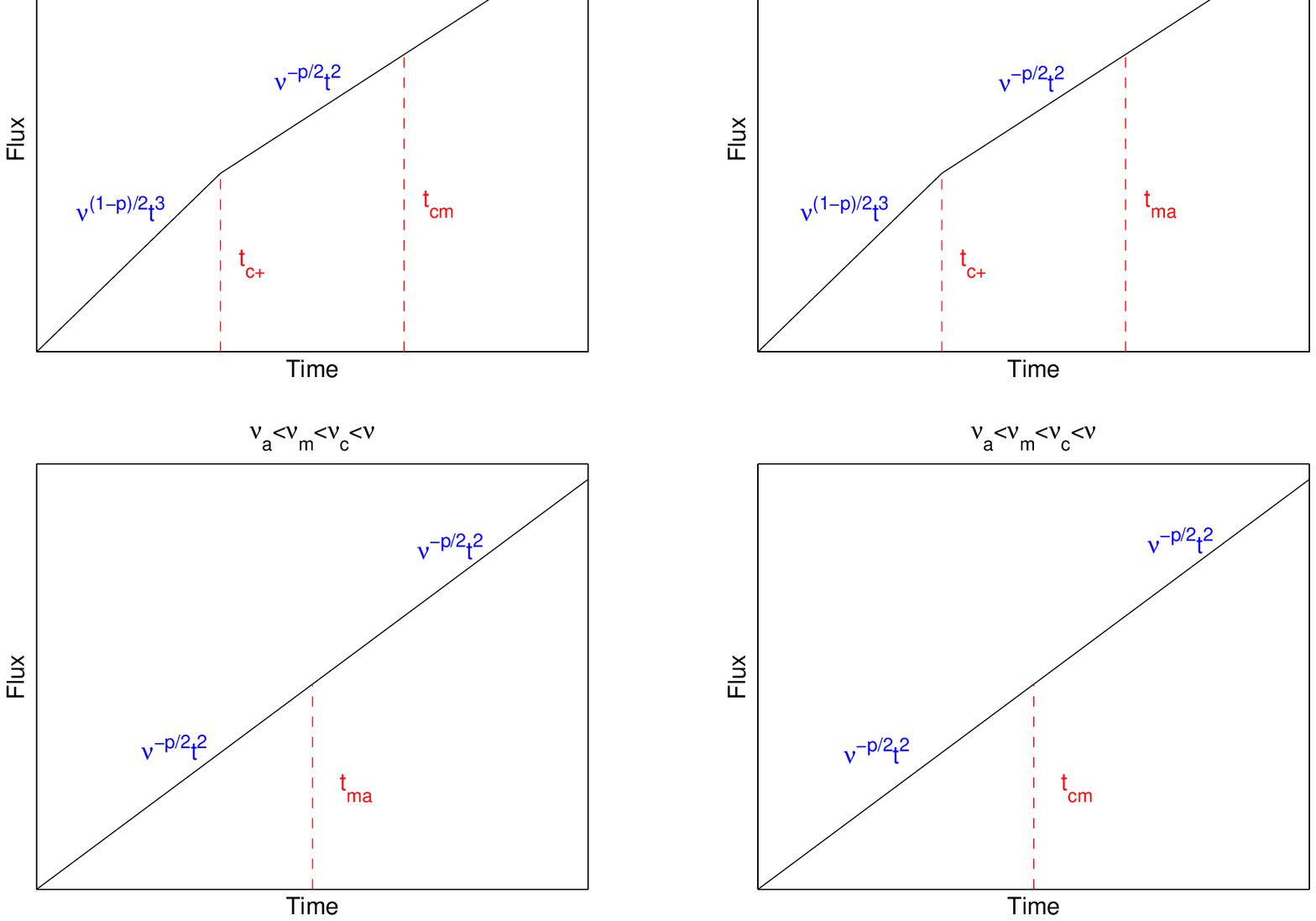}
    \caption{Figure 1 continued.}
  \label{fig11b}
\end{figure}

\begin{figure}
  \includegraphics[angle=0,width=1.0 \textwidth]{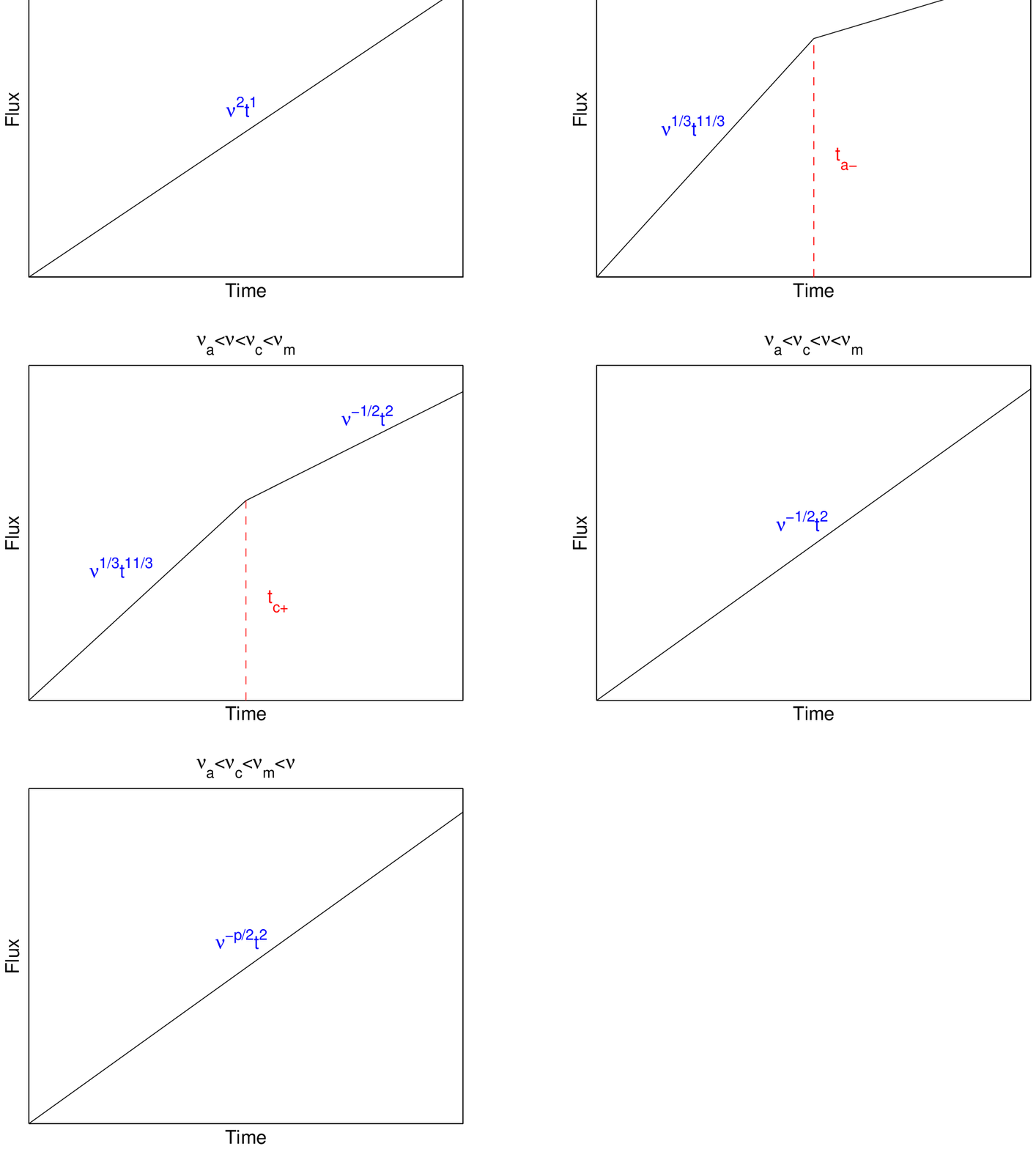}
    \caption{Same as Fig. 1, but with the initial characteristic frequency
order $\nu_a < \nu_c < \nu_m$.}
  \label{fig12}
\end{figure}

\begin{figure}
  \includegraphics[angle=0,width=1.0\textwidth]{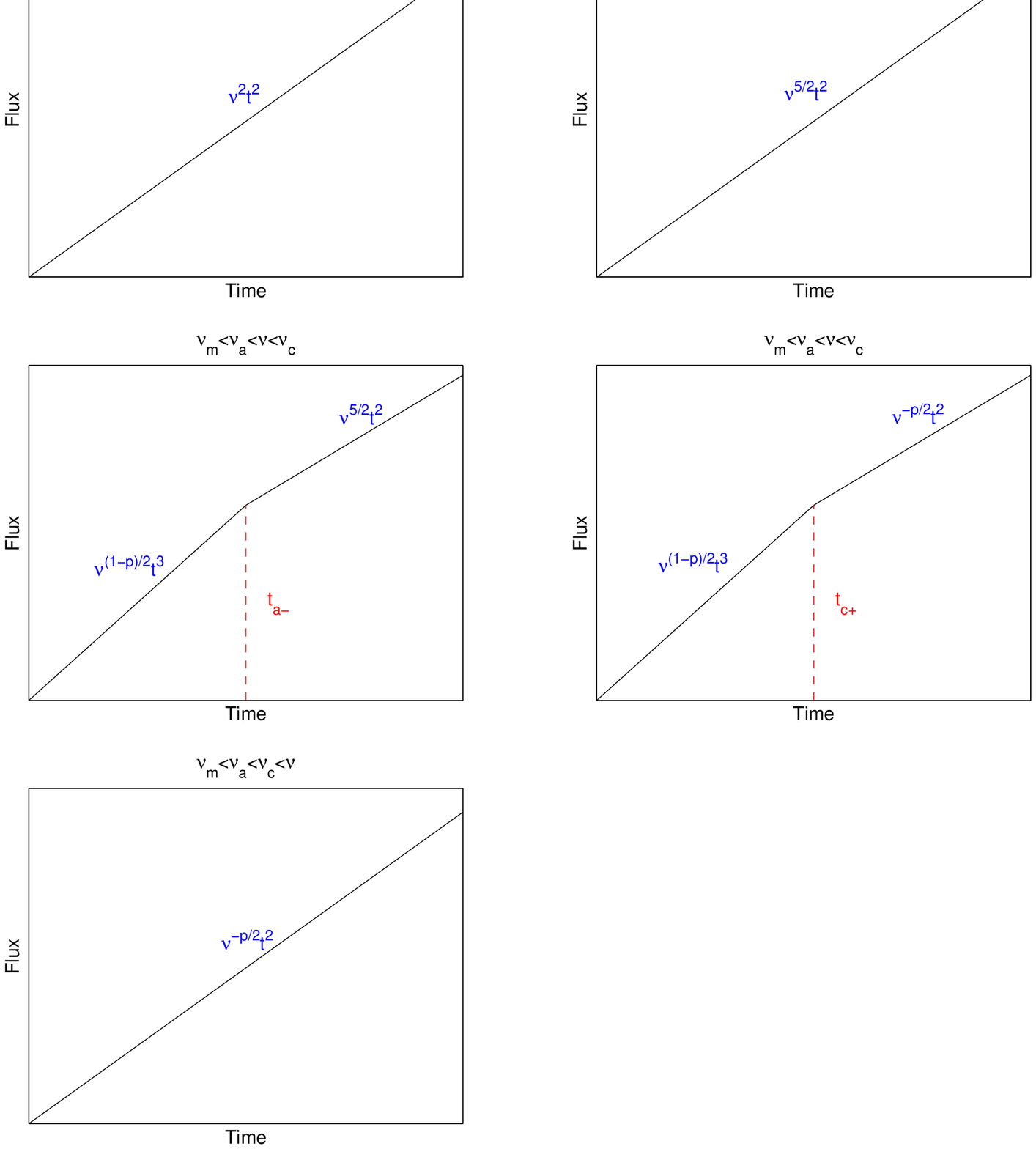}
    \caption{Same as Fig. 1, but with the initial characteristic frequency
order $\nu_m < \nu_a < \nu_c$.}
  \label{fig13}
\end{figure}

\begin{figure}
  \includegraphics[angle=0,width=1.0 \textwidth]{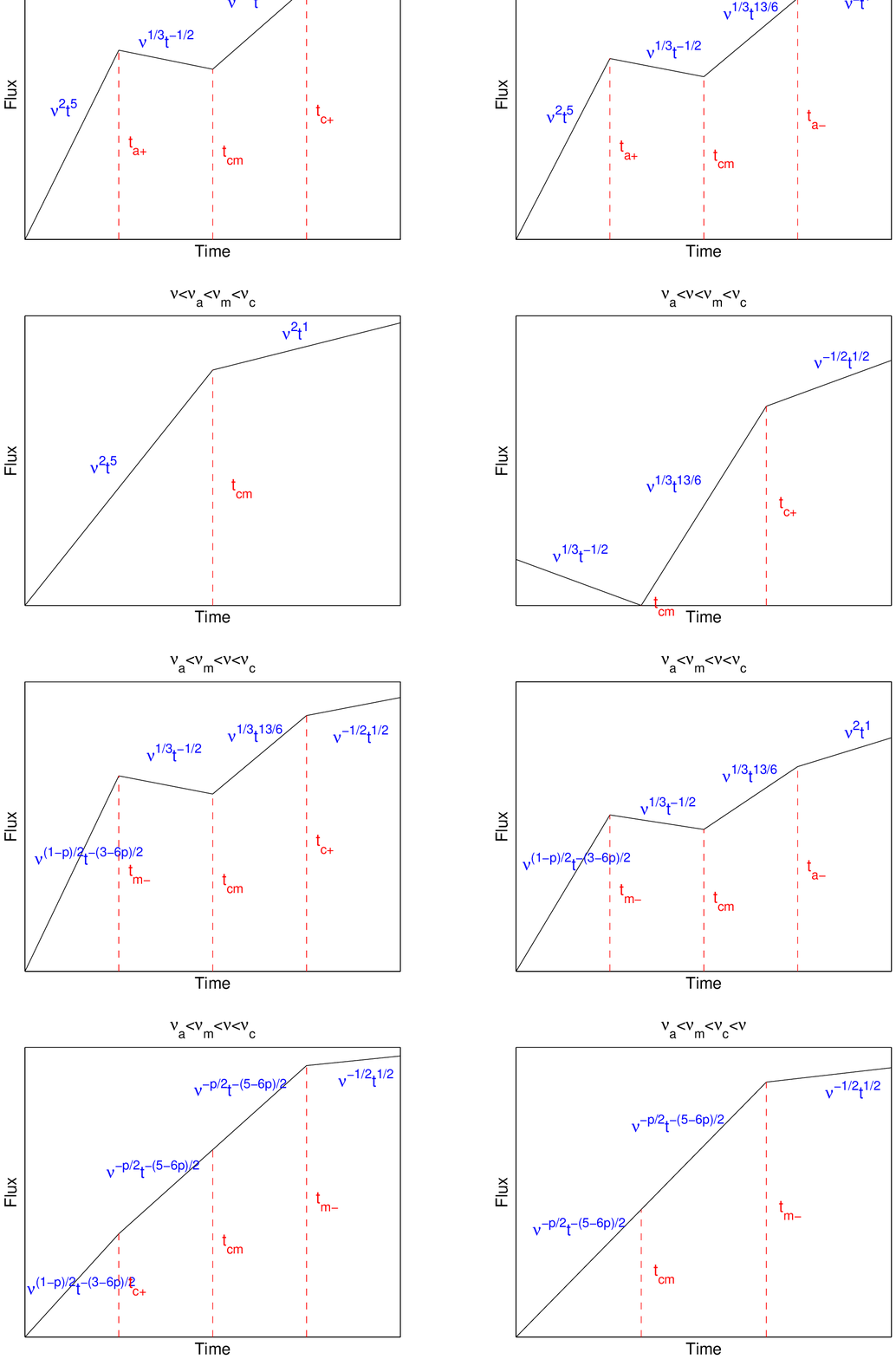}\\
    \caption{All possible reverse shock lightcurves during Phase 1
(reverse shock crossing phase), for thin shell ISM model
and the initial characteristic frequency order $\nu_a < \nu_m < \nu_c$.}
  \label{fig21}
\end{figure}

\begin{figure}
  \includegraphics[angle=0,width=1.0 \textwidth]{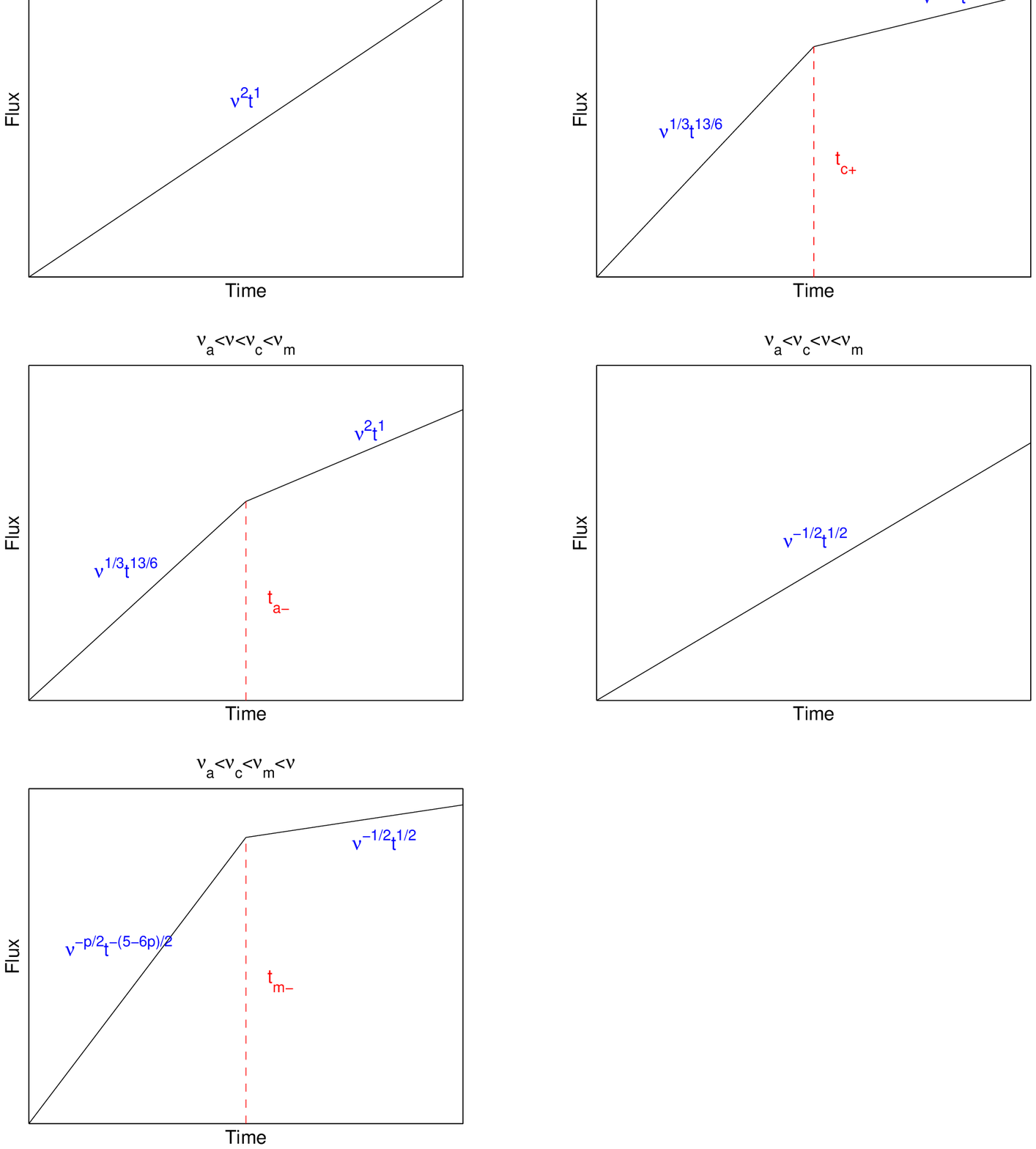}\\
    \caption{Same as Fig. \ref{fig21}, but with the initial characteristic
frequency order $\nu_a < \nu_c < \nu_m$.}
  \label{fig22}
\end{figure}

\begin{figure}
  \includegraphics[angle=0,width=1.0 \textwidth]{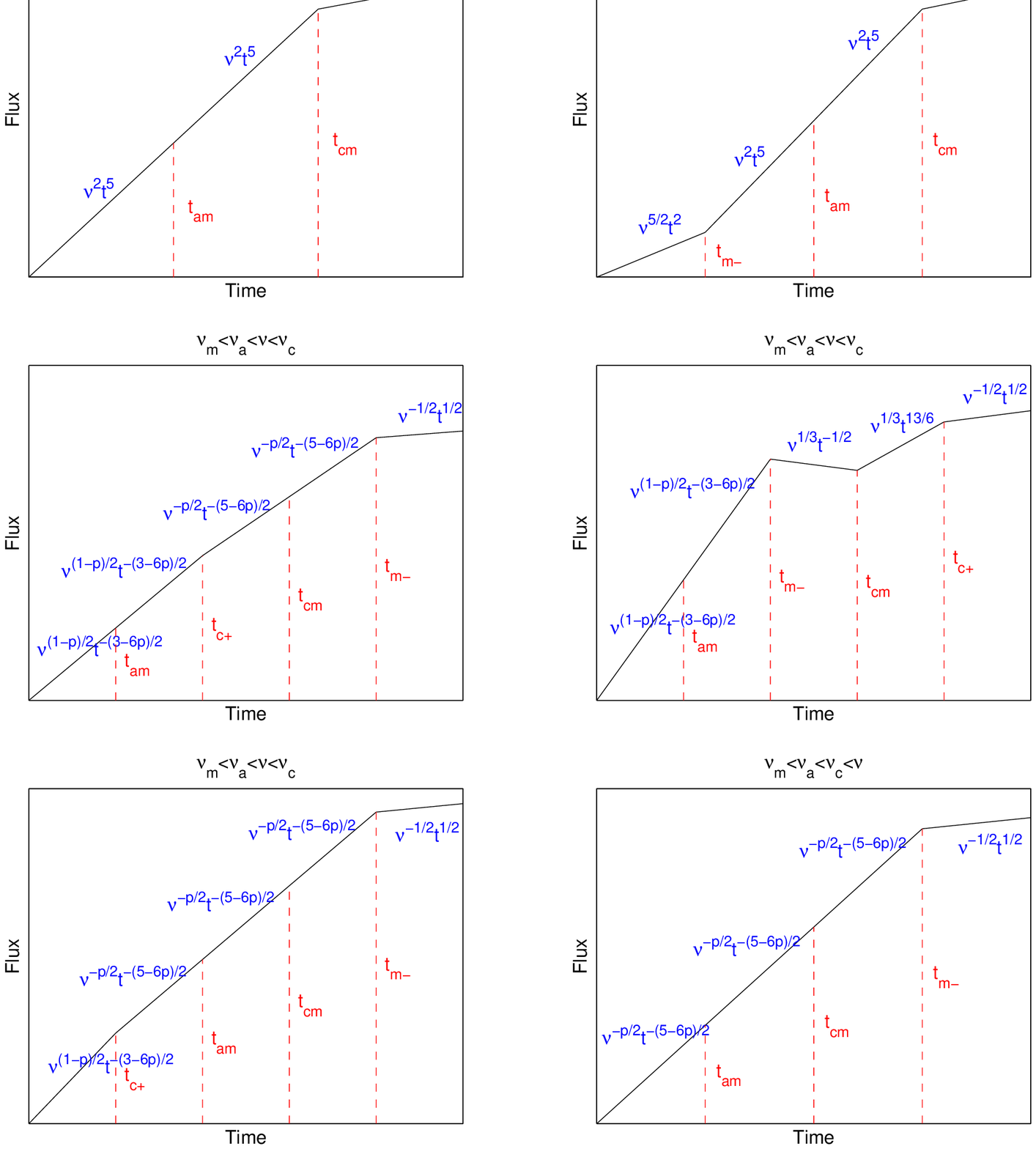}\\
    \caption{Same as Fig. \ref{fig21}, but with the initial characteristic
frequency order $\nu_m < \nu_a < \nu_c$.}
  \label{fig23}
\end{figure}

\begin{figure}
  \includegraphics[angle=0,width=1.0\textwidth]{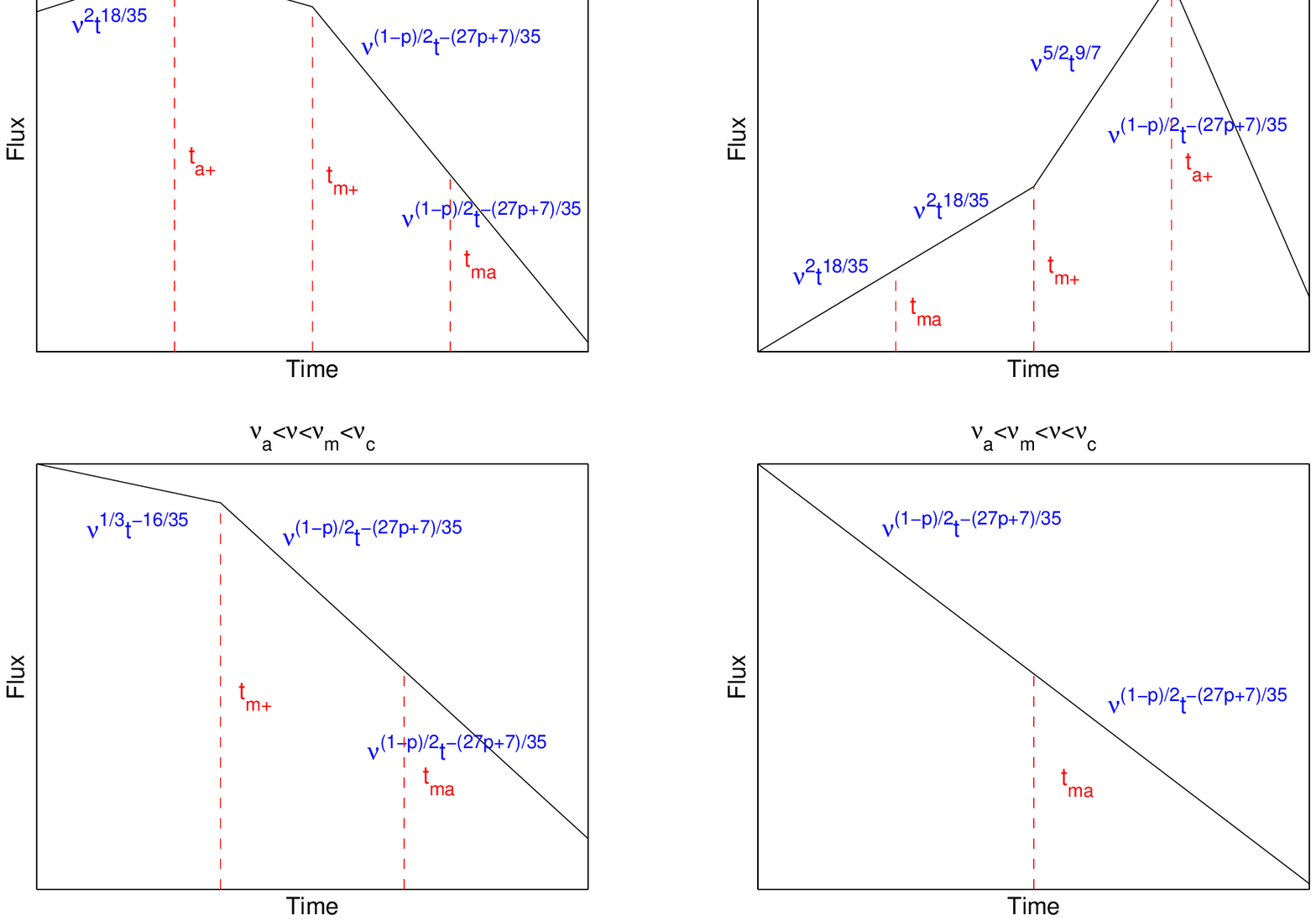}\\
    \caption{All possible reverse shock lightcurves after reverse
shock crossing the shell, for thin shell ISM model and the initial
characteristic frequency order $\nu_a < \nu_m < \nu_c$.}
  \label{fig31}
\end{figure}

\begin{figure}
  \includegraphics[angle=0,width=1.0\textwidth]{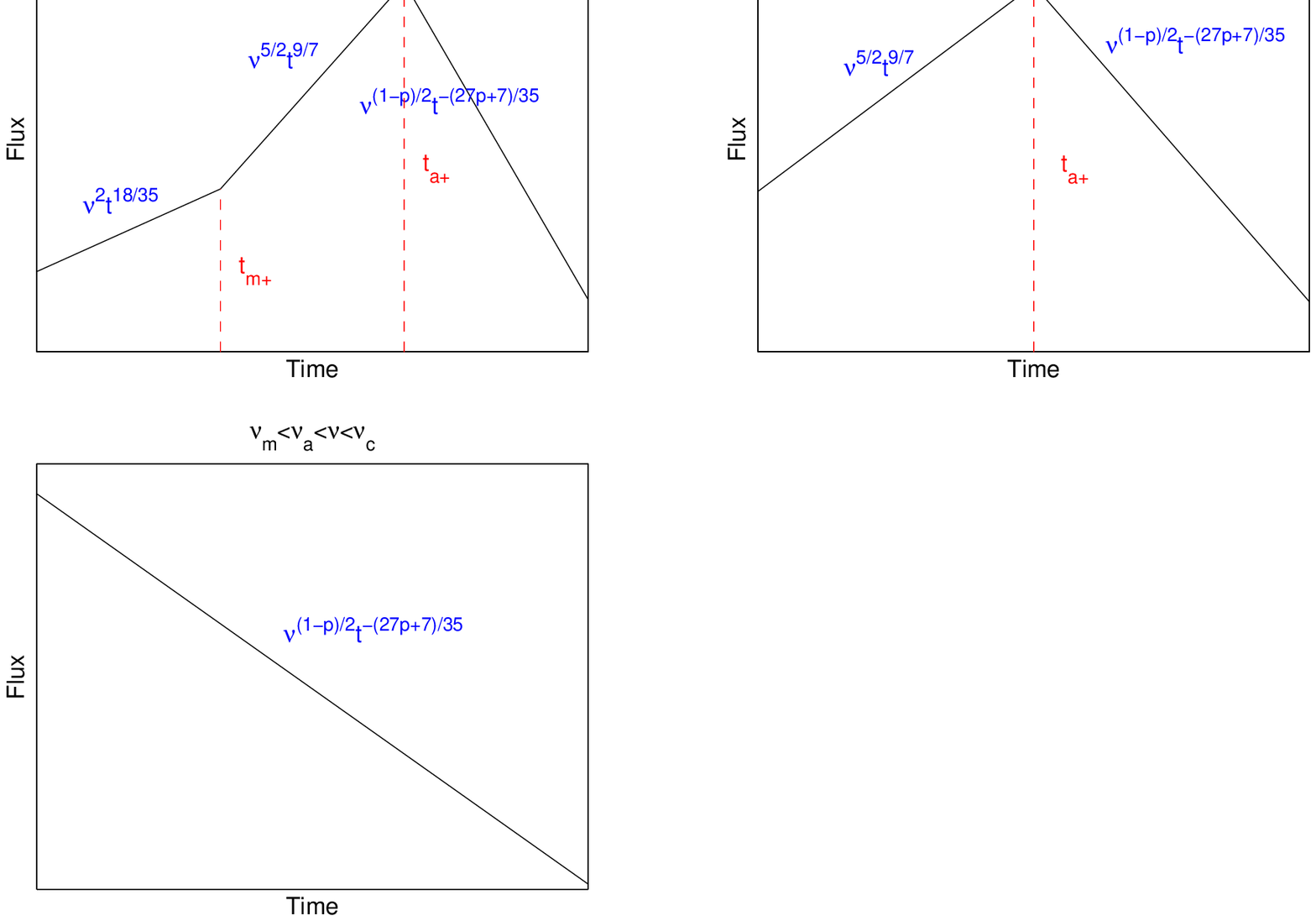}\\
    \caption{Same as Fig. \ref{fig31}, but with the
initial characteristic frequency order $\nu_m < \nu_a < \nu_c$.}
  \label{fig33}
\end{figure}

\begin{figure}
  \includegraphics[angle=0,width=1.0\textwidth]{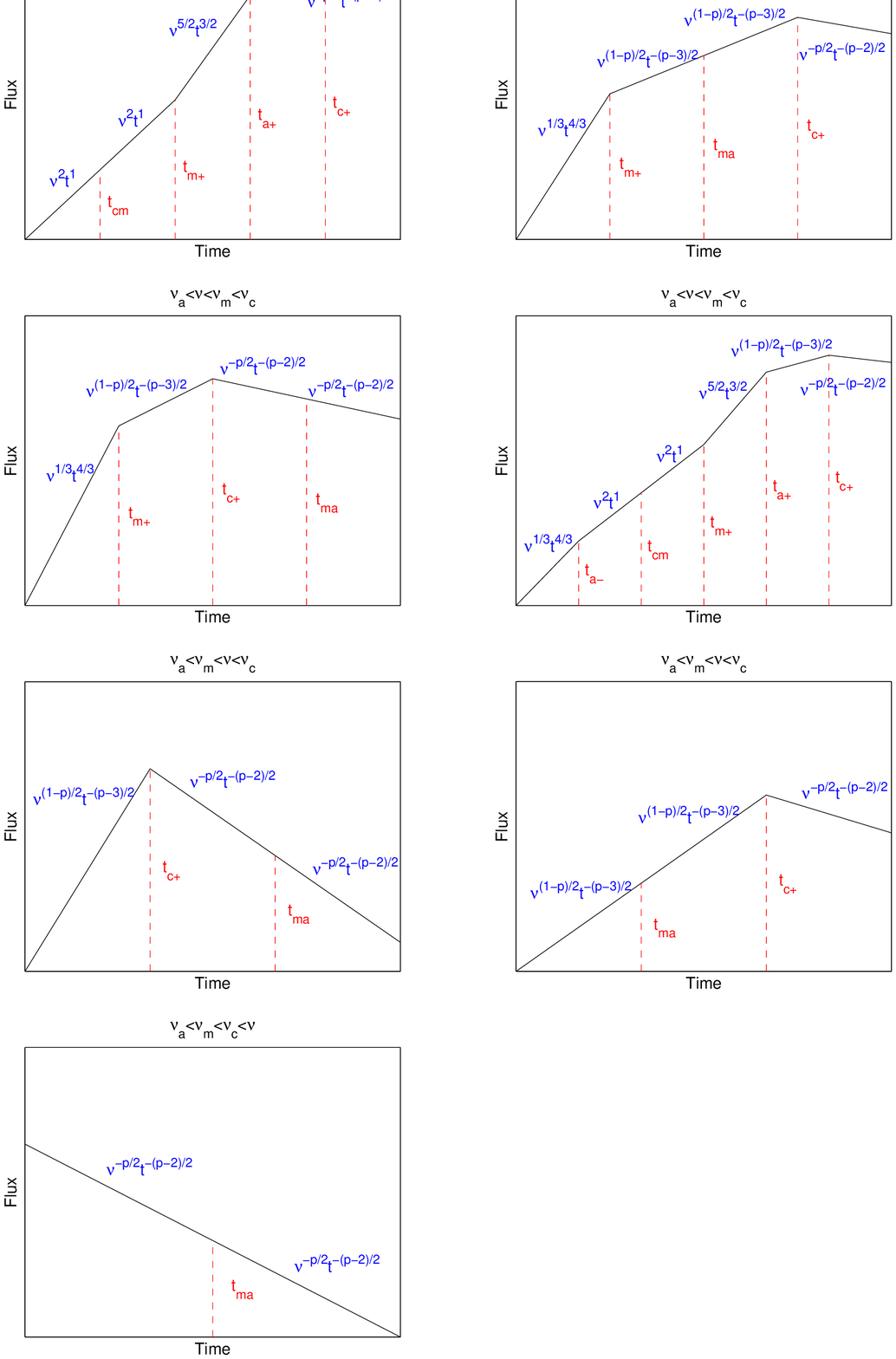}\\
    \caption{All possible forward shock lightcurves during Phase 1
(reverse shock crossing phase), for thick shell ISM model
and the initial characteristic frequency order $\nu_a < \nu_m < \nu_c$.}
  \label{fig61}
\end{figure}

\begin{figure}
  \includegraphics[angle=0,width=1.0\textwidth]{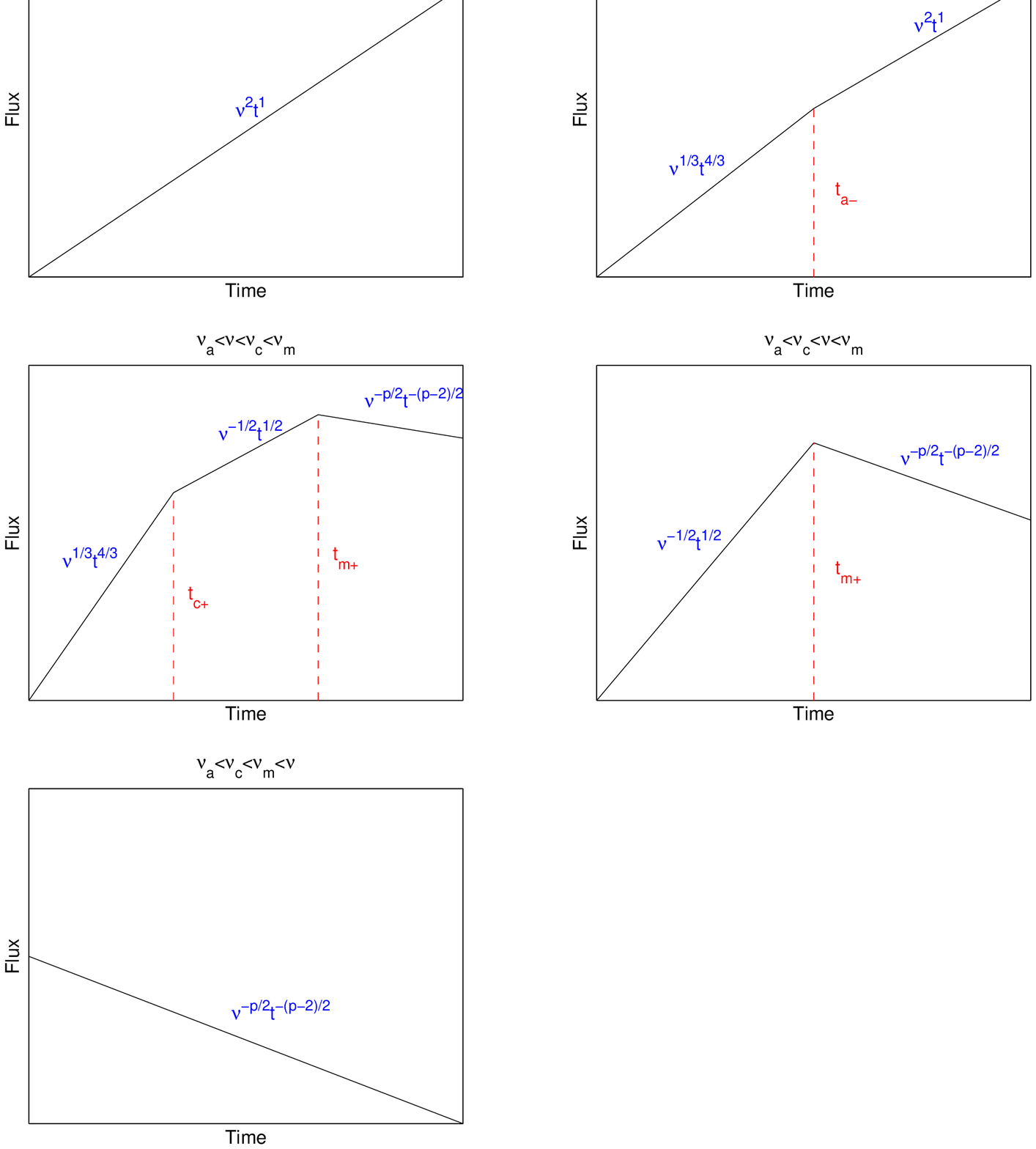}\\
    \caption{Same as Fig. \ref{fig61},  but with the initial characteristic
frequency order $\nu_a < \nu_c < \nu_m$.}
  \label{fig62}
\end{figure}

\begin{figure}
  \includegraphics[angle=0,width=1.0\textwidth]{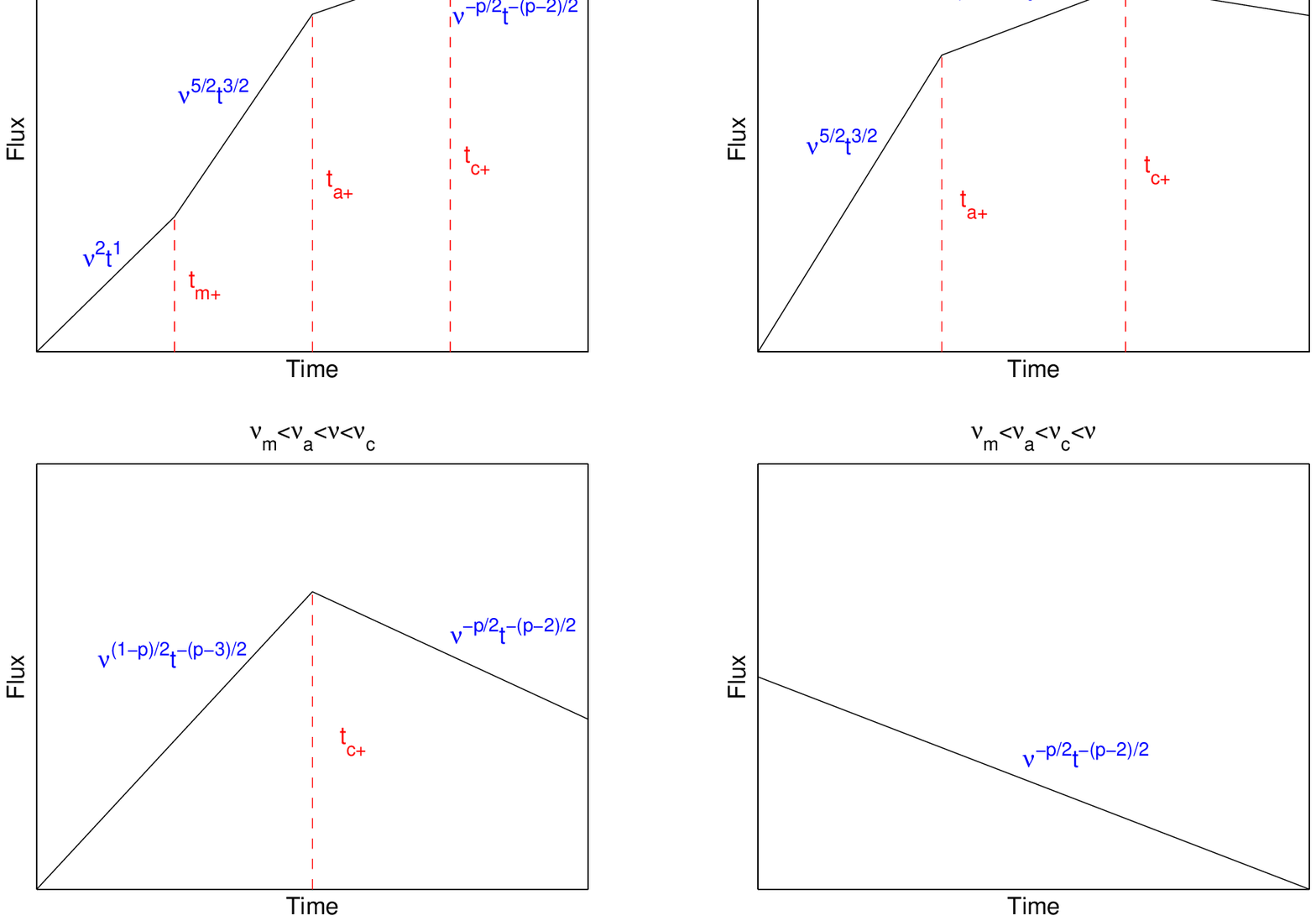}\\
    \caption{Same as Fig. \ref{fig61},  but with the initial characteristic
frequency order $\nu_m < \nu_a < \nu_c$.}
  \label{fig63}
\end{figure}

\begin{figure}
  \includegraphics[angle=0,width=1.0\textwidth]{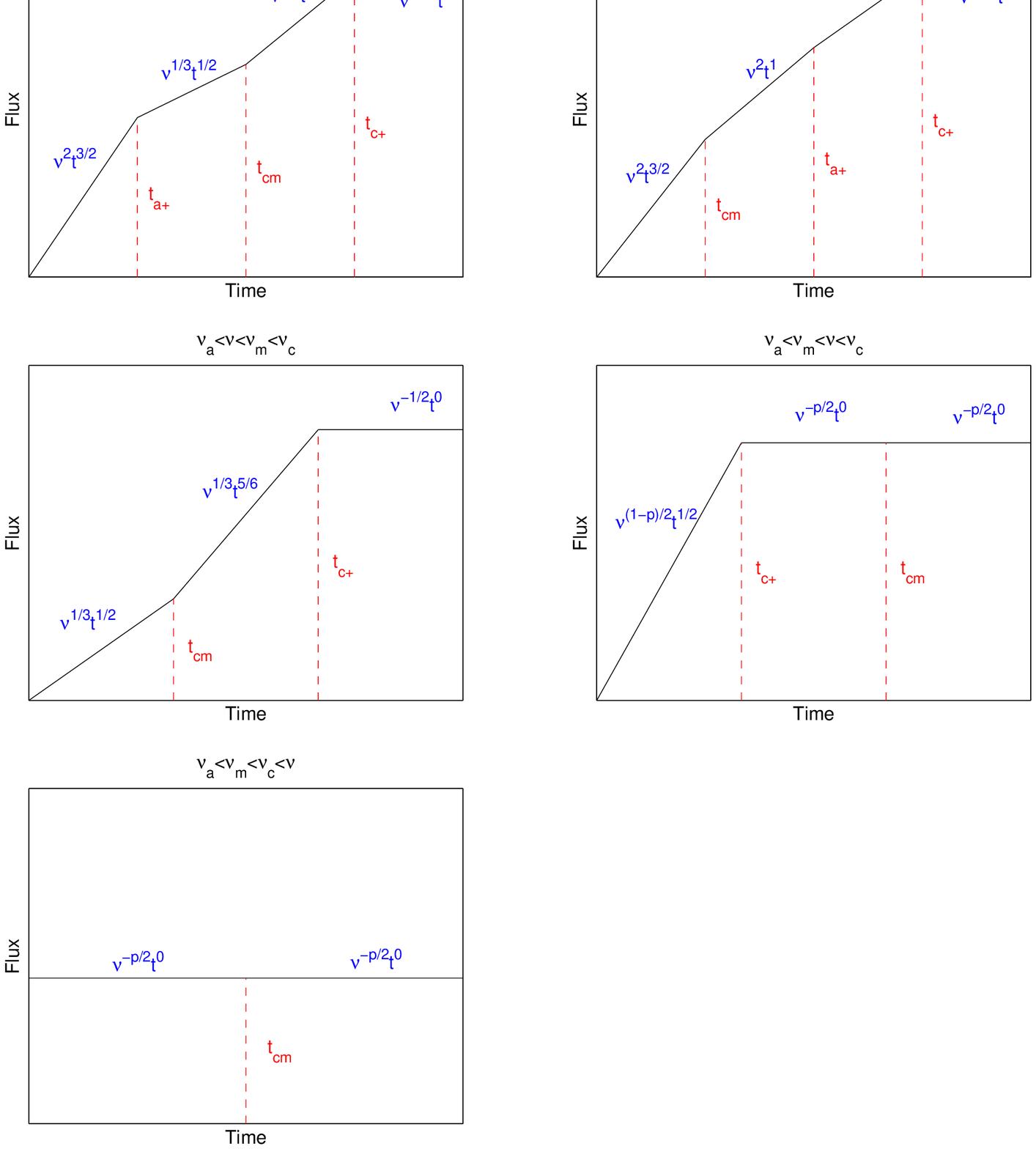}\\
    \caption{All possible reverse shock lightcurves during Phase 1
(reverse shock crossing phase), for thick shell ISM model and the
initial characteristic frequency order $\nu_a < \nu_m < \nu_c$.}
  \label{fig41}
\end{figure}

\begin{figure}
  \includegraphics[angle=0,width=1.0 \textwidth]{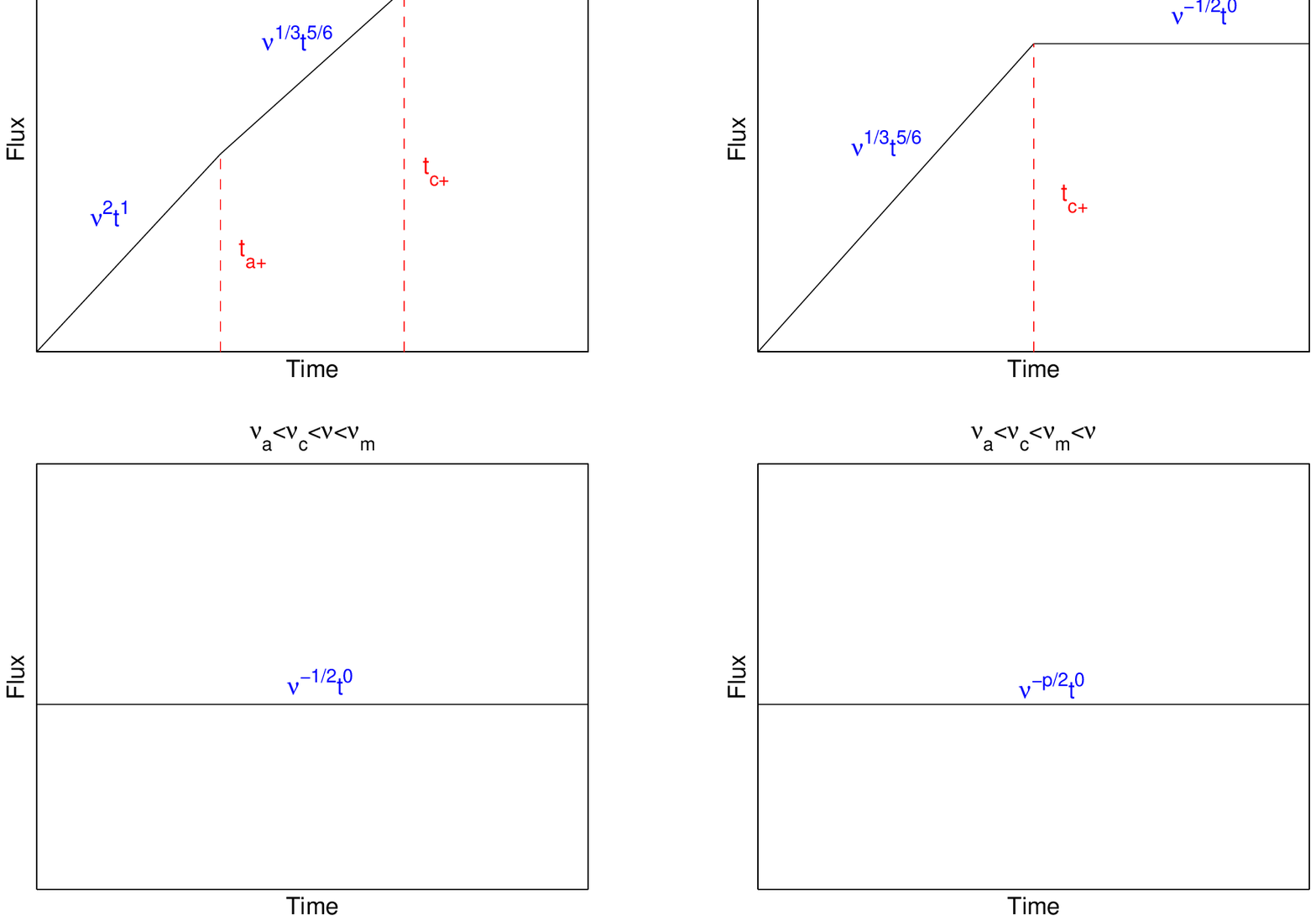}\\
    \caption{Same as Fig. \ref{fig41}, but
with the initial characteristic frequency order $\nu_a < \nu_c <
\nu_m$.}
  \label{fig42}
\end{figure}

\begin{figure}
  \includegraphics[angle=0,width=1.0 \textwidth]{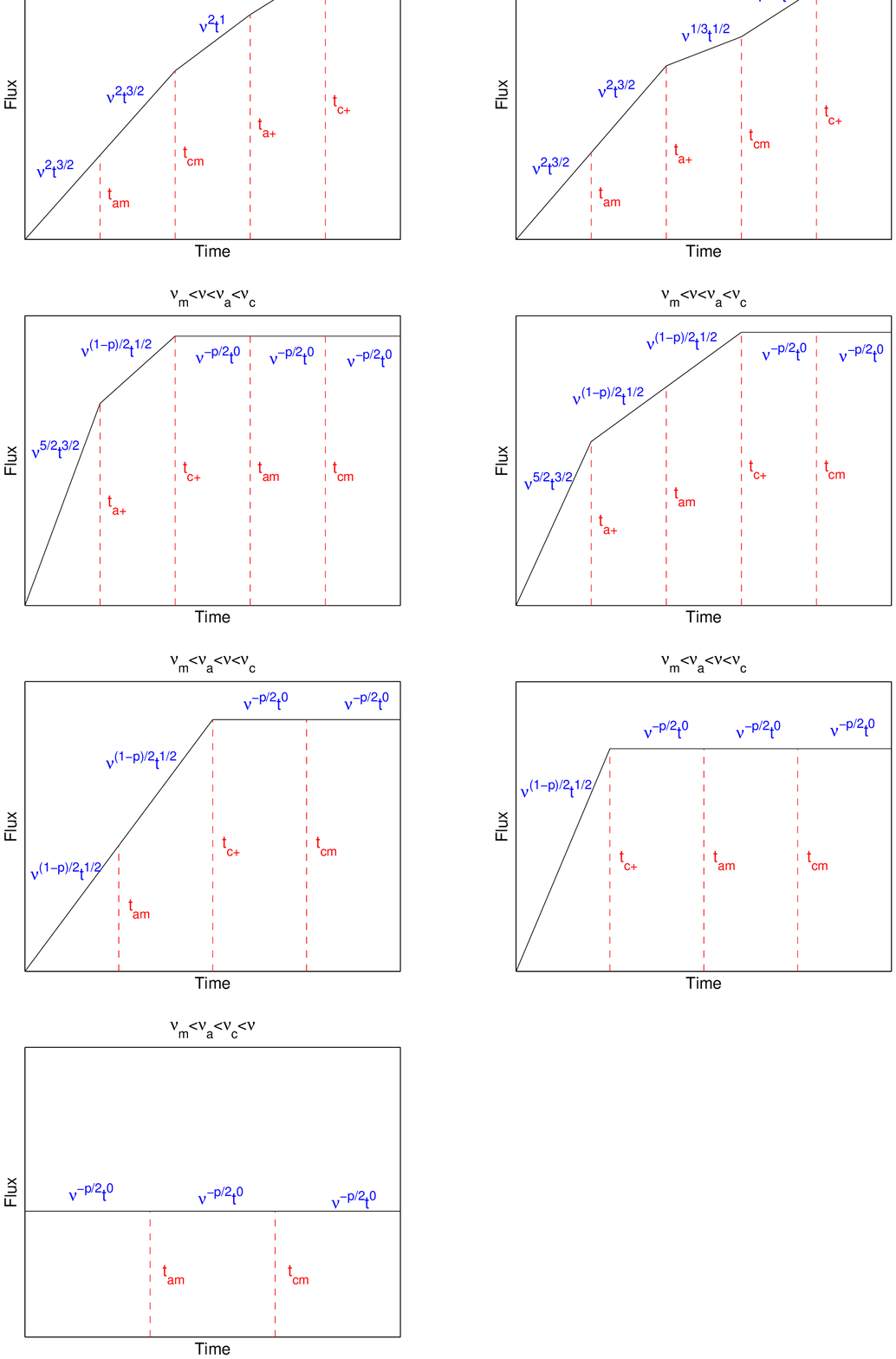}\\
    \caption{Same as Fig. \ref{fig41},  but with the initial characteristic
frequency order $\nu_m < \nu_a < \nu_c$.}
  \label{fig43}
\end{figure}

\begin{figure}
  \includegraphics[angle=0,width=1.0\textwidth]{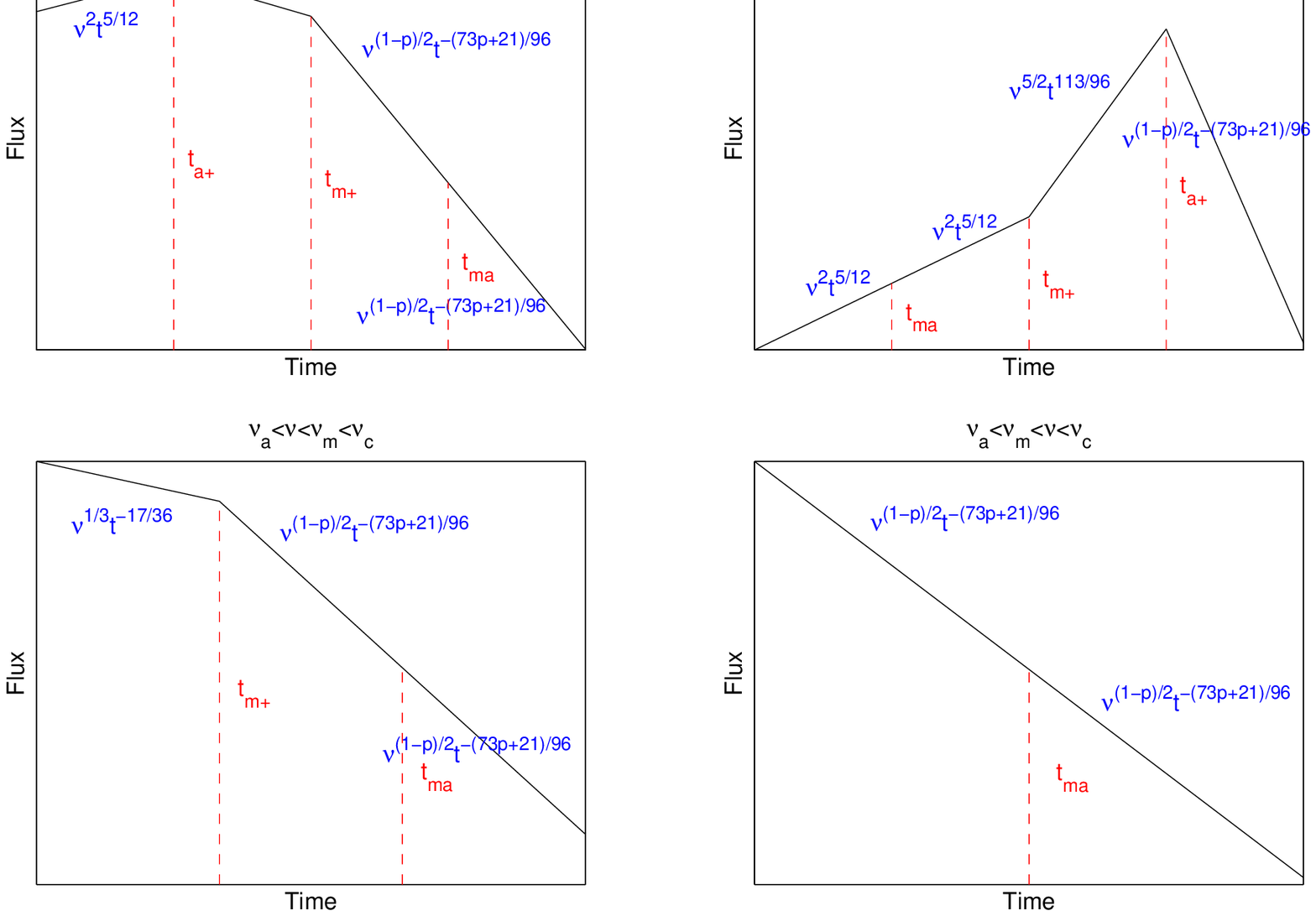}\\
    \caption{All possible reverse shock lightcurves after reverse shock
crosses the shell,  for thick shell ISM model
and the initial characteristic frequency order
$\nu_a < \nu_m < \nu_c$.}
  \label{fig51}
\end{figure}

\begin{figure}
  \includegraphics[angle=0,width=1.0\textwidth]{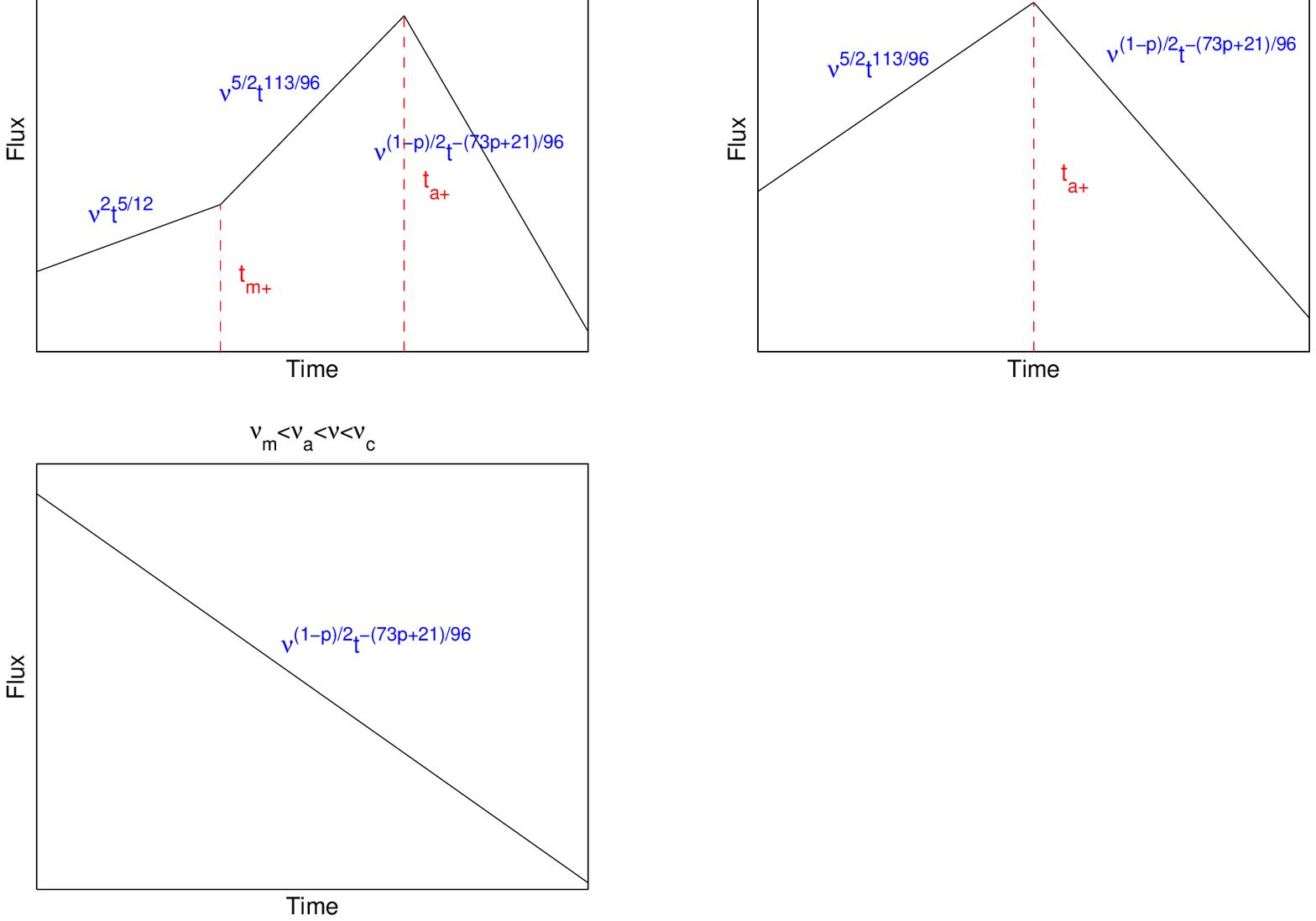}\\
    \caption{Same as Fig. \ref{fig51},  but with the initial characteristic
frequency order $\nu_m < \nu_a < \nu_c$.}
  \label{fig53}
\end{figure}

\begin{figure}
  \includegraphics[angle=0,width=1.0\textwidth]{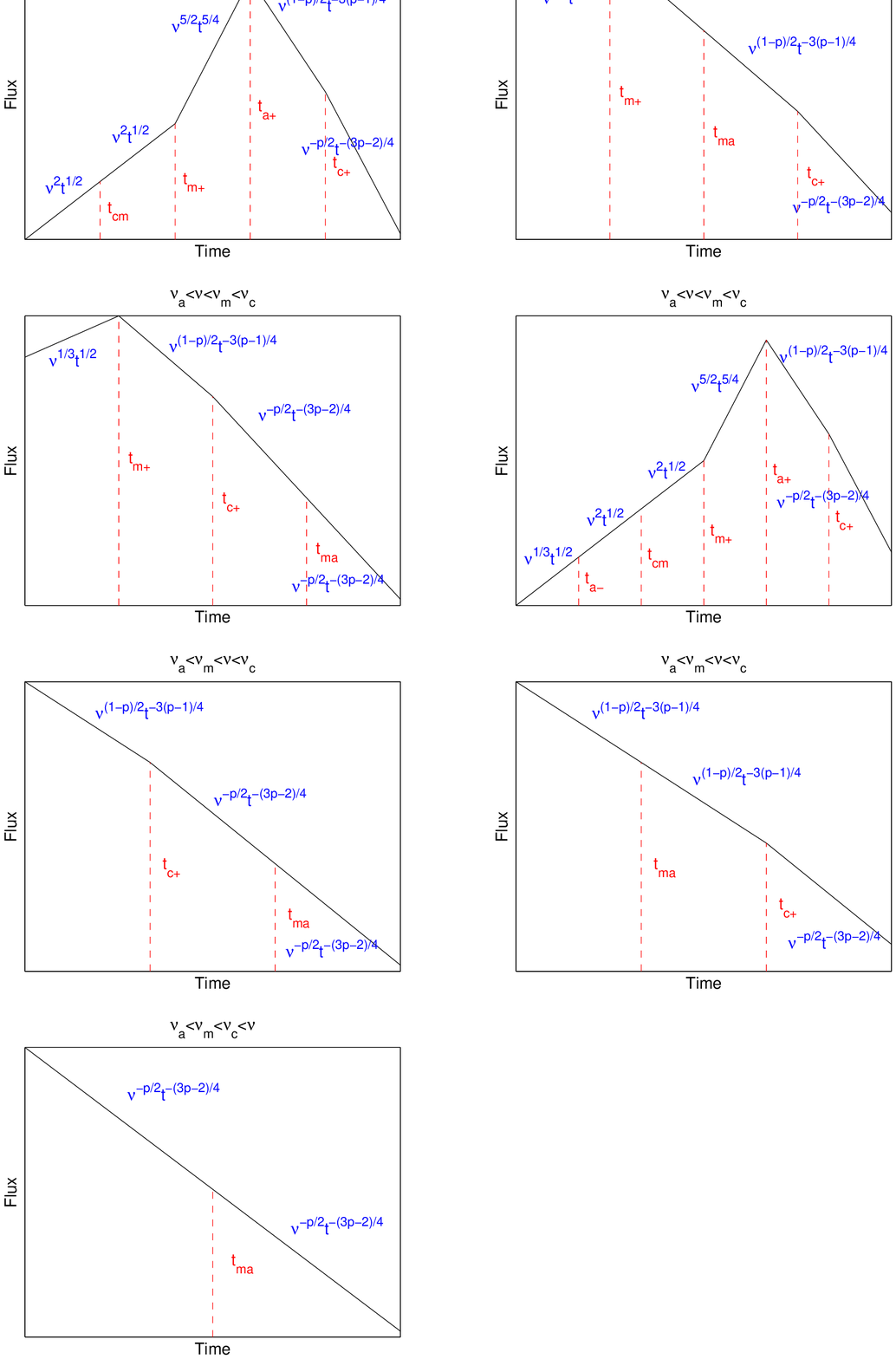}\\
    \caption{All possible forward shock lightcurves during Phase 2
(relativistic, isotropic, self-similar deceleration phase), with
an ISM medium and initial characteristic
frequency order $\nu_a < \nu_m < \nu_c$.}
  \label{fig71}
\end{figure}

\begin{figure}
  \includegraphics[angle=0,width=1.0 \textwidth]{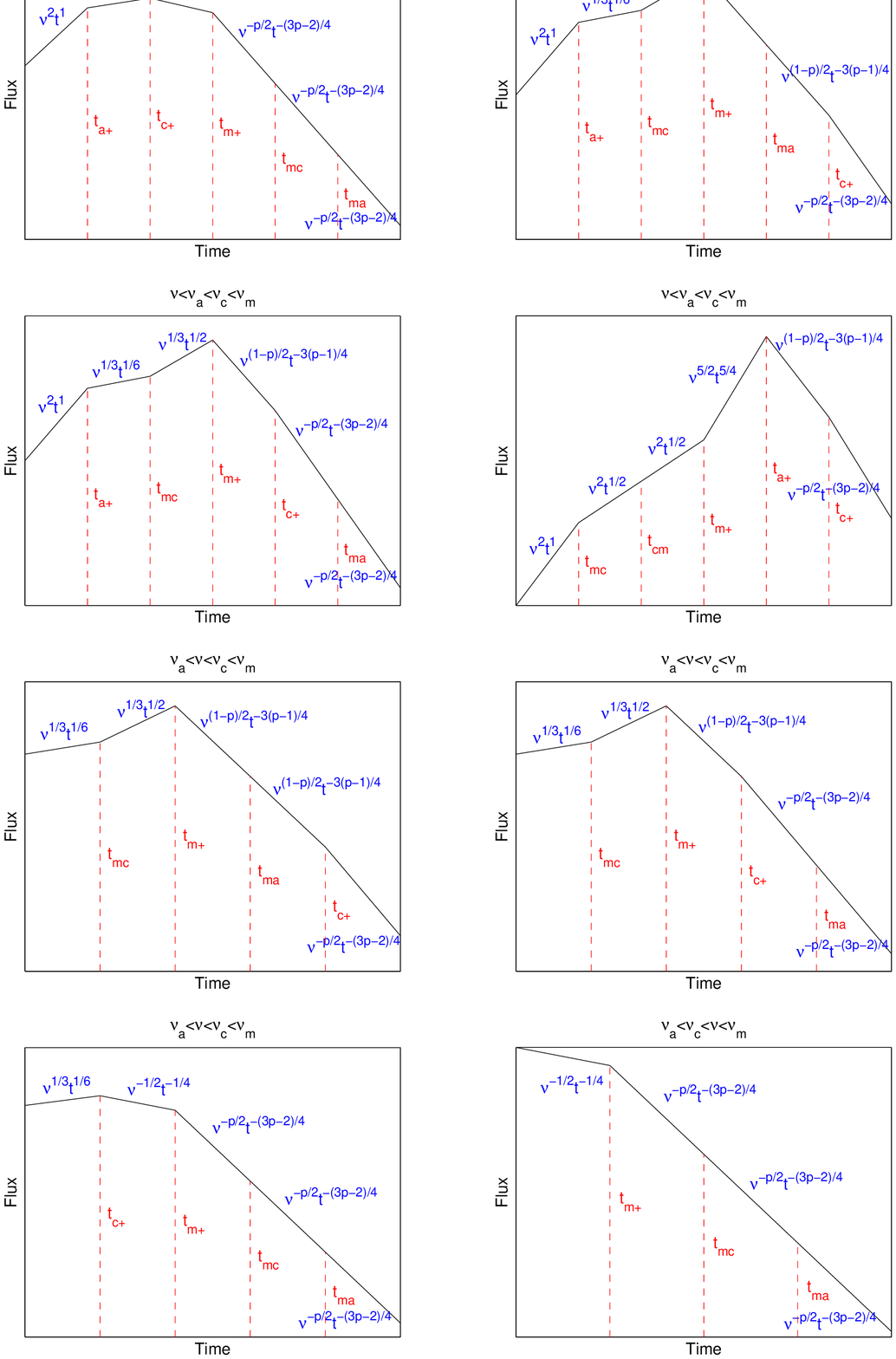}\\
    \caption{Same as Fig. \ref{fig71},  but with the initial characteristic
frequency order $\nu_a < \nu_c < \nu_m$.}
  \label{fig72}
\end{figure}

\begin{figure}
  \includegraphics[angle=0,width=1.0 \textwidth]{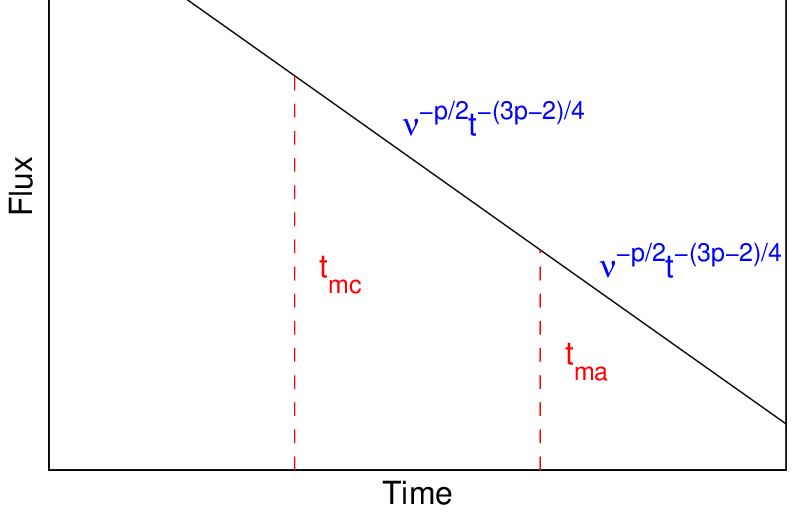}\\
    \caption{Figure \ref{fig72} continued.}
  \label{fig72b}
\end{figure}

\begin{figure}
  \includegraphics[angle=0,width=1.0 \textwidth]{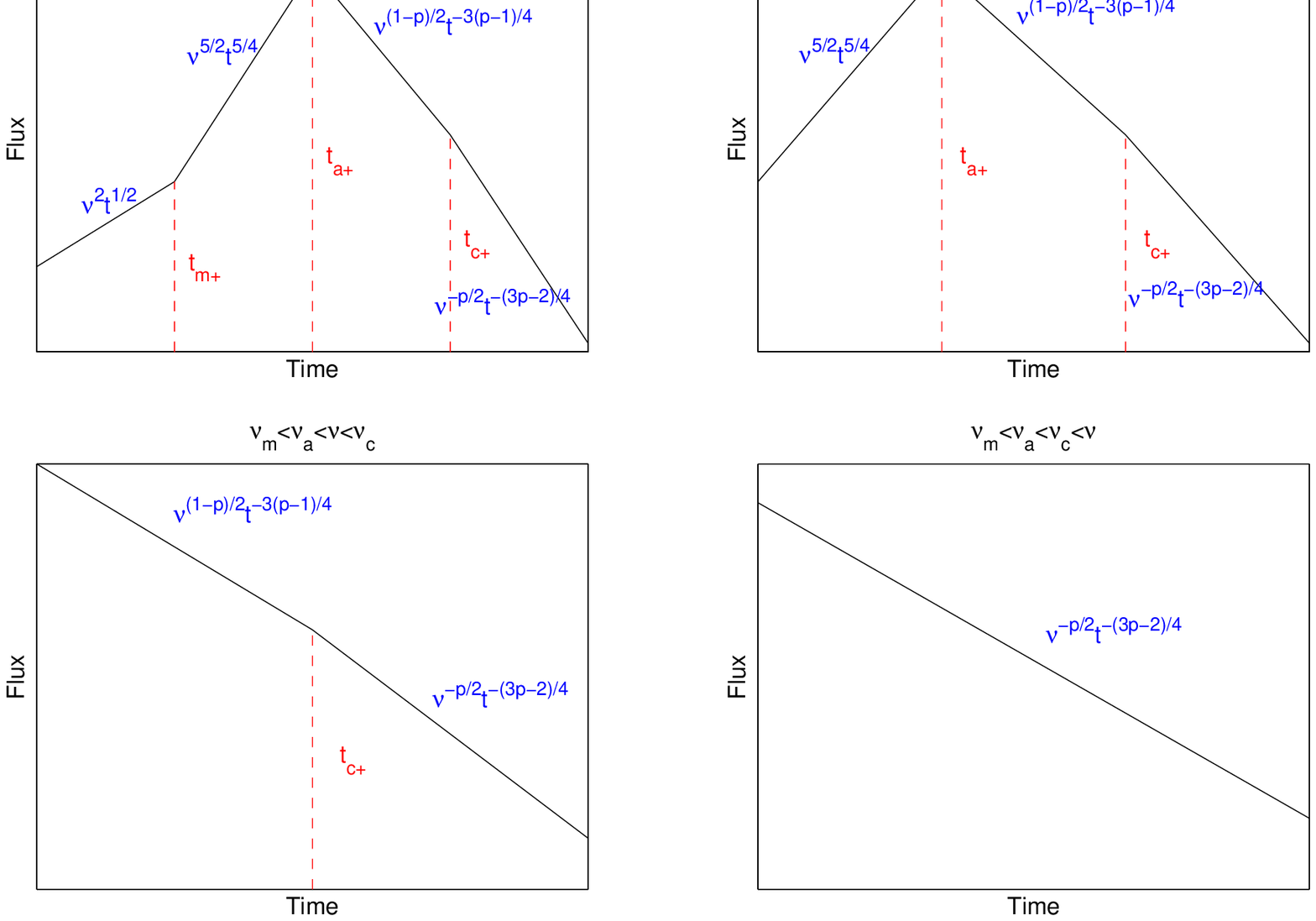}\\
    \caption{Same as Fig. \ref{fig71},  but with the initial characteristic
frequency order $\nu_m < \nu_a < \nu_c$.}
  \label{fig73}
\end{figure}

\begin{figure}
  \includegraphics[angle=0,width=1.0 \textwidth]{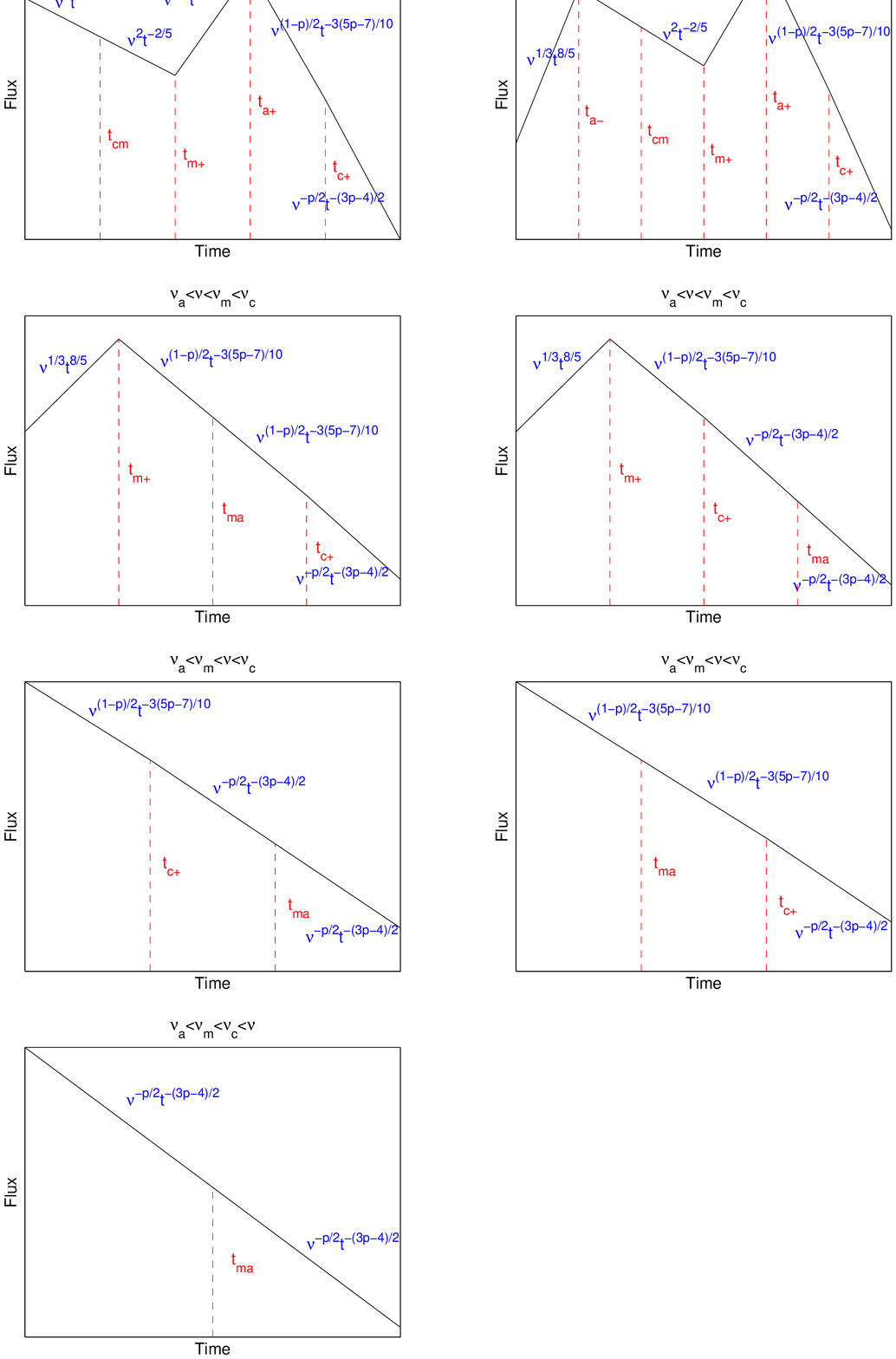}\\
    \caption{All possible forward shock lightcurves during Phase 4
(Newtonian phase), with an ISM medium and initial characteristic
frequency order $\nu_a < \nu_m < \nu_c$.}
  \label{fig81}
\end{figure}

\begin{figure}
  \includegraphics[angle=0,width=1.0 \textwidth]{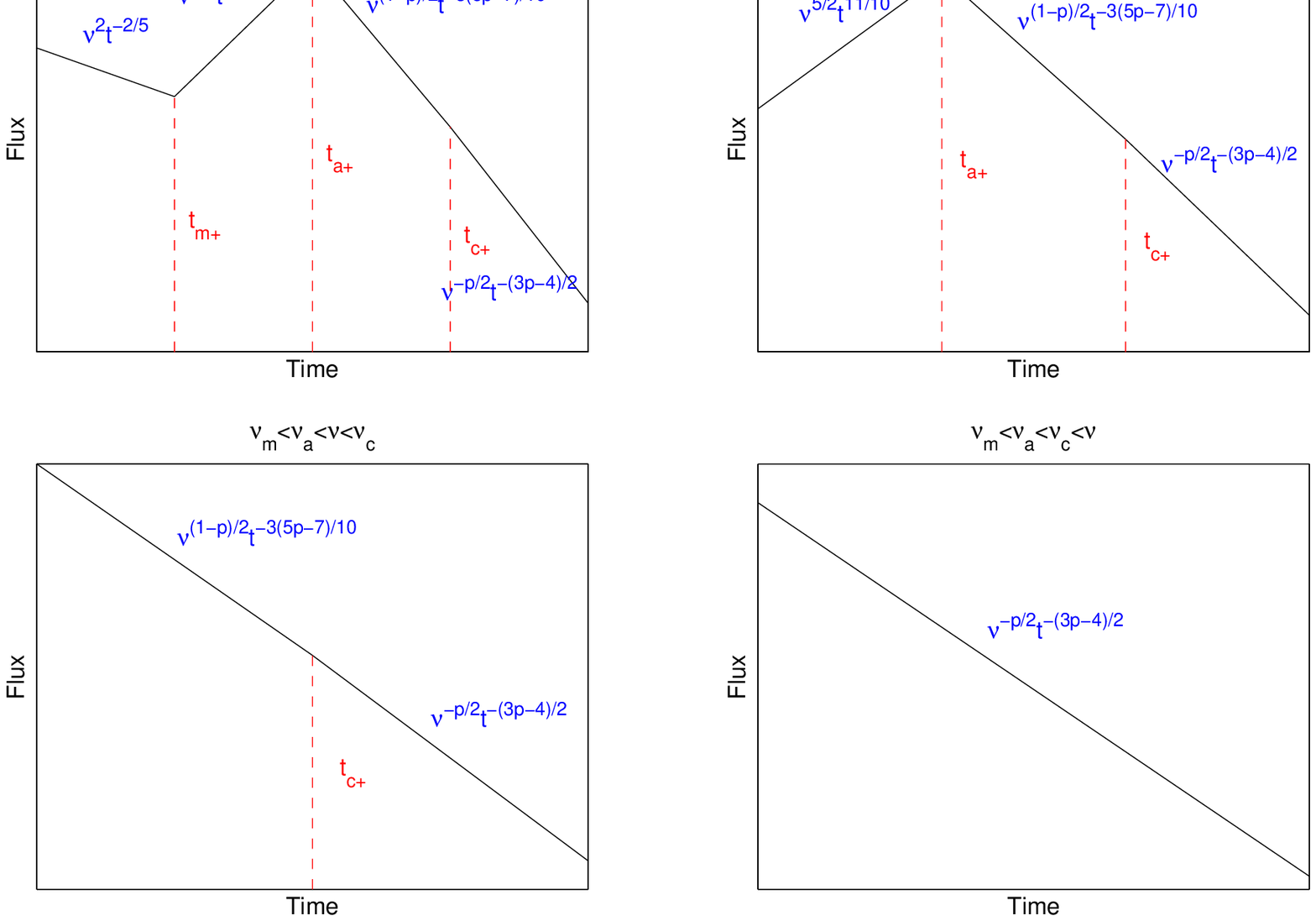}\\
    \caption{Same as Fig. \ref{fig81},  but with the initial characteristic
frequency order $\nu_m < \nu_a < \nu_c$.}
  \label{fig83}
\end{figure}

\begin{figure}
  \includegraphics[angle=0,width=1.0 \textwidth]{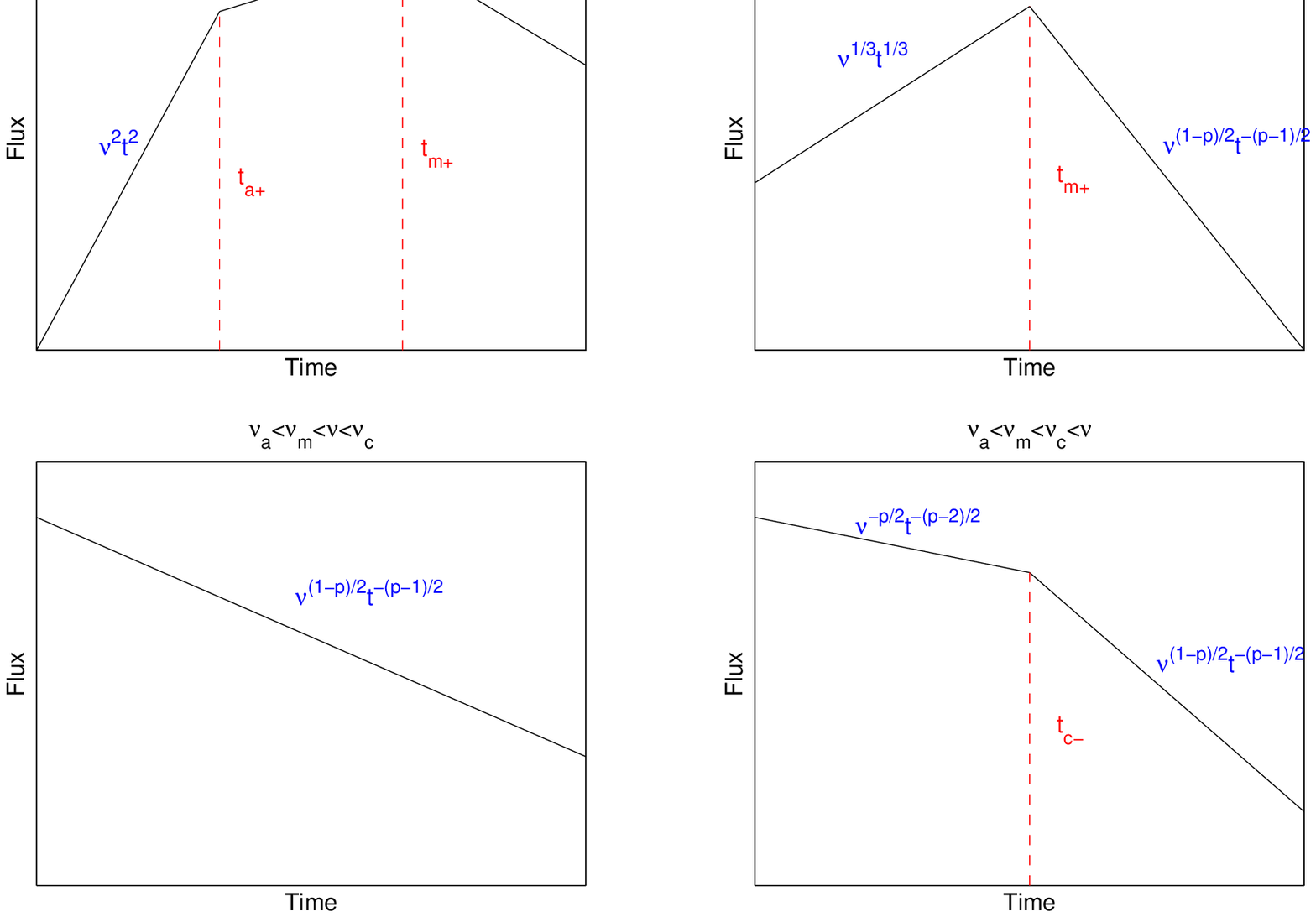}\\
    \caption{All possible forward shock lightcurves during Phase 1
(reverse shock crossing phase), for thin shell wind model
and the initial characteristic frequency order $\nu_a < \nu_m < \nu_c$.}
  \label{figwind11}
\end{figure}

\begin{figure}
  \includegraphics[angle=0,width=1.0\textwidth]{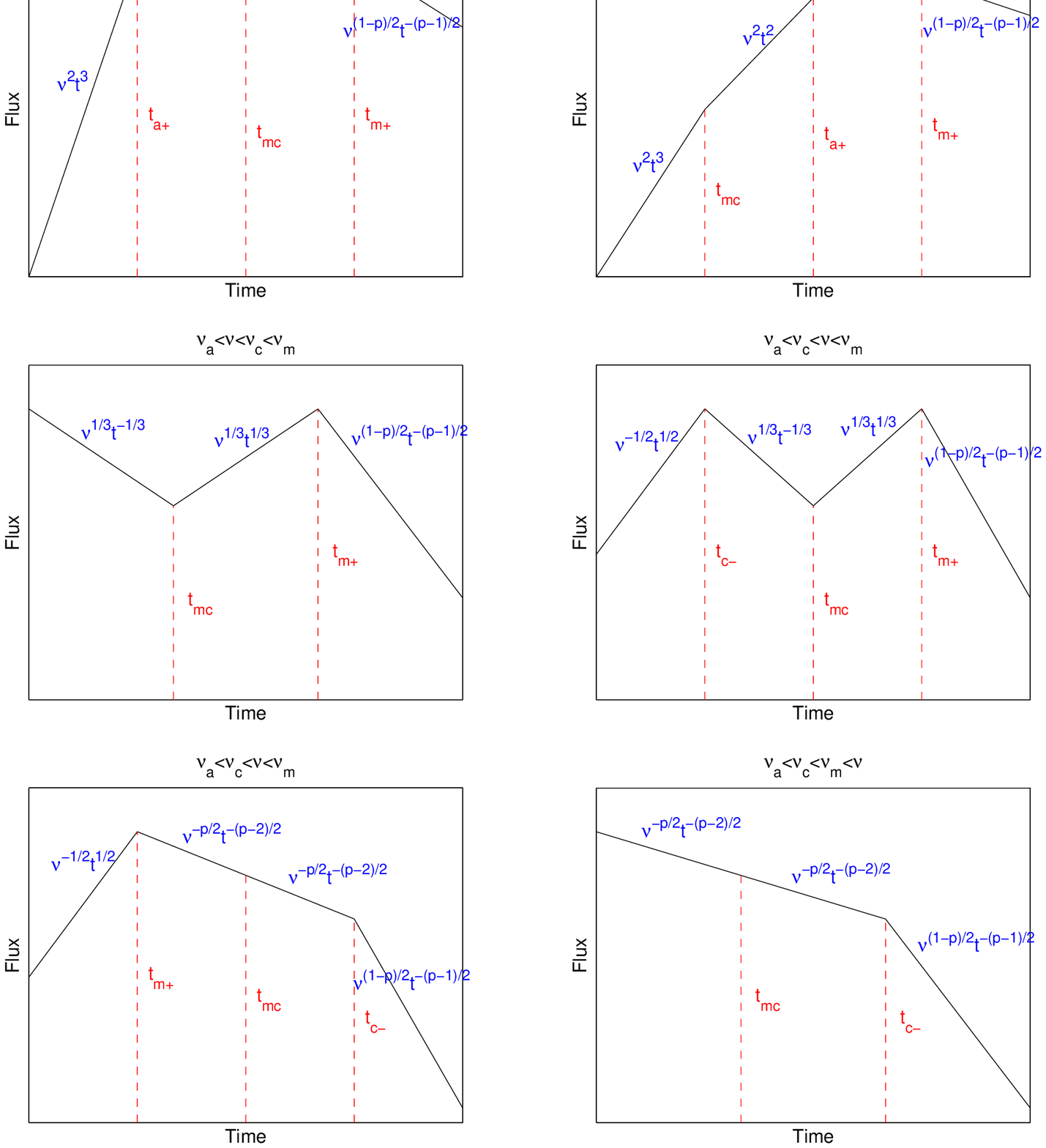}
    \caption{Same as Fig. \ref{figwind11},  but with the initial
characteristic frequency order $\nu_a < \nu_c < \nu_m$.}
  \label{figwind12}
\end{figure}

\begin{figure}
  \includegraphics[angle=0,width=1.0\textwidth]{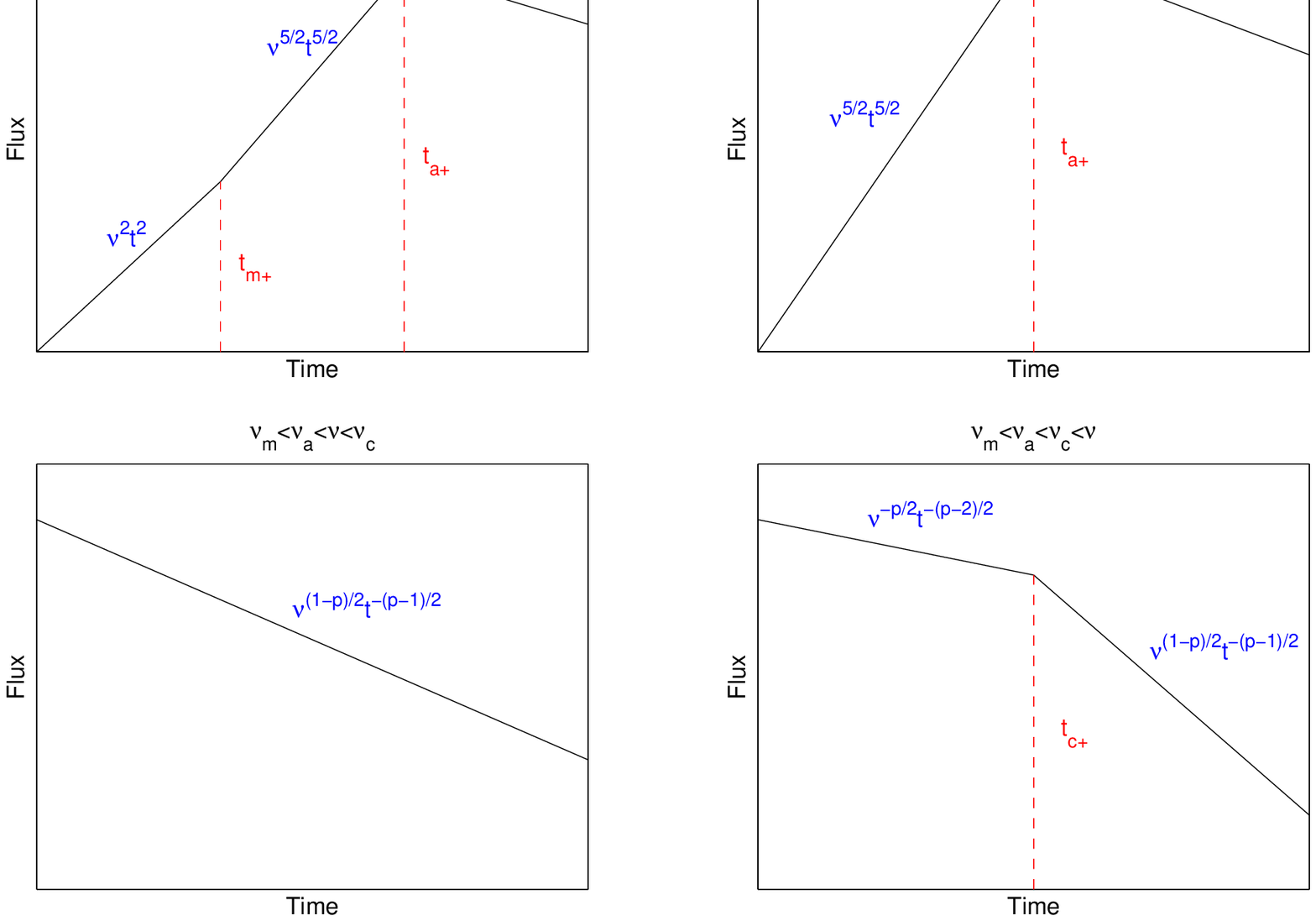}
    \caption{Same as Fig. \ref{figwind11},  but with the initial
characteristic frequency order $\nu_m < \nu_a < \nu_c$.}
  \label{figwind13}
\end{figure}

\begin{figure}
  \includegraphics[angle=0,width=1.0\textwidth]{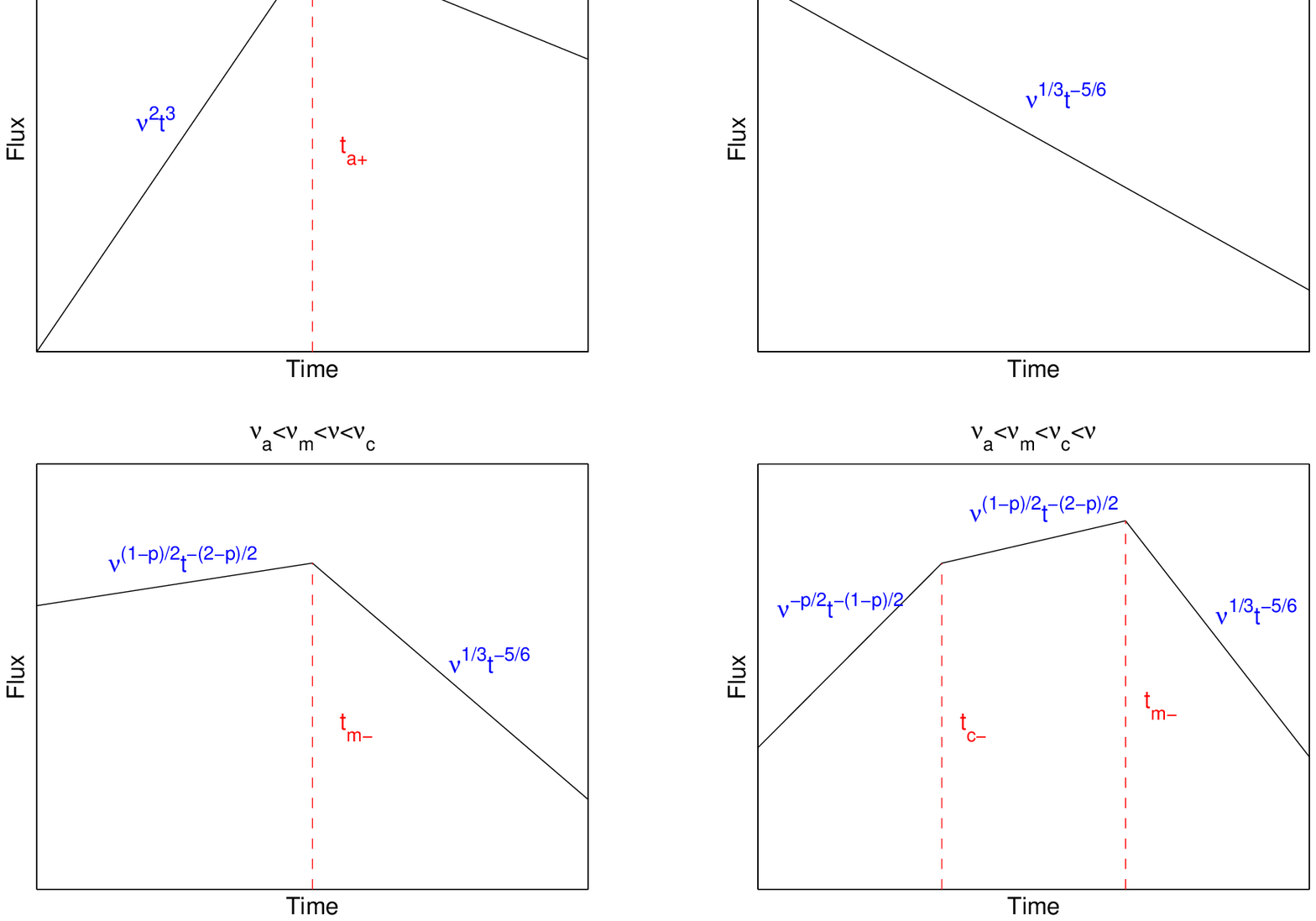}\\
    \caption{All possible reverse shock lightcurves during Phase 1
(reverse shock crossing phase), for thin shell wind model
and the initial characteristic frequency order $\nu_a < \nu_m < \nu_c$.}
  \label{figwind21}
\end{figure}

\begin{figure}
  \includegraphics[angle=0,width=1.0 \textwidth]{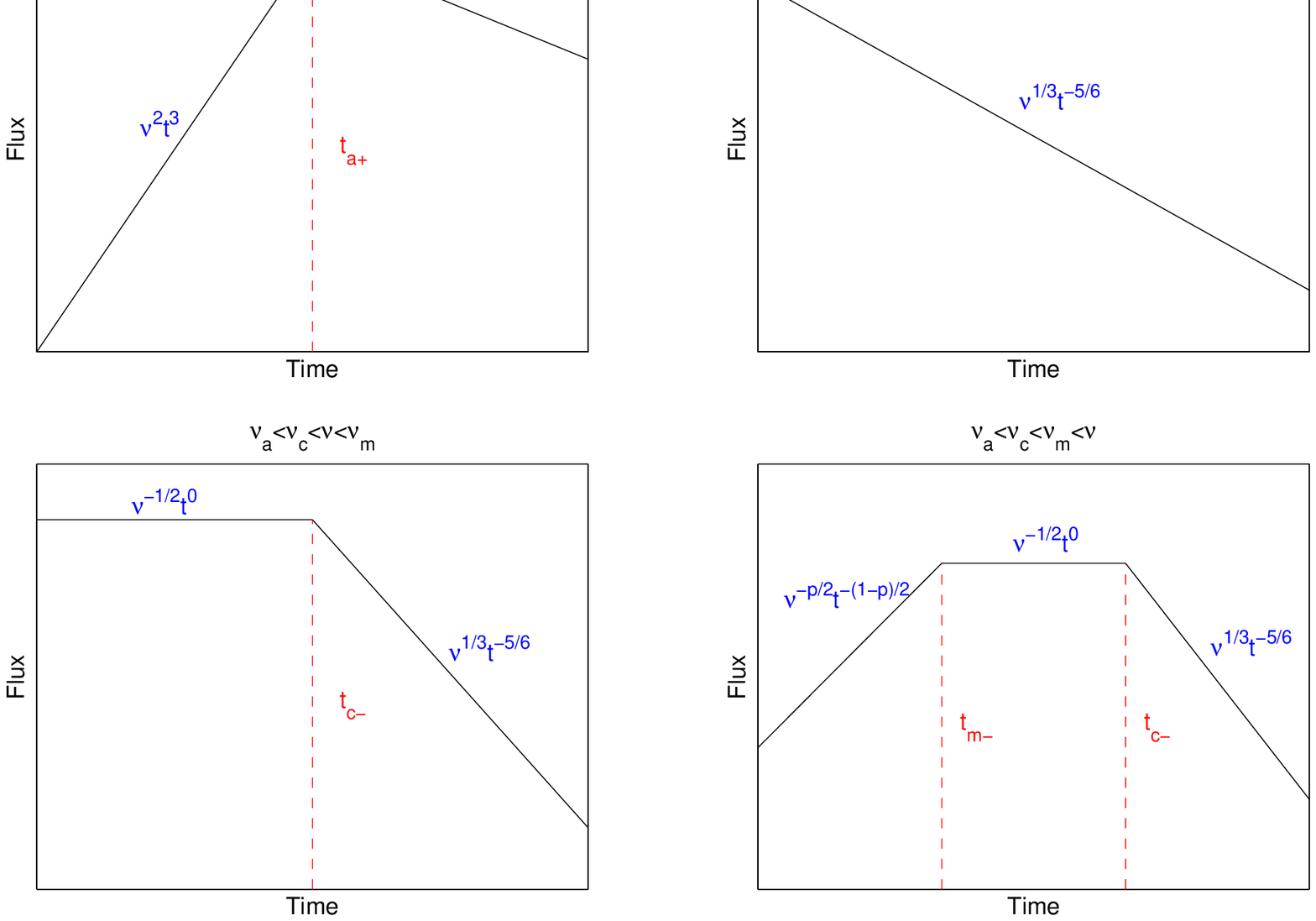}\\
    \caption{Same as Fig. \ref{figwind21},  but with the initial
characteristic frequency order $\nu_a < \nu_c < \nu_m$.}
  \label{figwind22}
\end{figure}

\begin{figure}
  \includegraphics[angle=0,width=1.0 \textwidth]{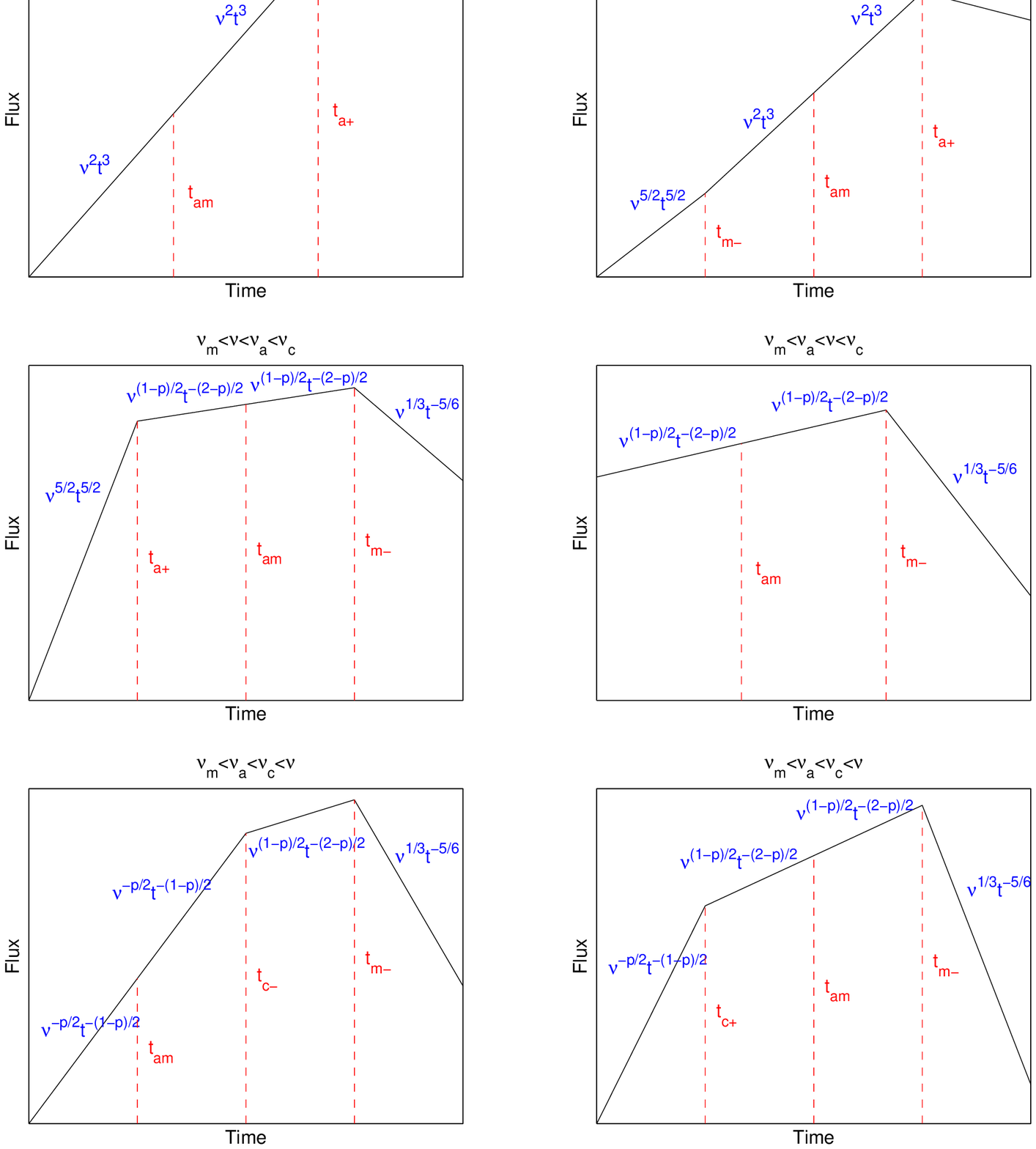}\\
    \caption{Same as Fig. \ref{figwind21},  but with the initial
characteristic frequency order $\nu_m < \nu_a < \nu_c$.}
  \label{figwind23}
\end{figure}

\begin{figure}
  \includegraphics[angle=0,width=1.0\textwidth]{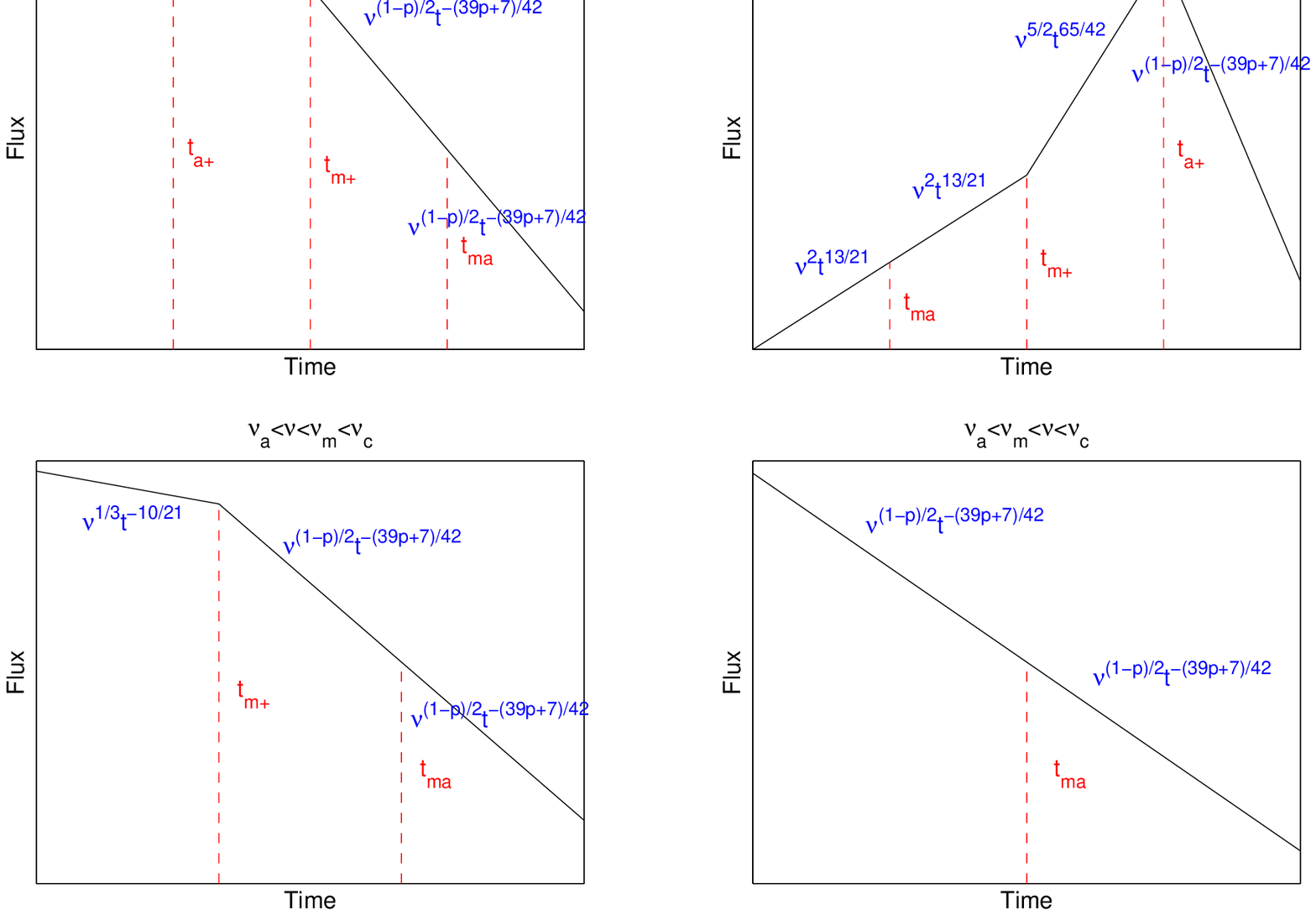}\\
    \caption{All possible reverse shock lightcurves after reverse
shock crossing, for thin shell wind model and the initial
characteristic frequency order $\nu_a < \nu_m < \nu_c$.}
  \label{figwind31}
\end{figure}

\begin{figure}
  \includegraphics[angle=0,width=1.0\textwidth]{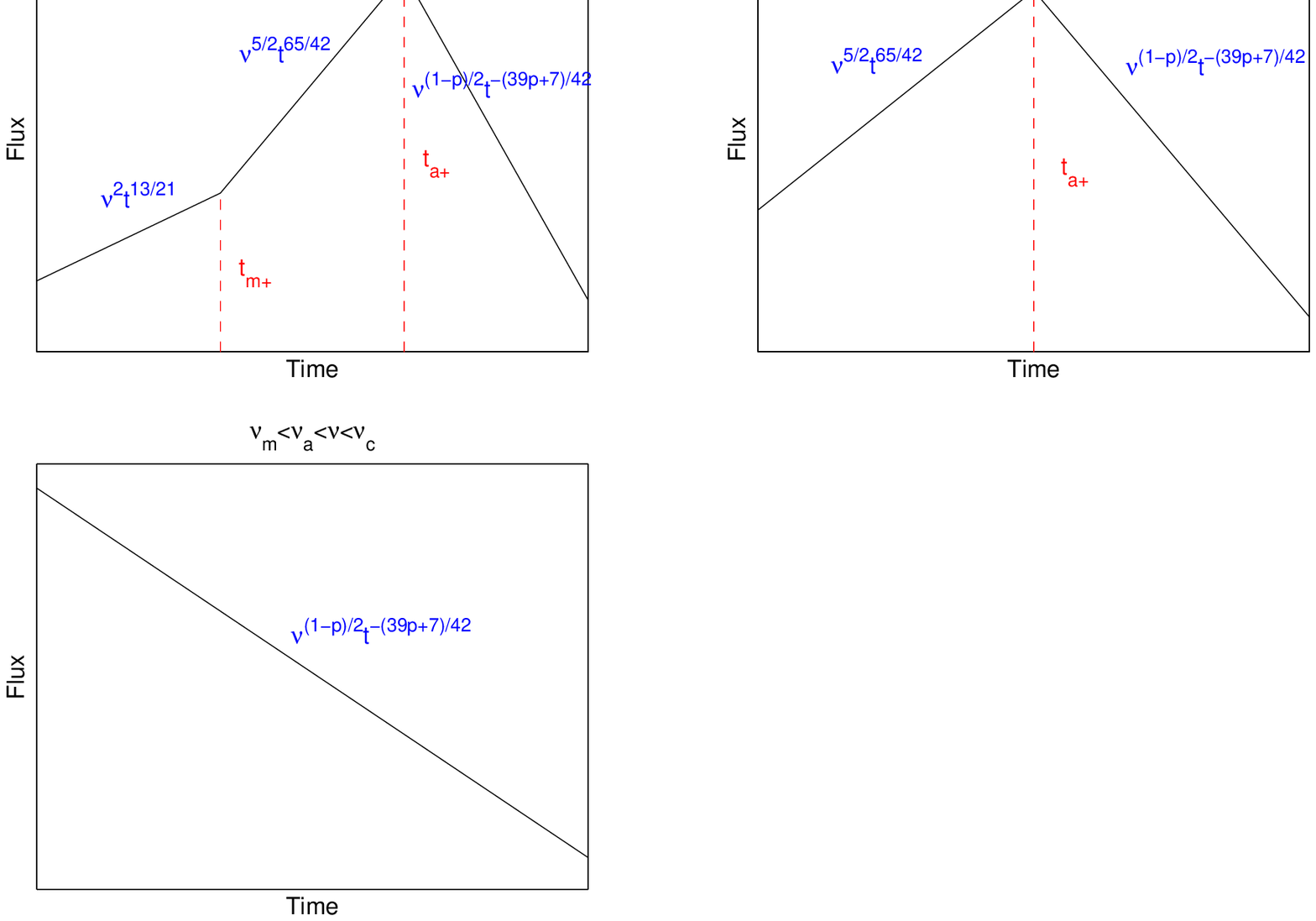}\\
    \caption{Same as Fig. \ref{figwind31},  but with the initial
characteristic frequency order $\nu_m < \nu_a < \nu_c$.}
  \label{figwind33}
\end{figure}

\begin{figure}
  \includegraphics[angle=0,width=1.0 \textwidth]{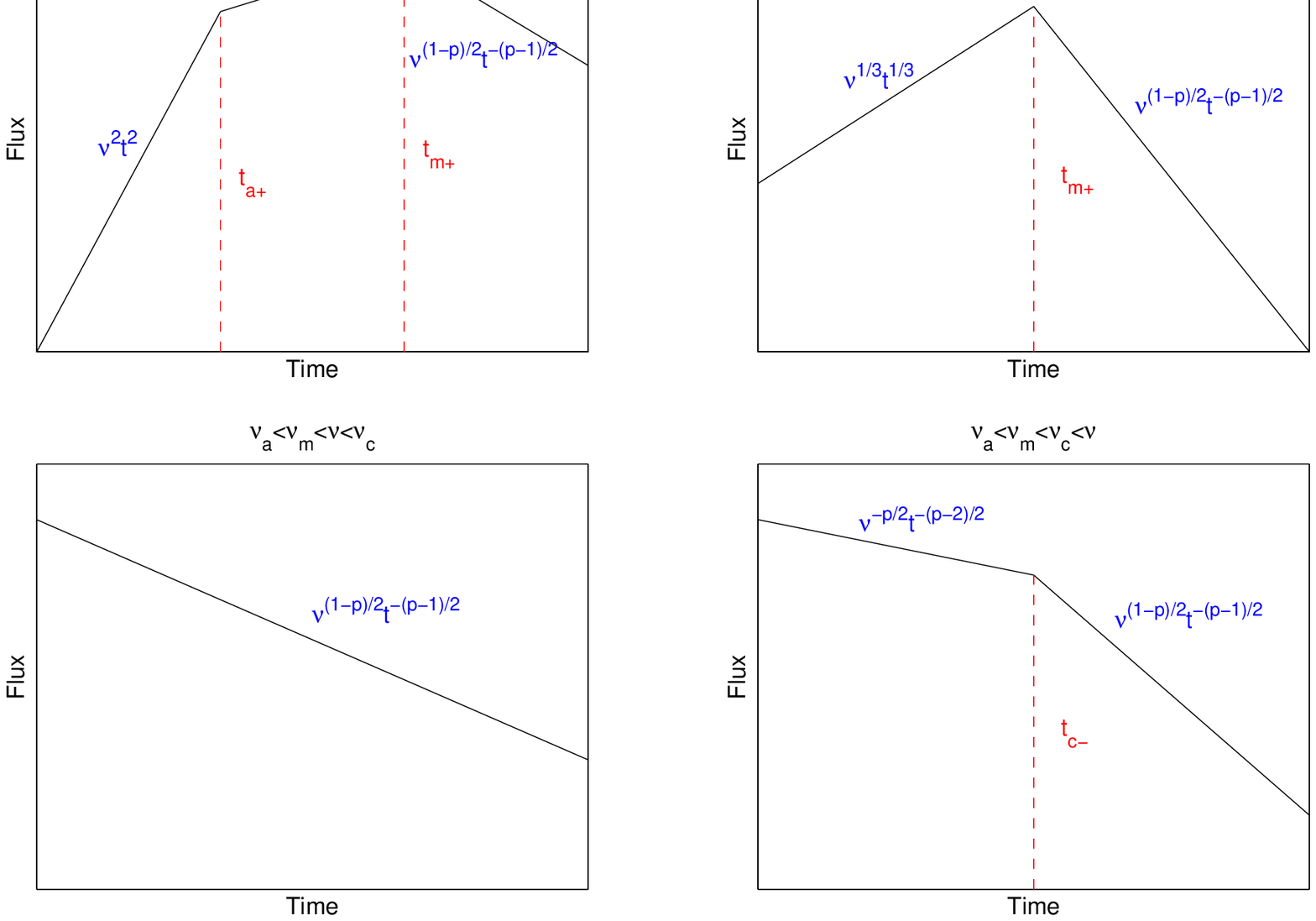}\\
    \caption{All possible forward shock lightcurves during Phase 1
(reverse shock crossing phase), for thick shell wind model and the initial
characteristic frequency order $\nu_a < \nu_m < \nu_c$.}
  \label{figwind61}
\end{figure}

\begin{figure}
  \includegraphics[angle=0,width=1.0\textwidth]{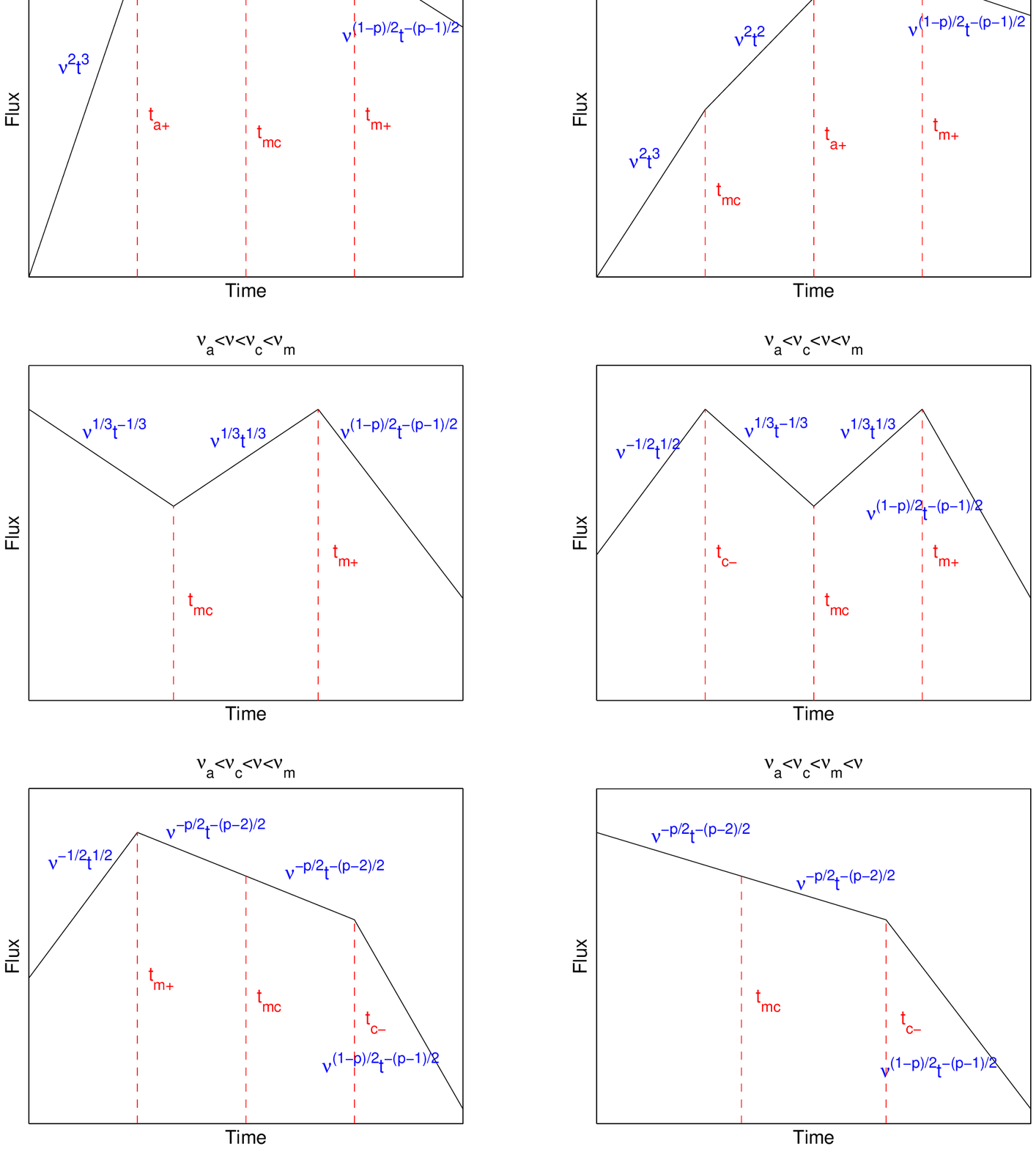}\\
    \caption{Same as Fig. \ref{figwind61},  but with the initial
characteristic frequency order $\nu_a < \nu_c < \nu_m$.}
  \label{figwind62}
\end{figure}

\begin{figure}
  \includegraphics[angle=0,width=1.0\textwidth]{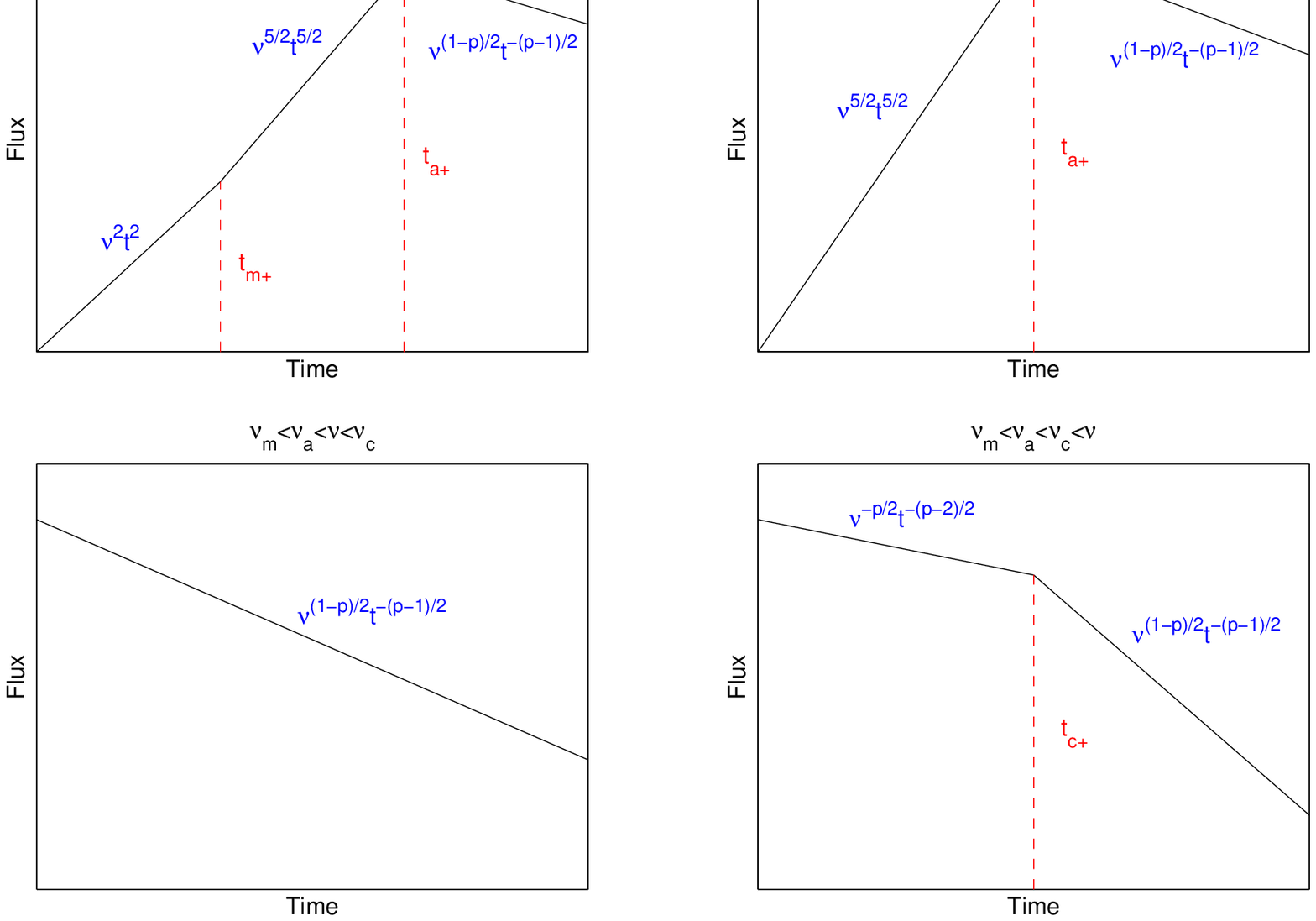}\\
    \caption{Same as Fig. \ref{figwind61},  but with the initial
characteristic frequency order $\nu_m < \nu_a < \nu_c$.}
  \label{figwind63}
\end{figure}

\begin{figure}
  \includegraphics[angle=0,width=1.0\textwidth]{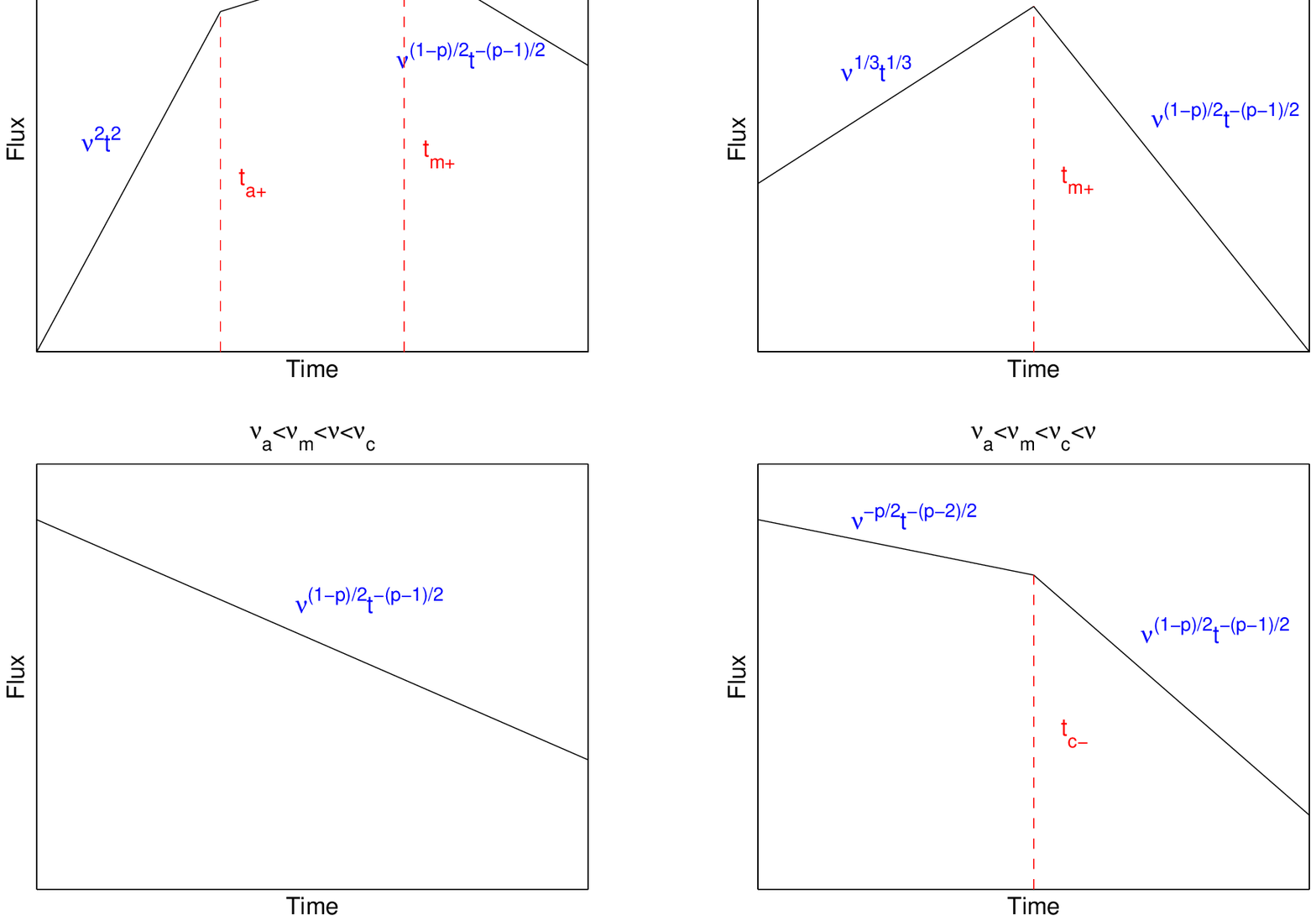}\\
    \caption{All possible reverse shock lightcurves during Phase 1
(reverse shock crossing phase), for thick shell wind model and the initial
characteristic frequency order $\nu_a < \nu_m < \nu_c$.}
  \label{figwind41}
\end{figure}

\begin{figure}
  \includegraphics[angle=0,width=1.0\textwidth]{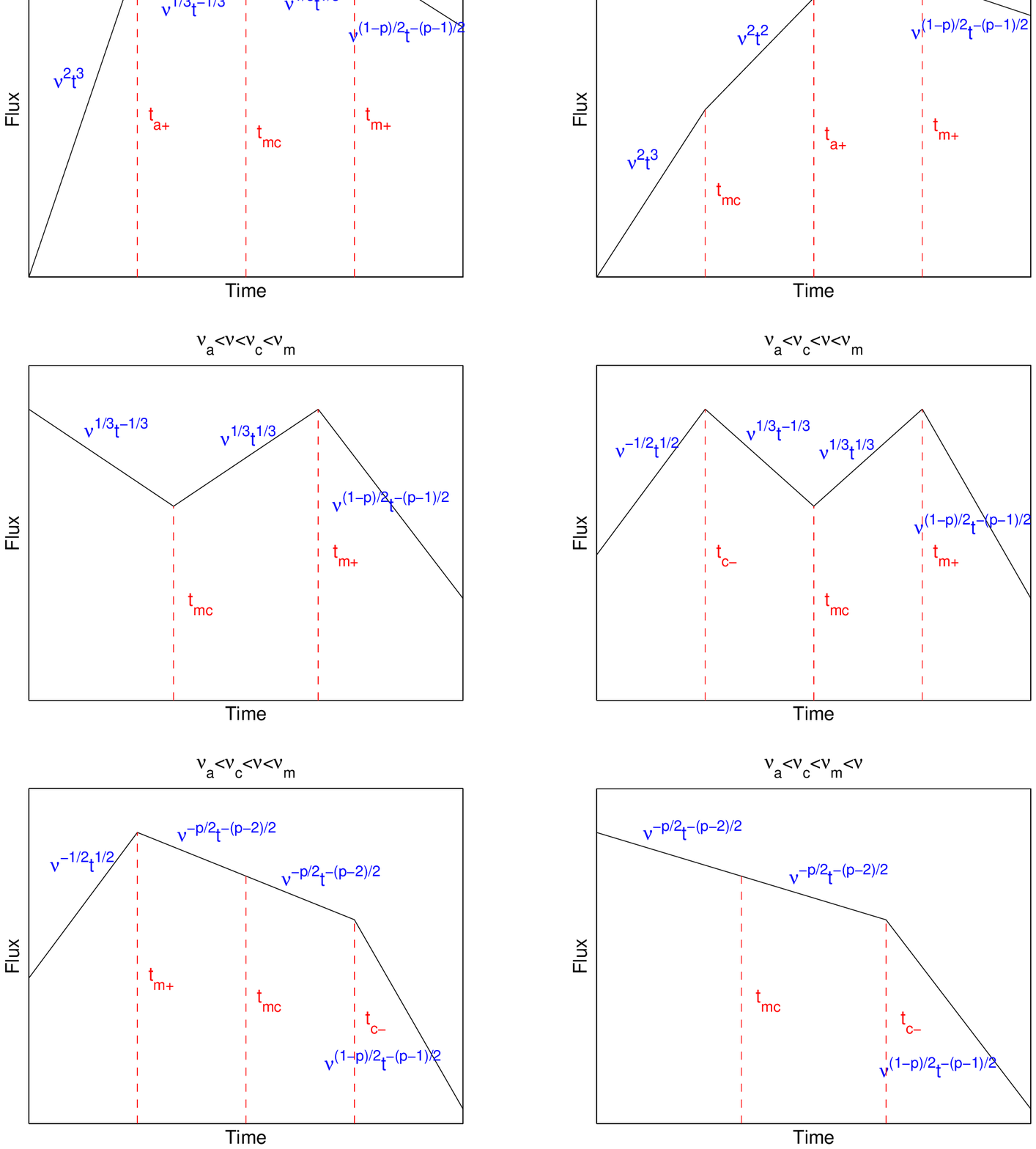}\\
    \caption{Same as Fig. \ref{figwind41},  but with the initial
characteristic frequency order $\nu_a < \nu_c < \nu_m$.}
  \label{figwind42}
\end{figure}

\begin{figure}
  \includegraphics[angle=0,width=1.0\textwidth]{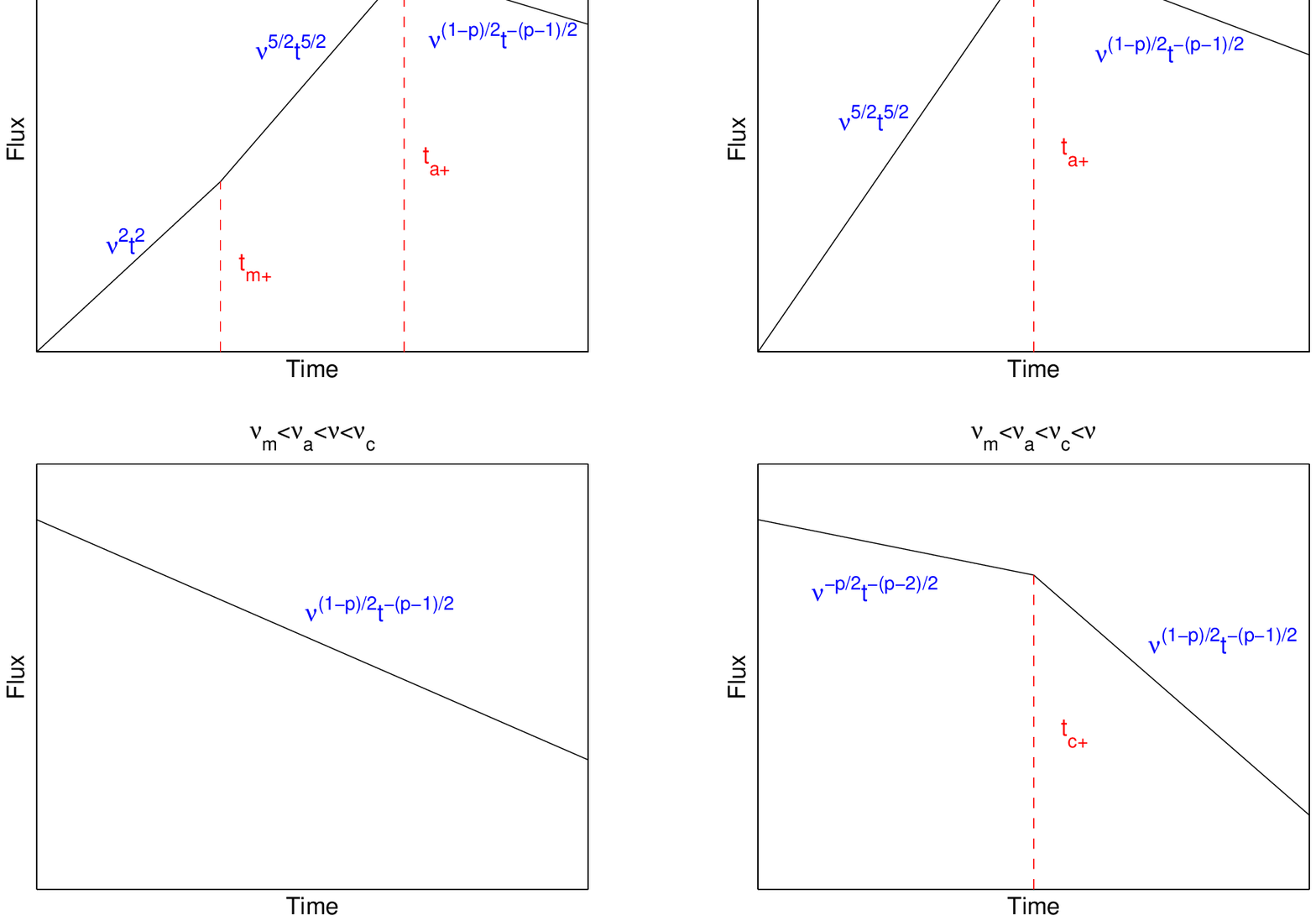}\\
    \caption{Same as Fig. \ref{figwind41},  but with the initial
characteristic frequency order $\nu_m < \nu_a < \nu_c$.}
  \label{figwind43}
\end{figure}

\begin{figure}
  \includegraphics[angle=0,width=1.0 \textwidth]{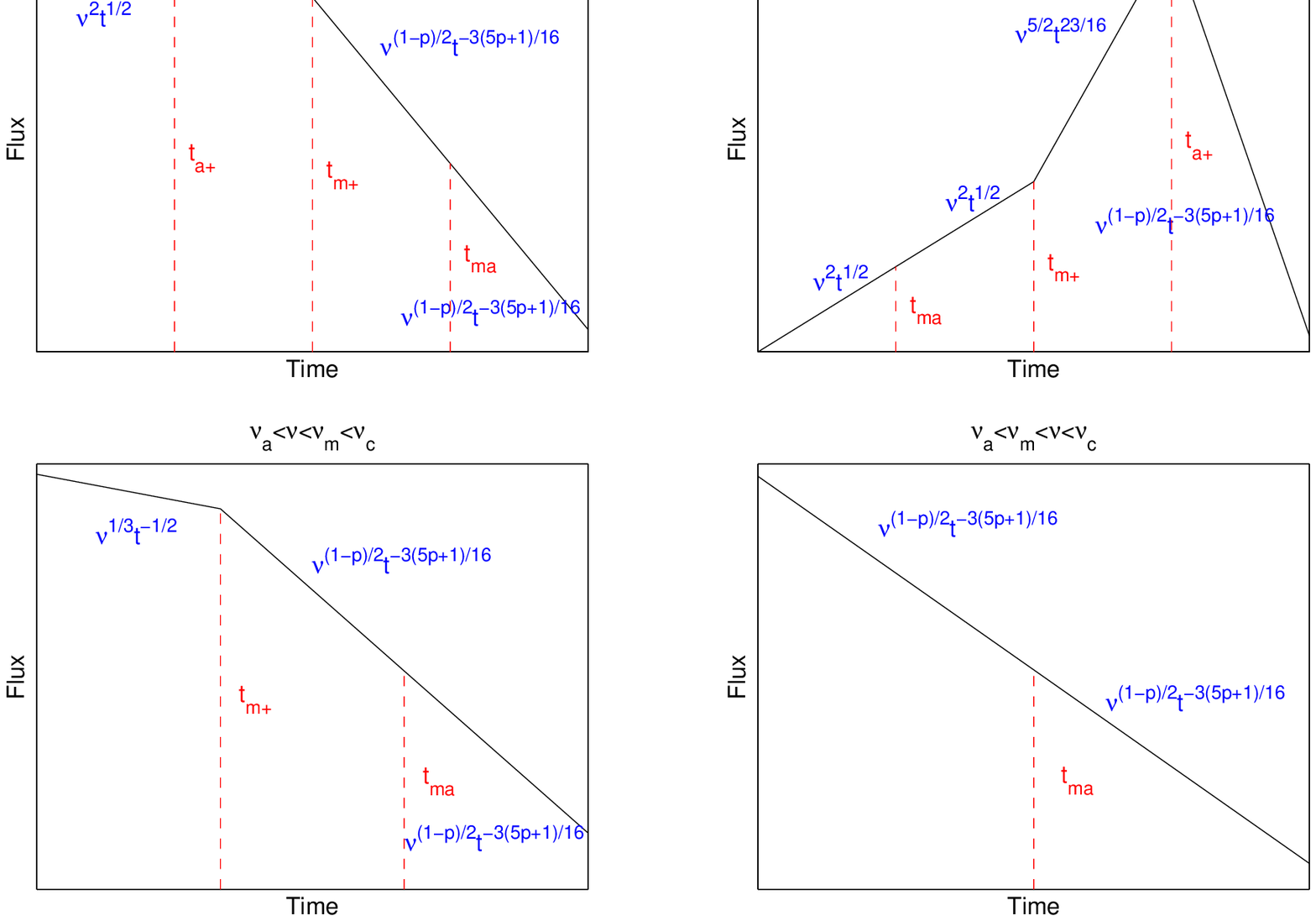}\\
    \caption{All possible reverse shock lightcurves after reverse
shock crossing, for thick shell wind model and the initial
characteristic frequency order $\nu_a < \nu_m < \nu_c$.}
  \label{figwind51}
\end{figure}

\begin{figure}
  \includegraphics[angle=0,width=1.0\textwidth]{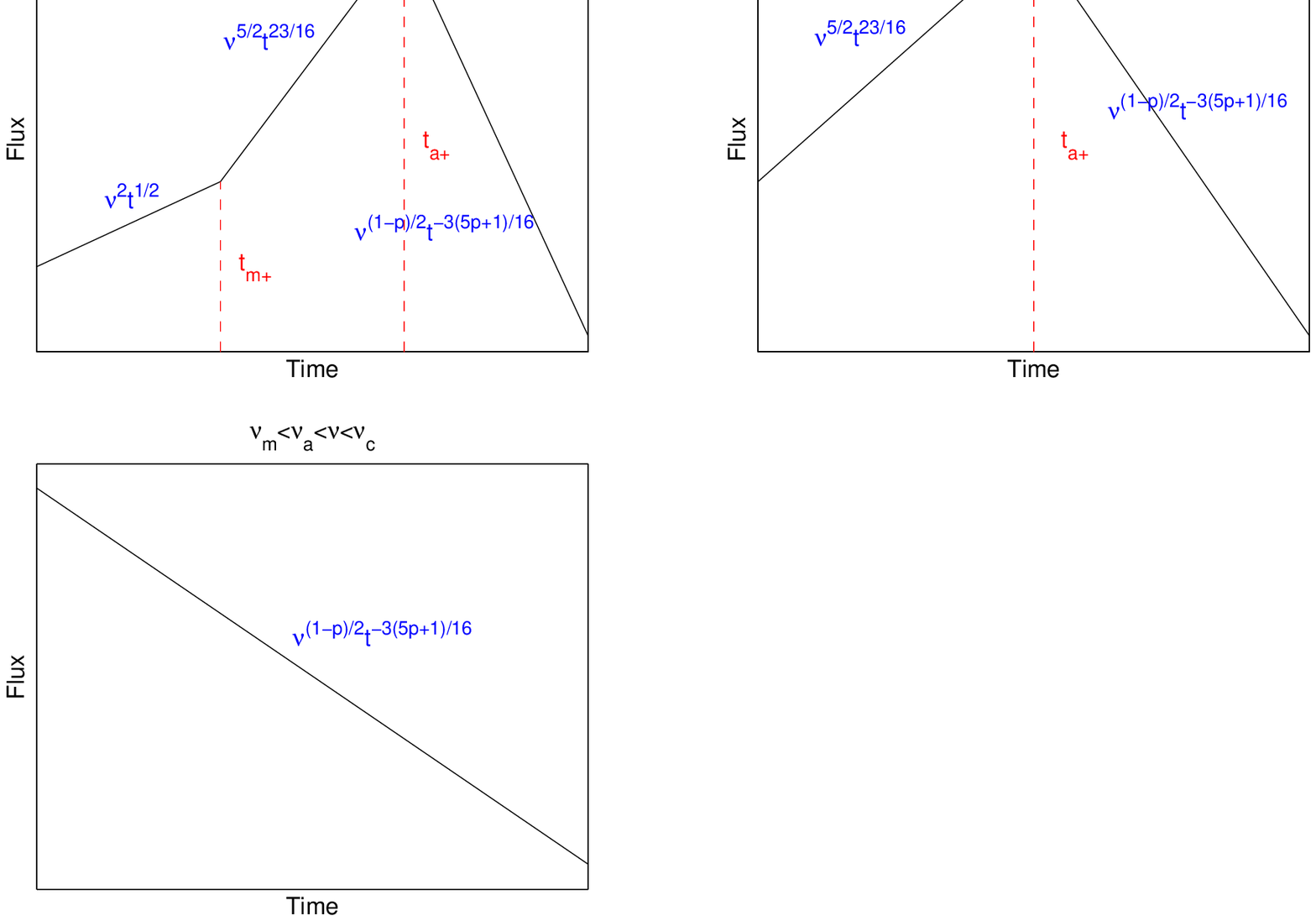}\\
    \caption{Same as Fig. \ref{figwind51},  but with the initial
characteristic frequency order $\nu_m < \nu_a < \nu_c$.}
  \label{figwind53}
\end{figure}

\begin{figure}
  \includegraphics[angle=0,width=1.0 \textwidth]{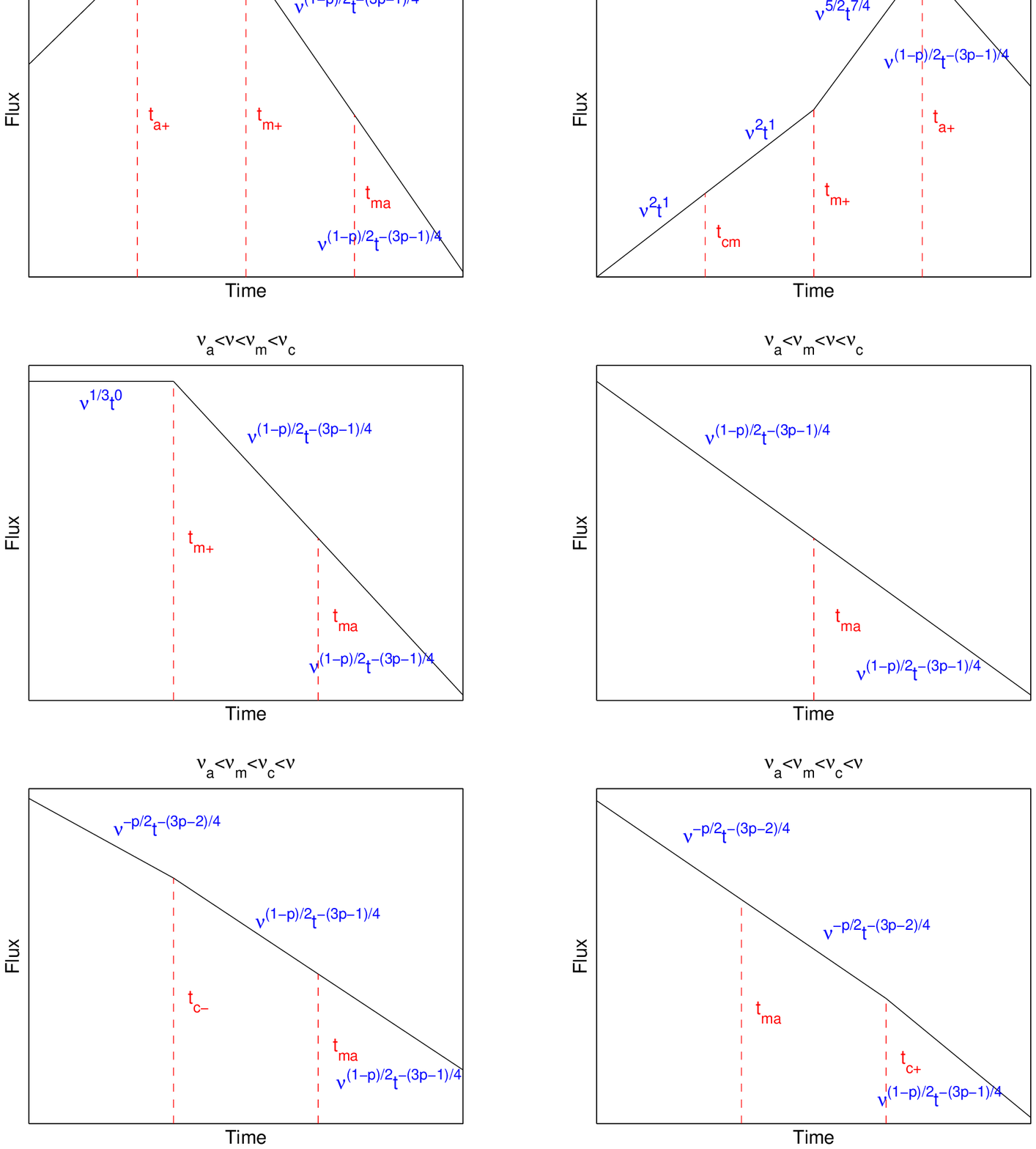}\\
    \caption{All possible forward shock lightcurves during Phase
2 (relativistic, isotropic, self-similar deceleration phase),
for a wind medium and the initial
characteristic frequency order $\nu_a < \nu_m < \nu_c$.}
  \label{figwind71}
\end{figure}

\begin{figure}
  \includegraphics[angle=0,width=1.0 \textwidth]{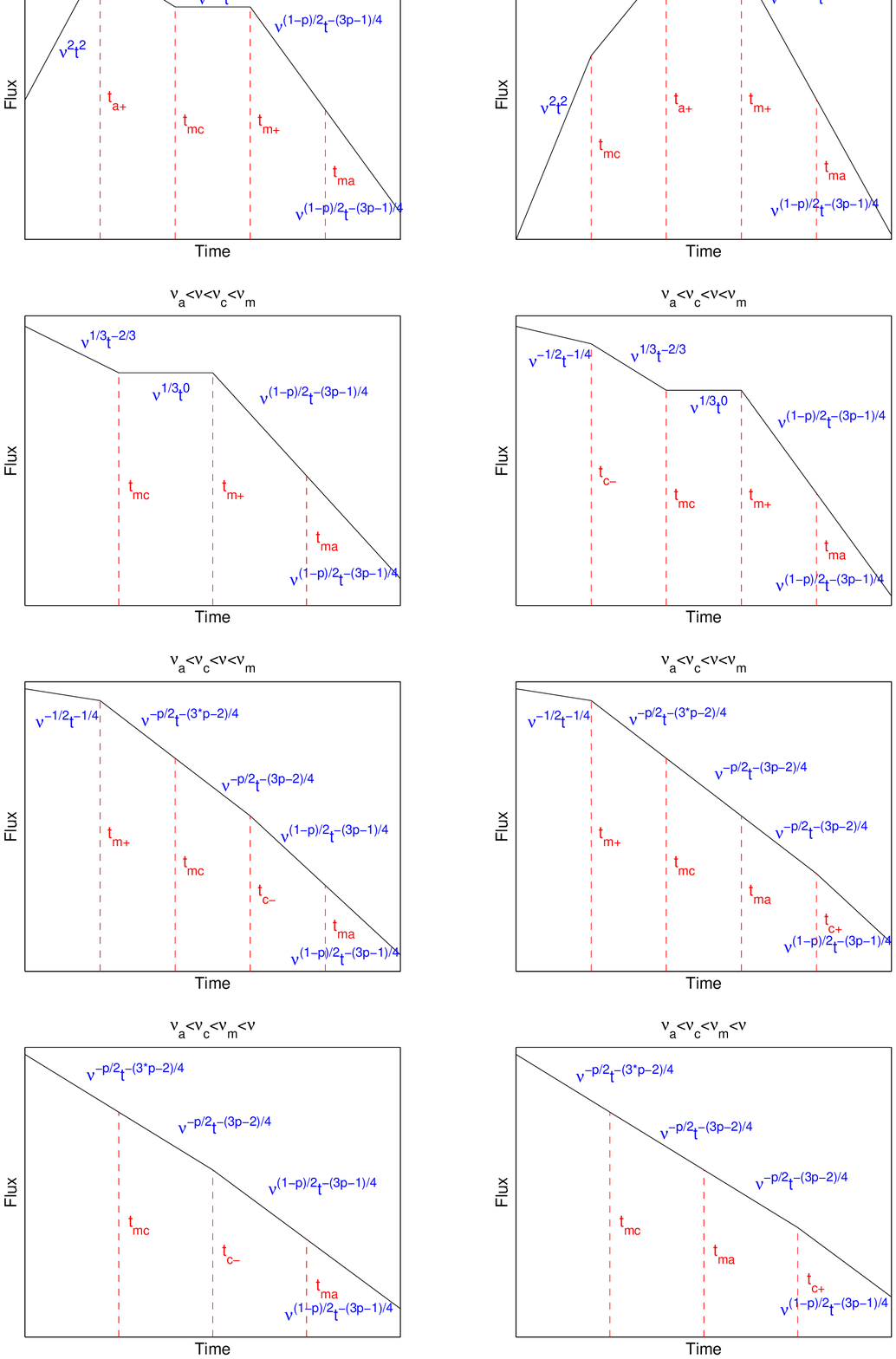}\\
    \caption{Same as Fig. \ref{figwind71},  but with the initial
characteristic frequency order $\nu_a < \nu_c < \nu_m$.}
  \label{figwind72}
\end{figure}

\begin{figure}
  \includegraphics[angle=0,width=1.0 \textwidth]{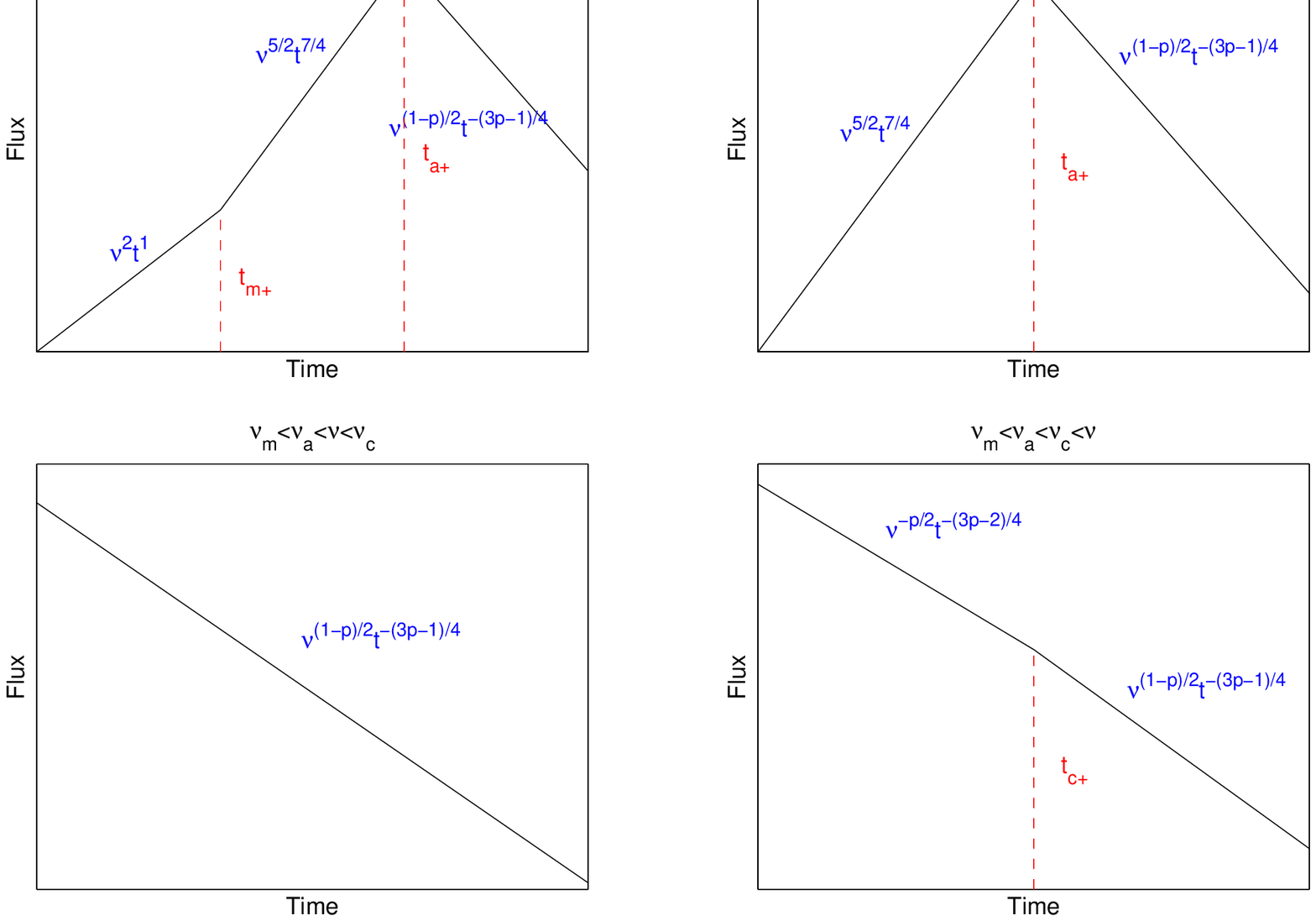}\\
    \caption{Same as Fig. \ref{figwind71},  but with the initial
characteristic frequency order $\nu_m < \nu_a < \nu_c$.}
  \label{figwind73}
\end{figure}

\begin{figure}
  \includegraphics[angle=0,width=1.0 \textwidth]{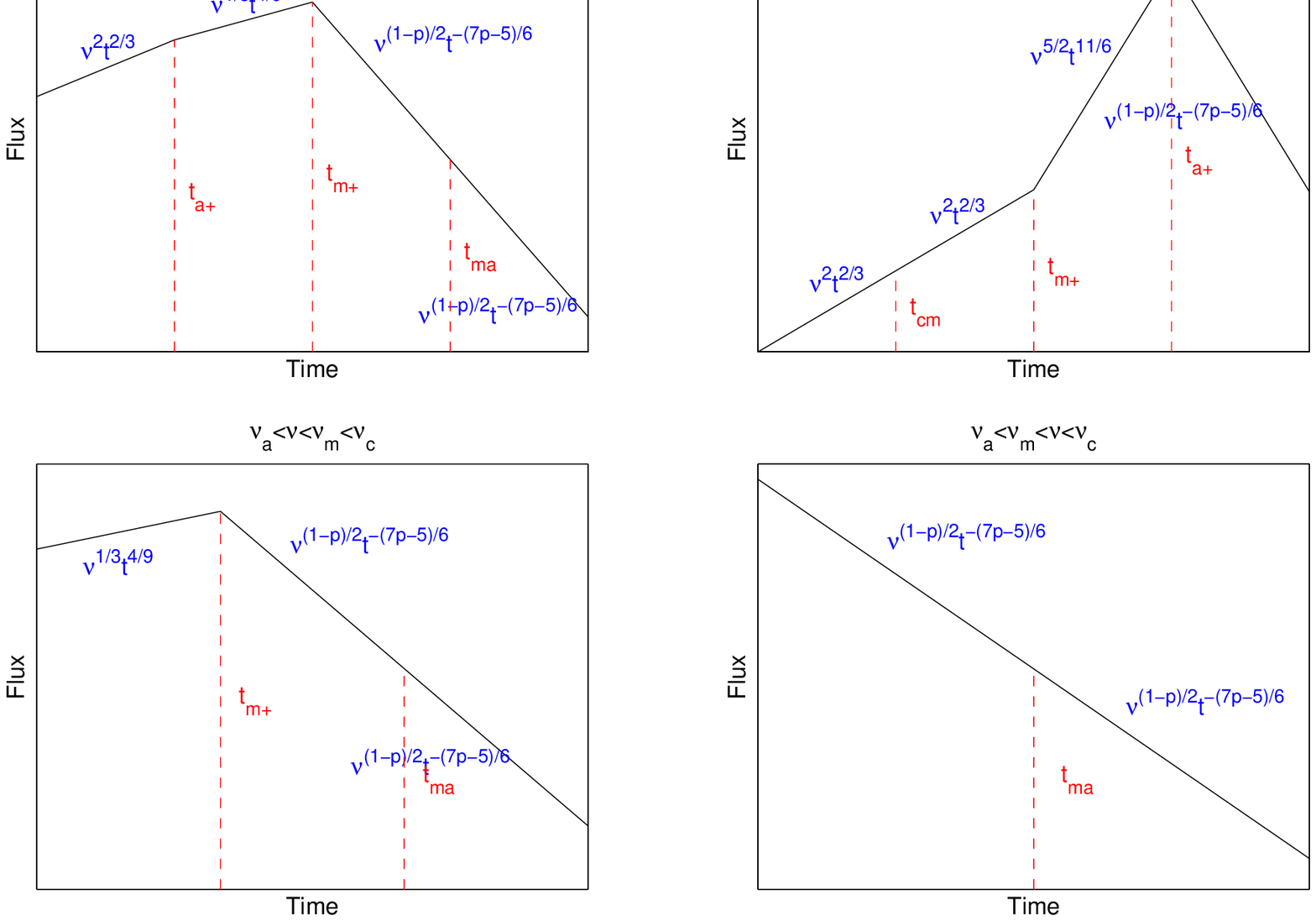}\\
    \caption{All possible forward shock lightcurves during Phase
4 (Newtonian phase), for a wind medium and the initial
characteristic frequency order $\nu_a < \nu_m < \nu_c$.}
  \label{figwind81}
\end{figure}

\begin{figure}
  \includegraphics[angle=0,width=1.0\textwidth]{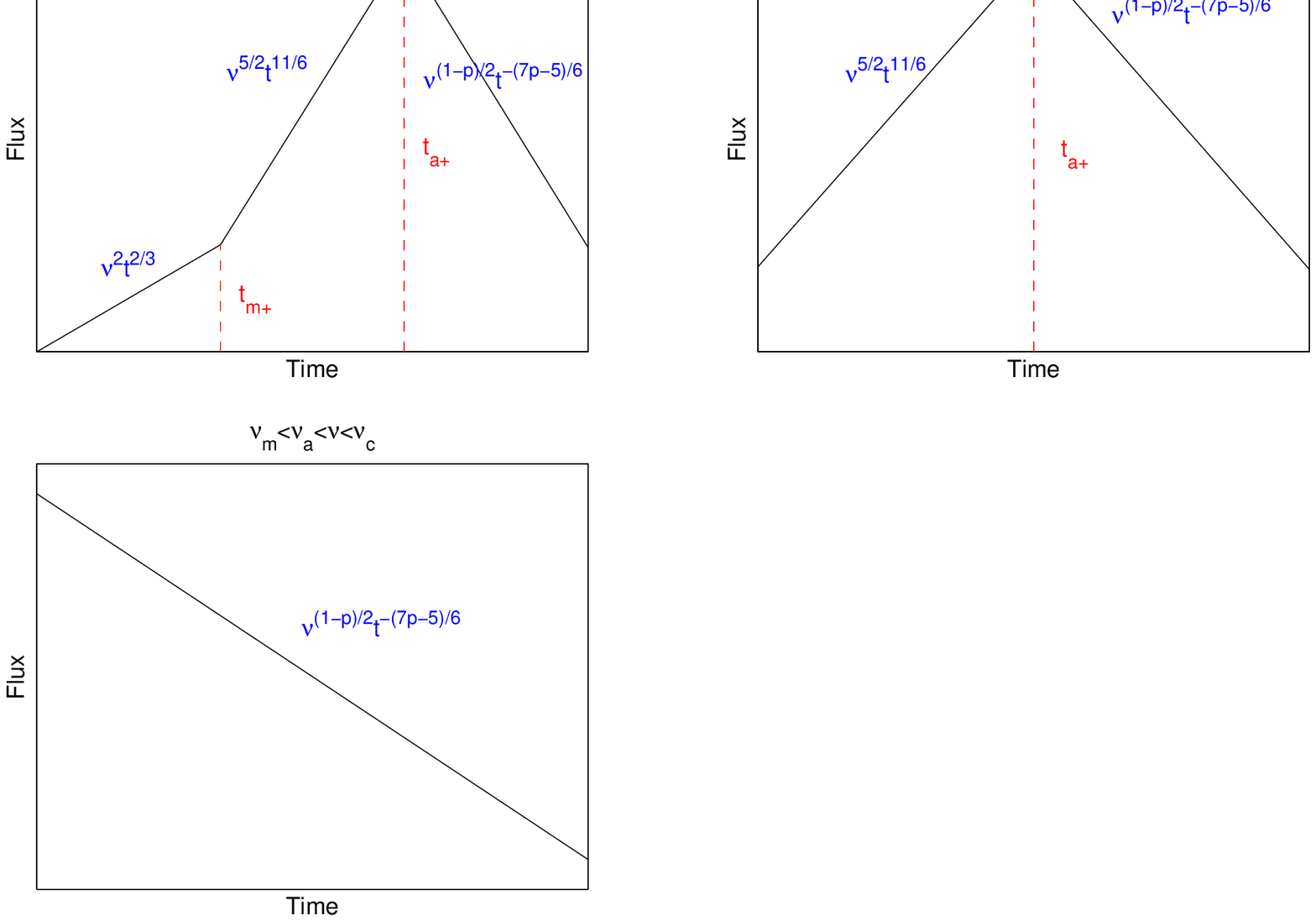}\\
    \caption{Same as Fig. \ref{figwind81},  but with the initial
characteristic frequency order $\nu_m < \nu_a < \nu_c$.}
  \label{figwind83}
\end{figure}

In order to make readers more conveniently use this reference tool,
we plot all the possible lightcurve shapes that can be derived
analytically\footnote{The only spectral regimes that are not
included are all the spectral orders that invoke $\nu_a > \nu_c$.
For such combinations, the power-law description of electron energy
distribution is no longer valid, and pile up of electrons near
$\gamma_a$ is expected \citep{kobayashi04,gao13}. Since the exact
shape of electron distribution cannot be obtained analytically, we
do not include these cases in the figures. Such electron pile-up
condition is usually not satisfied in most models reviewed in this
paper. The only relevant model is the reverse shock model during the
shock crossing phase for a wind medium, when $A_*$ is large enough
\citep{kobayashi04,gao13}.  }, and present spectral and temporal
indices for each temporal segment for all the phases discussed in
Section 3. These are presented in Figures \ref{fig11} to
\ref{figwind83}. Some of these lightcurves may demand extreme
afterglow parameters. However, since we aim at a complete reference
of the models and keep a wide open range of the observational
frequency and model parameters, we have included all the possible
frequency regime transitions for all the phases. In reality, one
could use the observational data to narrow down the possibilities to
identify the most relevant lightcurve segments. For easy
identification, Table 21 summarizes the corresponding figure numbers
for different dynamical models and spectral regimes.

\begin{figure}
\begin{minipage}[b]{0.5\textwidth}
\centering
\includegraphics[width=2.7in]{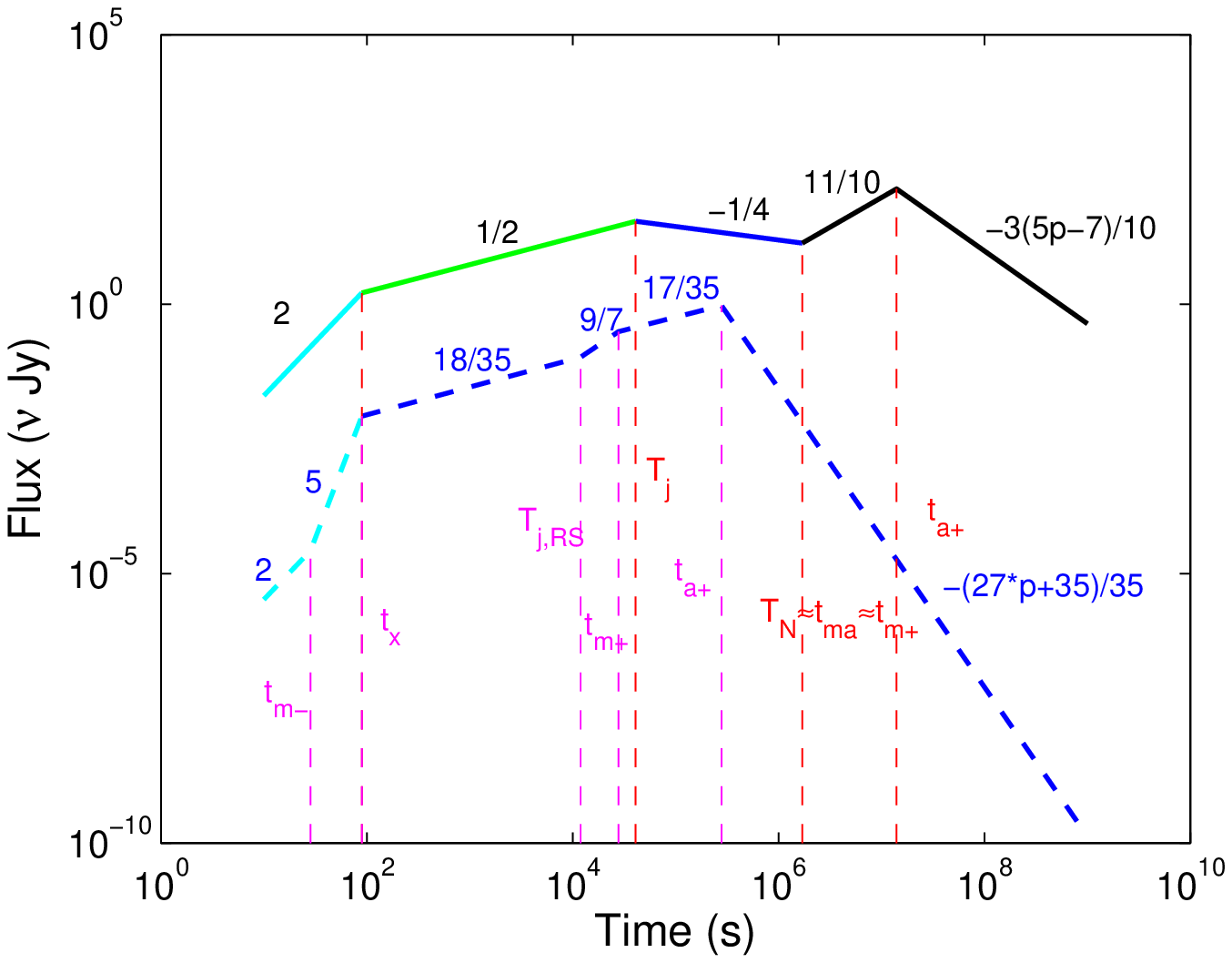}
\end{minipage}%
\begin{minipage}[b]{0.5\textwidth}
\centering
\includegraphics[width=2.7in]{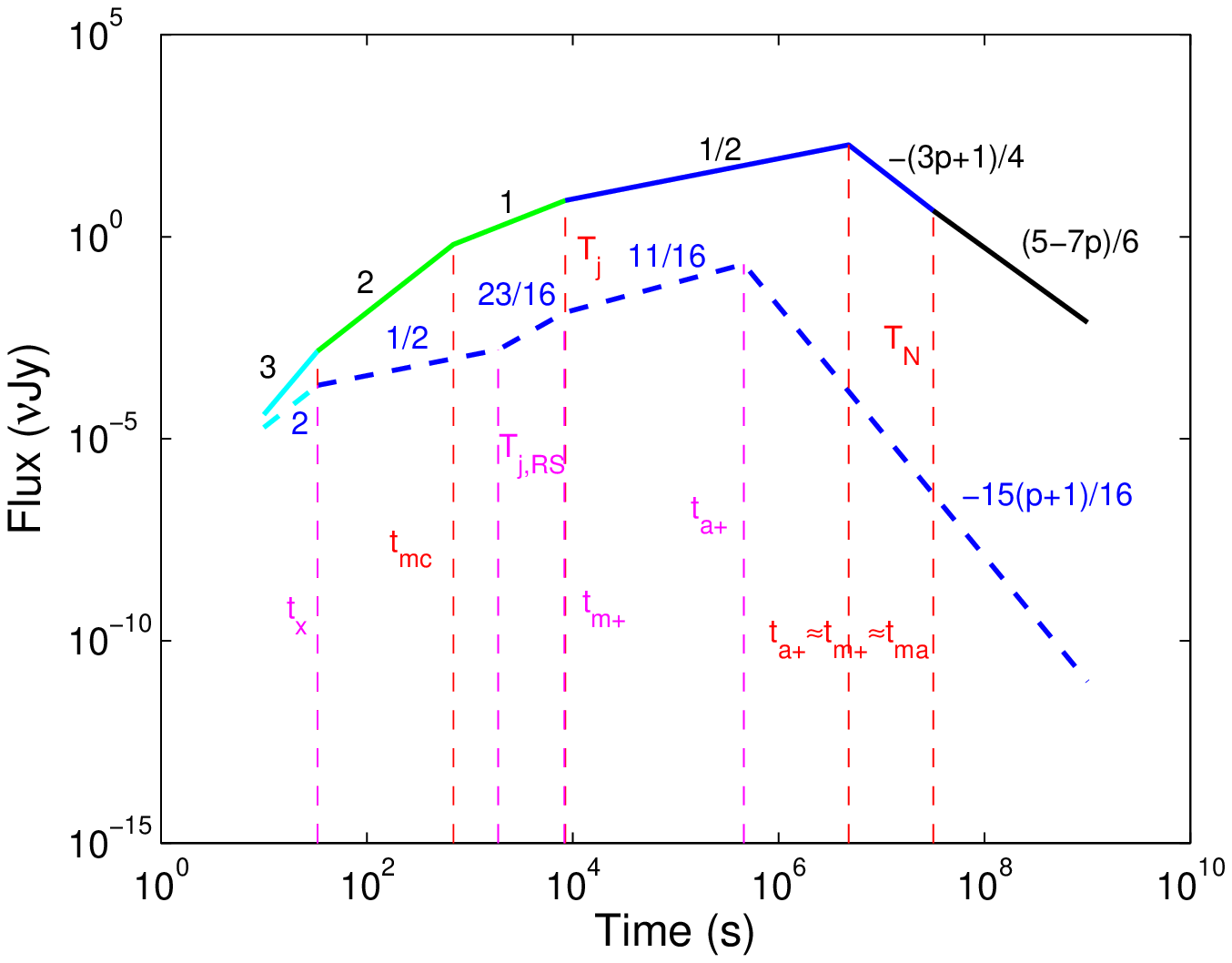}
\end{minipage}\\
\begin{minipage}[b]{0.5\textwidth}
\centering
\includegraphics[width=2.7in]{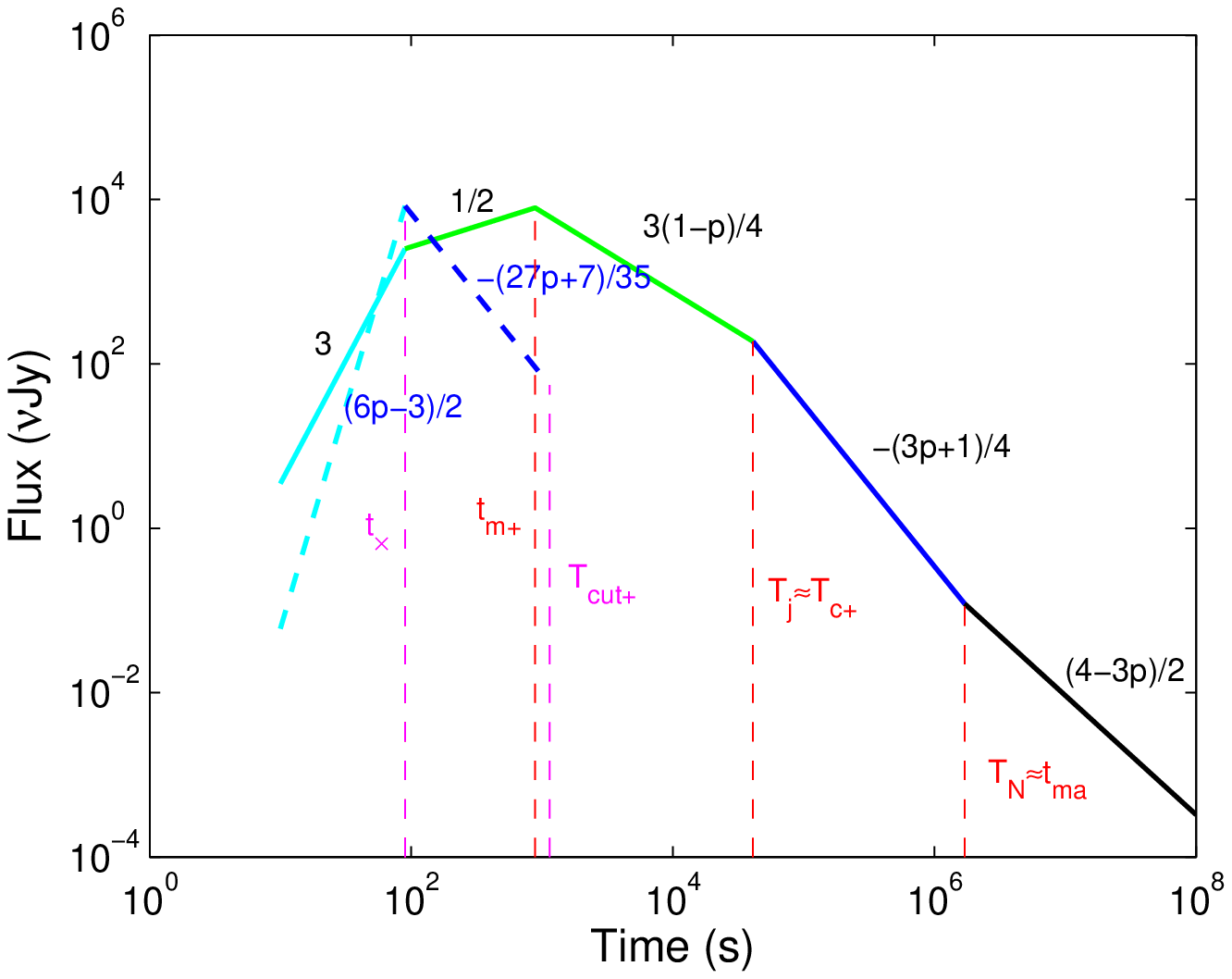}
\end{minipage}%
\begin{minipage}[b]{0.5\textwidth}
\centering
\includegraphics[width=2.7in]{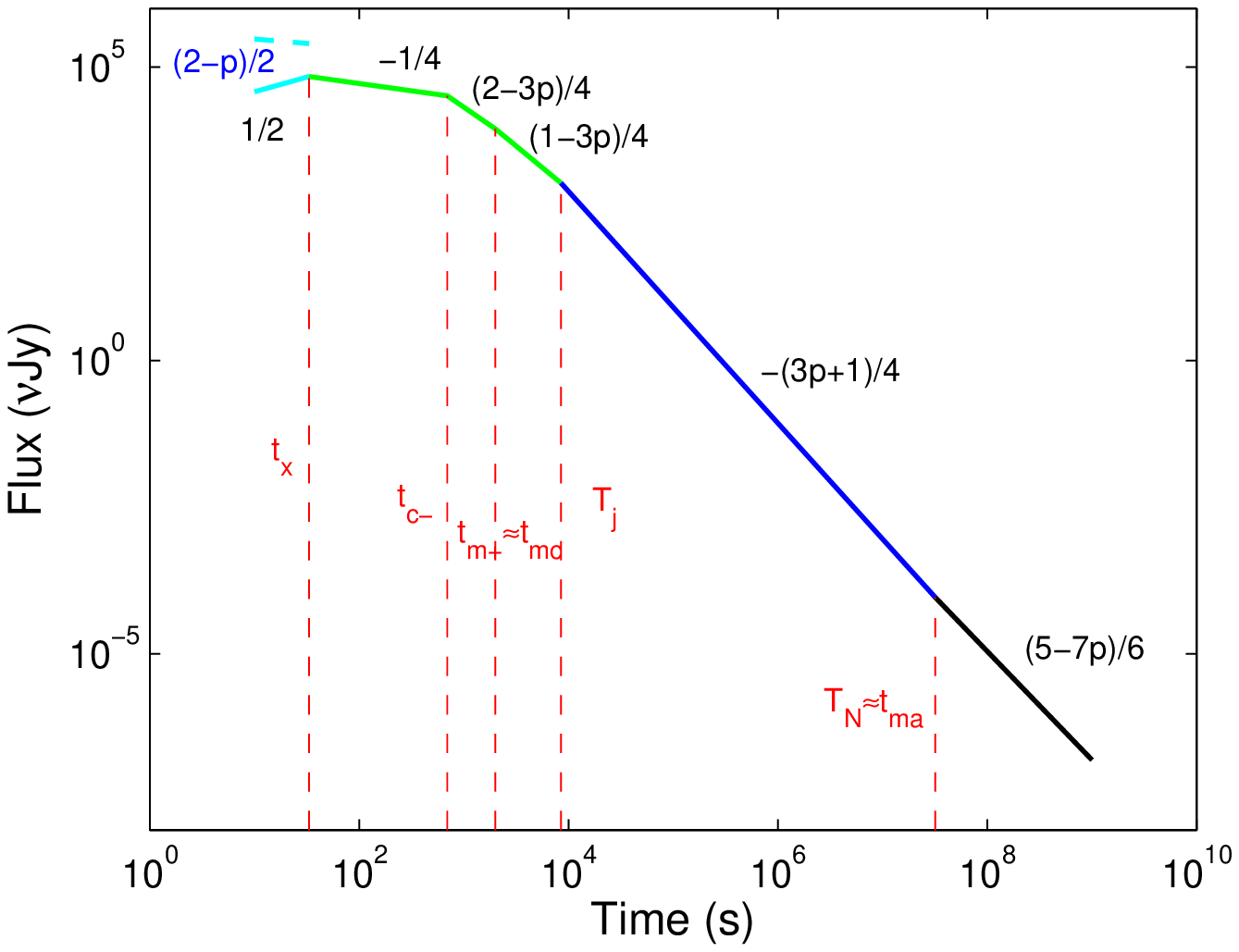}
\end{minipage}\\
\begin{minipage}[b]{0.5\textwidth}
\centering
\includegraphics[width=2.7in]{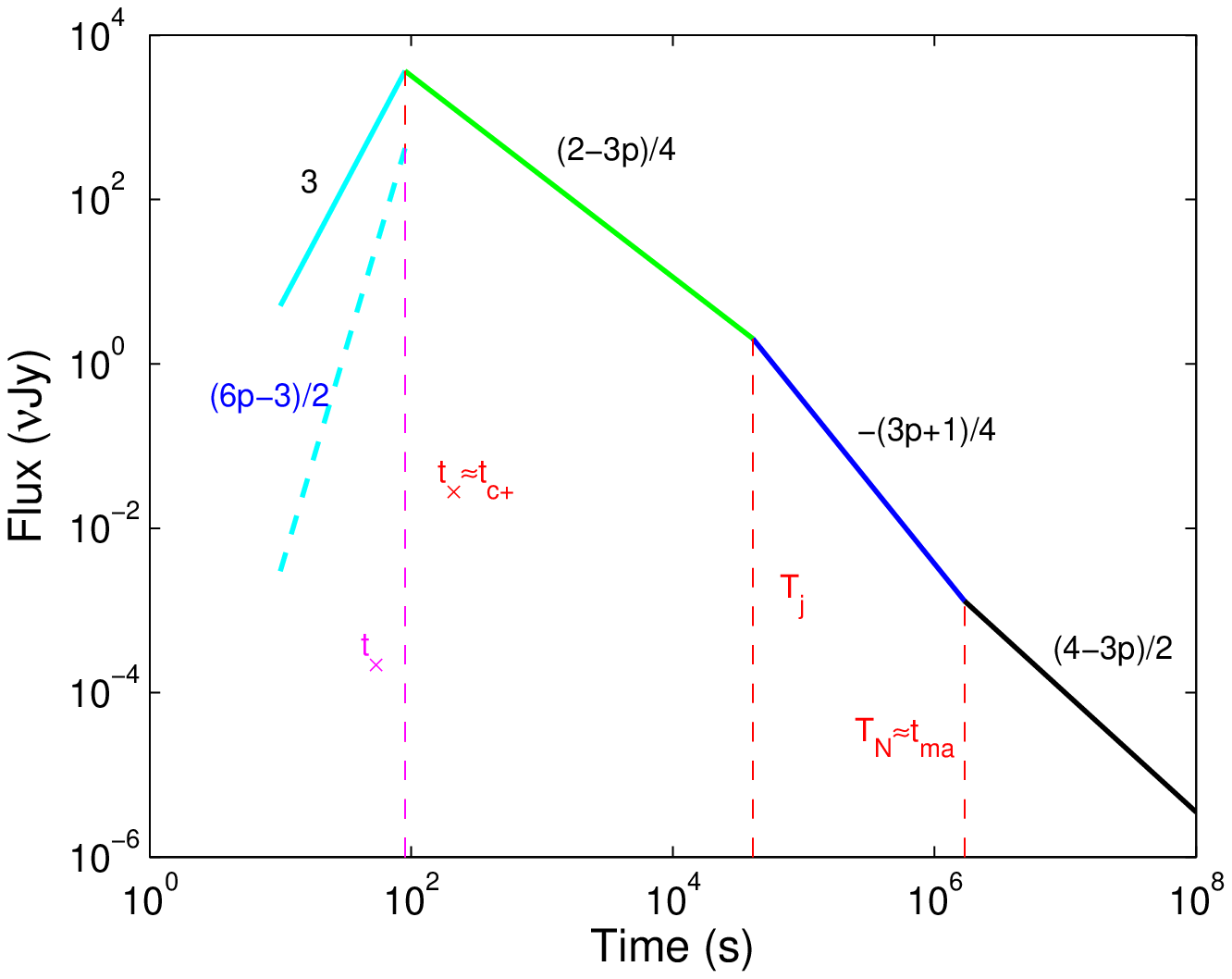}
\end{minipage}%
\begin{minipage}[b]{0.5\textwidth}
\centering
\includegraphics[width=2.7in]{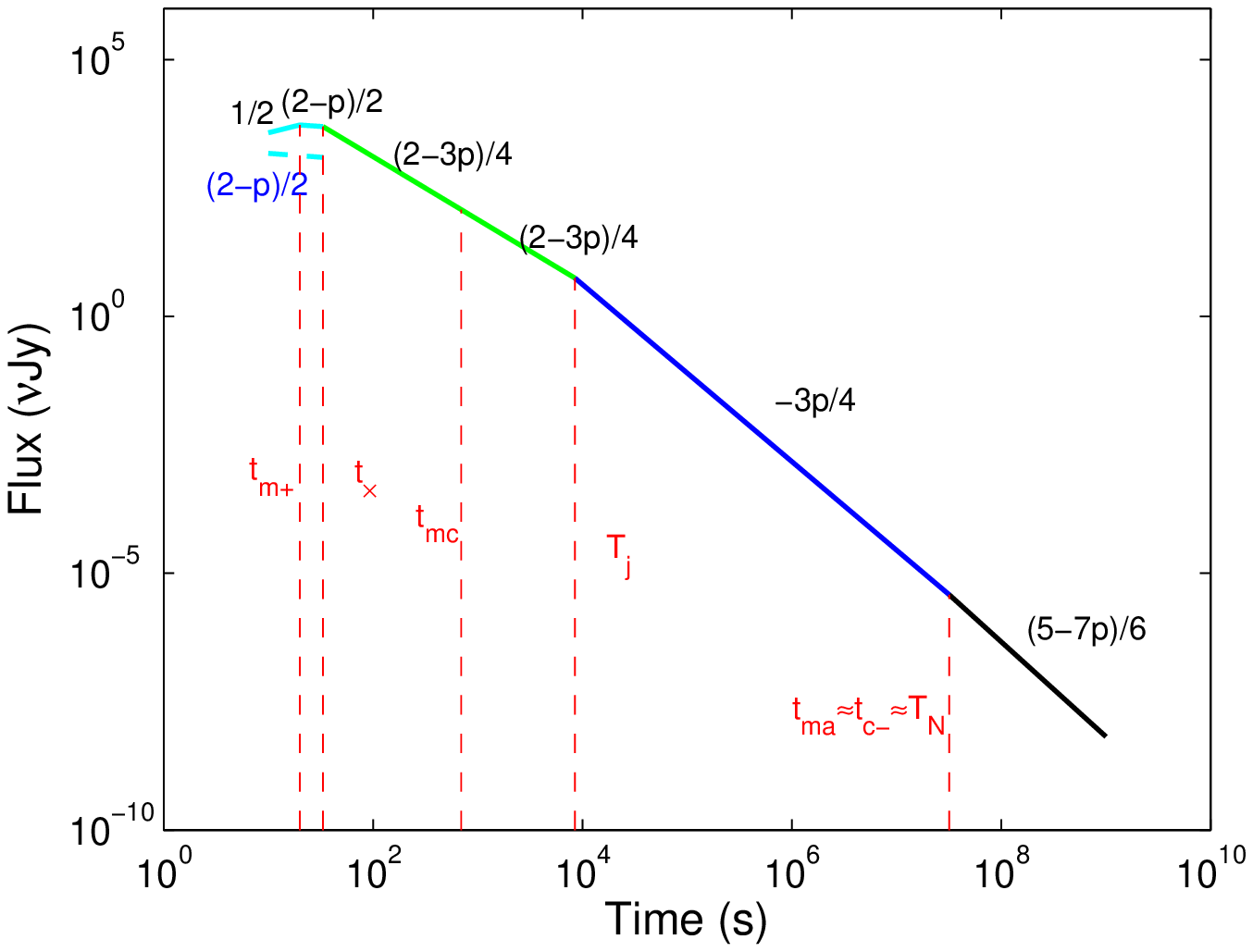}
\end{minipage}\\
       \caption{Example light curves in the radio, optical and X-ray bands
for a set of typical parameter values (see text). The left and right
panels are for the ISM and wind medium, respectively. In each panel,
from top to bottom are the lightcurves in the radio, optical and
X-ray band, respectively. Notations are the same with other Figures.
The parameters $T_{cut+}$,$t_{\times}$ $T_j$ and $T_{\rm N}$ denote
$\nu_{\rm cut}$ crossing time, the shock crossing time, jet break
time, and the transition time to the Newtonian phase, respectively.
The solid and dashed lightcurves denote contributions from the
forward and reverse shock, respectively. The 4 different phases of
forward shock emission are marked with 4 different colors. Notice
that the reverse shock light curves have a sharp ending, which
corresponds to time beyond which no on-axis electron radiation
contributes to the band (i.e. after shock crossing and $\nu >
\nu_{\rm cut}$. In reality, there should be emission from high
latitudes during these phases, so in these regimes there should be a
steeply-decaying lightcurve with slope $-(2+\beta)$, where $\beta$
is the flux density spectra index in the band
\citep{kumarpanaitescu00}.}
           \label{standard}
            \end{figure}

It is worth emphasizing that a critical time to separate Phase 1
(reverse shock crossing phase) and Phase 2 (self-similar
deceleration phase) is the shock crossing time $t_\times$
(Eq.\ref{tcross}). At $t_\times$, the ratios of the forward and
reverse shock quantities $F_{\nu,max}$, $\nu_m$, $\nu_c$ etc. can be
coasted into some simple forms \citep{zhang03}. Practically, one can
derive the forward shock scaling first (which is easier), and
extrapolate to $t_\times$. Then applying the reverse-to-forward
shock ratios of critical parameters \citep{zhang03,harrison13}, one
can derive the reverse shock parameters at $t_\times$. One can then
apply the reverse shock scaling laws to derive reverse shock
quantities. This approach would also lead to the same expressions
derived in \S3.1.2 and \S3.1.4. By comparing the reverse-to-forward
shock flux ratio at $t_\times$, one can determine which component
dominates for a specific frequency, see Figure \ref{standard} for
example.

The numerous possible lightcurves in each phase make it impossible
to draw all possible overall lightcurves. We therefore only draw a
set of example lightcurves based on a standard set of parameters. In
Figure \ref{standard}, we present the ``standard'' afterglow light
curves in radio ($10^9~ \rm{Hz}$), optical ($10^{15}~ \rm{Hz}$) and
X-ray ($10^{17}~ \rm{Hz}$) bands, by adopting a set of typical
parameter values: the total energy $E\sim10^{52}~ \rm{erg}$, initial
Lorentz factor $\Gamma_0=100$, width of ejecta $\Delta_0=10^{12}~
\rm{cm}$, jet opening angle $\theta_j = 0.1$, microphysics shock
parameters $\epsilon_e=0.1$, $\epsilon_B=0.01$ and electron index
$p=2.3$ for both forward and reverse shocks. For the ISM model, we
take $n_0=1 \rm~{cm^{-3}}$, so that the reverse shock is
non-relativistic and the system is in the thin-shell approximation.
For the wind model, we take $A_{*}=0.1$, the reverse shock is
relativistic and the system is in the thick-shell approximation.
More detailed studies on the standard models can be found in the
literature
\citep[e.g.][]{sari98,chevalier00,granotsari02,wu03,kobayashizhang03b,zou05}.

Several remarks regarding Fig.45 are worth addressing. 1. Only
external shock afterglow light curves are plotted. If one includes
the internal-origin ``prompt'' emission also, one would expect
another component before $t_\times$. There has been no observations
in the radio band in this time frame. In the optical and X-ray band,
this component is usually brighter than the external shock
component, and hence, would mask the early phase of the lightcurves.
After the cessation of the prompt emission, the lightcurve usually
transits to the afterglow emission through a ``steep decay'' likely
due to the high-latitude emission \citep[e.g. as observed in the
early X-ray afterglow detected with {\em
Swift},][]{tagliaferri05,zhang06,zhangbb07}. 2. The lightcurves are
plotted with identical microphysics parameters $\epsilon_e$ and
$\epsilon_B$ in the forward and reverse shocks. For the particular
set of parameters adopted, the reverse shock flux is usually lower
than that of forward shock in both radio and X-ray band, and it only
dominates the forward shock emission in the optical band early on
for a brief time. Observatinal data, on the other hand, require
different microphysics parameters in the two shocks, in particular,
a more magnetized reverse shock than the forward shock
\citep{fan02,zhang03,kumar03,harrison13}. This corresponds to the
ISM models with enhanced reverse shock peaks in the optical and
radio bands. Specifically, in the radio lightcurve (top-left panel),
the reverse shock flux at $t_{\rm a+}$ is much brighter than the
forward shock flux; in the optical band (mid-left panel), the
reverse shock flux at $t_\times$ way exceeds the forward shock flux,
and even at $t_{\rm m+}$ the reverse shock flux is higher than that
of forward shock, so that the optical flux shows a ``flattening''
behavior \citep{zhang03}. These are the ``radio flares'' and
``optical flashes'' as observed in some GRBs, such as GRB 990123
\citep{akerlof99,kulkarni99,kobayashisari00}. 3. Combining
lightcurve features and spectral properties is essential to diagnose
the physical origins of the afterglow emission. For example, the
peaks of the light curves could be due to a hydro-dynamical origin
(shock crossing or jet break) or crossing of a spectral break
($\nu_m$ or $\nu_a$). The former should not be accompanied by a
color change while the latter should. Taking spectral observations
before and after a certain break time is therefore crucial to
identify the correct model to interpret the data. The hydrodynamical
breaks are also expected to be ``achromatic'', i.e. occuring in all
wavelengths, while the frequency crossing breaks should be
chromatic. So simultaneous observations in all wavelengths are also
important to diagnose the physics of afterglow emission. 4. Some
light curve properties can be quickly applied to diagonose the
properties of the ambient medium. For example, in the pre-jet-break
phase, the wind model has a steeper slope than the ISM model. In the
optical band, a fast-rising optical flash would point towards an ISM
origin. In the radio band, a forward shock peak due to jet break
(achromatic break with other bands such as optical) would point
towards an ISM origin.

\section{Limitations of the analytical models}

Despite their great success, the analytical synchrotron external shock
models are known to have certain limitations that hinder a precise
description of GRB afterglows. In many situations, numerical
calculations are needed. In this section we itemize all the limitations
of the analytical approach, which serve as a caution to readers to
apply the analytical models reviewed in this paper.

\begin{itemize}
\item Swift observations suggest that X-ray flares observed in the
afterglow phase can be best modeled as internal emission of late
central engine activities
\citep{burrows05,zhang06,fanwei05,ioka05,lazzati07,maxham09}. It is
likely that some X-ray plateaus followed by steep decays (internal
plateaus) are also caused by late central engine activities
\citep{troja07,liang07b,lyons10}. A more extreme view interprets all
the X-ray afterglow as emission from the central engine
\citep{ghisellini07,kumar08a,kumar08b}. Therefore the external shock
model discussed in this review is not relevant to interpret X-ray
flares and internal X-ray plateaus, and possibly even the entire
X-ray emission.
\item A relativistic ejecta moving towards the observer has a complicated
equal arrival time effect
\citep{waxman97c,sari98b,panaitescu98b,granot99}, which smooths the
spectral and temporal breaks \citep{granotsari02}. The sharp
transition in the blastwave dynamics adopted in analytical models is
also an approximation. As a result, the sharp breaks predicted in
the analytical models usually do not exist.
\item Since the strength of the shock is continuously decreasing as the
blastwave decelerates, the magnetic field strengths continuously decay
in the shocked region. Electrons therefore cool in a varying
magnetic field, which leads to a very smooth or non-existence of
$\nu_c$ \citep{uhm13a}, see also \cite{vaneerten09}. In the fast
cooling regime, exactly the same effect makes the fast cooling
spectrum harder \citep{uhm13b} than $F_\nu \propto \nu^{-1/2}$
proposed by \cite{sari98}. In view of this, a sharp temporal or
spectral break observed in GRB afterglow lightcurve or spectrum must
not be associated with electron cooling \citep{uhm13a}.
\item All the analytical models reviewed in this article consider
synchrotron radiation only. Synchrotron self-Compton (SSC) effect
may be important in the afterglow phase \citep{weilu98,dermer00,zhangmeszaros01b}. Invoking synchrotron self-Compton (SSC) would
complicate the matter. In particular, it would enhance cooling
by a factor of $(1+Y)$, where $Y = L_{\rm IC}/L_{\rm syn} = U_{\rm ph}
/U_{\rm B}$, $L_{\rm IC}$ and $L_{\rm syn}$ are the luminosities
of the SSC and synchrotron components, respectively, and $U_{\rm ph}$
and $U_{\rm B}$ are the energy densities of the synchrotron photons
and magnetic fields, respectively. The detailed treatments of the
SSC effect can be found in \cite{sari01} and \cite{gao13}. During
the reverse shock crossing phase, besides SSC in the reverse shock
and forward shock regions, scattering of photons from the other shock
by electrons from both shocked regions can be also important,
which make more complicated spectra and lightcurves \citep{wang01,wang01b}.
\item Only adiabatic models are reviewed in the paper. In the literature,
radiative models have been also discussed
\citep[e.g.][]{sari97b,boettcher00}. However, since $\epsilon_e$ is usually
small, a GRB blastwave cannot be fully radiative even if electrons
are in the fast cooling regime. A partially radiative fireball and
its dynamical evolution have been discussed by various authors
\citep[e.g.][]{huang99,huang00,peer12,nava12} and the detailed
lightcurves of these cases have been calculated by \cite{wu05}.
\item Numerical simulations are needed to well describe the transitions
among various phases. For example, the analytical models in Phase 1
(reverse shock crossing phase) and Phase 2 (self-similar phase) do
not match exactly. After reverse shock crossing, how the blastwave
self-adjusts itself to the Blandford-McKee profile can be only addressed
by numerical simulations \citep[e.g.][]{kobayashisari00}. Sideway expansion
after the ``jet break'' phase and the transition from the ultra-relativistic
phase to deep Newtonian phase all need numerical simulations to resolve
the details \citep{cannizzo04,zhangmacfadyen09,vaneerten12}.
\item The lightcurves involving collimated jets are complicated and
usually require numerical treatments. Even for a uniform jet, the
shape of the jet break may depend on the viewing angle from the jet
axis \citep{granot02,vaneerten12}. If the viewing angle is outside
the jet cone, one expects a variety of lightcurves for the so-called
``orphan afterglows'', which cannot be properly addressed analytically.
More complicated jets invoke angular structure with decreasing
luminosity and Lorentz factor with respect to the jet axis
\citep{meszaros98}. The commonly discussed the jet structures
include power law \citep{meszaros98}, Gaussian \citep{zhang04},
and two-component conical jets \citep{berger03b,racusin08}.
An on-axis observer would see a steeper lightcurve
than the isotropic case \citep{meszaros98,daigou01,panaitescu05}.
For an off-axis observer \citep{rossi02,zhangmeszaros02b}, the
lightcurve may show a jet-break-like feature as the jet axis enters
the field of view, but the exact shape of the break depends on the
angular structure of the jet and the viewing angle
\citep{kumargranot03,granotkumar03}. The two-component jets
can show more complicated lightcurve behaviors \citep{huang04,peng05}.
\item It is possible that due to continuous energy injection or
ejecta Lorentz factor stratification, a long-lived reverse shock
may continue to exist, and the blastwave never enters the
Blandford-McKee phase. The long-lasting reverse shock can show
rich afterglow lightcurve features \citep{uhm12}, which may show
up above the forward shock contribution if the reverse shock
emission is enhanced. A more extreme view is that the entire
observed afterglow is of a reverse shock origin \citep{uhm07,genet07}.
\item Analyses of early afterglow data \citep{fan02,zhang03,kumar03}
and theoretical considerations
\citep{usov92,meszarosrees97b,metzger11,lei13} suggest that the GRB
central engine is likely magnetized. The GRB ejecta therefore likely
carries a certain degree of magnetization. The reverse shock models
presented here apply to low-magnetization cases. For moderate to
high magnetization, the shock jump conditions and the strength of
reverse shock are modified \citep{zhangkobayashi05,fan04b}, and
numerical simulations are needed to achieve precise results
\citep{mimica09}. Also numerical simulations \citep{sironi09}
suggest that electron acceleration becomes suppressed in a
magnetized shock, which would also affect the predicted synchrotron
radiation flux.
\item All the models invoke constant microphysics parameters
$\epsilon_e$ and $\epsilon_B$. In principle, these parameters
may evolve with time also, and some authors have considered such
more complicated models \citep[e.g.][]{ioka06,fanpiran06a}.
\item More complicated afterglow models invoke density bumps
\citep{dailu02b,daiwu03,nakargranot07}, violent energy injection into
the blastwave via collision from a fast shell ejected at late
times \citep{zhangmeszaros02a,geng13}, and patchy jets
\citep{kumarpiran00a,ioka05}.
\item Finally, in the early afterglow phase, additional physical processes
may modify the blastwave dynamics. These include pair loading effect
caused by interaction between radiation front and ambient medium
\citep{madau00,thompson00,meszaros01,beloborodov02} and neutron
decay effect from a neutron-rich ejecta
\citep{derishev01,beloborodov03b,fan05b}.
\end{itemize}

\section*{Acknowledgements}
We thank the referee Shiho Kobayashi for many constructive
suggestions, and L. Resmi for the helpful discussion on the
derivation details. This work is supported by NSF under Grant No.
AST-0908362,  by National Natural Science Foundation of China
(grants 11003004 ,11173011 and U1231101), and National Basic
Research Program (``973'' Program) of China under Grant No.
2009CB824800 and 2014CB845800. HG and WHL acknowledges a Fellowship
from China Scholarship Program for support, XFW acknowledges support
by the One-Hundred-Talents Program of Chinese Academy of Sciences.

\begin{table}[H]
 \begin{center}{\scriptsize
 \begin{tabular}{c|c|c|c|c|c|c|c|c} \hline\hline

& \multicolumn{6}{c|}{Phase 1} & \multicolumn{1}{c|}{}
& \multicolumn{1}{c}{}\\
\cline{2-3} \cline{4-5} \cline{6-7} Initial characteristic&
\multicolumn{3}{c|}{Thin shell} & \multicolumn{3}{c|}{Thick shell}
& \multicolumn{1}{c|}{Phase 2}& \multicolumn{1}{c}{Phase 4}\\
\cline{2-3} \cline{4-5} \cline{6-7}frequency
order                                 & FS               & $\rm RS_{pre}$        & $\rm RS_{post}$   &$\rm FS$     & $\rm RS_{pre}$  & $\rm RS_{post}$        &                  &\\
 \hline
 $\nu_a < \nu_m < \nu_c$ (ISM)    \ \ & \ \ $1-2$\ \     & \ \ $5$\ \            & \ \ $8$\ \        & \ \ $10$\ \ & \ \ $13$\ \     & \ \ $16$\ \            & \ \ $18$\ \      & \ \ $22$\ \ \\
\hline
  $\nu_a < \nu_c < \nu_m$ (ISM)   \ \ & \ \ $3$\ \       & \ \ $6$\ \            & \ \ $--$ \ \      & \ \ $11$\ \ & \ \ $14$\ \     & \ \ $--$\ \            & \ \ $19-20$\ \   & \ \ $--$\ \   \\
     \hline
  $\nu_m < \nu_a < \nu_c$  (ISM)  \ \ & \ \ $4$\ \       & \ \ $7$\ \            & \ \ $9$\ \        & \ \ $12$\ \ & \ \ $15$\ \     & \ \ $17$\ \            & \ \ $21$\ \      & \ \ $23$\ \   \\
  \hline
 $\nu_a < \nu_m < \nu_c$ (Wind)    \ \ & \ \ $24$\ \     & \ \ $27$\ \           & \ \ $30$\ \       & \ \ $32$\ \ & \ \ $35$\ \     & \ \ $38$\ \            & \ \ $40$\ \      & \ \ $43$\ \   \\
\hline
  $\nu_a < \nu_c < \nu_m$ (Wind)   \ \ & \ \ $25$\ \     & \ \ $28$\ \           & \ \ $--$ \ \      & \ \ $33$\ \ & \ \ $36$\ \     & \ \ $--$\ \            & \ \ $41$\ \      & \ \ $--$\ \     \\
     \hline
  $\nu_m < \nu_a < \nu_c$  (Wind)  \ \ & \ \ $26$\ \     & \ \ $29$\ \           & \ \ $31$\ \       & \ \ $34$\ \ & \ \ $37$\ \     & \ \ $39$\ \            & \ \ $42$\ \      & \ \ $44$\ \    \\
   \hline\hline
 \end{tabular}
 }
 \end{center}
 \caption{Collection of figure numbers corresponding to different dynamical models and initial spectra regimes.}
 \end{table}

\clearpage

\clearpage

\clearpage

\clearpage

\clearpage


\clearpage
\appendix{Appendix A: $p$-dependent coefficients in analytical solutions}

$G(p)=\left(\frac{p-2}{p-1}\right)^2$. For convenience, we define
$f(p)=\frac{\Gamma(\frac{3p+22}{12})\Gamma(\frac{3p+2}{12})}{\Gamma(\frac{3p+19}{12})\Gamma(\frac{3p-1}{12})}$.

\textbf{Thin Shell Forward shock}
\begin{eqnarray}
&&g^{I}(p)=  \left(\frac{p-1}{p-2}\right)^{}(p+1)^{3/5}f(p)^{3/5}  \nonumber\\
&&g^{II}(p)=  1.5\times10^{-\frac{30}{p+4}}\left(\frac{p-2}{p-1}\right)^{\frac{2(p-1)}{p+4}}(p+1)^{\frac{2}{p+4}}f(p)^{\frac{2}{p+4}}  \nonumber\\
&&g^{III}(p)=  (p+1)^{3/5}f(p)^{3/5}  \nonumber\\
&&g^{IV}(p)=  e^{\frac{38p-76}{p-1}}(3736-1868p)^{\frac{2}{p-1}}(p-1)^{-\frac{2}{p-1}} \nonumber\\
&&g^{V}(p)=  e^{\frac{38-19p}{p-1}}(3736-1868p)^{\frac{1}{1-p}}(p-1)^{\frac{1}{p-1}}(p+1)^{3/5}f(p)^{3/5}  \nonumber\\
&&g^{VI}(p)=  1.9\times10^{\frac{16(p-2)}{p+4}}e^{\frac{7}{p+4}}(3736-1868p)^{\frac{2}{p+4}}(p-1)^{-\frac{2}{p+4}}(p+1)^{\frac{2}{p+4}}f(p)^{\frac{2}{p+4}}  \nonumber\\
&&g^{VII}(p)=  (p+1)^{3/5}f(p)^{3/5}  \nonumber\\
&&g^{VIII}(p)=  \left(\frac{p-1}{p-2}\right)^{}(p+1)^{3/5}f(p)^{3/5}  \nonumber\\
&&g^{IX}(p)=  4.0\times10^{-\frac{16}{p+4}}\left(\frac{p-2}{p-1}\right)^{\frac{2(p-1)}{p+4}}(p+1)^{\frac{2}{p+4}}f(p)^{\frac{2}{p+4}}  \nonumber\\
&&g^{X}(p)=  (p+1)^{3/5}f(p)^{3/5}  \nonumber\\
&&g^{XI}(p)= 2^{\frac{11(p-2)}{p-1}}3^{\frac{p-2}{p-1}}e^{\frac{13p-27}{p-1}}(3736-1868p)^{\frac{2}{p-1}}(p-1)^{-\frac{2}{p-1}}\nonumber\\
&&g^{XII}(p)=  2^{-\frac{11(p-2)}{2(p-1)}}3^{\frac{2-p}{2(p-1)}}e^{\frac{13-7p}{p-1}}(3736-1868p)^{\frac{1}{1-p}}(p-1)^{\frac{1}{p-1}}(p+1)^{3/5}f(p)^{3/5}  \nonumber\\
&&g^{XIII}(p)=  2^{\frac{11(p-2)}{p+4}}3^{\frac{2-p}{p+4}}e^{\frac{41}{p+4}}(3736-1868p)^{\frac{2}{p+4}}(p-1)^{-\frac{2}{p+4}}(p+1)^{\frac{2}{p+4}}f(p)^{\frac{2}{p+4}}  \nonumber\\
&&g^{XIV}(p)=  (p+1)^{3/5}f(p)^{3/5}  \nonumber\\
\end{eqnarray}
\textbf{Thin Shell Reverse shock}
\begin{eqnarray}
&&g^{I}(p)=  \left(\frac{p-1}{p-2}\right)^{}(p+1)^{3/5}f(p)^{3/5}  \nonumber\\
&&g^{II}(p)=  4.1\times10^{-\frac{360}{p+4}}\left(\frac{p-2}{p-1}\right)^{\frac{2(p-1)}{p+4}}(p+1)^{\frac{2}{p+4}}f(p)^{\frac{2}{p+4}}  \nonumber\\
&&g^{III}(p)=  (p+1)^{3/5}f(p)^{3/5}  \nonumber\\
&&g^{IV}(p)=e^{\frac{38p-76}{p-1}}(3.0\times10^{33}-1.5\times10^{33}p)^{\frac{2}{p-1}}(p-1)^{-\frac{2}{p-1}}\nonumber\\
&&g^{V}(p)=  e^{\frac{38-19p}{p-1}}(3.0\times10^{33}-1.5\times10^{33}p)^{\frac{1}{1-p}}(p-1)^{\frac{1}{p-1}}(p+1)^{3/5}f(p)^{3/5}  \nonumber\\
&&g^{VI}(p)=  5.5\times10^{\frac{16p-60}{p+4}}(3.0\times10^{33}-1.5\times10^{33}p)^{\frac{2}{p+4}}(p-1)^{-\frac{2}{p+4}}(p+1)^{\frac{2}{p+4}}f(p)^{\frac{2}{p+4}}  \nonumber\\
&&g^{VII}(p)=  (p+1)^{3/5}f(p)^{3/5}  \nonumber\\
&&g^{VIII}(p)=  \left(\frac{p-1}{p-2}\right)^{}(p+1)^{3/5}f(p)^{3/5}  \nonumber\\
&&g^{IX}(p)=  1.3\times10^{-\frac{486}{p+4}}3^{-\frac{25}{p+4}}\pi^{-\frac{9}{p+4}}\left(\frac{p-2}{p-1}\right)^{\frac{2(p-1)}{p+4}}(p+1)^{\frac{2}{p+4}}f(p)^{\frac{2}{p+4}}  \nonumber\\
&&g^{X}(p)=  (p+1)^{3/5}f(p)^{3/5}  \nonumber\\
&&g^{XI}(p)=2^{\frac{11(p+6)}{p-1}}3^{\frac{p+4}{p-1}}e^{\frac{13p-27}{p-1}}(1.5\times10^{33}-7.6\times10^{32}p)^{\frac{1}{1-p}}(p-1)^{-\frac{2}{p-1}}\nonumber\\
&&g^{XII}(p)=  2^{-\frac{11(p+6)}{p-1}}3^{\frac{p+4}{2(1-p)}}e^{\frac{13-7p}{p-1}}(1.5\times10^{33}-7.6\times10^{32}p)^{\frac{1}{1-p}}(p-1)^{\frac{1}{p-1}}(p+1)^{3/5}f(p)^{3/5}  \nonumber\\
&&g^{XIII}(p)=  1.8\times10^{-\frac{30}{p+4}}787^{\frac{2(p-2)}{p+4}}2^{\frac{11(p+6)}{p+4}}3^{\frac{p+9}{p+4}}\pi^{\frac{1}{p+4}}(1.5\times10^{33}-7.6\times10^{32}p)^{\frac{1}{1-p}}\nonumber\\
&&~~~~~~~~~~~~~~(p-1)^{-\frac{2}{p+4}}(p+1)^{\frac{2}{p+4}}f(p)^{\frac{2}{p+4}}  \nonumber\\
&&g^{XIV}(p)=  (p+1)^{3/5}f(p)^{3/5}  \nonumber\\
&&g^{XV}(p)=  \left(\frac{p-1}{p-2}\right)^{}(p+1)^{3/5}f(p)^{3/5}  \nonumber\\
&&g^{XVI}(p)=  8.3\times10^{-\frac{22}{p+4}}\left(\frac{p-2}{p-1}\right)^{\frac{2(p-1)}{p+4}}(p+1)^{\frac{2}{p+4}}f(p)^{\frac{2}{p+4}}  \nonumber\\
&&g^{XVII}(p)=5.2\times10^{-10}e^{\frac{38p-76}{p-1}}(1068p-1068)^{\frac{2}{p-1}}(2-p)^{-\frac{2}{p-1}}\nonumber\\
&&g^{XVIII}(p)= 1.8\times10^{-5}e^{\frac{38-19p}{p-1}}(1068p-1068)^{\frac{1}{1-p}}(2-p)^{\frac{1}{p-1}}(p+1)^{3/5}f(p)^{3/5}  \nonumber\\
&&g^{XIX}(p)=  9.6\times10^{\frac{6p-72}{p+4}}e^{\frac{35}{p+4}}(1068p-1068)^{\frac{2}{p+4}}(2-p)^{-\frac{2}{p+4}}(p+1)^{\frac{2}{p+4}}f(p)^{\frac{2}{p+4}}  \nonumber\\
&&g^{XX}(p)=  \left(\frac{p-1}{p-2}\right)^{}(p+1)^{3/5}f(p)^{3/5}  \nonumber\\
&&g^{XXI}(p)=  1.8\times10^{-\frac{26}{p+4}}\pi^{\frac{6}{p+4}}\left(\frac{p-2}{p-1}\right)^{\frac{2(p-1)}{p+4}}(p+1)^{\frac{2}{p+4}}f(p)^{\frac{2}{p+4}}  \nonumber\\
&&g^{XXII}(p)=1.0\times10^{-25}2^{\frac{33(p-2)}{(1-p)}}e^{\frac{113-56p}{p-1}}\pi^{\frac{p-2}{(1-p)}}(1068p-1068)^{\frac{2}{p-1}}(2-p)^{-\frac{2}{p-1}}\nonumber\\
&&g^{XXIII}(p)= 3.6\times10^{\frac{47p-60}{p-1}}2^{\frac{33(p-2)}{2(p-1)}}\pi^{\frac{2-p}{2(p-1)}}(1068p-1068)^{\frac{1}{1-p}}(2-p)^{\frac{1}{p-1}}\nonumber\\
&&~~~~~~~~~~~~~~(p+1)^{3/5}f(p)^{3/5}  \nonumber\\
&&g^{XXIV}(p)=  1.8\times10^{-\frac{49p+52}{p+4}}2^{-\frac{33(p-2)}{p+4}}e^{\frac{445}{p+4}}\pi^{\frac{8-p}{p+4}}(1068p-1068)^{\frac{2}{p+4}}(2-p)^{-\frac{2}{p+4}}\nonumber\\
&&~~~~~~~~~~~~~~~~~(p+1)^{\frac{2}{p+4}}f(p)^{\frac{2}{p+4}}  \nonumber\\
\end{eqnarray}
\textbf{Thick Shell Forward shock}
\begin{eqnarray}
&&g^{I}(p)=  \left(\frac{p-1}{p-2}\right)^{}(p+1)^{3/5}f(p)^{3/5}  \nonumber\\
&&g^{II}(p)=  1.4\times10^{-\frac{10}{p+4}}\left(\frac{p-1}{p-2}\right)^{\frac{2(1-p)}{p+4}}(p+1)^{\frac{2}{p+4}}f(p)^{\frac{2}{p+4}}  \nonumber\\
&&g^{III}(p)=  (p+1)^{3/5}f(p)^{3/5}  \nonumber\\
&&g^{IV}(p)=e^{\frac{44p-88}{p-1}}(12-6p)^{\frac{2}{p-1}}(p-1)^{-\frac{2}{p-1}}\nonumber\\
&&g^{V}(p)=  e^{\frac{44-22p}{p-1}}(12-6p)^{\frac{1}{1-p}}(p-1)^{\frac{1}{p-1}}(p+1)^{3/5}f(p)^{3/5}  \nonumber\\
&&g^{VI}(p)=1.9\times10^{\frac{16(p-2)}{p+4}}0.003^{\frac{2-p}{p+4}}0.1^{\frac{2}{p+4}}(12-6p)^{\frac{2}{p+4}}(p-1)^{-\frac{2}{p+4}}(p+1)^{\frac{2}{p+4}}f(p)^{\frac{2}{p+4}}\nonumber\\
&&g^{VII}(p)=  (p+1)^{3/5}f(p)^{3/5}  \nonumber\\
&&g^{VIII}(p)=  \left(\frac{p-1}{p-2}\right)^{}(p+1)^{3/5}f(p)^{3/5}  \nonumber\\
&&g^{IX}(p)=  2^{\frac{105}{p+4}}e^{\frac{127}{p+4}}\pi^{\frac{3}{p+4}}\left(\frac{p-1}{p-2}\right)^{\frac{2(1-p)}{p+4}}(p+1)^{\frac{2}{p+4}}f(p)^{\frac{2}{p+4}}  \nonumber\\
&&g^{X}(p)=  (p+1)^{3/5}f(p)^{3/5}  \nonumber\\
&&g^{XI}(p)=2^{\frac{88-9p}{4(1-p)}}3^{\frac{p-4}{2(p-1)}}\pi^{\frac{p}{4(1-p)}}(0.009-0.005p)^{\frac{2}{p-1}}(p-1)^{-\frac{2}{p-1}}\nonumber\\
&&g^{XII}(p)= 0.005^{\frac{2}{p-1}}2^{\frac{88-9p}{8(p-1)}}3^{\frac{4-p}{4(p-1)}}\pi^{\frac{p}{8(p-1)}}(2-p)^{\frac{1}{1-p}}(p-1)^{\frac{1}{p-1}}(p+1)^{3/5}f(p)^{3/5}  \nonumber\\
&&g^{XIII}(p)=1.3^{\frac{2(p-2)}{p+4}}2^{\frac{9p-106}{2(p+4)}}3^{\frac{p-6}{2(p+4)}}25^{-\frac{10(p+3)}{(p+4)(p-1)}}e^{\frac{106p+23}{(p+4)(p-1)}}\pi^{\frac{2-p}{2(p+4)}}(2+p-p^2)^{\frac{2}{p+4}}\nonumber\\
&&~~~~~~~~~~~~~~(p-1)^{-\frac{2}{p+4}}(p+1)^{\frac{2}{p+4}}f(p)^{\frac{2}{p+4}}\nonumber\\
&&g^{XIV}(p)=  (p+1)^{3/5}f(p)^{3/5}  \nonumber\\
\end{eqnarray}
\textbf{Thick Shell Reverse shock}
\begin{eqnarray}
&&g^{I}(p)=  \left(\frac{p-1}{p-2}\right)^{}(p+1)^{3/5}f(p)^{3/5}  \nonumber\\
&&g^{II}(p)=  1.0\times10^{12}e^{-\frac{66}{p+4}}\left(\frac{p-2}{p-1}\right)^{\frac{2(p-1)}{p+4}}(p+1)^{\frac{2}{p+4}}f(p)^{\frac{2}{p+4}}  \nonumber\\
&&g^{III}(p)=  (p+1)^{3/5}f(p)^{3/5}  \nonumber\\
&&g^{IV}(p)=e^{\frac{44(p-2)}{p-1}}(5.8\times10^{5}-2.9\times10^{5}p)^{\frac{2}{p-1}}(p-1)^{-\frac{2}{p-1}}\nonumber\\
&&g^{V}(p)=  e^{\frac{22(2-p)}{p-1}}(5.8\times10^{5}-2.9\times10^{5}p)^{\frac{1}{1-p}}(p-1)^{\frac{1}{p-1}}(p+1)^{3/5}f(p)^{3/5}  \nonumber\\
&&g^{VI}(p)=  4.2\times10^{\frac{16p-44}{p+4}}0.003^{\frac{2-p}{p+4}}(5.8\times10^{5}-2.9\times10^{5}p)^{\frac{2}{p+4}}(p-1)^{-\frac{2}{p+4}}(p+1)^{\frac{2}{p+4}}f(p)^{\frac{2}{p+4}}  \nonumber\\
&&g^{VII}(p)=  (p+1)^{3/5}f(p)^{3/5}  \nonumber\\
&&g^{VIII}(p)=  \left(\frac{p-1}{p-2}\right)^{}(p+1)^{3/5}f(p)^{3/5}  \nonumber\\
&&g^{IX}(p)=  1.6\times10^{-\frac{100}{p+4}}2^{-\frac{47}{p+4}}\pi^{-\frac{1}{p+4}}\left(\frac{p-2}{p-1}\right)^{\frac{2(p-1)}{p+4}}(p+1)^{\frac{2}{p+4}}f(p)^{\frac{2}{p+4}}  \nonumber\\
&&g^{X}(p)=  (p+1)^{3/5}f(p)^{3/5}  \nonumber\\
&&g^{XI}(p)=2^{\frac{9p+44}{4(p-1)}}3^{\frac{p}{2(p-1)}}\pi^{\frac{4-p}{4(p-1)}}(1.5\times10^{9}-7.3\times10^{8})^{\frac{2}{p-1}}(p-1)^{-\frac{2}{p-1}}\nonumber\\
&&g^{XII}(p)= 2^{-\frac{9p+44}{8(p-1)}}3^{-\frac{p}{4(p-1)}}\pi^{\frac{p-4}{8(p-1)}}(1.5\times10^{9}-7.3\times10^{8})^{\frac{1}{1-p}}(p-1)^{\frac{1}{p-1}}(p+1)^{3/5}f(p)^{3/5}  \nonumber\\
&&g^{XIII}(p)=  2.9\times10^{\frac{114-43p-21p^2}{(p+4)(p-1)}}e^{\frac{15p-144}{(p+4)(p-1)}}2^{\frac{9p+166}{4(p+4)}}3^{\frac{p+2}{2(p+4)}}\pi^{\frac{10-p}{4(p+4)}}(2+p-p^2)^{\frac{2}{p+4}}\nonumber\\
&&~~~~~~~~~~~~~~(p-1)^{-\frac{2}{p+4}}(p+1)^{\frac{2}{p+4}}f(p)^{\frac{2}{p+4}}  \nonumber\\
&&g^{XIV}(p)=  (p+1)^{3/5}f(p)^{3/5}  \nonumber\\
&&g^{XV}(p)=  4.29\times10^{21}\left(\frac{p-1}{p-2}\right)^{}(p+1)^{3/5}f(p)^{3/5}  \nonumber\\
&&g^{XVI}(p)=  5.2\times10^{-12}e^{\frac{253}{p+4}}\left(\frac{p-2}{p-1}\right)^{\frac{2(p-1)}{p+4}}(p+1)^{\frac{2}{p+4}}f(p)^{\frac{2}{p+4}}  \nonumber\\
&&g^{XVII}(p)=3.4\times10^{-10}0.5^{\frac{p}{p-1}}e^{\frac{38p-77}{p-1}}(1321p-1321)^{\frac{2}{p-1}}(2-p)^{-\frac{2}{p-1}}\nonumber\\
&&g^{XVIII}(p)= 8.2\times10^{-5}e^{\frac{38-19p}{p-1}}(1321p-1321)^{\frac{1}{1-p}}(2-p)^{\frac{1}{p-1}}(p+1)^{3/5}f(p)^{3/5}  \nonumber\\
&&g^{XIX}(p)=  7.3\times10^{-\frac{24p+72}{p+4}}0.5^{\frac{p}{p+4}}e^{\frac{43}{p+4}}(1321p-1321)^{\frac{2}{p+4}}(2-p)^{-\frac{2}{p+4}}(p+1)^{\frac{2}{p+4}}f(p)^{\frac{2}{p+4}}  \nonumber\\
&&g^{XX}(p)= \left(\frac{p-1}{p-2}\right)^{}(p+1)^{3/5}f(p)^{3/5}  \nonumber\\
&&g^{XXI}(p)=  5.7\times10^{-\frac{82}{p+4}}2^{\frac{19}{2(p+4)}}3^{\frac{9}{4(p+4)}}5^{\frac{21}{2(p+4)}}\pi^{-\frac{1}{p+4}}\left(\frac{p-2}{p-1}\right)^{\frac{2(p-1)}{p+4}}(p+1)^{\frac{2}{p+4}}f(p)^{\frac{2}{p+4}}  \nonumber\\
&&g^{XXII}(p)=1.0\times10^{4}0.3^{\frac{2(p-2)}{p-1}}2^{\frac{33(4-p)}{4(p-1)}}625^{\frac{8-2p}{p-1}}\pi^{\frac{4-p}{4(p-1)}}(1321p-1321)^{\frac{2}{p-1}}(2-p)^{-\frac{2}{p-1}}\nonumber\\
&&g^{XXIII}(p)= 5.9\times10^{17}0.3^{\frac{2-p}{p-1}}2^{\frac{33(p-4)}{8(p-1)}}625^{\frac{p-4}{p-1}}\pi^{\frac{p-4}{8(p-1)}}(1321p-1321)^{\frac{1}{1-p}}(2-p)^{\frac{1}{p-1}}\nonumber\\
&&~~~~~~~~~~~~~~(p+1)^{3/5}f(p)^{3/5}  \nonumber\\
&&g^{XXIV}(p)=  1.6\times10^{\frac{4p+22}{p+4}}0.3^{\frac{2(p-2)}{p+4}}2^{\frac{33(4-p)}{4(p+4)}}3^{\frac{37-2p}{4(p+4)}}5^{\frac{245-16p}{2(p+4)}}\pi^{\frac{10-p}{4(p+4)}}(1321p-1321)^{\frac{2}{p+4}}\nonumber\\
&&~~~~~~~~~~~~~~(2-p)^{-\frac{2}{p+4}}(p+1)^{\frac{2}{p+4}}f(p)^{\frac{2}{p+4}}  \nonumber\\
\end{eqnarray}
\textbf{Adiabatic Deceleration With(or Without) Energy Injection}
\begin{eqnarray}
&&g^{I}(p)= \left(\frac{p-1}{p-2}\right)^{}(p+1)^{3/5}f(p)^{3/5}  \nonumber\\
&&g^{II}(p)= e^{\frac{11}{p+4}}\left(\frac{p-2}{p-1}\right)^{\frac{2(p-1)}{p+4}}(p+1)^{\frac{2}{p+4}}f(p)^{\frac{2}{p+4}}  \nonumber\\
&&g^{III}(p)= (p+1)^{3/5}f(p)^{3/5}  \nonumber\\
&&g^{IV}(p)=e^{\frac{47p-95}{p-1}}(0.3-0.15p)^{\frac{2}{p-1}}(p-1)^{-\frac{2}{p-1}}\nonumber\\
&&g^{V}(p)=e^{\frac{47-24p}{p-1}}(0.3-0.15p)^{\frac{1}{1-p}}(p-1)^{\frac{1}{p-1}}(p+1)^{3/5}f(p)^{3/5}  \nonumber\\
&&g^{VI}(p)= 1.8\times10^{\frac{2(p-2)}{p+4}}0.00008^{\frac{2-p}{p+4}}0.02^{\frac{2}{p+4}}e^{\frac{11}{p+4}}(0.3-0.15p)^{\frac{2}{p+4}}\nonumber\\
&&~~~~~~~~~~~~~(p-1)^{-\frac{2}{p+4}}(p+1)^{\frac{2}{p+4}}f(p)^{\frac{2}{p+4}}  \nonumber\\
&&g^{VII}(p)=  (p+1)^{3/5}f(p)^{3/5}  \nonumber\\
&&g^{VIII}(p)= \left(\frac{p-1}{p-2}\right)^{}(p+1)^{3/5}f(p)^{3/5}  \nonumber\\
&&g^{IX}(p)= e^{\frac{273}{p+4}}\left(\frac{p-2}{p-1}\right)^{\frac{2(p-1)}{p+4}}(p+1)^{\frac{2}{p+4}}f(p)^{\frac{2}{p+4}}  \nonumber\\
&&g^{X}(p)= (p+1)^{3/5}f(p)^{3/5}  \nonumber\\
&&g^{XI}(p)=0.3^{\frac{2(p-2)}{p-1}}2^{\frac{56-3p}{2(1-p)}}3^{\frac{8-3p}{4(1-p)}}5^{\frac{p+40}{2(1-p)}}\pi^{\frac{p}{4(1-p)}}(3736-1868p)^{\frac{2}{p-1}}(p-1)^{-\frac{2}{p-1}}\nonumber\\
&&g^{XII}(p)=0.3^{\frac{2-p}{p-1}}2^{\frac{56-3p}{4(p-1)}}3^{\frac{8-3p}{8(p-1)}}5^{\frac{p+40}{4(p-1)}}\pi^{\frac{p}{8(p-1)}}(3736-1868p)^{\frac{1}{1-p}}(p-1)^{\frac{1}{p-1}}(p+1)^{3/5}f(p)^{3/5}  \nonumber\\
&&g^{XIII}(p)= 0.3^{\frac{2(p-2)}{p+4}}2^{\frac{3(p-2)}{2(p+4)}}3^{\frac{3(p-2)}{4(p+4)}}5^{\frac{2-p}{2(p+4)}}e^{\frac{41}{p+4}}(3736-1868p)^{\frac{2}{p+4}}\nonumber\\
&&~~~~~~~~~~~~~~~(p-1)^{-\frac{2}{p+4}}(p+1)^{\frac{2}{p+4}}f(p)^{\frac{2}{p+4}}  \nonumber\\
&&g^{XIV}(p)=  (p+1)^{3/5}f(p)^{3/5}  \nonumber\\
\end{eqnarray}
\textbf{Newtonian Phase}
\begin{eqnarray}
&&g^{I}(p)=  \left(\frac{p-1}{p-2}\right)^{}(p+1)^{3/5}f(p)^{3/5}  \nonumber\\
&&g^{II}(p)=e^{\frac{219}{p+4}}\left(\frac{p-2}{p-1}\right)^{\frac{2(p-1)}{p+4}}(p+1)^{\frac{2}{p+4}}f(p)^{\frac{2}{p+4}}  \nonumber\\
&&g^{III}(p)=e^{\frac{53p-106}{p-1}}(1.6\times10^{-9}-8.3\times10^{-10}p)^{\frac{2}{p-1}}(p-1)^{-\frac{2}{p-1}}\nonumber\\
&&g^{IV}(p)=e^{\frac{53-26p}{p-1}}(1.6\times10^{-9}-8.3\times10^{-10}p)^{\frac{1}{1-p}}(p-1)^{\frac{1}{p-1}}(p+1)^{3/5}f(p)^{3/5}  \nonumber\\
&&g^{V}(p)= 5.4\times10^{\frac{26(p-2)}{p+4}}28245^{\frac{2-p}{p+4}}e^{\frac{10}{p+4}}(0.3-0.15p)^{\frac{2}{p+4}}(p-1)^{-\frac{2}{p+4}}(p+1)^{\frac{2}{p+4}}f(p)^{\frac{2}{p+4}}  \nonumber\\
&&g^{VI}(p)=  \left(\frac{p-1}{p-2}\right)^{}(p+1)^{3/5}f(p)^{3/5}  \nonumber\\
&&g^{VII}(p)=2^{\frac{842}{3(p+4)}}e^{\frac{509}{p+4}}\left(\frac{p-2}{p-1}\right)^{\frac{2(p-1)}{p+4}}(p+1)^{\frac{2}{p+4}}f(p)^{\frac{2}{p+4}}  \nonumber\\
&&g^{VIII}(p)=2^{\frac{3p+158}{3(1-p)}}3^{\frac{10-3p}{3(1-p)}}\pi^{\frac{p}{3(1-p)}}e^{\frac{22p-45}{p-1}}(5.6\times10^{-18}-2.8\times10^{-18}p)^{\frac{2}{p-1}}(p-1)^{-\frac{2}{p-1}}\nonumber\\
&&g^{IX}(p)=2.8\times10^{-\frac{36}{p-1}}2^{\frac{3p+158}{5(p-1)}}3^{\frac{10-3p}{6(p-1)}}e^{\frac{144-11p}{p-1}}\pi^{\frac{2}{3(p-1)}}(2-p)^{\frac{1}{1-p}}(p-1)^{\frac{1}{p-1}}(p+1)^{3/5}f(p)^{3/5}  \nonumber\\
&&g^{X}(p)= 2.8\times10^{-\frac{36}{p-1}}2^{-\frac{3p+136}{3(p+4)}}3^{\frac{3p-8}{3(p+4)}}73399^{\frac{2(p-2)}{p+4}}e^{\frac{104p+300}{(p+4)(p-1)}}\pi^{-\frac{2}{3(p+4)}}(2+p-p^2)^{\frac{2}{p+4}}\nonumber\\
&&~~~~~~~~~~~~~(p-1)^{-\frac{2}{p+4}}(p+1)^{\frac{2}{p+4}}f(p)^{\frac{2}{p+4}}  \nonumber\\
\end{eqnarray}

\end{document}